\documentclass[10pt]{article}
\IfFormatAtLeastTF{2026-06-01}{}{
\pdfglyphtounicode{angbracketleftBig}{27E8 FE02}
\pdfglyphtounicode{angbracketleftbig}{27E8 FE01}
\pdfglyphtounicode{angbracketleftbigg}{27E8 FE03}
\pdfglyphtounicode{angbracketrightBig}{27E9 FE02}
\pdfglyphtounicode{angbracketrightbig}{27E9 FE01}
\pdfglyphtounicode{angbracketrightbigg}{27E9 FE03}
\pdfglyphtounicode{arrowhookleft}{21AA}
\pdfglyphtounicode{braceleftbig}{007B FE01}
\pdfglyphtounicode{braceleftbigg}{007B FE03}
\pdfglyphtounicode{bracerightbig}{007D FE01}
\pdfglyphtounicode{bracerightbigg}{007D FE03}
\pdfglyphtounicode{bracketleftBig}{005B FE02}
\pdfglyphtounicode{bracketleftBigg}{005B FE04}
\pdfglyphtounicode{bracketleftbig}{005B FE01}
\pdfglyphtounicode{bracketleftbigg}{005B FE03}
\pdfglyphtounicode{bracketrightBig}{005D FE02}
\pdfglyphtounicode{bracketrightBigg}{005D FE04}
\pdfglyphtounicode{bracketrightbig}{005D FE01}
\pdfglyphtounicode{bracketrightbigg}{005D FE03}
\pdfglyphtounicode{circlemultiplydisplay}{2A02 FE02}
\pdfglyphtounicode{circleplusdisplay}{2A01 FE02}
\pdfglyphtounicode{circleplustext}{2A01 FE01}
\pdfglyphtounicode{contintegraldisplay}{222E FE02}
\pdfglyphtounicode{hatwide}{02C6 FE01}
\pdfglyphtounicode{hatwider}{02C6 FE02}
\pdfglyphtounicode{integraldisplay}{222B FE02}
\pdfglyphtounicode{integraltext}{222B FE01}
\pdfglyphtounicode{mapsto}{21A6}
\pdfglyphtounicode{parenleftBig}{0028 FE02}
\pdfglyphtounicode{parenleftBigg}{0028 FE04}
\pdfglyphtounicode{parenleftbig}{0028 FE01}
\pdfglyphtounicode{parenleftbigg}{0028 FE03}
\pdfglyphtounicode{parenleftbt}{239D}
\pdfglyphtounicode{parenlefttp}{239B}
\pdfglyphtounicode{parenrightBig}{0029 FE02}
\pdfglyphtounicode{parenrightBigg}{0029 FE04}
\pdfglyphtounicode{parenrightbig}{0029 FE01}
\pdfglyphtounicode{parenrightbigg}{0029 FE03}
\pdfglyphtounicode{parenrightbt}{23A0}
\pdfglyphtounicode{parenrighttp}{239E}
\pdfglyphtounicode{productdisplay}{220F FE02}
\pdfglyphtounicode{producttext}{220F FE01}
\pdfglyphtounicode{radicalBig}{221A FE02}
\pdfglyphtounicode{radicalBigg}{221A FE04}
\pdfglyphtounicode{radicalbig}{221A FE01}
\pdfglyphtounicode{slashbig}{002F FE01}
\pdfglyphtounicode{summationdisplay}{2211 FE02}
\pdfglyphtounicode{summationtext}{2211 FE01}
\pdfglyphtounicode{tildewide}{02DC FE01}
\pdfglyphtounicode{unionsqdisplay}{2A06 FE02}
\pdfglyphtounicode{uniontext}{22C3 FE01}
\pdfglyphtounicode{vextenddouble}{20E6}
\pdfglyphtounicode{vextendsingle}{20D3}
}
\usepackage[T1]{fontenc}
\usepackage{amsmath,mathtools,amssymb,amsfonts,amsthm,mathrsfs,bm}
\usepackage[dvipsnames]{xcolor}
\usepackage{graphicx,placeins,booktabs,tikz}
\usetikzlibrary{arrows.meta,decorations.pathmorphing,decorations.markings}
\definecolor{radialPairInk}{HTML}{263944}
\definecolor{radialPairLine}{HTML}{246697}
\definecolor{radialPairMark}{HTML}{A94B27}

\definecolor{noteblue}{HTML}{246697}
\definecolor{noterust}{HTML}{A75630}
\definecolor{noteink}{HTML}{263944}
\definecolor{notegreen}{HTML}{367665}
\definecolor{noterule}{HTML}{CFD8DC}
\tikzset{amplitude/.style={font=\small,text=noteink,line cap=round,line join=round,
          every node/.style={inner sep=1.6pt}},
         leg/.style={draw=noteblue,line width=1.05pt},
         wave/.style={leg,decorate,decoration={snake,amplitude=1.05pt,segment length=5.5pt,
                      pre length=1.5pt,post length=1.5pt}},
         oriented/.style={line width=.8pt,-{Stealth[length=4.5pt,width=3.8pt]}},
         direction mark/.style={postaction={decorate},decoration={markings,
           mark=at position #1 with {\arrow{Stealth[length=4.5pt,width=3.8pt]}}}},
         boundary/.style={draw=noteink!75,line width=.65pt},
         cut/.style={draw=noterust,line width=.9pt},
         vertex/.style={circle,draw=noteink,fill=noteink,inner sep=0pt,minimum size=4.5pt},
         external/.style={circle,draw=noteink,fill=white,line width=.8pt,inner sep=0pt,minimum size=4.8pt},
         head/.style={font=\small\bfseries,text=noteink}}

\newcommand{\SummaryRegionalAmplitudes}{%
\begin{tikzpicture}[amplitude]
  \node[head] at (0,3.56) {Usual $3+1$ quantization};
  \node[head] at (7.98,3.56) {Algebraic quantization with null cones};
  \draw[noteink!52,line width=.55pt] (-1.77,3.23)--(1.74,3.23);
  \draw[noteblue!65,line width=.55pt] (2.46,3.23)--(13.48,3.23);
  \draw[noteink!41,line width=.62pt] (2.20,-.72)--(2.20,3.74);
  \begin{scope}[yshift=-.24cm]
    \node at (0,2.83) {$|\Psi\rangle$};
    \draw[noteink!43,line width=.54pt]
      (-.21,2.59)--(-.84,2.14) (.21,2.59)--(.84,2.14);
    \foreach \cx/\pinch in {-.91/0,.91/.53}{
      \begin{scope}[shift={(\cx cm,1.29cm)},declare function={
          sliceradius(\t)=.70*sin(\t)*(1-\pinch*(sin(\t))^8);
          slatdepth(\t)=.17*sliceradius(\t)*sin(\t);}]
        \path[fill=noteink!6]
          plot[domain=0:180,samples=41,variable=\t] ({sliceradius(\t)},{.72*cos(\t)})
          --plot[domain=180:0,samples=41,variable=\t] ({-sliceradius(\t)},{.72*cos(\t)})--cycle;
        \foreach \lat in {48,90,132}{
          \draw[noteink!35,densely dotted,line width=.45pt]
            plot[domain=0:180,samples=25,variable=\t]
              ({sliceradius(\lat)*cos(\t)},{.72*cos(\lat)+slatdepth(\lat)*sin(\t)});
          \draw[noteink!65,line width=.49pt]
            plot[domain=180:360,samples=25,variable=\t]
              ({sliceradius(\lat)*cos(\t)},{.72*cos(\lat)+slatdepth(\lat)*sin(\t)});}
        \foreach \mer in {-41,41}{
          \draw[noteink!54,line width=.46pt]
            plot[domain=0:180,samples=41,variable=\t]
              ({sliceradius(\t)*sin(\mer)},{.72*cos(\t)-slatdepth(\t)*cos(\mer)});}
        \draw[noteink!85,line width=.83pt]
          plot[domain=0:180,samples=41,variable=\t] ({sliceradius(\t)},{.72*cos(\t)})
          --plot[domain=180:0,samples=41,variable=\t] ({-sliceradius(\t)},{.72*cos(\t)})--cycle;
      \end{scope}}
    \node at (-.91,.25) {$\Psi[h_{ij}]$};
    \node at (.91,.25) {$\Psi[h'_{ij}]$};
  \end{scope}
  \def\amplitudeRegion##1##2##3{%
    \foreach \direction/\shade in {-1/noteblue,1/notegreen}{
      \begin{scope}[shift={(0,{\direction*(##1+##2)})}]
        \path[draw=\shade!82,fill=\shade!14,line width=.70pt]
          (-.19,{\direction*.81})--(.19,{\direction*.81})
          --(.50,{\direction*.50})--(0,0)--(-.50,{\direction*.50})--cycle;
        \draw[\shade,line width=1.0pt] (-.19,{\direction*.81})--(.19,{\direction*.81});
        \fill[\shade] ({\direction*.33},{\direction*.67}) circle[radius=1.55pt];
        \ifnum##3=0
          \ifnum\direction<0
            \node[anchor=east,inner sep=1.8pt,text=\shade] at (-.41,-.73) {$F$};
          \else
            \node[anchor=west,inner sep=1.8pt,text=\shade] at (.41,.73) {$G$};
          \fi
        \fi
        \ifnum##3=1
          \ifnum\direction<0
            \node[font=\footnotesize] at (0,-.48) {$U_1$};
          \else
            \node[font=\footnotesize] at (0,.48) {$U_2$};
          \fi
        \fi
      \end{scope}
      \foreach \side in {-1,1}{
        \begin{scope}[shift={({\side*##1},{\direction*##2})}]
          \ifnum\side<0 \def\celltint{7}\else\def\celltint{21}\fi
          \path[draw=\shade!68,fill=\shade!\celltint,line width=.59pt]
            (0,0)--(\side,0)--({.50*\side},{.50*\direction})--cycle;
        \end{scope}}
      \ifnum##3<2
        \foreach \side in {-1,1}{
          \draw[noterust,line width=.98pt]
            ({\side*##1},{\direction*##2})--({\side*(1+##1)},{\direction*##2});
          \ifnum##3=0
            \node[external,draw=noterust,minimum size=3.0pt]
              at ({\side*##1},{\direction*##2}) {};
          \fi}
      \fi}
    \ifcase##3
      \node[font=\footnotesize,text=noterust] at (0,0) {$S$};
    \or
      \node[text=noterust,inner sep=.8pt] at (0,0) {$\Sigma$};
    \or
      \draw[noteink!75,line width=.99pt]
        (-.19,-.81)--(.19,-.81)--(1,0)--(.19,.81)--(-.19,.81)--(-1,0)--cycle;
      \draw[white,line width=.82pt] (-.985,0)--(.985,0);
      \draw[noterust!65,line width=.53pt,dash pattern=on 2.2pt off 1.6pt] (-.985,0)--(.985,0);
      \node[above,inner sep=2.3pt] at (0,.81) {$\langle a|$};
      \node[below,inner sep=2.3pt] at (0,-.81) {$|b\rangle$};
    \fi}
  \begin{scope}[shift={(4.04cm,1.43cm)},x=1.13cm,y=1.13cm]
    \amplitudeRegion{.25}{.24}{0}
  \end{scope}
  \draw[oriented,noteink] (5.57,1.43)--(6.82,1.43);
  \node[above,inner sep=2pt] at (6.195,1.48) {$\cup_S$};
  \begin{scope}[shift={(8.10cm,1.43cm)},x=1.13cm,y=1.13cm]
    \amplitudeRegion{0}{.24}{1}
  \end{scope}
  \draw[oriented,noteink] (9.39,1.43)--(10.77,1.43);
  \node[above,inner sep=2pt] at (10.08,1.48) {$\sum_\beta$};
  \begin{scope}[shift={(12.05cm,1.43cm)},x=1.13cm,y=1.13cm]
    \amplitudeRegion{0}{0}{2}
  \end{scope}
  \node at (7.98,-.50)
    {$\sum_\beta\langle a|U_2|e_\beta\rangle\langle e_\beta|U_1|b\rangle
        =\langle a|U_2U_1|b\rangle$};
\end{tikzpicture}%
}

\newcommand{\SummaryPreparation}{%
\begin{tikzpicture}[amplitude]
 \begin{scope}
  \path[fill=noteblue!10,draw=noteblue!68] (-2.68,.62)
   ..controls(-2.17,1.00)and(-1.88,1.27)..(-1.52,1.45)--(-1.52,-.18)
   ..controls(-1.97,-.01)and(-2.38,.35)..cycle;
  \path[fill=notegreen!10,draw=notegreen!72] (1.47,1.45)..controls(1.83,1.25)and(2.26,.95)..(2.66,.62)
   ..controls(2.30,.31)and(1.89,-.03)..(1.47,-.18)--cycle;
  \draw[leg] (-1.52,.62)--(1.47,.62);
  \foreach \x in {-1.52,1.47}{
   \draw[cut,line width=1.1pt] (\x,-.20)--(\x,1.46);
   \node[above,text=noterust] at (\x,1.48) {$P_{\mathscr D}$};}
  \node[vertex] at (0,.62) {}; \node[above] at (0,.75) {$X$};
  \node[left] at (-2.71,.62) {$|\Omega_{\mathscr A}\rangle$};
  \node[right] at (2.69,.62) {$\langle v|$};
 \end{scope}
 \draw[oriented,noteink] (3.47,.62)--(4.85,.62)
   node[midway,above,inner ysep=3pt] {$\sum_v$};
 \begin{scope}[shift={(7.27cm,0)}]
  \path[fill=noteblue!8,draw=noteblue!68] (-2.10,.62)--(-1.10,1.40)--(0,1.13)--(0,.09)--(-1.10,-.20)--cycle;
  \path[fill=notegreen!8,draw=notegreen!68] (0,1.13)--(1.10,1.40)--(2.10,.62)--(1.10,-.20)--(0,.09)--cycle;
  \draw[leg] (-2.10,.62)..controls(-1.48,.63)and(-1.40,1.04)..(-1.10,1.40)
     ..controls(-.67,.77)and(-.34,.62)..(0,.62)
     ..controls(.40,.62)and(.73,.89)..(1.10,1.40)
     ..controls(1.40,.93)and(1.71,.62)..(2.10,.62);
  \node[vertex] at (-1.10,1.40) {}; \node[vertex] at (1.10,1.40) {};
  \node[above] at (-1.10,1.53) {$X$};
  \node[above] at (1.10,1.53) {$X^\dagger$};
  \draw[cut,line width=1.1pt] (0,-.02)--(0,1.25);
  \node[above,text=noterust] at (0,1.31) {$P_{\mathscr D}$};
  \node[below] at (0,-.31) {$\bigl\|P_{\mathscr D}X|\psi\rangle\bigr\|^2$};
 \end{scope}
\end{tikzpicture}
}

\newcommand{\SummaryBoost}{%
\begin{tikzpicture}[amplitude]
 \foreach \cx/\step in {0/0,6.25/1}{
  \begin{scope}[shift={(\cx cm,0)}]
   \draw[boundary] (-2.36,.40)--(2.45,.40);
   \draw[cut] (0,.18)--(0,.68);
   \node[text=noterust,below] at (0,.16) {$S$};
   \node[below] at (-1.40,.32) {$-d$};
   \ifnum\step=0
    \node[head] at (0,2.08) {Before the boost};
    \draw[wave] (-1.40,.40)..controls(-.98,1.75)and(.98,1.75)..(1.40,.40);
    \node[external] at (1.40,.40) {}; \node[below] at (1.40,.32) {$d$};
   \else
    \node[head] at (.05,2.08) {After the boost};
    \draw[wave] (-1.40,.40)..controls(-.96,1.75)and(1.41,1.75)..(2.01,.40);
    \draw[noterust!75,densely dashed,line width=.65pt] (1.40,.22)--(1.40,.55);
    \node[external] at (2.01,.40) {}; \node[below] at (2.01,.32) {$e^\eta d$};
   \fi
   \node[external] at (-1.40,.40) {};
  \end{scope}
 }
\end{tikzpicture}
}

\newcommand{\SummaryNullData}{%
\begin{tikzpicture}[amplitude]
 \begin{scope}
  \node[head] at (0.10,3.02) {Neighboring null generators};
  \path[fill=noteblue!6] (-.84,.34)--(-1.66,2.27)
    arc[start angle=180,end angle=360,x radius=1.66cm,y radius=.34cm]--(.84,.34)
    arc[start angle=0,end angle=-180,x radius=.84cm,y radius=.19cm]--cycle;
  \draw[noteblue!71,line width=.68pt] (-.84,.34)--(-1.66,2.27) (.84,.34)--(1.66,2.27);
  \foreach \y/\w/\h in {.34/.84/.19,1.28/1.24/.25,2.27/1.66/.34}{
   \draw[noteblue!53,line width=.52pt] (0,\y) ellipse[x radius=\w cm,y radius=\h cm];}
  \foreach \a/\b/\c/\d/\e/\f in {-.43/.31/-.67/1.26/-1.01/2.26,.16/.22/.44/1.15/.59/2.06,.43/.36/.72/1.30/1.14/2.26,-.16/.45/-.39/1.41/-.46/2.46}{
   \draw[noteblue!80,line width=.68pt] (\a,\b)..controls(\a,.68)and(\c,.98)..(\c,\d)
    ..controls(\c,1.63)and(\e,1.99)..(\e,\f);}
  \path[fill=noterust!13,draw=noterust!80] (-.43,.31)--(.16,.22)--(.43,.36)--(-.16,.45)--cycle;
  \path[fill=noterust!22,draw=noterust] (-.67,1.26)--(.44,1.15)--(.72,1.30)--(-.39,1.41)--cycle;
  \path[fill=noterust!18,draw=noterust] (-1.01,2.26)--(.59,2.06)--(1.14,2.26)--(-.46,2.46)--cycle;
  \draw[oriented,noterust] (-1.01,2.26)--(.40,2.08);
  \draw[oriented,noterust] (-1.01,2.26)--(-.52,2.44);
  \draw[oriented,noteink!70] (-1.21,.68)--(-1.52,1.53);
  \node[anchor=north] at (0,.05) {$\partial\Sigma_0$};
  \node[anchor=west] at (1.66,2.30) {$S$};
 \end{scope}
 \begin{scope}[shift={(5.22cm,.01cm)},declare function={
   transversewidth(\x)=.155+.14*(\x+1.90);
   focussedwidth(\x)=transversewidth(\x)-.07488372/1.28*(max(0,\x+1.28)^2-max(0,\x+.64)^2);}]
  \node[head] at (-.02,3.01) {Time delay from shear};
  \draw[noteink!48,densely dashed,line width=.60pt] (.44,1.41)--(.44,.78);
  \node[anchor=west,font=\footnotesize,text=noteink!85] at (.50,1.12) {$u$};
  \foreach \y/\rad/\stop/\sheared in {1.96/transversewidth/.44/0,.36/focussedwidth/2.05/1}{
   \begin{scope}[shift={(0,\y cm)}]
    \path[fill=noteblue!7] plot[domain=-1.90:\stop,samples=35,variable=\v]
      (\v,{\rad(\v)}) -- plot[domain=\stop:-1.90,samples=35,variable=\v] (\v,{-\rad(\v)}) -- cycle;
    \ifnum\sheared=1
      \path[fill=noterust!20] plot[domain=-1.28:-.64,samples=14,variable=\v] (\v,{\rad(\v)})
       -- plot[domain=-.64:-1.28,samples=14,variable=\v] (\v,{-\rad(\v)}) -- cycle;
      \draw[noterust!65,line width=.55pt] (-1.26,.04)..controls(-1.09,.16)and(-.87,-.03)..(-.65,.05);
      \node[anchor=south,text=noterust,font=\footnotesize] at (-1.00,.46) {$\sigma_{AB}\ne 0$};
    \else
      \node[anchor=south,text=noteblue,font=\footnotesize] at (-1.00,.46) {$\sigma_{AB}=0$};
    \fi
    \foreach \sgn in {-1,1}
      \draw[noteblue!73,line width=.70pt] plot[domain=-1.90:\stop,samples=38,variable=\v] (\v,{\sgn*\rad(\v)});
    \draw[noteblue!90,line width=.62pt] (-1.90,0) ellipse[x radius=.065cm,y radius=.155cm];
    \draw[noteblue!72,line width=.55pt] (-1.30,0) ellipse[x radius=.084cm,y radius=.239cm];
    \ifnum\sheared=1
      \draw[noteblue,line width=.65pt] (.44,0) ellipse[x radius=.131cm,y radius=.378cm];
    \fi
    \filldraw[fill=noterust!9,draw=noterust,line width=.95pt] (\stop,0) ellipse[x radius=.163cm,y radius=.4826cm];
   \end{scope}}
  \draw[noteink!53,line width=.55pt] (.44,-.18)--(.44,-.29)--(2.05,-.29)--(2.05,-.18);
  \node[below,font=\footnotesize] at (1.245,-.30) {$u_*-u$};
 \end{scope}
 \begin{scope}[shift={(10.05cm,0)}]
  \node[head] at (0,3.02) {Transport from the corner};
  \coordinate (lowerNear) at (-1.65,.04);
  \coordinate (jointNear) at (.45,1.30);
  \coordinate (upperNear) at (-1.65,2.50);
  \begin{scope}[shift={(1.17cm,.02cm)}]
    \coordinate (lowerFar) at (-1.65,.04);
    \coordinate (jointFar) at (.45,1.30);
    \coordinate (upperFar) at (-1.65,2.50);
  \end{scope}
  \path[draw=noteblue!75,fill=noteblue!11,line width=.77pt]
    (lowerNear)--(jointNear)--(jointFar)--(lowerFar)--cycle;
  \path[draw=notegreen!87,fill=notegreen!12,line width=.77pt]
    (jointNear)--(upperNear)--(upperFar)--(jointFar)--cycle;
  \foreach \t in {.24,.50,.76}{
   \draw[noteblue!65,line width=.52pt] ({-1.65+1.17*\t},{.04+.02*\t})
     --({.45+1.17*\t},{1.30+.02*\t});
   \draw[notegreen!69,line width=.52pt] ({.45+1.17*\t},{1.30+.02*\t})
     --({-1.65+1.17*\t},{2.50+.02*\t});}
  \foreach \t in {.29,.59,.83}{
   \draw[noteblue!58,line width=.49pt] ({-1.65+2.10*\t},{.04+1.26*\t})--++(1.17,.02);}
  \foreach \t in {.15,.22,.68}{
   \draw[notegreen!73,line width=.49pt] ({.45-2.10*\t},{1.30+1.20*\t})--++(1.17,.02);}
  \draw[noterust,line width=1.14pt] (jointNear)--(jointFar);
  \node[anchor=west] at (1.68,1.34) {$S$};
  \coordinate (cornerDatum) at (1.035,1.31);
  \coordinate (innerCut) at (-.351,.4784);
  \draw[noterust,line width=1.24pt,direction mark=.36] (cornerDatum)--(innerCut);
  \draw[oriented,noteblue,line width=1.26pt] (innerCut)--++(.52,.312);
  \draw[oriented,notegreen,line width=1.24pt] (cornerDatum)--++(-.70,.40);
  \draw[oriented,notegreen,line width=1.24pt] (innerCut)--++(-.70,.40);
  \node[external,draw=noterust,minimum size=4pt] at (cornerDatum) {};
  \node[vertex,minimum size=4pt] at (innerCut) {};
  \node[anchor=east,text=notegreen] at (-1.12,.90) {$n$};
  \draw[noteblue!75,line width=.42pt] (.12,.74)--(.39,.45);
  \node[anchor=north west,inner sep=.9pt,text=noteblue] at (.39,.45) {$\ell$};
  \node[anchor=east,text=notegreen] at (-1.71,2.47) {$\mathscr H_q$};
  \node[anchor=east,text=noteblue] at (-1.72,.07) {$\mathscr H_p$};
 \end{scope}
\end{tikzpicture}%
}

\newcommand{\SummarySpacetimeCover}{%
\begin{tikzpicture}[amplitude,declare function={
  phTime(\t)=atan(exp(3*sqrt(3)*\t/4))/45-1;
  singularTime(\x)=acos((exp(1.5)/2-1)/(exp(1.5)/2+1)*cos(90*\x))/90;
  saRad(\a)=sin(deg(\a)); caRad(\a)=cos(deg(\a));}]
 \def\oldPast{-.90}\def\oldFuture{0}
 \def\oldLow{-.80}\def\oldHigh{-.16}
 \def\newPast{-.44}\def\newFuture{.44}
 \def\newLow{-.37}\def\newHigh{.37}
 \def\separation{.16}
 \node[head] at (0,2.34) {Schwarzschild spacetime};
 \node[head] at (6.45,2.34) {On the photon sphere};
 \begin{scope}[x=1.69cm,y=1.69cm]
  \def\radialDiamond##1##2##3##4##5##6{%
   \pgfmathsetmacro{\halfspan}{((##3)-(##2))/2}
   \pgfmathsetmacro{\middle}{((##3)+(##2))/2}
   \pgfmathsetmacro{\bottomspan}{(##4)-(##2)}
   \pgfmathsetmacro{\topspan}{(##3)-(##5)}
   \path[draw=##6,fill=##6,fill opacity=.105,line width=.73pt]
     ({##1-\bottomspan},{##4})--({##1-\halfspan},\middle)
     --({##1-\topspan},{##5})--({##1+\topspan},{##5})
     --({##1+\halfspan},\middle)--({##1+\bottomspan},{##4})--cycle;
   \draw[##6,line width=1.03pt]
     ({##1-\bottomspan},{##4})--({##1+\bottomspan},{##4})
     ({##1-\topspan},{##5})--({##1+\topspan},{##5});}
  \radialDiamond{-.94}{-.56}{.56}{-.47}{.47}{noteink!39}
  \radialDiamond{-.10}{-.70}{.70}{-.59}{.59}{noteink!52}
  \radialDiamond{.46}{-.70}{.44}{-.61}{.35}{noteink!43}
  \draw[noteink!68,dash pattern=on 2.6pt off 1.8pt,line width=.70pt]
    (-1,-1)--(1,1) (-1,1)--(1,-1);
  \radialDiamond{1}{phTime(\newPast)}{phTime(\newFuture)}
                   {phTime(\newLow)}{phTime(\newHigh)}{notegreen}
  \radialDiamond{1}{phTime(\oldPast)}{phTime(\oldFuture)}
                   {phTime(\oldLow)}{phTime(\oldHigh)}{noteblue}
  \draw[noterust,line width=1.30pt] ({1+phTime(\oldHigh)},{phTime(\oldHigh)})
                                 --({1-phTime(\oldHigh)},{phTime(\oldHigh)});
  \draw[noteblue,line width=.83pt] (1,{phTime(-.45)})--(1,{phTime(\oldHigh)});
  \draw[noteblue,dash pattern=on 2.2pt off 1.35pt,line width=.85pt]
    (1,{phTime(\oldHigh)})--(1,{phTime(.07)});
  \draw[noteink,line width=.95pt] (-1,1)--(-2,0)--(-1,-1) (1,1)--(2,0)--(1,-1);
  \foreach \side in {-1,1}{
    \draw[noteink,line width=.8pt] (-1,\side)
      \foreach \i in {0,...,27}{
       --({-1+2*(\i+.5)/28},{\side*(singularTime(-1+2*(\i+.5)/28)+.018)})
       --({-1+2*(\i+1)/28},{\side*singularTime(-1+2*(\i+1)/28)})};
    \node[font=\footnotesize] at (0,{\side*(singularTime(0)+.137)}) {$r=0$};
    \node[font=\footnotesize] at ({1.64*\side},.61) {$\mathscr I^+$};
    \node[font=\footnotesize] at ({1.64*\side},-.61) {$\mathscr I^-$};
    \foreach \x in {1.66,1.73,1.80}
      \fill[noteink!52] ({\side*\x},0) circle[radius=.44pt];
    \foreach \t in {.636,.674,.712}
      \fill[noteink!52] (-.10,{\side*\t}) circle[radius=.40pt];
  }
 \end{scope}
 \begin{scope}[shift={(6.18cm,.69cm)},x=4.32cm,y=3.36cm]
  \def\depth{.23}
  \def\angularRing##1##2##3##4##5{%
    \draw[##4,line width=##5]
      plot[domain=0:360,samples=65,variable=\p]
      ({saRad(##1)*caRad(##3)+caRad(##1)*saRad(##3)*cos(\p)},
       {(##2)+\depth*saRad(##3)*sin(\p)}); }
  \def\angularDiamond##1##2##3##4##5##6{%
   \pgfmathsetmacro{\middle}{((##2)+(##3))/2}
   \pgfmathsetmacro{\halfspan}{((##3)-(##2))/2}
   \path[fill=##6,fill opacity=.09]
    plot[domain=##4:\middle,samples=17,variable=\t] ({saRad((##1)-(\t-(##2)))},\t)
    --plot[domain=\middle:##5,samples=17,variable=\t] ({saRad((##1)-((##3)-\t))},\t)
    --plot[domain=##5:\middle,samples=17,variable=\t] ({saRad((##1)+((##3)-\t))},\t)
    --plot[domain=\middle:##4,samples=17,variable=\t] ({saRad((##1)+(\t-(##2)))},\t)--cycle;
   \foreach \sign in {-1,1}{
    \draw[##6,line width=.9pt]
     plot[domain=##4:\middle,samples=17,variable=\t] ({saRad((##1)+\sign*(\t-(##2)))},\t)
     --plot[domain=\middle:##5,samples=17,variable=\t] ({saRad((##1)+\sign*((##3)-\t))},\t);}
   \angularRing{##1}{##4}{(##4)-(##2)}{##6}{.85pt}
   \angularRing{##1}{\middle}{\halfspan}{##6!70}{.62pt}
   \angularRing{##1}{##5}{(##3)-(##5)}{##6}{.85pt}}
  \angularDiamond{\separation}{\newPast}{\newFuture}{\newLow}{\newHigh}{notegreen}
  \angularDiamond{0}{\oldPast}{\oldFuture}{\oldLow}{\oldHigh}{noteblue}
  \foreach \plane in {212,270,332}{
   \def\photonX{-saRad(\t)*cos(\plane)}
   \def\photonY{\t-\depth*saRad(\t)*sin(\plane)}
   \draw[noteblue,line width=.90pt]
     plot[domain=-.45:\oldHigh,samples=20,variable=\t] ({\photonX},{\photonY});
   \draw[noteblue,line width=.94pt,dash pattern=on 2.4pt off 1.4pt]
     plot[domain=0:\oldHigh,samples=16,variable=\t] ({\photonX},{\photonY});
   \draw[noteblue,line width=.94pt,dash pattern=on 2.4pt off 1.4pt]
     plot[domain=0:.07,samples=8,variable=\t] ({\photonX},{\photonY});}
  \angularRing{0}{\oldHigh}{-\oldHigh}{noterust}{1.26pt}
  \draw[noterust!80,line width=.43pt] ({saRad(\oldHigh)},\oldHigh)--(-.246,-.145);
  \node[anchor=east,inner sep=1pt,text=noterust] at (-.247,-.145) {$\Sigma$};
  \node[anchor=east,inner sep=2pt,text=noteblue] at ({-saRad(.45)},-.45) {$S$};
 \end{scope}
\end{tikzpicture}%
}

\newcommand{\SummaryHamiltonian}{%
\begin{tikzpicture}[amplitude]
 \begin{scope}
  \node[head,anchor=base] at (0,3.31) {Lapse along the null cone};
  \foreach \dx/\sheared in {-1.18/0,1.18/1}{
   \begin{scope}[shift={(\dx cm,0)}]
    \path[fill=noteblue!6] (-.69,.51)--(.56,.51)--(.82,2.42)--(-.47,2.42)--cycle;
    \foreach \x in {-.65,-.26,.13,.52}{
      \draw[noteblue!54,line width=.55pt] (\x,.52)--({\x+.24},2.42);}
    \ifnum\sheared=1
      \path[fill=noterust!16,draw=noterust!45,line width=.45pt]
         (-.61,1.13)--(.62,1.13)--(.70,1.78)--(-.53,1.78)--cycle;
      \foreach \x in {-.39,-.12,.15,.42}
       \draw[noterust!90,line width=.65pt] (\x,1.20)..controls({\x+.17},1.38)and({\x-.04},1.57)..({\x+.10},1.71);
      \def\cutlapse{.29}
      \node[below] at (0,.34) {$\sigma_{AB}\ne 0$};
    \else
      \def\cutlapse{.66}
      \node[below] at (0,.34) {$\sigma_{AB}=0$};
    \fi
    \draw[boundary] (-.69,.51)--(.57,.51) (-.47,2.42)--(.83,2.42);
    \draw[oriented,noteblue,line width=1.18pt] (.85,2.42)--(.94,3.08);
    \draw[oriented,noteblue,line width=1.18pt] (.59,.51)--({.59+.09*\cutlapse/.66},{.51+\cutlapse});
   \end{scope}}
  \node[anchor=east] at (-1.72,2.40) {$S$};
 \end{scope}
 \node[head,anchor=base] at (6.80,3.31) {Time evolution between slices};
 \begin{scope}[shift={(4.15cm,.63cm)},declare function={
      tiltch(\a)=(exp(\a)+exp(-\a))/2;
      tiltsh(\a)=(exp(\a)-exp(-\a))/2;}]
   \def\rapidity{.47}
   \def\slicelength{4.50}
   \def\normallength{1.80}
   \coordinate (corner) at (0,0);
   \coordinate (slice0) at (\slicelength,0);
   \coordinate (slice1) at ({\slicelength*tiltch(\rapidity)},
                            {\slicelength*tiltsh(\rapidity)});
   \coordinate (normal0) at (0,\normallength);
   \coordinate (normal1) at ({\normallength*tiltsh(\rapidity)},
                             {\normallength*tiltch(\rapidity)});
   \path[fill=notegreen!8] (corner)
     -- plot[domain=0:\rapidity,samples=16,variable=\a]
       ({\slicelength*tiltch(\a)},{\slicelength*tiltsh(\a)})--cycle;
   \draw[noteblue,line width=1.05pt] (corner)--
     coordinate[pos=.79] (data0)
     node[pos=.79,below=1.2pt,inner sep=1pt,text=noteblue] {$(q,p)_t$} (slice0);
   \draw[notegreen,line width=1.05pt] (corner)--
     coordinate[pos=.79] (data1)
     node[pos=.79,anchor=south east,xshift=7pt,yshift=5pt,inner sep=1pt,text=notegreen] {$(q,p)_{t+\dd t}$} (slice1);
   \node[anchor=west,text=noteblue] at (slice0) {$\Sigma_t$};
   \node[anchor=west,text=notegreen] at (slice1) {$\Sigma_{t+\dd t}$};

   \foreach \distance in {.96,1.93}{
    \fill[noteblue] (\distance,0) circle[radius=1.15pt];
    \draw[noteink!65,line width=.64pt,-{Stealth[length=3.5pt,width=2.9pt]},
          shorten <=1.7pt,shorten >=1.6pt]
      plot[domain=0:\rapidity,samples=16,variable=\a]
        ({\distance*tiltch(\a)},{\distance*tiltsh(\a)});
    \fill[notegreen] ({\distance*tiltch(\rapidity)},
                      {\distance*tiltsh(\rapidity)}) circle[radius=1.15pt];}

   \draw[oriented,noteink!85,line width=.82pt]
     (data0)--node[midway,fill=notegreen!8,inner xsep=1.4pt,inner ysep=1pt]
       {$\{\cdot,H_c\}\,\dd t$} (data1);

   \draw[oriented,noteblue,line width=1.05pt] (corner)--(normal0);
   \draw[oriented,notegreen,line width=1.05pt] (corner)--(normal1);
   \node[anchor=east,text=noteblue,inner sep=0pt] at (-.12,1.32) {$n_t^a$};
   \node[anchor=west,text=notegreen,inner sep=0pt] at (.77,1.46) {$n_{t+\dd t}^a$};
   \draw[oriented,noterust,line width=.92pt]
     plot[domain={.15*\rapidity}:{.85*\rapidity},samples=16,variable=\a]
       ({\normallength*tiltsh(\a)},{\normallength*tiltch(\a)});
   \path (normal0)--node[midway,above=4pt,inner sep=.6pt,text=noterust]
     {$\{\cdot,\mathscr K\}\,\dd\gamma$} (normal1);
   \draw[noteink,fill=white,line width=.55pt] (corner) circle[radius=2.5pt];
   \fill[noteink] (corner) circle[radius=1.25pt];
   \node[anchor=east,inner sep=0pt] at (-.12,.02) {$S$};
 \end{scope}
\end{tikzpicture}%
}

\newcommand{\SummarySources}{%
\begin{tikzpicture}[amplitude]
 \begin{scope}
  \node[head] at (0,2.68) {One deformation};
  \path[fill=noteblue!6] (-2.04,1.68)--(1.46,1.68)..controls(2.11,1.68)and(2.11,.55)..(1.46,.55)
    --(-2.04,.55)--cycle;
  \draw[noteblue,line width=.95pt] (-2.04,1.68)--(1.46,1.68);
  \draw[oriented,noteblue] (-1.59,1.68)--(-.97,1.68);
  \node[vertex] at (.02,1.68) {}; \node[above,text=noterust] at (.02,1.80) {$F$};
  \draw[boundary] (1.46,1.68)..controls(2.12,1.68)and(2.12,.55)..(1.46,.55);
  \draw[notegreen,line width=.95pt] (1.46,.55)--(-2.04,.55);
  \draw[oriented,notegreen] (-.93,.55)--(-1.55,.55);
  \node[external] at (-2.04,1.68) {}; \node[external] at (-2.04,.55) {};
  \node[anchor=west,text=noteblue] at (.42,1.95) {$U_{I+F}$};
  \node[anchor=west,text=notegreen] at (.42,.18) {$U_I^\dagger$};
 \end{scope}
 \begin{scope}[shift={(6.10cm,0)}]
  \node[head] at (0,2.68) {Two ordered deformations};
  \path[fill=noteblue!8] (-2.36,1.02)--(-1.12,1.99)--(0,1.53)--(0,.68)--(-1.12,.17)--cycle;
  \path[fill=notegreen!8] (0,1.53)--(1.12,1.99)--(2.36,1.02)--(1.12,.17)--(0,.68)--cycle;
  \draw[boundary] (-2.36,1.02)--(-1.12,1.99)--(0,1.53)--(1.12,1.99)--(2.36,1.02);
  \draw[boundary] (-2.36,1.02)--(-1.12,.17)--(0,.68)--(1.12,.17)--(2.36,1.02);
  \draw[cut] (0,.57)--(0,1.64); \node[below,text=noterust] at (0,.52) {$\Sigma$};
  \draw[leg] (-2.36,1.02)..controls(-1.79,1.02)and(-1.51,1.52)..(-1.12,1.99)
    ..controls(-.79,1.49)and(-.47,1.02)..(0,1.02)
    ..controls(.49,1.02)and(.79,1.49)..(1.12,1.99)
    ..controls(1.51,1.52)and(1.79,1.02)..(2.36,1.02);
  \node[vertex] at (-1.12,1.99) {}; \node[vertex] at (1.12,1.99) {};
  \node[above,text=noterust] at (-1.12,2.11) {$G$}; \node[above,text=noterust] at (1.12,2.11) {$F$};
  \draw[oriented,noteblue] (-.41,1.12)--(-.12,1.02);
  \draw[oriented,noteblue] (1.80,1.23)--(2.02,1.07);
  \node[external] at (-2.36,1.02) {}; \node[external] at (2.36,1.02) {};
 \end{scope}
\end{tikzpicture}%
}

\newcommand{\DrawBasisCycle}[2]{%
  \foreach \angle in {90,18,-54,-126,-198}{
    \draw[oriented,noteink!73] (\angle-12:#1)
      arc[start angle=\angle-12,end angle=\angle-60,radius=#1cm];
    \fill[noteblue] (\angle:#1) circle (2.55pt);
  }
  \ifnum#2=1
    \node[rectangle,draw=noterust,line width=1pt,fill=white,
          inner sep=0pt,minimum size=5.35pt] at (90:#1) {};
  \fi
}

\newcommand{\SummaryCycles}{%
\begin{tikzpicture}[amplitude]
  \begin{scope}
    \node[head] at (-.05,2.38) {Allowed areas};
    \path[fill=noteblue!18] (-.77,.91) rectangle (-.17,1.26);
    \draw[boundary] (-1.97,.91) rectangle (1.65,1.26);
    \foreach \x in {-1.37,-.77,-.17,.44,1.04}
      \draw[noteink!55,line width=.5pt] (\x,.91)--(\x,1.26);
    \draw[noteblue,line width=1pt] (-.77,.91) rectangle (-.17,1.26);
    \node[above,text=noteblue] at (-.47,1.26) {$I$};
    \node[below] at (-1.97,.87) {$0$}; \node[below] at (1.65,.87) {$s_*$};
    \fill[noteblue] (-1.30,.04) circle (2.55pt);
    \node[anchor=west] at (-1.11,.04) {in $I$};
    \node[rectangle,draw=noterust,fill=white,line width=1pt,
          inner sep=0pt,minimum size=5.35pt] at (.23,.04) {};
    \node[anchor=west] at (.42,.04) {outside $I$};
  \end{scope}
  \foreach \dx/\outside in {3.82/1,7.72/0}{
    \begin{scope}[shift={(\dx cm,.84cm)}]
      \ifnum\outside=1
        \node[head] at (0,1.54) {One state outside $I$};
      \else
        \node[head] at (0,1.54) {All states in $I$};
      \fi
      \DrawBasisCycle{.80}{\outside}
      \ifnum\outside=1
        \node[above] at (0,.91) {$i$};
        \node[right] at (.84,.26) {$V(i)$};
      \fi
    \end{scope}
  }
\end{tikzpicture}%
}

\newcommand{\SummaryContradiction}{%
\begin{tikzpicture}[amplitude]
  \foreach \dx/\cutopen in {0/0,6.50/1}{
    \begin{scope}[shift={(\dx cm,0)}]
      \ifnum\cutopen=0
        \node[head] at (0,3.05) {Uncut preparation};
      \else
        \node[head] at (0,3.05) {Cut at $S$};
      \fi
      \path[fill=noteblue!6] (-1.92,1.17)..controls(-1.65,2.13)and(-.75,2.32)..(0,2.32)
        --(0,.63)..controls(-.91,.63)and(-1.77,.76)..cycle;
      \path[fill=notegreen!6] (0,2.32)..controls(.75,2.32)and(1.65,2.13)..(1.92,1.17)
        ..controls(1.77,.76)and(.91,.63)..(0,.63)--cycle;
      \draw[leg] (-1.92,1.17)..controls(-1.65,2.13)and(-.75,2.32)..(0,2.32);
      \draw[notegreen,line width=1.05pt] (0,2.32)..controls(.75,2.32)and(1.65,2.13)..(1.92,1.17);
      \node[above,inner sep=2pt] at (0,2.32) {$\Sigma_1$};
      \ifnum\cutopen=0
        \draw[boundary] (-1.92,1.17)..controls(-1.52,.45)and(1.52,.45)..(1.92,1.17);
        \node[below] at (0,.55) {$S$};
        \node[font=\normalsize] at (0,-.47) {$Z[X]>0$};
      \else
        \draw[boundary] (-1.92,1.17)..controls(-1.53,.79)and(-.95,.64)..(-.66,.62);
        \draw[boundary] (.66,.62)..controls(.95,.64)and(1.53,.79)..(1.92,1.17);
        \draw[cut] (-.66,.45)--(-.66,.80) (.66,.45)--(.66,.80);
        \begin{scope}[shift={(0,.67cm)}]
          \fill[white] (0,0) circle (.64cm);
          \DrawBasisCycle{.54}{1}
        \end{scope}
        \node[font=\normalsize] at (0,-.47) {$Z_P[X]\leq0$};
      \fi
      \node[vertex] at (-1.42,.82) {};
      \node[vertex] at (1.42,.82) {};
    \end{scope}
  }
  \draw[oriented,noterust] (4.67,-.47)--(1.83,-.47);
  \node[above,text=noterust] at (3.25,-.43) {all cycles};
\end{tikzpicture}%
}

\newcommand{\SummaryFock}{%
\begin{tikzpicture}[amplitude]
  \begin{scope}
    \node[head] at (0,3.72) {Two radiative modes};
    \path[fill=noteblue!6] (-.40,.28)--(-1.17,2.57)
       arc[start angle=180,end angle=360,x radius=1.17cm,y radius=.23cm]--(.40,.28)
       arc[start angle=0,end angle=-180,x radius=.40cm,y radius=.09cm]--cycle;
    \draw[boundary] (-.40,.28)--(-1.17,2.57) (.40,.28)--(1.17,2.57);
    \draw[noteink!46,densely dashed] (1.17,2.57) arc[start angle=0,end angle=180,x radius=1.17cm,y radius=.23cm];
    \draw[cut] (-1.17,2.57) arc[start angle=180,end angle=360,x radius=1.17cm,y radius=.23cm];
    \draw[boundary] (0,.28) ellipse[x radius=.40cm,y radius=.09cm];
    \foreach \sgn in {-1,1}{\draw[noteink!35,densely dashed] ({.15*\sgn},.28)--({.54*\sgn},2.37);}
    \draw[noteblue,line width=1.12pt] plot[domain=0:1,samples=40,variable=\v]
       ({-.15-.39*\v+.13*sin(360*\v)}, {.28+2.09*\v});
    \draw[noterust,line width=1.10pt] plot[domain=0:1,samples=60,variable=\v]
       ({.15+.39*\v+.13*sin(720*\v)}, {.28+2.09*\v});
    \node[text=noteblue,above] at (-.53,2.91) {$j$};
    \node[text=noterust,above] at (.53,2.91) {$k$};
    \node[anchor=west,text=noterust] at (1.23,2.57) {$\mathscr A$};
    \node[anchor=north,inner sep=1pt] at (0,.10) {$|\Omega_{\mathscr A}\rangle$};
  \end{scope}
  \begin{scope}[shift={(6.72cm,0)}]
    \node[head] at (0,3.72) {Null cone Fock space};
    \node[anchor=east] at (.20,2.89)
      {$\mathfrak a_j^\dagger\mathfrak a_k^\dagger|\Omega_{\mathscr A}\rangle\ = $};
    \begin{scope}[shift={(.70,2.89)}]
      \draw[boundary,line width=.85pt] (1.15,-.45)--(1.15,.45);
      \draw[wave] (0,.24)--(1.15,.24);
      \draw[wave,draw=noterust] (0,-.24)--(1.15,-.24);
      \fill[noteink] (1.15,.24) circle[radius=1.65pt]
                     (1.15,-.24) circle[radius=1.65pt];
      \node[anchor=east,text=noteblue] at (-.04,.24) {$j$};
      \node[anchor=east,text=noterust] at (-.04,-.24) {$k$};
    \end{scope}
    \begin{scope}[yshift=1.30cm]
      \node[anchor=base east,inner xsep=0pt,yshift=-.25em] at (-.95,0)
        {$\langle\Omega_{\mathscr A}|\mathfrak a_{k'}\mathfrak a_{j'}
          \mathfrak a_j^\dagger\mathfrak a_k^\dagger|\Omega_{\mathscr A}\rangle ={}$};
      \node[anchor=base,yshift=-.25em] at (1.92,0) {$+$};
      \foreach \cx/\braJ/\braK/\weight in {
        .40/braTop/braBottom/{\delta_{j'j}\delta_{k'k}},
         3.44/braBottom/braTop/{\delta_{j'k}\delta_{k'j}}}{
        \begin{scope}[shift={(\cx,0)}]
          \coordinate (braTop) at (-.68,.29);
          \coordinate (braBottom) at (-.68,-.29);
          \coordinate (ketJ) at (.68,.29);
          \coordinate (ketK) at (.68,-.29);
          \draw[wave] (ketJ)--(\braJ);
          \draw[wave,draw=noterust,preaction={draw=white,line width=3.5pt}]
            (ketK)--(\braK);
          \foreach \point in {braTop,braBottom,ketJ,ketK}
            \fill[noteink] (\point) circle[radius=1.65pt];
          \node[anchor=east,xshift=-2pt] at (braTop) {$j'$};
          \node[anchor=east,xshift=-2pt] at (braBottom) {$k'$};
          \node[anchor=west,xshift=2pt,text=noteblue] at (ketJ) {$j$};
          \node[anchor=west,xshift=2pt,text=noterust] at (ketK) {$k$};
          \node at (0,-.77) {$\weight$};
        \end{scope}}
    \end{scope}
  \end{scope}
\end{tikzpicture}%
}

\newcommand{\SummaryOpenState}{%
\begin{tikzpicture}[amplitude]
 \begin{scope}
  \node[head] at (0,3.94) {Preparing the state};
  \path[fill=noteblue!9] (-.71,.33)--(-1.60,1.91)
    arc[start angle=180,end angle=360,x radius=1.60cm,y radius=.29cm]--(.71,.33)
    arc[start angle=0,end angle=-180,x radius=.71cm,y radius=.13cm]--cycle;
  \path[fill=notegreen!8] (-1.60,1.91)--(-.71,3.42)
    arc[start angle=180,end angle=360,x radius=.71cm,y radius=.16cm]--(1.60,1.91)
    arc[start angle=0,end angle=-180,x radius=1.60cm,y radius=.29cm]--cycle;
  \draw[noteblue,line width=.85pt] (-.71,.33)--(-1.60,1.91) (.71,.33)--(1.60,1.91);
  \draw[notegreen,line width=.85pt] (-1.60,1.91)--(-.71,3.42) (1.60,1.91)--(.71,3.42);
  \foreach \a in {212,245,290,326}{
    \draw[noteblue!58,line width=.58pt] ({.71*cos(\a)},{.33+.13*sin(\a)})
     --({1.6*cos(\a)},{1.91+.29*sin(\a)});
    \draw[notegreen!58,line width=.58pt] ({1.6*cos(\a)},{1.91+.29*sin(\a)})
     --({.71*cos(\a)},{3.42+.16*sin(\a)});}
  \draw[noteink!55,densely dashed,line width=.55pt] (1.60,1.91) arc[start angle=0,end angle=180,x radius=1.60cm,y radius=.29cm];
  \draw[boundary] (-1.60,1.91) arc[start angle=180,end angle=360,x radius=1.60cm,y radius=.29cm];
  \filldraw[draw=noteblue,fill=noteblue!15,line width=.7pt] (0,.33) ellipse[x radius=.71cm,y radius=.13cm];
  \filldraw[draw=noterust,fill=white,line width=.95pt] (0,3.42) ellipse[x radius=.71cm,y radius=.16cm];
  \draw[cut,line width=.88pt] (-.71,3.42) arc[start angle=180,end angle=360,x radius=.71cm,y radius=.16cm];
  \node[vertex] at (-.86,1.20) {};
  \draw[noteink!65,line width=.52pt] (-.91,1.18)--(-1.74,.99);
  \node[anchor=east] at (-1.80,.99) {$F$};
  \node[anchor=east] at (-1.65,1.95) {$S$};
  \node[anchor=west,text=noterust] at (.77,3.43) {$\Sigma_1$};
  \node[below] at (0,.08) {$\Psi_{\Sigma_0}(q_i)$};
 \end{scope}
 \begin{scope}[shift={(6.30cm,0)}]
  \node[head] at (0,3.94) {Fields on the open cut};
  \foreach \yc/\squash/\width in {2.55/0/1,.98/1/.83}{
   \begin{scope}[shift={(0,\yc cm)},xscale=\width]
    \path[draw=noterust,fill=notegreen!6,line width=.78pt] (0,0) ellipse[x radius=1.34cm,y radius=.52cm];
    \begin{scope}
    \clip (0,0) ellipse[x radius=1.34cm,y radius=.52cm];
    \foreach \k in {-3,-2,...,3}{
      \ifnum\squash=0
       \draw[noteblue!73,line width=.64pt] ({.42*\k-.20},-.53)--({.42*\k+.20},.53);
       \draw[notegreen!73,line width=.64pt] (-1.38,{.18*\k-.04})--(1.38,{.18*\k+.04});
      \else
       \draw[noteblue!73,line width=.64pt] ({.41*\k-.19},-.53)..controls({.25*\k+.12},-.15)and({.30*\k+.40},.17)..({.46*\k+.24},.54);
       \draw[notegreen!73,line width=.64pt] (-1.38,{.17*\k-.06})..controls(-.61,{.24*\k+.24})and(.54,{.16*\k-.27})..(1.38,{.16*\k+.06});
      \fi}
    \end{scope}
    \draw[noteink!72,line width=.78pt] (-.69,.02)--(.69,.02);
    \foreach \mark in {-.69,.69}{\fill[noteink] (\mark,.02) circle (1.7pt);}
   \end{scope}}
  \coordinate (openBranch) at (-2.43,1.78);
  \draw[noterust,densely dashed,line width=.62pt] (-4.71,3.38)
    ..controls(-3.66,3.38)and(-3.20,2.35)..(openBranch);
  \draw[noterust,densely dashed,line width=.62pt] (openBranch)..controls(-2.10,2.02)and(-1.68,2.49)..(-1.34,2.55);
  \draw[noterust,densely dashed,line width=.62pt] (openBranch)..controls(-2.10,1.55)and(-1.63,1.10)..(-1.12,.98);
 \end{scope}
\end{tikzpicture}%
}

\usepackage{footmisc}
\usepackage[colorlinks=true,citecolor=Blue,linkcolor=Blue]{hyperref}
\usepackage[labelfont={small,bf,sf},textfont={small,sf}]{caption}
\usepackage{authblk}
\usepackage[in]{fullpage}
\usepackage{cite,microtype}
\hypersetup{breaklinks=true,pdftitle={An algebraic quantization of gravity from the null cone},pdfauthor={Venkatesa Chandrasekaran}}
\numberwithin{equation}{section}
\allowdisplaybreaks[1]
\newcommand{\dd}{\mathrm d}
\newcommand{\Tr}{\operatorname{Tr}}

\newcommand{\GN}{G_N}

\newcommand{\Lie}{\pounds}
\newcommand{\supp}{\operatorname{supp}}
\newcommand{\normalorder}[1]{\mathopen{\vcentcolon}\,#1\,\mathclose{\vcentcolon}}
\newcommand{\rank}{\operatorname{rank}}

\title{\Large An algebraic quantization of gravity from the null cone}

\author{Venkatesa Chandrasekaran\thanks{venc@stanford.edu}}
\affil{Leinweber Institute for Theoretical Physics, Stanford University, Stanford, CA 94305, USA}
\date{\normalsize November 23, 2026}

\newcommand{\RevisionNeedspace}[1]{%
  \par\begingroup
  \dimen0=\dimexpr\pagegoal-\pagetotal\relax
  \ifdim\dimen0<#1\relax\newpage\fi
  \endgroup}
\hypersetup{urlcolor=Blue}

\usepackage{needspace}

\begin{document}
\flushbottom
\makeatletter\def\@textbottom{\vskip 0pt plus 6pt}\makeatother
\setlength{\skip\footins}{8pt plus 1pt minus 1pt}
\maketitle
\setlength{\abovedisplayskip}{9.5pt plus 3pt minus 4pt}
\setlength{\belowdisplayskip}{9.5pt plus 3pt minus 4pt}

\begin{abstract}
We propose a background independent quantization of vacuum general relativity on four dimensional Lorentzian spacetimes. Starting from first order gravity, we solve the constraints on intersecting null cones and obtain the reduced phase space at finite $G_N$. At fixed area we construct a Fock space from Gaussian wavefunctions of canonical pairs and prepare states whose span is dense in its radiative sector. Gravitational focusing renders their overlaps finite as we include arbitrarily short wavelengths at fixed $G_N$ and $\hbar$, even though expanding first in $G_N$ yields a loop divergence. We choose an extension to the full algebra by requiring the sum over intermediate states to reproduce amplitudes computed from the reduced action. This choice, which we check is consistent with our construction, yields a representation in which the algebra of observables generates a type~I factor. We use diffeomorphism invariance of the Lorentzian path integral under gluing to show that a macroscopic causal diamond of bounded size has a finite dimensional Hilbert space. From the resulting Lorentzian trace formula, we obtain the Bekenstein Hawking entropy for marginally trapped surfaces and its logarithmic correction for large Schwarzschild black holes. We explicitly compute this correction to one loop using the Lorentzian replica trick, from which we predict the distribution of area eigenvalues and recover the coefficient Sen obtained from the Euclidean path integral. We end by deriving the semiclassical Einstein equations to one loop in the low energy limit.
\end{abstract}

\begingroup
\setlength{\parskip}{0.6pt}
\tableofcontents
\endgroup

\section{Introduction}
\label{sec:bf-introduction}
\label{sec:bf-overview}

In perturbative quantum gravity we typically choose a fixed background spacetime and quantize small fluctuations of its metric. In a locally covariant formulation we can nevertheless make the observables independent, order by order, of how we split the metric into a background and a fluctuation~\cite{BrunettiFredenhagenRejzner2016}. Nonperturbatively, background independence requires us to treat all spacetimes dynamically, so we must specify observables independently of any particular background. One way to do this is to dress a bulk field to a reference frame defined by gravitational data on the boundary of a finite subregion. Relative to this frame we can specify where to evaluate the field in different spacetimes even as the metric fluctuates strongly. We would like to use these observables to define finite transition amplitudes between gravitational states on initial and final spacelike slices, and to understand when a classical spacetime satisfying the Einstein equations emerges from these amplitudes.

For scattering about Minkowski spacetime we usually specify momenta and polarizations of incoming and outgoing gravitons at the asymptotic boundary, but loop corrections to their scattering amplitudes in pure general relativity lead to UV divergences that make the theory perturbatively nonrenormalizable~\cite{GoroffSagnotti1986}. String theory addresses this difficulty by replacing pointlike interactions with the splitting and joining of strings. At long wavelengths string amplitudes reproduce general relativity~\cite{ScherkSchwarz1974}, but the expansion in string coupling does not fix them nonperturbatively even for flat space scattering, let alone for transitions between states on the fluctuating boundary of a finite subregion.

Loop quantum gravity on the other hand begins with a densitized triad on a spacelike slice that determines its metric, while the conjugate Ashtekar Barbero connection encodes its extrinsic curvature~\cite{Ashtekar1986,Barbero1995}. In these variables one can impose invariance under diffeomorphisms within the slice by group averaging~\cite{AshtekarEtAl1995}. For general boundary states, it remains an open problem how to then solve the Hamiltonian constraint so that the physical inner product is independent of how one slices spacetime and its semiclassical continuum limit recovers the Einstein equations~\cite{ThiemannGiesel2023}.

In AdS/CFT we can construct gravitational observables through their representations in a boundary conformal field theory, whose inner product and Hamiltonian already define amplitudes in the bulk dual nonperturbatively~\cite{Maldacena9711200}. Observables in a boundary subregion can even tell us about gravitational subregions in the bulk. For example, the Ryu Takayanagi formula relates the entanglement entropy of such a subregion to the area of a bulk minimal surface~\cite{RyuTakayanagi0603001}, while quantum error correction explains how different boundary subregions can represent the same bulk observable~\cite{AlmheiriDongHarlow14117041,Harlow160703901}.

Motivated by the above, we seek in this paper a nonperturbative algebra of observables for a finite bulk subregion directly from gravitational data on its own boundary, with the gravitational action determining the time evolution of its observables~\cite{PeningtonWitten2023}. In a causal diamond we can dress these observables to a dynamical reference frame at the spacelike corner where the future null cone from its past tip intersects the past null cone from its future tip. This frame specifies where we evaluate an observable as the metric fluctuates, so we include its position and orientation as gravitational edge modes in the reduced phase space. This dressing is invariant under a diffeomorphism acting on the bulk metric and boundary edge modes together, while surface symmetries transform edge modes relative to the metric and act on the observables through gravitational surface charges \cite{DF,Speranza}. Describing two subregions independently introduces a reference frame on each side of the boundary between them, so gluing requires invariance under the surface symmetry acting on both sides together. On the boundary Hilbert spaces this condition defines the entangling product, wherein one restricts their tensor product to states annihilated by the sum of charges on the two boundaries, each defined using its outward normal \cite{DF}.\footnote{This is in the spirit of existing proposals in the literature on quantizing gravitational subregions through representations of the corner symmetry algebra~\cite{Freidel:2015gpa,Freidel:2016bxd,FreidelGeillerPranzetti2020,Freidel:2023bnj}. In two dimensions one can explicitly glue representations of the corner symmetry group in this way~\cite{Ciambelli:2024qgi}.}

On solutions of the gravitational constraints, the Hamiltonian reduces to a surface charge \cite{ReggeTeitelboim,IW}, so a boundary algebra containing it also contains time translations of its observables. This property, together with the Einstein equations and the equations of motion for matter, allows perturbative reconstruction of bulk observables from boundary data in asymptotically flat and AdS spacetimes \cite{Marolf08082842}. In perturbative quantum gravity, the bulk subregion algebras of black hole exteriors and static regions of de Sitter spacetime likewise contain a boundary (or observer) Hamiltonian \cite{CPW220910454,CLPW220610780}. For a black hole exterior, adjoining the unitaries that implement modular flow yields a type~II crossed product~\cite{Witten2022CrossedProduct,JensenSorceSperanza2023,Kudler-Flam:2023qfl}, and one can also include the charges for angle dependent boosts of the horizon~\cite{KlingerKudlerFlamSatishchandran2026}. For subregions of a black hole horizon, the algebra of dressed fields and surface charges can be derived directly by canonically quantizing metric perturbations and gravitational edge modes \cite{Subregion260107915}.

A description of the interior in terms of boundary observables is particularly natural when the boundary is null. To specify a classical solution inside a null cone, we can choose two functions along its generators together with data at its cuts and solve the null constraints \cite{ChoquetBruhat10064467,Chrusciel12091971}. Expressing the symplectic form in terms of this data yields the Poisson brackets of freely specifiable fields \cite{Reisenberger07122541,Reisenberger12113880,CF}. We can thus quantize the gravitational data of one subregion without first expanding the metric about a chosen classical solution. On a boundary shared by two such subregions, the same data specifies which quantum states we must match when we join them. Composing their amplitudes in this way allows us to define a gravitational path integral on a spacetime covered by subregions. It has likewise been proposed to compute transition amplitudes in the bulk by gluing amplitudes for gravitational edge modes at the corners of null hypersurfaces~\cite{WielandNullInitialData2019}.

In pure Jackiw Teitelboim (JT) gravity with two asymptotic boundaries one can readily complete this program to obtain the Hilbert space and Hamiltonian~\cite{HarlowJafferis2020} on the disk. In three dimensional pure gravity, one can actually do the same for causal diamonds~\cite{AndradeESilvaJacobson2023}. To see how boundary states can determine the amplitude for a larger region, we can turn to gravity with negative cosmological constant, where quantization of the reduced phase space leads to Virasoro TQFT~\cite{CollierEberhardtZhang2023}. With geodesic lengths or dihedral angles fixed on the outer boundary of a triangulation~\cite{Hartman2025CTV}, we glue the amplitudes for individual tetrahedra by summing over labels on interior edges to obtain the gravitational partition function at fixed topology.

Extending this description to four dimensions requires us to account for propagating degrees of freedom of the gravitational field in each subregion when we sum over states on its boundary. Our starting point is first order gravity, in which the tetrad determines the metric and defines a local Lorentz frame, while the spin connection $\omega^{\mu\nu}$ specifies parallel transport in that frame. Upon recasting the Einstein Hilbert action in terms of these variables, the curvature of $\omega^{\mu\nu}$ couples to a spacetime two-form $B^{\mu\nu}$ with antisymmetric indices. If this two-form is varied freely, the theory is topological and has no local propagating degrees of freedom. To obtain general relativity, we impose the simplicity constraint that $B^{\mu\nu}$ be the Hodge dual of $e^\mu\curlywedge e^\nu$, where $e^\mu$ are the tetrad one-forms and $\curlywedge$ denotes their spacetime wedge product.\footnote{The theory with an unrestricted two-form is called BF theory. The Pleba\'nski action for constrained BF theory reproduces general relativity for this choice of $B^{\mu\nu}$ with an invertible tetrad~\cite{Plebanski1977}. Other deformations of BF theory allow for dynamical torsion~\cite{Cattaneo:2023cnt}.} Varying the connection then sets torsion to zero, while varying $B^{\mu\nu}$ yields the Einstein equations. Although the simplicity constraint restricts the fields at the corner~\cite{Freidel:2020ayo}, it contains no derivatives and thus adds no boundary term to the variation of the bulk action. The same boundary variables can therefore be used for gluing, now with the two local propagating degrees of freedom of general relativity included. In four dimensional spin foam models one likewise imposes simplicity constraints in BF theory and glues amplitudes for individual simplices~\cite{EngleLivinePereiraRovelli2008,FreidelKrasnov2008}.

Once the null constraints are solved, the action depends only on boundary fields and edge modes. To quantize them we follow the algebraic formulation of quantum field theory, in which observables and their relations can be specified before choosing a state~\cite{HaagKastler1964,Haag1996}. This is important even for QFT in curved spacetime, where there need not be a preferred vacuum~\cite{Wald1994QFT}. Here the spacetime itself can fluctuate, so it becomes critical. We use the first order action to determine relational observables, including boundary and corner terms in order for composition in time to agree with gluing. Coupling a source $j$ to an observable in the action then lets us insert that observable in an amplitude by differentiating with respect to $j$. A source coupled to the shear of the null generators measures the radiative part of a gravitational excitation.\footnote{The conformal metric along the null cone specifies the two polarizations of gravitational radiation. The shear measures how this metric varies along the null generators, much as the Bondi news at future null infinity measures how the asymptotic shear varies with retarded time~\cite{Bishop2005Bondi}, hence we refer to the shear as a radiative observable, and upon quantization, a radiative insertion or operator. Upon linearization about a classical solution, these radiative modes correspond to gravitons.} To prepare the excitation itself we act on the incoming state with the creation operator of a reduced canonical pair. The conjugate momentum entering this operator can also depend on the edge modes at the cuts, so exciting a single metric polarization can change their fluctuations.

To construct the gravitational path integral over all of spacetime, we identify the outgoing boundary of one subregion with the incoming boundary of another and sum over physical states on their shared boundary. A cobordism $\mathscr M:\Sigma_0\to\Sigma_1$ is a spacetime subregion bounded by an incoming hypersurface $\Sigma_0$, an outgoing hypersurface $\Sigma_1$, and possibly additional boundary components on which states are specified (Figure~\ref{fig:bf-composition}). In order to specify incoming and outgoing states for a causal diamond $\mathscr D$, we cut off its past and future tips with small spacelike hypersurfaces $\Sigma_0$ and $\Sigma_1$, which we refer to throughout as the \emph{caps} of $\mathscr D$.\footnote{One might worry that cutting off the tips might discard Planck scale physics near either vertex. Each cap bounds only one causal diamond, however, and another can cover the subregion we excise (Fig.~\ref{fig:bf-local-reconstruction}). A process localized near the vertex then contributes to the amplitude for this second causal diamond and hence to the amplitude for the full region when we sum over physical states on the cap as in~\eqref{bfgeo:sewing-main}.} Following the null generators from either cap, we reach the corner $S$ where the two null cones intersect (Figure~\ref{fig:bf-causal-diamond-3d}). With the affine parameter of an incoming null generator set to zero at $\Sigma_0$ and its scale fixed by the gravitational boundary data there, its value at $S$ defines the reading of a gravitational clock. To specify bulk fields on either null cone we must say where in spacetime they are evaluated and in which local Lorentz frame their components are measured, hence the boundary edge modes enter the sum over states along with bulk fields when we join two causal diamonds. 

\subsection{Outline of paper}
\label{sec:outline-of-paper}

In Section~\ref{jt:prelude-classical} we derive the momentum conjugate to the gravitational clock in JT gravity and show that it ranges over the real line with the matter field and both corner dilatons fixed. In Section~\ref{jt:prelude-quantum} we quantize this pair and vary the phase of the clock wavefunction to construct infinitely many orthogonal states with identical clock probabilities in a causal diamond of bounded size. The observables of the causal diamond distinguish these states and generate a type~$\mathrm I_\infty$ factor.

In Sections~\ref{bfgeo:null-initial-data} and~\ref{bfgeo:symplectic-edge-modes} we solve the null constraints in four dimensions, where shear changes the clock reading even at fixed corner area. From the reduced action we derive the Poisson brackets of the radiative and corner data and, in Section~\ref{bfgeo:reduced-hamiltonian-main}, the Hamiltonian that evolves them. In Sections~\ref{sec:metric-bulk-observables} and~\ref{sec:pi-matching} we match fields to identify the same dressed observable in overlapping subregions.

In Sections~\ref{bfq:observable-algebra} and~\ref{bfq:bulk-diffeomorphisms} we quantize these brackets and deform the action to define an algebra of observables whose cocycle vanishes under bulk diffeomorphisms. In Section~\ref{bfq:algebra-states} we construct a Fock space at fixed corner area from Gaussian wavefunctions of the canonical pairs. In Section~\ref{bfq:fixed-area-uv} we show that gravitational focusing renders overlaps UV finite for a dense set of radiative states, even though expanding first in $G_N$ can yield a divergence at two loops.

In Section~\ref{bfq:state-extension} we choose a pure extension of the Gaussian to the full algebra that assigns each pair of source operators the overlap of the states they prepare from the same incoming wavefunction using the reduced gravitational action. We check this prescription by summing over intermediate area eigenstates in the radiative two point function across the corner. In Section~\ref{bfq:full-area-uv} we show that focusing bounds the change in area fluctuations under an infinitesimal null translation, whereas in $g\phi^6$ theory and QED at fixed bare couplings we find UV divergent corrections to fluctuations of the boost generator, which is the field theory analogue of the area operator. We then show finiteness of the energy flux through two null cones glued at $S$ (Section~\ref{bfq:glued-null-energy}) and glue amplitudes by summing over the states on an intermediate cut (Section~\ref{bfuv:completeness}).

In Sections~\ref{bfq:canonical-quantization} and~\ref{bfq:diamond-state-existence} we construct the gravitational clock operator and project onto its allowed readings to prepare a state of a macroscopic causal diamond of bounded size. In Section~\ref{bfent:subregion-boost-main} we use diffeomorphism invariance of the path integral to show that this causal diamond has a finite dimensional Hilbert space and a discrete area spectrum. In Section~\ref{sec:bf-entropy} we trace over interior states to obtain the Bekenstein Hawking entropy for marginally trapped surfaces. In Section~\ref{bfent:one-loop-clock} we use the Lorentzian replica trick to compute the logarithmic correction at one loop for large Schwarzschild black holes. The same calculation predicts the distribution of gravitational clock readings in the interior and recovers Sen's result at fixed mass~\cite{SenSchwarzschild2013}.

In Section~\ref{sec:spatial-topology} we represent a state on a connected slice as an entangled state on two disconnected components with the same boundary data, realizing the algebraic criterion for $\mathrm{ER}=\mathrm{EPR}$~\cite{McNamaraWang2026}. In Sections~\ref{bfq:classical-solution} and~\ref{bfq:graviton-backreaction} we derive the semiclassical Einstein equations through one loop for coherent initial states. In Section~\ref{sec:open-questions} we discuss open questions. In Appendix~\ref{sec:lorentzian-algebras} we recover the Hilbert space and trace from amplitudes on closed spacetimes using Lorentzian axioms equivalent to those of~\cite{ColafranceschiDongMarolfWang2024} when the Euclidean cylinder has finite trace.

\section[A prologue in JT gravity]{A prologue in JT gravity}
\label{sec:jt-prelude}

In a background independent theory of gravity, the clocks and reference frames with which we compare events belong to the same quantum system as spacetime itself. Given a classical metric, two observers can follow worldlines $\gamma_i$ between intersections $p,q$ and compare the proper times $\tau_i=\int^{\gamma_i}(-g_{ab}\,\mathrm dx^a\,\mathrm dx^b)^{1/2}$ recorded by their clocks. When no single metric accounts for a quantum state, however, neither the worldlines nor their readings can be fixed in advance. The observers can still start and stop their clocks where their worldlines intersect. We likewise define a gravitational clock for a causal diamond by following an incoming null generator from the spacelike cap $\Sigma_0$ that cuts off the past tip to the corner $S$ where it meets the outgoing null cone. Fixing the tangent vector in a local Lorentz frame on $\Sigma_0$ sets the scale of the affine parameter, so if we take it to vanish there its value at the corner defines the clock reading.

To understand what a bound on this reading tells us about the state of the region, we start with JT gravity coupled to a massless scalar field. Here the proper width of a causal diamond, and hence its clock reading, can fluctuate even though the metric has the same curvature in every solution. At each width we can solve for the dilaton sourced by the matter field along both null generators.

\begin{figure}[!htbp]
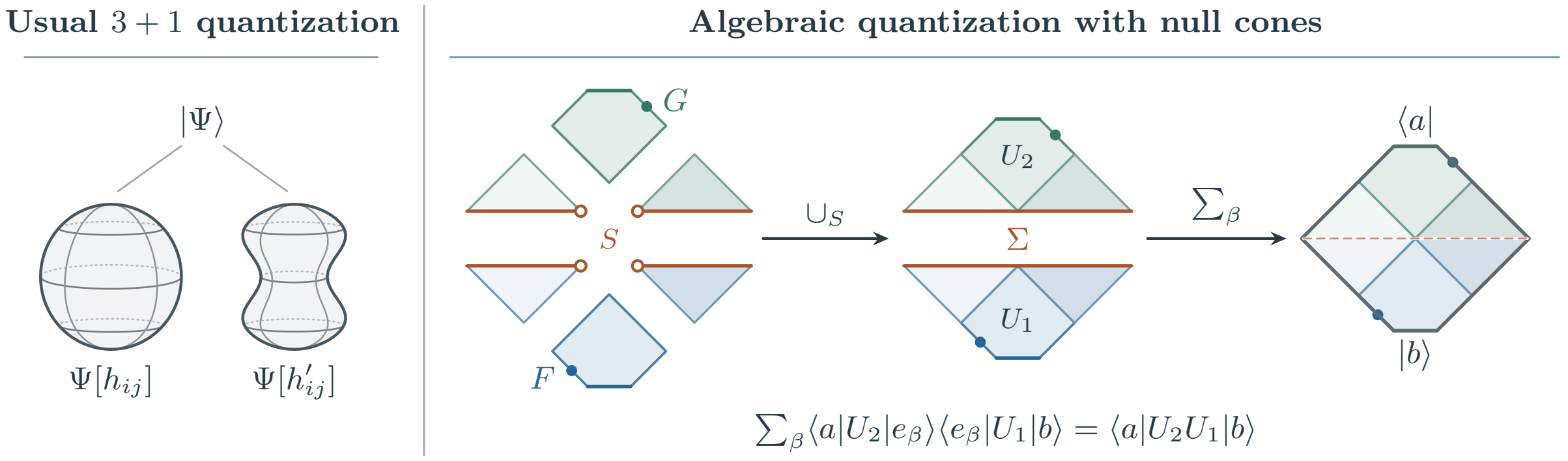

\centering
\resizebox{.96\linewidth}{!}{\SummaryRegionalAmplitudes}
\caption[States on a Cauchy slice and amplitudes on null cones]{On the left one evaluates the wavefunction on two spatial metrics of the same Cauchy slice, shown with one spatial dimension suppressed. On the right we match the tetrad and spin connection at $S$ to join the two parts of $\Sigma$. Summing over its basis states $|e_\beta\rangle$ composes the time evolution operators $U_1$ and $U_2$, with the sources $G$ and $F$ on opposite sides of the slice.}
\label{fig:summary-regional-amplitudes}
\end{figure}

In the usual $3+1$ quantization of general relativity one specifies a wavefunction of the spatial metric on an entire Cauchy slice of fixed topology~\cite{DeWitt1967I}. Even for a clock that runs only from $\Sigma_0$ to $S$, the argument of this wavefunction includes the metric outside the causal diamond. Two preparations that differ only outside $S$ can therefore have vanishing overlap on the full slice. To determine how many independent states the causal diamond admits, we need an inner product for preparations in the region itself. Since characteristic data on the incoming null cone, together with compatible data on $\Sigma_0$, determines the solution in the domain of dependence of the two surfaces~\cite{ChoquetBruhat10064467}, we can quantize this data to prepare a joint state of the clock and the matter field before adjoining an exterior. To obtain the amplitude for a larger region, we glue neighboring null cone preparations by matching their fields at the shared corner $S$. We then compose the amplitudes to the past and future of the joined slice $\Sigma$ by summing their product over a basis of states on that slice (Figure~\ref{fig:summary-regional-amplitudes}).

In short, we show that in JT gravity a causal diamond of bounded size admits infinitely many orthogonal states with identical clock probabilities, even when we fix the dilaton at both corners and prepare the matter field in the same state, and despite us finding that the associated algebra is type~$\mathrm I_\infty$. In four dimensions on the other hand, shear changes the affine parameter at which a null generator reaches a given area element at the corner, so a bound on the clock reading also restricts gravitational radiation. In Section~\ref{sec:diamond-hilbert-entropy} we show that it is precisely this fact which, unlike in JT gravity, yields a finite dimensional Hilbert space, including states prepared on different spatial topologies.

\subsection[A gravitational clock on a finite null cone]{A gravitational clock on a finite null cone}
\label{jt:prelude-classical}
\label{sec:space-and-time}

In JT gravity we can vary the flux of matter across a null generator while holding the metric and hence the clock reading fixed. To compare boost charges at that reading, we solve for the dilaton sourced by the flux, since the value of the dilaton at the corner generates the boost.

To that aim, null generators of the incoming null cone $\mathscr H_p$ of a causal diamond $\mathscr D$ run from $\Sigma_0$ to the corners $S=\{S_L,S_R\}$ where they meet the outgoing null cone $\mathscr H_q$. On either null generator the affine parameter specifies where to insert a field. The embedding $X[\mathfrak h]$ into spacetime depends on the gravitational boundary data $\mathfrak h$, including the local Lorentz frame $\mathfrak e_{\Sigma_0}$, so under a diffeomorphism $f$,
\begin{align}
&X[f^*\mathfrak h]=f^{-1}\circ X[\mathfrak h],\
X[f^*\mathfrak h]^*(f^*g)=X[\mathfrak h]^*g .
\label{prologue:relational-metric}
\end{align}
The metric $X^*g$ can therefore distinguish field configurations by what an observer would find, while assigning the same reading to those related by a diffeomorphism~\cite{DF,DonnellyGiddings2016}.

To determine the metric that enters this clock reading, we write $g_{ab}=\eta_{\mu\nu}e^\mu{}_a e^\nu{}_b$, where $\mu,\nu=0,1$, and vary $e^\mu{}_a$ independently of the spin connection $\omega_a{}^{\mu\nu}$, whose only independent component $\omega^{01}$ enters the torsion two-form $T^\mu=\dd e^\mu+\omega^\mu{}_\nu\curlywedge e^\nu$. Absorbing the overall gravitational coupling into $\Phi$ and $Y_\mu$, we write the first order action~\cite{IliesiuPufuVerlindeWang2019} with a minimally coupled scalar field $\varphi$ as
\begin{align}
&I[e,\omega,Y,\Phi,\varphi]=\int_{\mathscr D}\!\left[Y_\mu T^\mu+\Phi\big(\dd\omega^{01}+L_{\rm AdS}^{-2}\bm\epsilon_g\big)-\frac{1}{2}\dd\varphi\curlywedge\star\dd\varphi\right]+I_{\partial\mathscr D},
\label{jt:prelude-action}
\end{align}
where $L_{\rm AdS}$ is the AdS radius, $\bm\epsilon_g=e^0\curlywedge e^1$ the spacetime volume two-form, and $\star$ the Hodge dual defined by $g_{ab}$. Varying the Lagrange multipliers $Y_\mu$ imposes $T^\mu=0$ and fixes the spin connection in terms of $e^\mu{}_a$, and varying the dilaton $\Phi$ yields $R=-2/L_{\rm AdS}^2$.

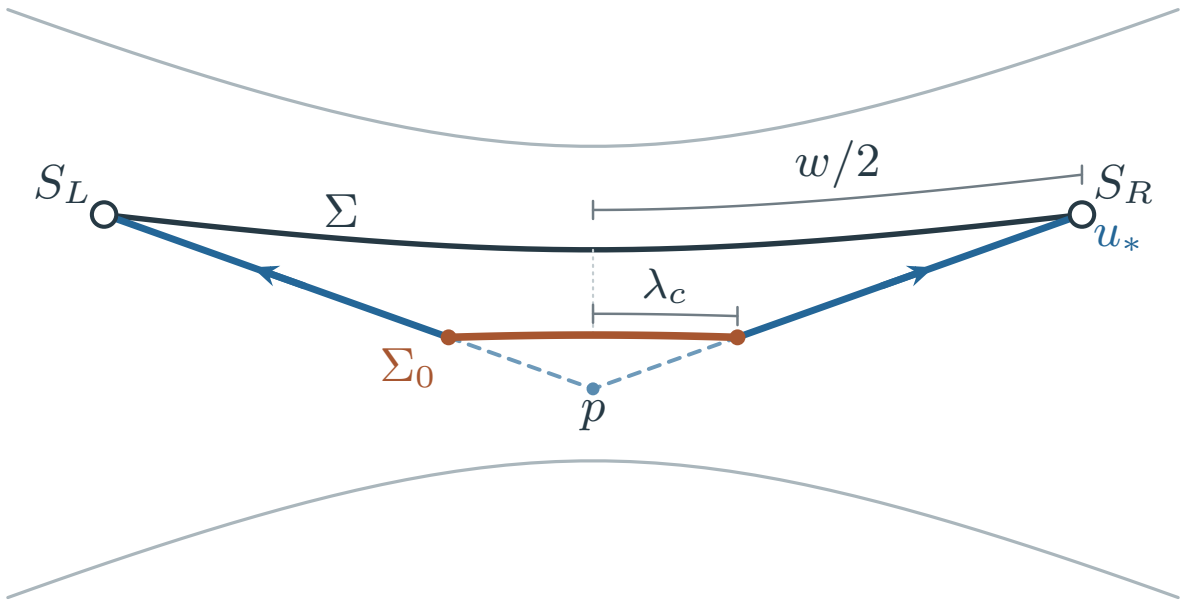
\begin{figure}[!htb]
\centering
\begingroup
\definecolor{jtInk}{HTML}{263944}
\definecolor{jtBlue}{HTML}{246697}
\definecolor{jtCap}{HTML}{A75630}
\definecolor{jtOutline}{HTML}{AAB6BC}
\def\spatialScale{2.45}
\def\timeScale{1.04}
\def\viewAngle{20}
\def\extent{1.58}
\def\atCorner{1.32}
\def\atCap{.39}
\begin{tikzpicture}[font=\small,text=jtInk,line cap=round,line join=round,
  every node/.style={inner sep=1.9pt},
  guide/.style={draw=jtInk!66,line width=.47pt},
  future/.style={jtBlue,-{Stealth[length=4.8pt,width=4.0pt]},line width=1.30pt},
  corner/.style={circle,fill=white,draw=jtInk,line width=.77pt,inner sep=0pt,minimum size=4.6pt},
  declare function={
    jtx(\r)=\spatialScale*\r;
    jty(\r,\t)=\timeScale*sqrt(1+\r*\r)*sin(\t+\viewAngle);}]
\pgfmathsetmacro{\tipTime}{-atan(\atCorner)}
\pgfmathsetmacro{\capTime}{atan(\atCap)+\tipTime}
\def\slice#1#2#3{plot[variable=\r,domain=#1:#2,samples=38]
  ({jtx(\r)},{jty(\r,#3)})}

\foreach \side in {-1,1}{
  \draw[jtOutline,line width=.60pt]
    plot[variable=\r,domain=-\extent:\extent,samples=48]
      ({jtx(\r)},{\side*\timeScale*sqrt(1+\r*\r)});}

\coordinate (tip) at (0,{jty(0,\tipTime)});
\coordinate (SL) at ({jtx(-\atCorner)},{jty(-\atCorner,0)});
\coordinate (SR) at ({jtx(\atCorner)},{jty(\atCorner,0)});
\coordinate (leftcap) at ({jtx(-\atCap)},{jty(-\atCap,\capTime)});
\coordinate (rightcap) at ({jtx(\atCap)},{jty(\atCap,\capTime)});

\draw[jtInk,line width=1.02pt] \slice{-\atCorner}{\atCorner}{0};
\draw[jtOutline!70,densely dotted,line width=.41pt]
  (0,{jty(0,0)})--(0,{jty(0,\capTime)});
\draw[jtBlue!66,line width=.74pt,dash pattern=on 2.2pt off 1.8pt]
  (leftcap)--(tip)--(rightcap);
\foreach \start/\finish in {leftcap/SL,rightcap/SR}{
  \draw[jtBlue,line width=1.30pt] (\start)--(\finish);
  \path (\start)--(\finish) coordinate[pos=.42] (before) coordinate[pos=.57] (after);
  \draw[future] (before)--(after);}
\draw[jtCap,line width=1.43pt] \slice{-\atCap}{\atCap}{\capTime};
\foreach \pt in {leftcap,rightcap}{\fill[jtCap] (\pt) circle[radius=1.5pt];}
\foreach \pt in {SL,SR}{\node[corner] at (\pt) {};}
\fill[jtBlue!76] (tip) circle[radius=1.22pt];

\node[above left] at (SL) {$S_L$};
\node[above right] at (SR) {$S_R$};
\node[below right,text=jtBlue] at (SR) {$u_*$};
\node[above] at ({jtx(-.68)},{jty(-.68,0)}) {$\Sigma$};
\node[below left,text=jtCap] at (leftcap) {$\Sigma_0$};
\node[below] at (tip) {$p$};

\foreach \limit/\time/\offset in {\atCorner/0/7.5pt,\atCap/\capTime/4pt}{
  \begin{scope}[yshift=\offset]
    \draw[guide] \slice{0}{\limit}{\time};
    \foreach \r in {0,\limit}{
      \draw[guide] ({jtx(\r)},{jty(\r,\time)}) ++(0,-1.7pt)--++(0,3.4pt);}
  \end{scope}}
\node[above,yshift=7.5pt] at ({jtx(.66)},{jty(.66,0)}) {$w/2$};
\node[above,yshift=4pt,font=\footnotesize] at ({jtx(.195)},{jty(.195,\capTime)}) {$\lambda_c$};
\end{tikzpicture}
\endgroup
\caption[The gravitational clock in $\mathrm{AdS}_2$]{In this side view of the $\mathrm{AdS}_2$ hyperboloid (gray), the null generators from $p$ meet $\Sigma_0$ a proper distance $\lambda_c=\epsilon w/2$ from its midpoint. We normalize their tangents there and follow the solid blue segments to the corners of $\Sigma$, each a proper distance $w/2$ from its midpoint, to obtain the clock reading $u_*(w)$ in~\eqref{jt:prelude-clock:c}.}
\label{fig:jt-null-geometry}
\end{figure}

With the curvature fixed, we use global coordinates $\tau,\sigma$ centered at the midpoint of the spacelike geodesic $\Sigma$ of length $w$ joining the corners. Its ends lie at $(0,\pm\sigma_S)$. In these coordinates the metric and the incoming null generators take the form
\begin{equation}
 \dd s^2=L_{\rm AdS}^2\sec^2\!\sigma\,(-\dd\tau^2+\dd\sigma^2),\ \tau=|\sigma|-\sigma_S.\label{jt:global-null-generators}
\end{equation}
Along the right sided null generator the affine parameter $u$ vanishes at the past tip $p$ and satisfies $(\partial_u\lambda)|_p=1$, where $\lambda$ denotes proper distance from the center on a slice of constant $\tau$. $\Sigma_0$ lies on such a slice with its ends at $\lambda_c=\epsilon w/2$ for fixed $0<\epsilon<1$. At the right sided corner $\lambda=w/2$ (Figure~\ref{fig:jt-null-geometry}), so the clock reads
\begin{subequations}\label{jt:prelude-clock}
\begin{align}
&\lambda=L_{\rm AdS}\int_0^\sigma\sec s\,\dd s,\label{jt:prelude-clock:a}\\
&u=L_{\rm AdS}\tan\sigma=L_{\rm AdS}\sinh(\lambda/L_{\rm AdS}),\label{jt:prelude-clock:b}\\
&u_*(w)=(\partial_u\lambda)|_{\Sigma_0}[u(w/2)-u(\lambda_c)]=L_{\rm AdS}\,\frac{\sinh[w/(2L_{\rm AdS})]-\sinh(\lambda_c/L_{\rm AdS})}{\cosh(\lambda_c/L_{\rm AdS})}.\label{jt:prelude-clock:c}
\end{align}
\end{subequations}

Because $(\partial_u\lambda)|_{\Sigma_0}$ in~\eqref{jt:prelude-clock:c} compensates any rescaling of $u$, the reflection about the midpoint of $\Sigma$ that exchanges the two null generators preserves the clock reading. We can therefore infer the width $w$ from a clock on either side, since its reading increases strictly from zero.

To compare the flux of matter at fixed clock reading, we take $u=0$ and $\partial_u\lambda=1$ on $\Sigma_0$, so $u=u_*$ at $S_R$. After eliminating the spin connection in~\eqref{jt:prelude-action}, variation with respect to $g^{ab}$ yields the null constraint along the right sided null generator~\cite{BakKimYi2023}. For initial data $\Phi(0),\partial_u\Phi(0)$, integrating this constraint determines the change relative to a solution with constant $\varphi$,
\begin{equation}
 \partial_u^2\Phi=-(\partial_u\varphi)^2,\ 
 \Delta\Phi(u_*)=-\int_0^{u_*}(u_*-u)(\partial_u\varphi)^2\,\dd u.\label{jt:prelude-null}
\end{equation}

\begin{figure}[!htb]
\centering
\begingroup
\definecolor{noteblue}{HTML}{246697}
\definecolor{noterust}{HTML}{A75630}
\definecolor{noteink}{HTML}{263944}
\definecolor{notegreen}{HTML}{367665}
\begin{tikzpicture}[font=\small,text=noteink,line cap=round,line join=round,
  every node/.style={inner sep=1.6pt},
  leg/.style={draw=noteblue,line width=1.05pt},
  oriented/.style={line width=.8pt,-{Stealth[length=4.5pt,width=3.8pt]}},
  boundary/.style={draw=noteink!75,line width=.65pt},
  external/.style={circle,draw=noteink,fill=white,line width=.8pt,inner sep=0pt,minimum size=4.8pt}]
  \foreach \x in {-2.85,2.85}{
    \draw[noteink!70,line width=1.0pt] (\x,-1.67)--(\x,1.67);
    \draw[noteink!55,densely dotted,line width=.72pt] (\x,-1.67)--(\x,-1.98)
      (\x,1.67)--(\x,1.98);
  }
  \node[below] at (-2.85,-2.00) {$\sigma=-\pi/2$};
  \node[below] at (2.85,-2.00) {$\sigma=\pi/2$};
  \path[fill=noteblue!8] (-1.53,0)--(-.30,-1.23)--(.30,-1.23)--(1.53,0)--cycle;
  \path[fill=notegreen!8] (-1.53,0)--(-.30,1.23)--(.30,1.23)--(1.53,0)--cycle;
  \draw[noteink!40,densely dashed,line width=.55pt] (-.30,-1.23)--(0,-1.53)--(.30,-1.23)
    (-.30,1.23)--(0,1.53)--(.30,1.23);
  \draw[leg] (-1.53,0)--(-.30,-1.23) (.30,-1.23)--(1.53,0);
  \draw[notegreen,line width=1.05pt] (-1.53,0)--(-.30,1.23) (.30,1.23)--(1.53,0);
  \draw[boundary,line width=.92pt] (-.30,-1.23)--(.30,-1.23) (-.30,1.23)--(.30,1.23);
  \node[anchor=east] at (-.46,-1.27) {$\Sigma_0$};
  \node[anchor=east] at (-.46,1.27) {$\Sigma_1$};
  \draw[noteink!73,densely dashed,line width=.74pt] (-1.53,0)--(1.53,0);
  \node[above] at (.78,.02) {$\Sigma$};
  \node[external,minimum size=4.6pt] at (-1.53,0) {};
  \node[external,minimum size=4.6pt] at (1.53,0) {};
  \node[anchor=east] at (-1.61,0) {$S_L$};
  \node[anchor=west] at (1.61,0) {$S_R$};
  \draw[noterust,line width=1.18pt] (.83,-.70)--(-.70,.83);
  \draw[oriented,noterust,line width=1.18pt] (.13,0)--(-.16,.29);
  \fill[noterust] (.83,-.70) circle (2.2pt);
  \node[anchor=west,text=noterust] at (.97,-.76) {$u_0$};
  \node[anchor=west,text=noteblue,font=\footnotesize] at (.39,-1.33) {$u=0$};
\end{tikzpicture}
\endgroup
\caption{Penrose diagram of the finite causal diamond between the timelike boundaries $\sigma=\pm\pi/2$ of global $\mathrm{AdS}_2$. The spacelike geodesic $\Sigma$ of proper length $w$ joins the two corners. For the matter shock in~\eqref{jt:prelude-null} we set $u=0$ on $\Sigma_0$ and $u=u_*$ at $S_R$. The shock crosses the right sided incoming null generator at $u_0$ and lowers $\Phi(S_R)$ at fixed $u_*$.}
\label{fig:jt-matter-shock}
\end{figure}
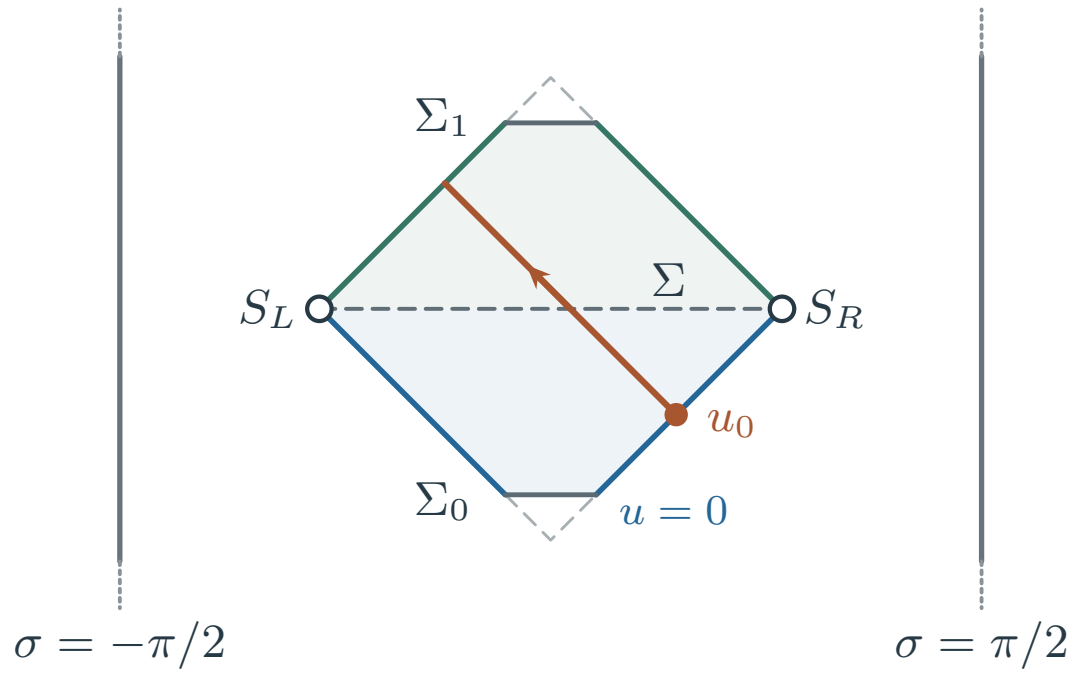

The two solutions can therefore have the same affine parameter $u_*$ at $S_R$ and different values of $\Phi(S_R)$ (Figure~\ref{fig:jt-matter-shock}). Even in vacuum we can change the corner values of the dilaton by rescaling $\Phi$ on the same spacetime.

In order to then quantize the field $\varphi$ whose flux changes the corner dilaton in~\eqref{jt:prelude-null}, we label cuts on each null generator by $v=2\lambda/w$, from $v=\epsilon$ at the ends of $\Sigma_0$ to $v=1$ at the corner. The wave equation $\partial_+\partial_-\varphi=0$ determines the field in the causal diamond from its two profiles $\varphi_i(v)$ and the field $\varphi_0(x)$ and momentum density $\pi_\varphi(x)$ in a fixed coordinate $x$ on $\Sigma_0$. To find their brackets and the momentum conjugate to the clock, we integrate the bulk symplectic potential of~\eqref{jt:prelude-action},
\begin{align}
&\bm\theta_{\mathscr M}=Y_\mu\,\delta e^\mu+\Phi\,\delta\omega^{01}-(\star\dd\varphi)\,\delta\varphi.\label{jt:prelude-potential-current}
\end{align}
Integrating from $S_L$ to $S_R$, we traverse the left sided null generator toward $\Sigma_0$ and the right sided null generator away from it. The pullback of $\star\dd\varphi$ changes sign between them, so the two scalar contributions add, while matching the scalar field and the local Lorentz frame cancels the terms at the two ends of $\Sigma_0$. Furthermore, a variation of $u_*$ displaces the null generators and $\Sigma_0$ together in $(\tau,\sigma)$, with conjugate momentum $p_{u_*}$. Therefore, including the rapidity $\Gamma_i$ of the local Lorentz frame at each corner yields, up to an exact variation,
\begin{align}
&\Theta=p_{u_*}\,\delta u_*+\sum_{i=L,R}\int_\epsilon^1(\partial_v\varphi_i)\,\delta\varphi_i\,\dd v
+\int_{\Sigma_0}\pi_\varphi\,\delta\varphi_0\,\dd x+\sum_{i=L,R}\mathscr K_i\,\delta\Gamma_i.\label{jt:prelude-reduced}
\end{align}
We note that in the symplectic form $\Omega=\delta\Theta$, the corner dilaton $\mathscr K_i=\Phi(S_i)$ generates boosts at $S_i$.

Now, with the scalar field and both corner charges fixed, the dilaton on $\Sigma_0$ can still vary. For two such solutions on the same spacetime the matter stress tensor cancels when we subtract the equations obtained by varying $g^{ab}$. Eliminating $\nabla^2\Delta\Phi$ with the trace then yields
\begin{equation}
 \nabla_a\nabla_b\Delta\Phi=\frac{g_{ab}\Delta\Phi}{L_{\rm AdS}^2}.\label{jt:dilaton-homogeneous-equation}
\end{equation}
Of its three independent solutions $\cos\tau/\cos\sigma$, $\sin\tau/\cos\sigma$ and $\tan\sigma$, only $\sin\tau/\cos\sigma$ vanishes at both corners~\cite{BakKimYi2023}. Hence, with the tetrad $e^0=L_{\rm AdS}\sec\sigma\,\dd\tau$ and $e^1=L_{\rm AdS}\sec\sigma\,\dd\sigma$, varying $\omega^{01}$ determines $\Delta Y_\mu$ from $\Delta\Phi$. The resulting constraint and its substitution in~\eqref{jt:prelude-potential-current} at fixed $v,x$ yield
\begin{subequations}
\label{jt:prelude-null-momentum}
\begin{align}
&\dd\Phi+Y_0e^1+Y_1e^0=0,\label{jt:prelude-null-momentum:a}\\
&\Delta\Phi=c\,\frac{\sin\tau}{\cos\sigma},\label{jt:prelude-null-momentum:b}\\
&\Delta\bm\theta_{\mathscr M}|_{\mathscr H_p\cup\Sigma_0}=-c\,\dd\!\left(\frac{\cos\tau}{\cos\sigma}\,\delta\sigma\right)\Big|_{\mathscr H_p\cup\Sigma_0},\label{jt:prelude-null-momentum:c}\\
&\Delta p_{u_*}\,\delta u_*=-2c\sec\sigma_S\,\delta\sigma_S=-\frac{c}{L_{\rm AdS}}\,\delta w,\label{jt:prelude-null-momentum:d}
\end{align}
\end{subequations}
where the boundary displacements $\delta\sigma=\pm\delta\sigma_S$ add, with $\delta w=2L_{\rm AdS}\sec\sigma_S\,\delta\sigma_S$ and, since $c$ is arbitrary, $p_{u_*}$ ranges over the real line at each $u_*>0$.

\subsection{Quantization and gluing}
\label{jt:prelude-quantum}

The dilaton mode in~\eqref{jt:prelude-null-momentum} allows us to vary the momentum conjugate to $u_*$ even with the matter field and both corner charges fixed. Once we quantize this pair, two wavefunctions with the same modulus yield the same distribution of clock readings, yet we can make their overlap vanish by choosing how their relative phase varies with $u_*$.

To see this, note that at fixed corner charges the terms for $\varphi$ in~\eqref{jt:prelude-reduced} are independent of $u_*$ and $p_{u_*}$, so we can quantize the clock and the matter field separately. For smooth $f$ supported away from the ends of null generator $i$, integration by parts in $\Omega$ shows that the smeared current $j_i(f)$ shifts $\varphi_i$ by $f/2$. With $u_*$ represented on $L^2(\mathbb R_+,\dd u_*)$ and a boost of rapidity $\eta$ at $S_i$ generated by $\Phi(S_i)$, we obtain
\begin{subequations}
\label{jt:prelude-operators}
\begin{align}
&j_i(f)=\int_\epsilon^1 f\,\partial_v\varphi_i\,\dd v,\label{jt:prelude-operators:a}\\
&[\widehat j_i(f),\widehat j_j(g)]=\frac{i\hbar}{2}\delta_{ij}\int_\epsilon^1 f(v)\,\partial_vg(v)\,\dd v\,\widehat 1,\label{jt:prelude-current-algebra}\\
&(\widehat{u}_*\psi)(u_*)=u_*\psi(u_*),\label{jt:prelude-operators:c}\\
&\big(e^{i\eta\widehat{\mathscr K}_i/\hbar}\psi\big)(u_*)=e^{i\eta\Phi(S_i)/\hbar}\psi(u_*).\label{jt:prelude-operators:d}
\end{align}
\end{subequations}

For the scalar field we choose a pure Gaussian state $|\Omega_\varphi\rangle$ of canonical pairs $(q_j,p_j)$ for the two profiles $\varphi_i$ on the null generators and the fields $\varphi_0,\pi_\varphi$ on $\Sigma_0$. The scalar field has a single constant mode on $\mathscr H_p\cup\Sigma_0$, and its conjugate momentum follows from~\eqref{jt:prelude-reduced},
\begin{equation}
 p_1=\int_{\Sigma_0}\pi_\varphi\,\dd x+\sum_{i=L,R}[\varphi_i(1)-\varphi_i(\epsilon)].\label{jt:constant-field-momentum}
\end{equation}
 Averaging $\varphi_0$ with a smooth weight that integrates to unity defines $q_1$, for which the Gaussian wavefunction lies in $L^2(\mathbb R,\dd q_1)$. For centers $q_j^0,p_j^0$ and widths $\Delta q_j>0$, the operators $\mathfrak a_j$ annihilate $|\Omega_\varphi\rangle$ and satisfy $[\mathfrak a_j,\mathfrak a_k^\dagger]=\delta_{jk}\widehat 1$, hence their adjoints generate the occupation number basis,
\begin{subequations}
\label{jt:prelude-scalar-fock}
\begin{align}
&\mathfrak a_j=\frac{\widehat q_j-q_j^0}{2\Delta q_j}
+\frac{i\Delta q_j}{\hbar}(\widehat p_j-p_j^0),\ j=1,2,\ldots,\label{jt:prelude-ladder}\\
&|\boldsymbol n\rangle=\prod_{j:\,n_j>0}\frac{(\mathfrak a_j^\dagger)^{n_j}}{\sqrt{n_j!}}\,|\Omega_\varphi\rangle,\label{jt:prelude-occupations}\\
&\mathcal H_\varphi\simeq\bigoplus_{m=0}^{\infty}\operatorname{Sym}^{m}\ell^2(\mathbb N).\label{jt:prelude-scalar-fock:c}
\end{align}
\end{subequations}
The vectors with finite total occupation $m=\sum_j n_j$ are orthonormal and span a dense subspace of $\mathcal H_\varphi$.

\begin{figure}[!htb]
\centering
\begingroup
\definecolor{jtMatter}{HTML}{246697}
\definecolor{jtClock}{HTML}{A75630}
\definecolor{jtInk}{HTML}{263944}
\normalsize\color{jtInk}
\tikzset{jt matter/.style={draw=jtMatter,line width=.9pt,line cap=round},
  jt clock/.style={draw=jtClock,line width=.9pt,line cap=round}}

\newcommand{\jtPairing}[2]{%
  \begin{tikzpicture}[baseline=-.52ex,every node/.style={inner sep=0pt}]
    \coordinate (braJ) at (-.64,.21);
    \coordinate (braK) at (-.64,-.21);
    \coordinate (ketJ) at (.64,.21);
    \coordinate (ketK) at (.64,-.21);
    \node[anchor=east,xshift=-2.7pt] at (braJ) {$j'$};
    \node[anchor=east,xshift=-2.7pt] at (braK) {$k'$};
    \node[anchor=west,xshift=2.7pt] at (ketJ) {$j$};
    \node[anchor=west,xshift=2.7pt] at (ketK) {$k$};
    #1
    \node[anchor=north] at (0,-.43) {$#2$};
  \end{tikzpicture}}
\newcommand{\jtClockLine}{u_*'\,\tikz[baseline=-.52ex]{%
    \draw[jt clock] (0,0)--(1.03,0);
    \fill[jtClock] (.515,0) circle[radius=1.55pt];
    \node[above,text=jtClock,inner sep=1pt] at (.515,.055) {$P_{\mathscr D}$};}\,u_*}
\begin{tabular}{@{}l@{\hspace{.58em}}l@{}}
(a) &
  $j\,\tikz[baseline=-.52ex]{%
    \draw[jt matter] (0,0)--(.55,0);
    \draw[jtInk,line width=.85pt] (.55,-.14)--(.55,.14);}
  =\mathfrak a_j^\dagger|\Omega_\varphi\rangle,\ 
  j'\,\tikz[baseline=-.52ex]{\draw[jt matter] (0,0)--(.69,0);}\,j
  =\delta_{j'j},\ 
  \jtClockLine=\mathbf1_{[0,\ell(\mathscr D)]}(u_*)\,\delta(u_*'-u_*)$\\[9pt]
(b) &
  $\langle u_*',\Omega_\varphi|\mathfrak a_{j'}\mathfrak a_{k'}\,
    P_{\mathscr D}\,\mathfrak a_j^\dagger\mathfrak a_k^\dagger
    |u_*,\Omega_\varphi\rangle
  =\jtClockLine\,\Biggl(\jtPairing{\draw[jt matter] (braJ)--(ketJ) (braK)--(ketK);}
              {\delta_{j'j}\delta_{k'k}}
   +\jtPairing{\draw[jt matter] (braJ)--(ketK);
               \draw[jt matter,preaction={draw=white,line width=2.8pt}] (braK)--(ketJ);}
              {\delta_{j'k}\delta_{k'j}}\Biggr)$
\end{tabular}
\endgroup

\caption[Contractions in the Hilbert space of JT gravity]{(a) An open line adds one excitation to the Gaussian state $|\Omega_\varphi\rangle$. Joining it to a line in the adjoint state identifies their mode indices. At fixed corner charges the clock line identifies the two readings, and $P_{\mathscr D}$ restricts their common value~\eqref{jt:prelude-projection}. (b) For a product of a clock state and the Gaussian matter state, the clock kernel in (a) multiplies both matter field contractions. The $n_j!$ Wick contractions in mode $j$ account for the normalization $(n_j!)^{-1/2}$ in~\eqref{jt:prelude-occupations}.}
\label{fig:jt-fock-rules}
\end{figure}
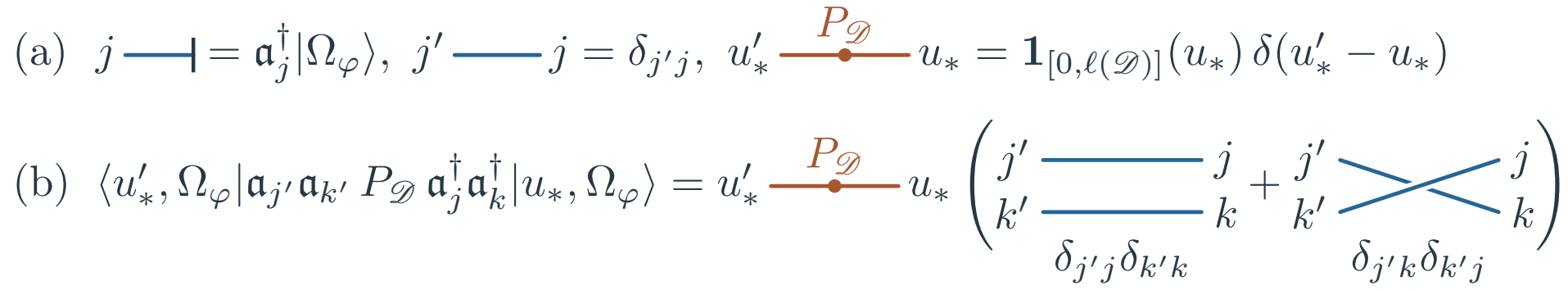

Having quantized the clock and the matter field on $\mathscr H_p\cup\Sigma_0$, we can prepare a state on the geodesic $\Sigma$ joining the corners. At fixed corner charges the coefficient of each occupation number state of $\varphi$ is a wavefunction of $u_*$. Bounding the clock reading by $\ell(\mathscr D)$ restricts each of these wavefunctions to the same interval,
\begin{subequations}
\label{jt:prelude-clock-spaces}
\begin{align}
&\mathcal H_\Sigma=L^2(\mathbb R_+,\dd u_*)\otimes\mathcal H_\varphi,\label{jt:prelude-clock-spaces:a}\\
&(P_{\mathscr D}\psi)(u_*)=\mathbf1_{[0,\ell(\mathscr D)]}(u_*)\,\psi(u_*),\label{jt:prelude-projection}\\
&\mathcal H_{\mathscr D}=P_{\mathscr D}\mathcal H_\Sigma
\simeq L^2([0,\ell(\mathscr D)],\dd u_*)\otimes\mathcal H_\varphi.\label{jt:prelude-projected-space}
\end{align}
\end{subequations}
We take the inner product by pairing the matter excitations in the bra and ket and integrating over their common clock reading (Figure~\ref{fig:jt-fock-rules}).

\begin{figure}[!htb]
\centering
\begingroup
\definecolor{jtBlue}{HTML}{246697}
\definecolor{jtGreen}{HTML}{367665}
\definecolor{jtRust}{HTML}{A75630}
\definecolor{jtInk}{HTML}{263944}
\begin{tikzpicture}[font=\small,text=jtInk,line cap=round,line join=round,
  every node/.style={inner sep=1.5pt},
  sew/.style={jtInk,line width=.85pt,-{Stealth[length=4.5pt,width=3.8pt]}},
  rim/.style={jtRust,line width=1.05pt},
  dot/.style={circle,inner sep=0pt,minimum size=3.8pt,draw=jtRust,fill=white,line width=.7pt}]
\newcommand{\jtHalf}[2]{%
  \fill[#1!12] (-1.12,0)--(-.30,{#2*.82})--(.30,{#2*.82})--(1.12,0)--cycle;
  \draw[#1,line width=.88pt] (-1.12,0)--(-.30,{#2*.82}) (.30,{#2*.82})--(1.12,0);
  \draw[jtInk!70,line width=.72pt] (-.30,{#2*.82})--(.30,{#2*.82});}
\node[anchor=east] at (-1.37,1.24) {(a)};
\begin{scope}[shift={(0,1.24cm)}]
  \begin{scope}[yshift=-.23cm]
    \jtHalf{jtBlue}{-1}
    \draw[rim] (-1.12,0)--(1.12,0);
    \draw[sew,jtRust] (-.28,0)--(.26,0);
    \node[text=jtBlue] at (0,-.37) {$|b\rangle$};
  \end{scope}
  \begin{scope}[yshift=.23cm]
    \jtHalf{jtGreen}{1}
    \draw[rim] (1.12,0)--(-1.12,0);
    \draw[sew,jtRust] (.28,0)--(-.26,0);
    \node[text=jtGreen] at (0,.37) {$\langle a|$};
  \end{scope}
  \foreach \side in {-1.12,1.12}{\draw[jtRust!70,densely dotted] (\side,-.23)--(\side,.23);}
\end{scope}
\draw[sew] (1.85,1.24)--(4.57,1.24);
\node[above,inner ysep=3pt] at (3.21,1.24)
  {$\displaystyle \sum_{\boldsymbol n}\int_0^{\ell(\mathscr D)}\!\dd u_*$};
\begin{scope}[shift={(6.44cm,1.24cm)}]
  \jtHalf{jtBlue}{-1}\jtHalf{jtGreen}{1}
  \draw[rim,densely dashed] (-1.12,0)--(1.12,0);
  \node[above,text=jtRust] at (0,.03) {$\Sigma$};
  \node[anchor=north] at (0,-.91) {$\langle a|P_{\mathscr D}|b\rangle$};
\end{scope}
\node[anchor=east] at (-1.37,-.64) {(b)};
\begin{scope}[yshift=-.64cm]
  \draw[jtBlue,line width=1.12pt] (-1.23,0)--(-.19,0);
  \draw[jtGreen,line width=1.12pt] (.19,0)--(1.23,0);
  \draw[sew,jtBlue] (-.89,0)--(-.51,0);
  \draw[sew,jtGreen] (.89,0)--(.51,0);
  \draw[jtRust!65,densely dotted] (-.19,0)--(.19,0);
  \node[dot] at (-.19,0) {}; \node[dot] at (.19,0) {};
  \node[anchor=south east] at (-.19,.07) {$\Gamma_L$};
  \node[anchor=south west] at (.19,.07) {$\Gamma_R$};
  \draw[sew] (1.85,0)--(4.57,0);
  \node[above,inner ysep=3pt] at (3.21,0) {$\Phi_L(S)=\Phi_R(S)$};
  \begin{scope}[xshift=6.44cm]
    \draw[jtBlue,line width=1.12pt] (-1.12,0)--(0,0);
    \draw[jtGreen,line width=1.12pt] (0,0)--(1.12,0);
    \fill[jtRust] (0,0) circle[radius=1.7pt];
    \node[above] at (0,.07) {$S$};
    \node[below] at (0,-.12) {$\Phi(S)\,\delta(\Gamma_R-\Gamma_L)$};
  \end{scope}
\end{scope}
\end{tikzpicture}
\endgroup
\caption[Gluing on an interval and at a corner]{(a) On $\Sigma$ we multiply the wavefunction of $|b\rangle$ by the conjugate wavefunction of $|a\rangle$ at the same clock reading and matter field occupation numbers. We sum over these mode occupations and integrate $u_*$ over the range in~\eqref{jt:prelude-projection}. For $\Psi_n$ and $\Psi_m$ the overlap of matter states is one and the clock integral equals $\delta_{mn}$. (b) Joining neighboring geodesic intervals identifies their dilatons at $S$. Since the two copies of $S$ have opposite orientations, the corner term in~\eqref{jt:prelude-reduced} depends only on the relative boost.}
\label{fig:jt-interval-sewing}
\end{figure}
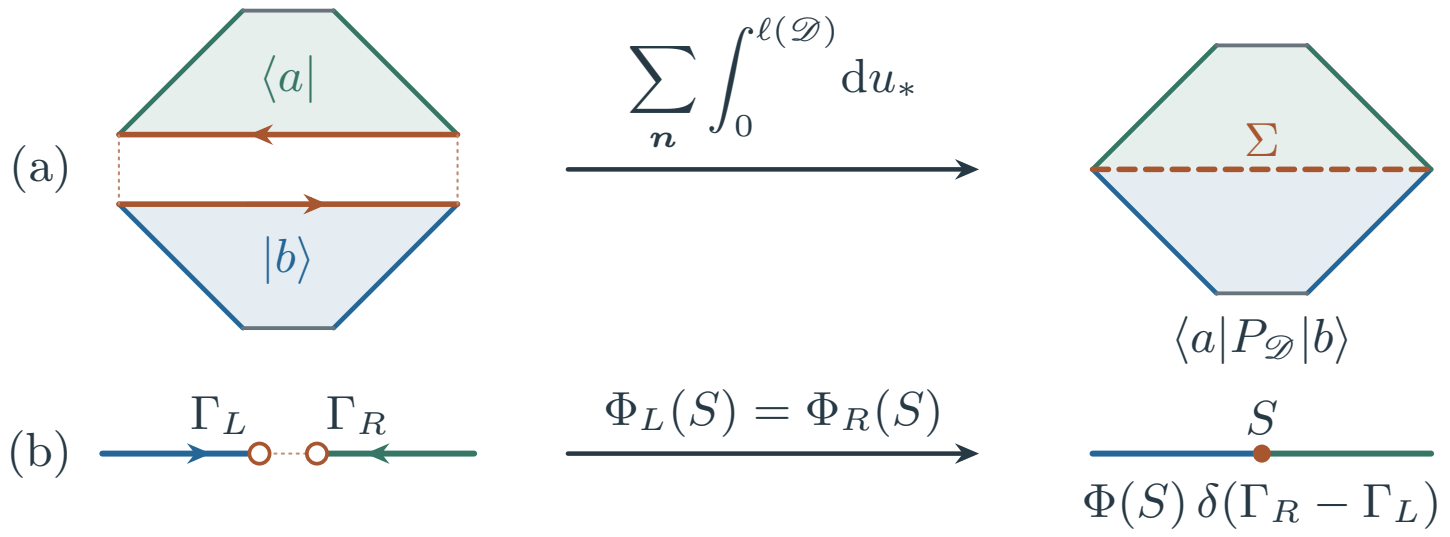

In Section~\ref{bfq:diamond-state-existence} we construct $\mathcal H_{\mathscr D}$ in four dimensions from a Gaussian state centered on a spherical null cone by projecting onto the allowed clock readings before and after acting with boundary observables~\eqref{bfq:diamond-cyclic-packet}. In JT gravity the projector acts on the clock alone, so we can hold the field in a normalized state $|\xi\rangle\in\mathcal H_\varphi$ and vary the phase of a normalized smooth clock wavefunction $h(u_*)$ supported in $0<u_*<\ell(\mathscr D)$. The states $\Psi_n$, $n\in\mathbb Z$, all have the same cumulative probability $F$ for the clock reading. We evaluate their overlap by gluing on $\Sigma$ (Figure~\ref{fig:jt-interval-sewing}a) and using the variable $s=F(u_*)$,
\begin{subequations}\label{jt:clock-state-overlap}
\begin{align}
&\Psi_n(u_*)=h(u_*)e^{2\pi i n F(u_*)}|\xi\rangle,\label{jt:clock-state-overlap:a}\\
&F(u_*)=\int_0^{u_*}|h(v)|^2\,\dd v,\label{jt:clock-state-overlap:b}\\*
&\langle\Psi_m|\Psi_n\rangle=\int_0^1 e^{2\pi i(n-m)s}\,\dd s=\delta_{mn}.\label{jt:clock-state-overlap:c}
\end{align}
\end{subequations}

To distinguish these states we can measure the generator of dilations of $u_*$, which differentiates the phase $e^{2\pi i nF(u_*)}$. Its symmetrically ordered form is $\mathcal Q(u_*p_{u_*})=-i\hbar(u_*\partial_{u_*}+1/2)$, where $\mathcal Q$ denotes Weyl quantization (Appendix~\ref{bfq:canonical-representation}). Since $F'(u_*)=|h(u_*)|^2$, we find
\begin{align}
&\langle\Psi_n|\mathcal Q(u_*\,p_{u_*})|\Psi_n\rangle-\langle\Psi_0|\mathcal Q(u_*\,p_{u_*})|\Psi_0\rangle
=2\pi n\hbar\int_0^\infty u_*|h(u_*)|^4\,\dd u_*.
\label{jt:prelude-momentum}
\end{align}
For any one choice of $|\xi\rangle$, the states $\Psi_n$ thus span an infinite dimensional subspace of $\mathcal H_{\mathscr D}$.

Having projected the states of a single interval onto the allowed clock readings, we can now join two such intervals to prepare a state on a larger spatial slice. On the time symmetric slice of the eternal black hole in pure JT gravity~\cite{HarlowJafferis2020}, we cut the geodesic at the bifurcation surface $S$. When we match the dilaton there, the opposite orientations make the two corner terms in~\eqref{jt:prelude-reduced} sum to $\Phi(S)\,\delta(\Gamma_R-\Gamma_L)$ (Figure~\ref{fig:jt-interval-sewing}b), so a common boost has zero charge, though of course the sum of the two proper lengths can still fluctuate in the joint state. In four dimensions, matching the area elements at $S$ likewise cancels the charge associated to a common boost~\eqref{bfgeo:glued-potential-main}, while a collar of fixed proper width can enclose different volumes (Figure~\ref{fig:prologue-space-time}).

In a similar vein, to extend a preparation $|b\rangle$ in time at fixed spatial topology, we glue it to a cobordism $\mathscr M:\Sigma_0\to\Sigma_1$ by summing over a basis on $\Sigma_0$. Partitioning the possible readings of the outgoing clock into intervals $I_\alpha$, we obtain
\begin{align}
&|\psi\rangle=U_{\mathscr M}|b\rangle
=\sum_\alpha P_\alpha U_{\mathscr M}|b\rangle,\
P_\alpha=\mathbf1_{I_\alpha}(\widehat{u}_*) .
\label{prologue:time-extension}
\end{align}
The state on $\Sigma_1$ can therefore be a coherent superposition of different elapsed times. To distinguish outgoing states with the same probabilities for $I_\alpha$, we combine bounded functions of $\widehat u_*$ and dilations with smearings of $\varphi_i$ on the null generators and of $\varphi_0,\pi_\varphi$ on $\Sigma_0$. At fixed $\Phi(S_L),\Phi(S_R)$ these observables generate all rank one operators under weak closure (Appendix~\ref{bfq:fixed-area-source-algebra}). Restricting both external states to $\mathcal H_{\mathscr D}$ then yields $\mathcal B(\mathcal H_{\mathscr D})$, of type~$\mathrm I_\infty$ since the $\Psi_n$ form an infinite orthonormal family. Intuitively, this is because we prepare a joint state $|u\rangle$ of the clock and the matter field on $\mathscr H_p\cup\Sigma_0$ before specifying an exterior, such that gluing its adjoint to $\Sigma_1$ yields the probability $|\langle u|U_{\mathscr M}|b\rangle|^2$ for normalized states.

\begin{figure}[!htb]
\centering
\begingroup
\definecolor{jtBlue}{HTML}{246697}
\definecolor{jtGreen}{HTML}{367665}
\definecolor{jtRust}{HTML}{A75630}
\definecolor{jtInk}{HTML}{263944}
\begin{tikzpicture}[font=\small,text=jtInk,line cap=round,line join=round,
  every node/.style={inner sep=1.5pt},
  sew/.style={jtInk,line width=.85pt,-{Stealth[length=4.5pt,width=3.8pt]}}]
\foreach \side/\shade in {-1/jtBlue,1/jtGreen}{
  \fill[\shade!13] (-.39,0)--(-1.23,{.84*\side})--(1.23,{.84*\side})--(.39,0)--cycle;
  \draw[\shade,line width=.95pt] (-.39,0)--(-1.23,{.84*\side}) (.39,0)--(1.23,{.84*\side});
  \draw[jtInk!72,line width=.80pt] (-1.23,{.84*\side})--(1.23,{.84*\side});
  \foreach \edge in {-1,1}{
    \begin{scope}[shift={({1.23*\edge},{.84*\side})},xscale=\edge,yscale=\side]
      \fill[jtRust] (0,0) circle[radius=1.65pt];
      \draw[jtRust,line width=.84pt] (-.12,0) arc[start angle=180,end angle=360,radius=.12cm];
    \end{scope}
  }
}
\draw[jtRust,densely dashed,line width=.92pt] (-.39,0)--(.39,0);
\node[anchor=east,text=jtRust] at (-.50,0) {$\Sigma_1$};
\node[text=jtBlue] at (0,-.49) {$C_{\mathscr D}$};
\node[text=jtGreen] at (0,.49) {$C_{\mathscr D}^{\dagger}$};
\node[below] at (0,-.92) {$|i\rangle$};
\node[above] at (0,.92) {$\langle j|$};
\node[anchor=north,text=jtRust,font=\footnotesize] at (-1.23,-.97) {$e^{-\pi\mathscr K_L/\hbar}$};
\node[anchor=north,text=jtRust,font=\footnotesize] at (1.23,-.97) {$e^{-\pi\mathscr K_R/\hbar}$};
\draw[sew] (1.92,0)--(4.06,0);
\node[above,inner ysep=3pt] at (2.99,0) {$\sum_i$};
\node[below,inner ysep=3pt] at (2.99,0) {$j=i$};
\begin{scope}[xshift=5.91cm]
  \foreach \angle/\shade in {180/jtBlue,0/jtGreen}{
    \path[fill=\shade!13] (\angle:1.05cm) arc[start angle=\angle,delta angle=180,radius=1.05cm]
      --({\angle+180}:.36cm) arc[start angle={\angle+180},delta angle=-180,radius=.36cm]--cycle;
    \draw[\shade,line width=.96pt] (\angle:1.05cm) arc[start angle=\angle,delta angle=180,radius=1.05cm];
    \draw[\shade,line width=.80pt] (\angle:.36cm) arc[start angle=\angle,delta angle=180,radius=.36cm];
  }
  \draw[jtRust,densely dashed,line width=.88pt] (-1.05,0)--(-.36,0);
  \draw[jtInk,densely dashed,line width=.85pt] (.36,0)--(1.05,0);
  \foreach \x in {.36,1.05}{\fill[jtRust] (\x,0) circle[radius=1.65pt];}
  \node[above] at (.70,.05) {$i$};
  \node at (0,0) {$S_L$};
  \node[anchor=west] at (1.12,0) {$S_R$};
  \node[anchor=north] at (0,-1.12) {$\Tr_{\mathcal H_{\mathscr D}}W$};
\end{scope}
\end{tikzpicture}
\endgroup
\caption[Gluing the JT gravity preparation to its time reflection]{Gluing the preparation to its time reflection sums over a complete basis on $\Sigma_1$, so the time evolution cancels by unitarity. On each copy the orange half turns multiply the state by $e^{-\pi\mathscr K_L/\hbar}$ at $S_L$ and $e^{-\pi\mathscr K_R/\hbar}$ at $S_R$. Identifying the two remaining intervals sets $j=i$ on the radial cut of the annulus, so the trace sums over $i$. Each orthogonal $\Psi_n$ enters with the same weight in~\eqref{jt:prelude-trace:a}.}
\label{fig:jt-reflected-trace}
\end{figure}
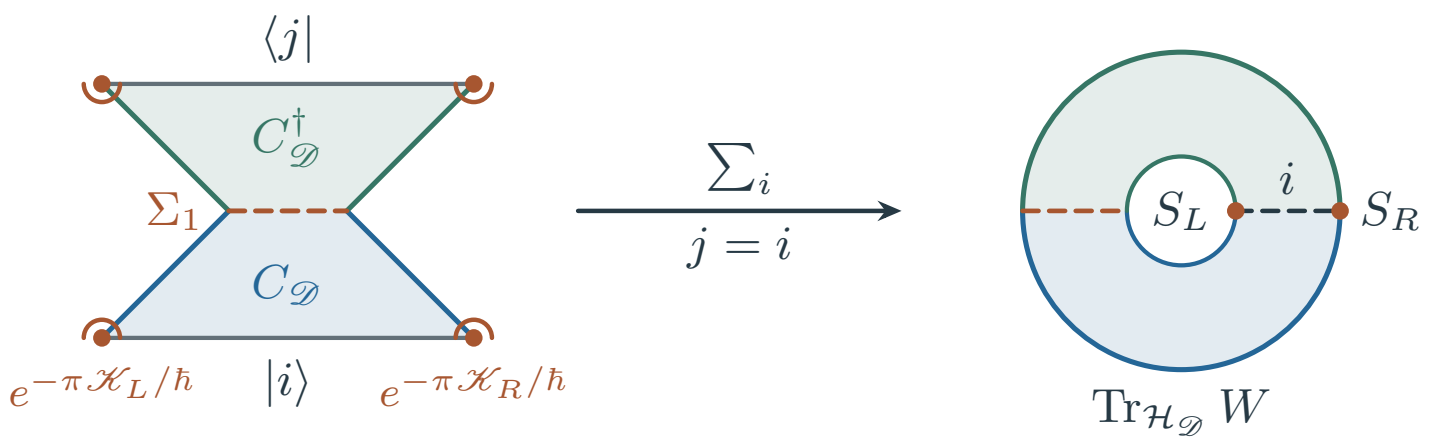

To compute the contribution of each $\Psi_n$ to the trace, we prepare the future half of $\mathscr D$. Analytically continuing the boost rapidity $\eta$ to $i\pi$ in~\eqref{jt:prelude-operators:d} multiplies $\Psi_n$ by $\exp[-\pi\Phi(S_i)/\hbar]$ at each $S_i$. After time evolution $U_{\mathscr M}$ to $\Sigma_1$, we glue this preparation $C_{\mathscr D}$ to its time reflection, summing over all states of the clock and $\varphi$. Its unnormalized density operator $W=C_{\mathscr D}^\dagger C_{\mathscr D}$ (Figure~\ref{fig:jt-reflected-trace}) then satisfies
\begin{subequations}\label{jt:prelude-trace}
\begin{align}
&\langle\Psi_m|W|\Psi_n\rangle=e^{-2\pi(\mathscr K_L+\mathscr K_R)/\hbar}\langle\Psi_m|U_{\mathscr M}^\dagger U_{\mathscr M}|\Psi_n\rangle=e^{-2\pi(\mathscr K_L+\mathscr K_R)/\hbar}\delta_{mn},\label{jt:prelude-trace:a}\\
&\Tr_{\mathcal H_{\mathscr D}}W\geq\sum_{n=1}^N\langle\Psi_n|W|\Psi_n\rangle=N e^{-2\pi(\mathscr K_L+\mathscr K_R)/\hbar}\label{jt:prelude-trace:b}
\end{align}
\end{subequations}
for arbitrary $N$. Thus $\Tr W=\infty$ and $W/\Tr W$ cannot define a density matrix.

Fundamentally, every occupation number state of $\varphi$ in JT gravity can have the same clock wavefunction, since $u_*$ depends only on the geodesic length~\eqref{jt:prelude-clock:c},~\eqref{jt:prelude-projection}. But this is not true in four dimensions, where, with the initial area element and expansion fixed, shear increases the clock reading at a cut of a given area element (Figure~\ref{fig:outgoing-clock-readings}), so projecting onto $\bar u_*\le\ell(\mathscr D)$ also restricts the radiative states at fixed corner area~\eqref{bfq:fixed-area-fock:c},~\eqref{bfq:clock-selected-states:b}.

Additionally, gluing across a time reflection weights a state by the two corner dilatons in JT gravity~\eqref{jt:prelude-trace:a} and by the corner area $\mathscr A$ in four dimensions~\eqref{bfent:full-gluing-main}. But it is only in four dimensions that the null constraints also bound $\mathscr A$ for a given $\ell(\mathscr D)$ and thereby constrain the relative boost generator $\mathscr A/(8\pi G_N)$. In Section~\ref{bfent:subregion-boost-main} we use the change in the graviton two point function across $S$ under a relative boost to show that a causal diamond of bounded size admits only finitely many independent states.

\FloatBarrier


\section{Reduced phase space of intersecting null cones}
\label{sec:bf-action}

In JT gravity the matter shock lowered the dilaton at the corner without changing the clock reading, since at fixed proper width the null generator followed the same geodesic. We could thus prepare any state of the matter field with the same clock wavefunction. In four dimensions the radiation we wish to quantize changes the transverse metric on the null cone itself.

As gravitational radiation shears neighboring null generators, their transverse separations grow at different rates in different directions. This change of shape affects the clock because the square of the shear reduces the expansion through the Raychaudhuri equation, even in vacuum~\cite{Reisenberger12113880}. With the same initial area element and expansion at $\partial\Sigma_0$, the null generators therefore enclose less area at a given affine parameter than they would without shear. Reaching the same area element at $S$ then takes a larger clock reading, so a bound on the average reading restricts the radiation on the null cone.

\begin{figure}[!htbp]
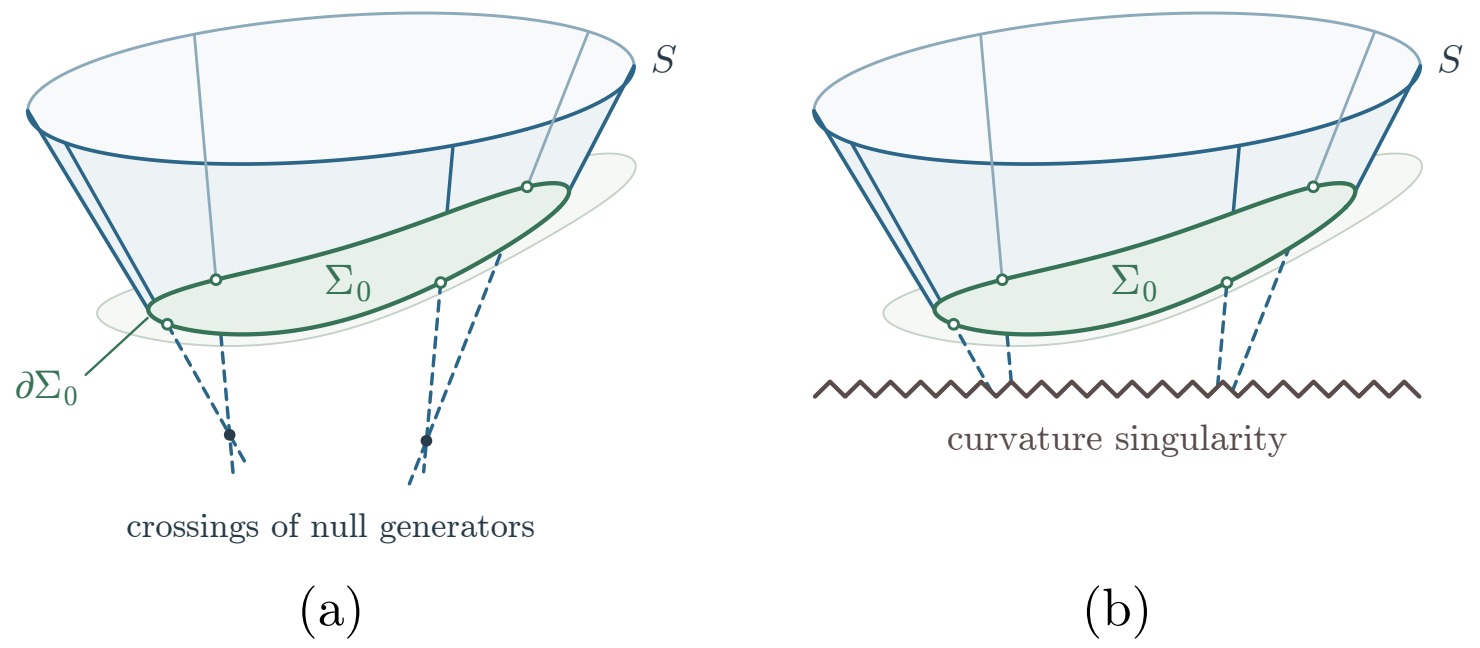

\centering
\begin{tabular}{@{}c@{\hspace{.015\linewidth}}c@{}}
\includegraphics[viewport=130 21 301 152,clip,width=.30\linewidth]{figures/causal-diamond-cap-and-singularity-panels.pdf} &
\includegraphics[viewport=304 21 475 152,clip,width=.30\linewidth]{figures/causal-diamond-cap-and-singularity-panels.pdf} \\
(a) & (b)
\end{tabular}
\caption{A smooth spacelike slice meets the incoming null generators at different affine parameters measured from $S$. The resulting cut $\partial\Sigma_0$ bounds the past cap $\Sigma_0$, shown in green. The dashed curves follow the null generators to the past of this cut, where they cross in pairs at distinct regular points in (a) and terminate at a spacelike curvature singularity in (b).}
\label{fig:bf-capped-null-generators}
\end{figure}

The same focusing can bring neighboring null generators to a caustic before they reach $S$. In solving the null constraints at finite $G_N$, we must then determine how much of the spacetime can be recovered from one null cone and compatible data on its spacelike cap~\cite{ChoquetBruhat10064467}. Where the spacetime itself remains smooth, we can place a spacelike cap before the null generators focus (Figure~\ref{fig:bf-capped-null-generators}) and choose a second causal diamond whose null boundaries are regular near the caustic~\cite{Penrose1972Topology}. Matching tetrads and spin connections on the overlap then allows us to continue the same solution using the characteristic data on the second causal diamond (Figure~\ref{fig:bf-local-reconstruction}). Since every point of a smooth globally hyperbolic vacuum spacetime admits such a neighborhood, we can choose a cover for each solution and recover the full reduced phase space $\mathcal P(\mathscr M)$ by gluing overlapping causal diamonds. In~\eqref{bfgeo:patched-phase-space} we identify different covers of the same solution while holding fixed the data on the boundary of $\mathscr M$.

\begin{figure}[t]
\centering
\includegraphics[viewport=2 21 126 163,clip,width=.30\linewidth]{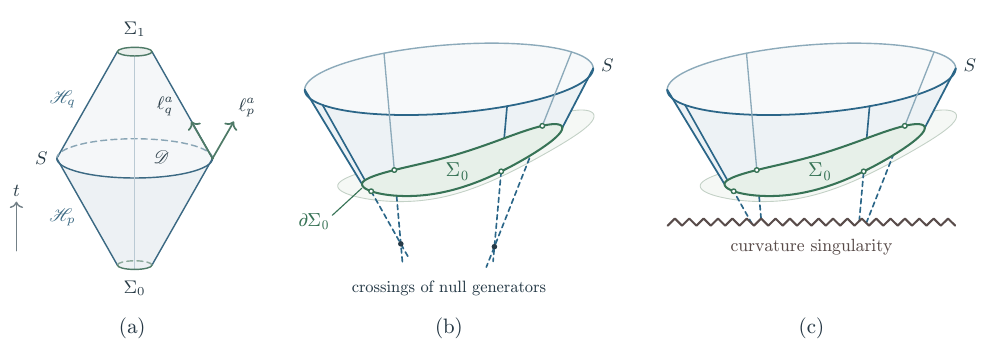}
\caption{A causal diamond $\mathscr D$. Small spacelike caps $\Sigma_0$ and $\Sigma_1$ replace the vertices and make the region a cobordism $\mathscr D:\Sigma_0\to\Sigma_1$. The null cones $\mathscr H_p$ and $\mathscr H_q$ extend from their respective caps and intersect at the corner $S$. The fields $B^{\mu\nu},\omega^{\mu\nu}$ and the boundary frames determine the tetrad through the simplicity and null constraints. The arrows $\ell_p^a$ and $\ell_q^a$ are the future directed null generators of these boundaries at $S$, normalized by the choice of affine parameter on $\mathscr H_p$ and $\mathscr H_q$, respectively. Time $t$ increases upwards. Dashed arcs lie behind the visible part of $\mathscr D$.}
\label{fig:bf-causal-diamond-3d}
\end{figure}

To describe a single causal diamond $\mathscr D$, we fix a smooth cobordism between the spacelike caps $\Sigma_0$ and $\Sigma_1$, with truncated null cones $\mathscr H_p$ and $\mathscr H_q$ meeting transversely at $S$ (Figure~\ref{fig:bf-causal-diamond-3d}). We write $\mathfrak e_\Sigma$ for the embedding $X$ and local Lorentz frame of each boundary component $\Sigma$. A displacement of $\Sigma$ or a boost of its local Lorentz frame relative to the bulk fields can carry a surface charge, so the variables on which these charges act become edge modes~\cite{HarlowWu2020,CF}. We therefore vary the first order action with its boundary terms included. The boundary variation determines the symplectic form, and translating a cap determines the Hamiltonian generating time evolution of both the radiative modes and the edge modes.

\suppressfloats[t]

\subsection{First order gravity}

When we joined two geodesic intervals in JT gravity, matching the dilaton at their shared corner canceled the charge of a common boost. In four dimensions we can tilt a spacelike slice about $S$ by a rapidity that varies across the surface, so we must match the boost charge at each point when we join the two sides. The component of the spin connection in the plane normal to $S$ transports the local Lorentz frame by a boost. We therefore vary $\omega^{\mu\nu}$ independently in the first order action, with $B^{\mu\nu}$ subject to the simplicity constraint, to identify the area element that weights the change of rapidity in the corner term.\footnote{The indices $\mu,\nu$ refer to an oriented vector bundle with metric $\eta_{\mu\nu}$ of signature $(-+++)$, while Latin indices $a,b$ serve as abstract spacetime indices.}

Using a torsion free spacetime derivative $\nabla_a$, we write the curvature two form in a local Lorentz frame as
\begin{equation}
F^{\mu\nu}{}_{ab}=2\nabla_{[a}\omega_{b]}{}^{\mu\nu}
+2\omega_{[a}{}^\mu{}_\rho\,\omega_{b]}{}^{\rho\nu}.
\label{bfgeo:spin-curvature}
\end{equation}
The action is then~\cite{Plebanski1977}
\begin{equation}
I[B,\omega,\Phi]
=\frac{1}{16\pi\GN}\int_{\mathscr M}
\left(B_{\mu\nu}\curlywedge F^{\mu\nu}
-\frac12\Phi_{\mu\nu\rho\sigma}B^{\mu\nu}\curlywedge B^{\rho\sigma}\right)
+I_{\partial \mathscr M},
\label{bfgeo:action}
\end{equation}
where the Lagrange multiplier satisfies $\Phi_{\mu\nu\rho\sigma}=\Phi_{[\mu\nu][\rho\sigma]}=\Phi_{\rho\sigma\mu\nu}$ and $\epsilon^{\mu\nu\rho\sigma}\Phi_{\mu\nu\rho\sigma}=\nobreak4\Lambda$.
We use $\epsilon_{0123}=1$, so $\epsilon^{\mu\nu\rho\sigma}\epsilon_{\mu\nu\rho\sigma}=-24$. The trace part of $\Phi_{\mu\nu\rho\sigma}$ is therefore $-(\Lambda/6)\epsilon_{\mu\nu\rho\sigma}$ and is fixed by $\Lambda$.

Varying the tracefree part of $\Phi$ imposes
\begin{equation}
\mathfrak S^{\mu\nu\rho\sigma}
=B^{\mu\nu}\curlywedge B^{\rho\sigma}
+\frac1{24}\epsilon^{\mu\nu\rho\sigma}\epsilon_{\alpha\beta\gamma\delta}
B^{\alpha\beta}\curlywedge B^{\gamma\delta}=0.
\label{bfgeo:simplicity-constraint}
\end{equation}
With an invertible tetrad $e^\mu{}_a$ and fixed orientation, we choose the following solution of $\mathfrak S=0$
\begin{equation}
B^{\mu\nu}=\frac12\epsilon^{\mu\nu}{}_{\rho\sigma}e^{\rho}\curlywedge e^{\sigma},
\label{bfgeo:simplicity}
\end{equation}
with $g_{ab}=\eta_{\mu\nu}e^{\mu}{}_ae^{\nu}{}_b$.
Thus $B^{\mu\nu}$ fixes $e^\mu{}_a$ up to local Lorentz transformations and hence fixes $g_{ab}$. Varying $\omega^{\mu\nu}$ and $B^{\mu\nu}$ then yields
\begin{subequations}
\label{bfgeo:field-equations}
\begin{align}
&\nabla_{[a}B^{\mu\nu}{}_{bc]}=0,\label{bfgeo:spin-connection-field-equation}\\
&F_{\mu\nu}=\Phi_{\mu\nu\rho\sigma}B^{\rho\sigma}.
\end{align}
\end{subequations}
With~\eqref{bfgeo:simplicity}, equations~\eqref{bfgeo:field-equations} imply $\nabla_{[a}e^\mu{}_{b]}=0$ and $R_{ab}=\Lambda g_{ab}$. These equations constrain the Ricci tensor, while the Weyl tensor carries the two local radiative degrees of freedom as usual. On a null boundary one can identify these degrees of freedom by separating the first and second class constraints on the tetrad and spin connection~\cite{Canepa:2020hol}. 

To see how \eqref{bfgeo:action} arises, we start from the usual Einstein Hilbert action 
\begin{align}
\frac{1}{16\pi\GN}\int_{\mathscr M}(R-2\Lambda)\,\boldsymbol\epsilon_g,
\end{align}
where $\boldsymbol\epsilon_g=\sqrt{-g}\,\dd^4x$ is the spacetime volume form. For the torsion free spin connection $\omega(e)$ of the tetrad, it can be shown that
\begin{subequations}
\label{bfgeo:tetrad-identities}
\begin{align}
&R\,\boldsymbol\epsilon_g
 =\frac{1}{2}\epsilon_{\mu\nu\rho\sigma}e^\mu\curlywedge e^\nu\curlywedge F^{\rho\sigma}[\omega(e)],\\
&\boldsymbol\epsilon_g
 =\frac{1}{24}\epsilon_{\mu\nu\rho\sigma}e^\mu\curlywedge e^\nu\curlywedge e^\rho\curlywedge e^\sigma .
\end{align}
\end{subequations}
On the solution~\eqref{bfgeo:simplicity}, $B^{\mu\nu}\curlywedge B^{\rho\sigma}$ is proportional to $\epsilon^{\mu\nu\rho\sigma}$, so only the fixed trace of $\Phi$ enters the volume term. Using the two tetrad identities above, we can thus write the bulk action with $\omega$ varied independently as
\begin{equation}
I-I_{\partial \mathscr M}
=\frac{1}{32\pi\GN}\int_{\mathscr M}
\epsilon_{\mu\nu\rho\sigma}e^{\mu}\curlywedge e^{\nu}\curlywedge
\left(F^{\rho\sigma}-\frac{\Lambda}{6}e^{\rho}\curlywedge e^{\sigma}\right).
\label{bfgeo:einstein-action}
\end{equation}

Before reduction the boundary data $\mathfrak h_\Sigma$ consists of $\mathfrak e_\Sigma$ and the pullbacks of $B^{\mu\nu}$ and $\omega^{\mu\nu}$ to $\Sigma$.\footnote{For the reduced phase space in tetrad variables on a spacelike or timelike boundary see~\cite{Cattaneo:2017hus,Canepa:2020ujx}. Extending the Batalin Vilkovisky formulation of metric gravity~\cite{Cattaneo:2015xca} to tetrad variables requires additional boundary conditions~\cite{Cattaneo:2017kkv}.} Varying the spin connection at fixed $B$ and $\Phi$ and integrating by parts yields
\begin{equation}
\delta_\omega\bigl(I-I_{\partial\mathscr M}\bigr)
=\frac{1}{16\pi\GN}\left[
-\frac12\int_{\mathscr M}(\bm\epsilon_g)^{dabc}\,\delta\omega_{d\mu\nu}
\,\nabla_{[a}B^{\mu\nu}{}_{bc]}\,\bm\epsilon_g
+\int_{\partial\mathscr M}B_{\mu\nu}\curlywedge\delta\omega^{\mu\nu}
\right].
\label{bfgeo:bulk-spin-variation}
\end{equation}
The bulk term vanishes by~\eqref{bfgeo:spin-connection-field-equation}. To cancel the boundary term we use the covariant derivatives of the future null generators $\ell_s^\mu$ of $\mathscr H_p,\mathscr H_q$ and the future unit normals $n_i^\mu$ to $\Sigma_0,\Sigma_1$. Holding fixed the tetrad and null normals, we vary only the spin connection to obtain
\begin{subequations}\label{bfgeo:spin-variation-of-normals}
\begin{align}
&\delta_\omega(\nabla_a\ell_s^\mu)=\delta\omega_a{}^\mu{}_\nu\,\ell_s^\nu,\label{bfgeo:spin-variation-of-normals:a}\\*
&\delta_\omega(\nabla_a n_i^\mu)=\delta\omega_a{}^\mu{}_\nu\,n_i^\nu.\label{bfgeo:spin-variation-of-normals:b}
\end{align}
\end{subequations}

Where the two null cones meet at $S$, their normals also enter the boundary action through $\ell_p\cdot\ell_q$. This inner product depends on the normalization of the null generators, which we fix by choosing a parameter on each null cone. On a cut of $\mathscr H_s$ we choose the other null normal $n_s$ so that $n_s\cdot\ell_s=-1$. At $S$ this yields
\begin{equation}
n_p^\mu=\frac{\ell_q^\mu}{-\ell_p\cdot\ell_q}.
\label{bfgeo:transverse-normal-at-corner}
\end{equation}
Truncating the vertices introduces a joint at each $\partial\Sigma_i$. There the unit normal to the spacelike cap lies in the plane of $\ell_s^\mu$ and $n_s^\mu$, so its inner product with the null generator specifies its direction,
\begin{equation}
n_i^\mu=\frac{\ell_s^\mu}{2(-n_i\cdot\ell_s)}-(n_i\cdot\ell_s)n_s^\mu,\ \
(s,i)=(p,0),(q,1).
\label{bfgeo:cap-normal-at-joint}
\end{equation}

Writing $\ell_s^a=e_\mu{}^a\ell_s^\mu$, we use the inverse tetrad $E_\mu{}^a$ projected onto a cut to express the inaffinity and expansion along a null boundary~\cite{Parattu:2015gga}. On a spacelike cap the corresponding term depends on the trace of its extrinsic curvature, with $n_i^a=e_\mu{}^an_i^\mu$ and $h_i^{ab}$ the inverse induced metric. In these conventions,
\begin{subequations}\label{bfgeo:boundary-curvatures}
\begin{align}
&\kappa_s=-n_{s\mu}\ell_s^a\nabla_a\ell_s^\mu,\label{bfgeo:boundary-curvatures:a}\\*
&\Theta_{(\ell_s)}=E_\mu{}^a\nabla_a\ell_s^\mu,\label{bfgeo:boundary-curvatures:b}\\*
&K_i=h_i^{ab}e_{b\mu}\nabla_a n_i^\mu.\label{bfgeo:boundary-curvatures:c}
\end{align}
\end{subequations}
From the area element $\bm\mu=E^2\curlywedge E^3$ determined by $B^{\mu\nu}$ at the joints, we obtain both the corner area and the volume form on each null boundary,
\begin{equation}
\mathscr A=\int_S\bm\mu,\ \bm\eta_s=\dd u_s\curlywedge\bm\mu.
\label{bfgeo:corner-area-volume}
\end{equation}
With $\bm\epsilon_\Sigma$ the volume form on a cap, the boundary action combines these curvature terms with the logarithms of the normal inner products at the three joints,
\begin{subequations}\label{bfgeo:boundary-action}
\begin{align}
&\begin{aligned}
&I_{\partial \mathscr M}=\sum_{s=p,q} I_{\mathscr H_s}
+\frac{1}{8\pi\GN}\bigg[
\int_{\Sigma_0}\bm\epsilon_\Sigma K_0
-\int_{\Sigma_1}\bm\epsilon_\Sigma K_1
\\
&+\int_{\partial\Sigma_0}\bm\mu\log(-n_0\cdot \ell_p)
+\int_{\partial\Sigma_1}\bm\mu\log(-n_1\cdot \ell_q)
-\int_S\bm\mu\log(-\ell_p\cdot \ell_q/2)\bigg],
\end{aligned}\label{bfgeo:boundary-action:a}\\
&I_{\mathscr H_s}=\frac{\epsilon_s}{8\pi\GN}\biggl[\int_{\mathscr H_s}\bm\eta_s(\kappa_s+\Theta_{(\ell_s)})-\int_{C_{s,+}}\bm\mu+\int_{C_{s,-}}\bm\mu\biggr],\label{bfgeo:boundary-action:b}
\end{align}
\end{subequations}
where the null contributions have signs $\epsilon_p=1$ and $\epsilon_q=-1$. The cuts $C_{s,-}$ and $C_{s,+}$ are at the lower and upper ends of $\mathscr H_s$ in $u_s$, respectively.
The $\kappa_s+\Theta_{(\ell_s)}$ terms cancel variations of the normal and transverse components of $\omega$ in $B_{\mu\nu}\curlywedge\delta\omega^{\mu\nu}$~\cite{JubbSamuelSorkinSurya161200149}. After imposing torsionlessness, $\Lie_{\ell_s}\bm\mu=\Theta_{(\ell_s)}\bm\mu$ integrates to the difference of the areas of the two cuts, so only the inaffinity enters the null boundary action~\cite{Lehner},
\begin{equation}
 \int_{\mathscr H_s}\bm\eta_s\Theta_{(\ell_s)}=\int_{C_{s,+}}\bm\mu-\int_{C_{s,-}}\bm\mu,\ I_{\mathscr H_s}=\frac{\epsilon_s}{8\pi\GN}\int_{\mathscr H_s}\bm\eta_s\kappa_s.\label{bfgeo:null-action-cut-areas}
\end{equation}
 This cancellation is made after variation: $\delta\omega$ and the terms at $C_{s,\pm}$ remain in the symplectic potential. In Appendix~\ref{app:bf-geometry}, we derive the signs from the orientations in~\eqref{bfgeo:boundary-action:a}. The integrals over $\Sigma_0,\Sigma_1,\partial\Sigma_0,\partial\Sigma_1$ vanish as the caps shrink to smooth vertices.

\subsection{Null initial data}
\label{bfgeo:null-initial-data}

The integral of the expansion in~\eqref{bfgeo:null-action-cut-areas} yields the change in area between two cuts of the incoming null cone $\mathscr H_p$. To determine the area element on each cut, we freely specify the two polarizations of gravitational radiation through its conformal metric. Since the null normal is tangent to $\mathscr H_p$, the variation of this metric along each null generator fixes the shear. Given the area element and expansion at $\partial\Sigma_0$, we can then integrate the focusing equation to determine the area element as a function of affine parameter~\cite{Reisenberger12113880}.

On either null boundary $\mathscr H$ we omit the label $s$ and write $\ell^\mu$ for the future directed, affinely parametrized null normal and $n^\mu$ for the transverse null normal, with $\ell\cdot n=-1$. The affine parameter $u$ has tangent $\ell^\mu=e^\mu{}_a(\partial_u)^a$, and the coordinates $y^A$ label the null generators. Together with $\ell^a$, the degenerate metric induced on $\mathscr H$ defines a Carrollian structure~\cite{Duval:2014lpa,Ciambelli:2019lap,Ciambelli:2025unn}. Since $\star_\eta^2=-1$ on two forms, the simplicity constraints determine the tetrad $E_A^\mu$ pulled back to a cut and hence its induced metric $q_{AB}=\eta_{\mu\nu}E_A^\mu E_B^\nu$. To separate the bulk degrees of freedom\footnote{We use ``bulk'' both for data along a null boundary, in contrast to corner data, and for the spacetime interior. It should be clear from context which we mean.} from the area, we choose the conformal metric $\bar q_{AB}$ to have unit determinant relative to the fixed area two form $\bm\epsilon_0$ and write
\begin{subequations}\label{bfgeo:null-cut-geometry}
\begin{align}
&(\star_\eta B_{uA})^{\mu\nu}=-2\ell^{[\mu}E_A^{\nu]},\label{bfgeo:null-cut-geometry:a}\\
&E_A^\mu=n_\nu(\star_\eta B_{uA})^{\nu\mu},\label{bfgeo:null-cut-geometry:b}\\
&q_{AB}=\varrho^2\bar q_{AB},\label{bfgeo:null-cut-geometry:c}\\
&\mathrm dA=\varrho^2\bm\epsilon_0.\label{bfgeo:null-cut-geometry:d}
\end{align}
\end{subequations}

On each null cone let $\gamma_y$ and $\gamma_{y'}$ denote the segments of any two distinct null generators between $\partial\Sigma_i$ and $S$. To specify $\partial\Sigma_i$ and keep the null cone regular up to $S$, we impose conditions on the tetrad $e$ and the embeddings $X$ in $\mathfrak e_\Sigma$. The allowed pairs form the set
\begin{equation}
\mathscr B_\epsilon=\bigl\{(e,\mathfrak e_\Sigma)\mid
\mathrm dA|_{\partial\Sigma_i}=\epsilon^2\,\mathrm dA|_S,\ 
\varrho|_{\gamma_y}>0,\ \gamma_y\cap\gamma_{y'}=\varnothing\bigr\}.
\label{bfgeo:diamond-boundary-conditions}
\end{equation}
At fixed $0<\epsilon<1$, both the corner area and the affine parameter at $S$ can vary. For smooth invertible tetrads the vacuum field equations~\eqref{bfgeo:field-equations} then yield the reduced phase space
\begin{equation}
\mathcal P(\mathscr D)=
\bigl\{(e,\omega(e),\mathfrak e_\Sigma)\mid R_{ab}=\Lambda g_{ab},\ (e,\mathfrak e_\Sigma)\in\mathscr B_\epsilon\bigr\}\big/\!\sim,
\label{bfgeo:single-diamond-phase-space}
\end{equation}
where $\sim$ identifies configurations related by a diffeomorphism or a local Lorentz transformation acting on the bulk fields and gravitational boundary data together.\footnote{By the Gauss and Codazzi equations, the metric and extrinsic curvature induced on each spacelike cap by a vacuum solution in $\mathcal P(\mathscr D)$ satisfy the Hamiltonian and momentum constraints. We thus impose these constraints before choosing the Gaussian wavefunction of the reduced radiative and corner variables in Section~\ref{bfq:algebra-states}.} The data on $\mathscr H_p$ and $\Sigma_0$, together with compatible data at their joint $\partial\Sigma_0$, determines the solution in their domain of dependence and hence the metric induced there on the outgoing null cone $\mathscr H_q$~\cite{Rendall1990,ChoquetBruhat10064467}.

To determine how the cut expands and shears, we need the change in separation of neighboring null generators, measured by
\begin{equation}
K_{AB}=E_{B\mu}\nabla_A\ell^\mu,\ E_A=\frac{\partial}{\partial y^A},\ [\ell,E_A]=0.
\label{bfgeo:optical-tensor}
\end{equation}
Since $\ell$ is normal to the cut, the torsionless condition yields
\begin{equation}
\nabla_{[a}e^\mu{}_{b]}=0\ \Longrightarrow\ \ell^a\nabla_a E_A{}^b=E_A{}^a\nabla_a\ell^b,\ K_{[AB]}=0.
\label{bfgeo:optical-transport}
\end{equation}
The trace of $K_{ab}$ is the expansion and its tracefree part is the shear,\footnote{Recall that lowercase Latin indices are abstract spacetime indices and capital indices denote components on a cut, projected with $q^a{}_b=\delta^a{}_b+\ell^an_b+n^a\ell_b$.}
\begin{equation}
\Theta_{(\ell)}=q^{ab}K_{ab},\ \sigma_{ab}=K_{ab}-\frac{1}{2}\Theta_{(\ell)}q_{ab}.
\label{bfgeo:optical-expansion-shear}
\end{equation}

We use the Raychaudhuri equation to determine how the cross sectional area changes along a null generator. Our curvature conventions are
\begin{equation}
2\nabla_{[c}\nabla_{d]}v^a=R^a{}_{bcd}v^b,\ R_{bd}=R^a{}_{bad}.
\label{bfgeo:curvature-conventions}
\end{equation}
For an affinely parametrized null generator in a solution of the vacuum field equations~\eqref{bfgeo:field-equations},
\begin{equation}
\ell^a\nabla_a\ell^\mu=0,\ R_{ab}\ell^a\ell^b=0,
\label{bfgeo:affine-vacuum-focusing}
\end{equation}
so $K_{ab}$ determines focusing along each null generator.

To determine the area from the shear, we use the inverse conformal metric to raise indices on its variations, so
$(\partial_u\bar q)^{AB}=\bar q^{AC}\bar q^{BD}\partial_u\bar q_{CD}$.
Expressing the expansion and shear in these variables puts the Raychaudhuri equation in the form
\begin{subequations}
\label{bfgeo:focusing}
\label{bfgeo:null-coframe-evolution-main}
\begin{align}
&\Lie_\ell q_{AB}=2K_{AB}=2\sigma_{AB}+\Theta_{(\ell)}q_{AB},\label{bfgeo:focusing:a}\\
&\Lie_\ell\Theta_{(\ell)}=-K_{ab}K^{ab}-R_{ab}\ell^a\ell^b,\label{bfgeo:focusing:b}\\
&\Theta_{(\ell)}=\frac{2\Lie_\ell\varrho}{\varrho},\label{bfgeo:focusing:c}\\
&\sigma_{AB}=\frac{\varrho^2}{2}\Lie_\ell\bar q_{AB},\label{bfgeo:focusing:d}\\
&\mathfrak C_\ell\equiv\Lie_\ell\Theta_{(\ell)}+\frac12\Theta_{(\ell)}^2+\sigma_{ab}\sigma^{ab}=0,\label{bfgeo:focusing:e}\\
&\frac{\dd^2\varrho}{\dd u^2}
+\frac{\varrho}{8}(\partial_u\bar q)^{AB}\,\partial_u\bar q_{AB}=0.\label{bfgeo:focusing:f}
\end{align}
\end{subequations}
The Raychaudhuri equation $\mathfrak C_\ell=0$ determines $\varrho$ from $\bar q$, $\varrho(0)$ and $\partial_u\varrho(0)$~\cite{Reisenberger12113880,CF}. At a smooth vertex we measure solid angle in a local Lorentz frame with unit time direction $t^a$. Setting $u=0$ there and normalizing each null tangent by $|t_a\ell^a|=1$ then yields, for the solid angle element $\bm\epsilon_0$ of this local Lorentz frame, $\varrho(0)=0$ and $\partial_u\varrho(0)=1$. For another $\bm\epsilon_0$, $\partial_u\varrho(0)$ corresponds to a positive function of $y^A$.

\begin{figure}[!htbp]
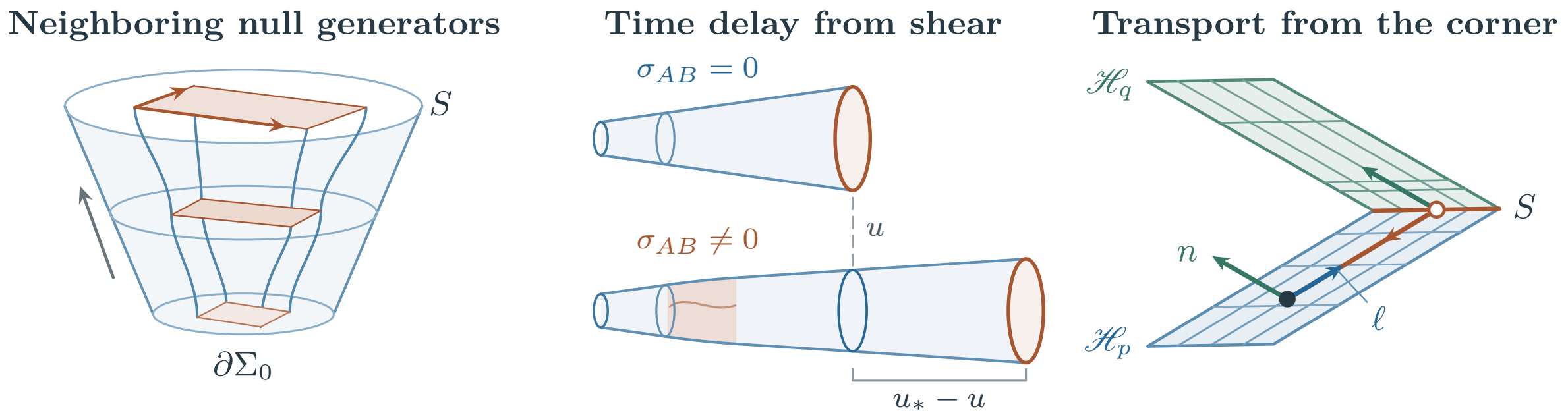

\centering
\resizebox{.95\linewidth}{!}{\SummaryNullData}
\caption[Shear and focusing on intersecting null cones]{The orange quadrilaterals mark the same four null generators on successive cuts. In the center the shear free congruence reaches the corner area element at the dotted line. The sheared congruence reaches it at the later clock reading $u_*$. On the right the intersecting null cone $\mathscr H_q$ supplies the transverse expansion and shear at $S$. The orange arrow transports this data along $\mathscr H_p$ to the marked cut, where $\ell$ is tangent to the null generator and $n$ is its transverse null normal.}
\label{fig:summary-data}
\end{figure}

To compare values of the affine parameter at a given area element on an expanding null generator, we use $v=\varrho/\varrho_S(y)$, where $\varrho_S(y)$ is the value of $\varrho$ at $S$. The cap condition in~\eqref{bfgeo:diamond-boundary-conditions} puts $\partial\Sigma_i$ at $v=\epsilon$ and $S$ at $v=1$. We can therefore integrate~\eqref{bfgeo:focusing:f} from $\partial\Sigma_i$. For initial slope $p_{\epsilon,y}=\left.\partial_u\varrho\right|_{v=\epsilon}>0$ we obtain
\begin{subequations}
\label{bfgeo:integrated-focusing-main}
\begin{align}
&\partial_v\log(\partial_u\varrho)=-\frac{v}{8}(\partial_v\bar q)^{AB}\partial_v\bar q_{AB},\label{bfgeo:integrated-focusing-main:a}\\
&\partial_u\varrho=p_{\epsilon,y}\exp[-\Xi_y(v)],\label{bfgeo:integrated-focusing-main:b}\\
&\Xi_y(v)=\frac18\int_\epsilon^v s\,(\partial_s\bar q)^{AB}\,\partial_s\bar q_{AB}\,\dd s,\label{bfgeo:integrated-focusing-main:c}\\
&p_{\epsilon,y}[u(v,y)-u(\epsilon,y)]=\varrho_S(y)\int_\epsilon^v e^{\Xi_y(s)}\dd s.\label{bfgeo:integrated-focusing-main:d}
\end{align}
\end{subequations}
For a smooth vertex $\epsilon=0$ and $p_{0,y}=1$, so at the corner the clock reads $u_*(y)=u(1,y)$. Since the focusing observable $\Xi_y(v)$ is nonnegative, shear increases the affine parameter needed to reach a given $\varrho_S(y)$ from the vertex. Shear encountered near the initial cut enters $\Xi_y(v)$ at every subsequent cut, so the delay continues to grow even after the shear has vanished (Figure~\ref{fig:summary-data}). On a capped null cone, rescaling $u$ by a positive factor multiplies the change in affine parameter from $\partial\Sigma_i$ to $S$ by that factor and divides $p_{\epsilon,y}$ by it. Their product therefore defines a clock independent of this normalization,
\begin{equation}
\label{bfgeo:capped-clock}
u_*(y)=p_{\epsilon,y}[u(1,y)-u(\epsilon,y)]
=\varrho_S(y)\int_\epsilon^1e^{\Xi_y(v)}\,\mathrm dv.
\end{equation}
We can also express the clock reading in terms of the dressing time $V$ of~\cite{CiambelliFreidelLeigh2024DressingTime}, since $\partial_u V\propto\varrho$ along each null generator by~\eqref{bfgeo:area-constraint} and~\eqref{bfgeo:clock-reduction}.

With $\int_S\bm\epsilon_0=4\pi$, the area and averaged clock are
\begin{equation}
 \mathscr A=\int_S\varrho_S(y)^2\bm\epsilon_0,\ \bar u_*=\frac{\int_S\varrho_S u_*\bm\epsilon_0}{\int_S\varrho_S\bm\epsilon_0}.\label{bfgeo:area-and-clock-average}
\end{equation}
 Each solution thus determines its own reading $\bar u_*$.\footnote{In Section~\ref{bfent:subregion-boost-main} we impose $\bar u_*\le\ell(\mathscr D)$ to count the states of a causal diamond of bounded size. Both $\bar u_*$ and $\mathscr A$ vary on this subset of~\eqref{bfgeo:single-diamond-phase-space}, with $(1-\epsilon)\sqrt{\mathscr A/(4\pi)}\le\bar u_*$ by~\eqref{bfgeo:capped-clock} and Cauchy Schwarz on $S$.}

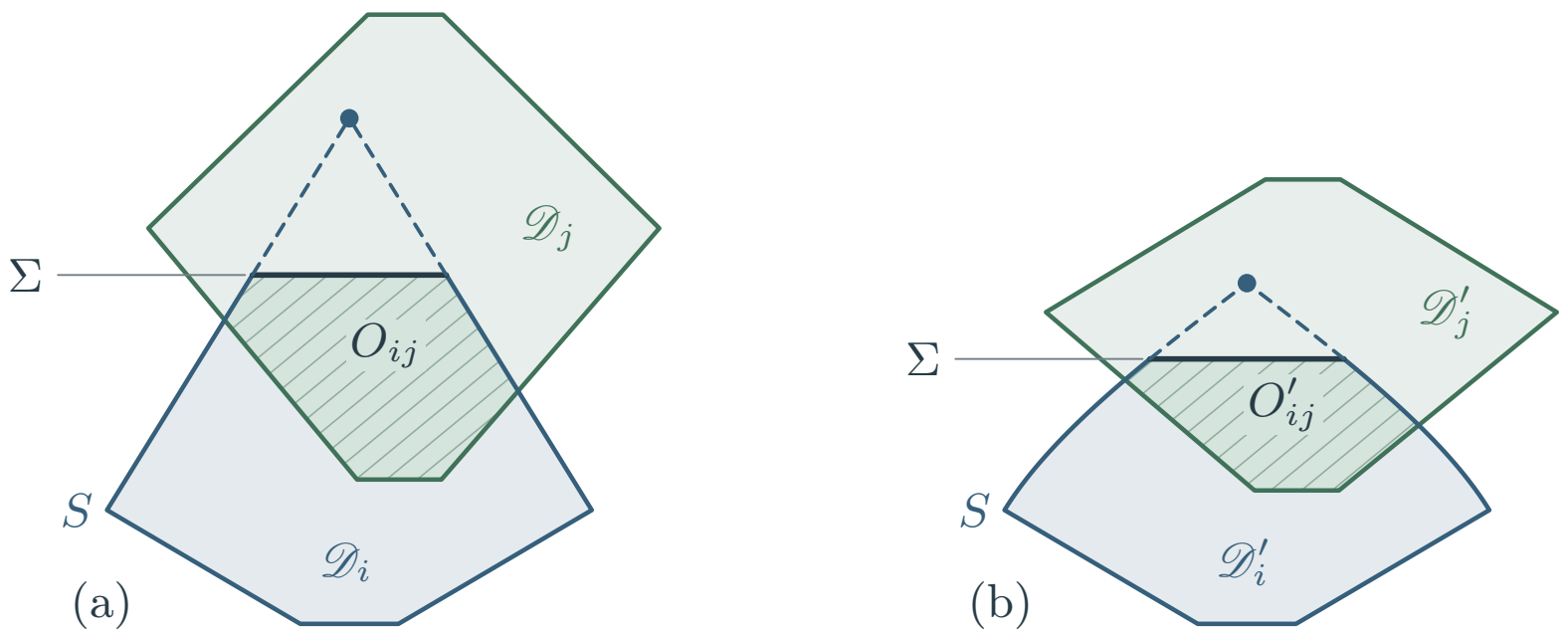
\begin{figure}[!htb]
\centering
\begin{tikzpicture}[x=1.02cm,y=.84cm,font=\small,line cap=round,line join=round]
  \definecolor{patchBlue}{HTML}{355F7B}
  \definecolor{patchGreen}{HTML}{3F7259}
  \definecolor{patchBoth}{HTML}{D5E4DC}
  \definecolor{patchHatch}{HTML}{92ADA0}
  \definecolor{patchInk}{HTML}{263B45}
  \def\straightWidth#1{(1-(#1)/3.1)}
  \def\curvedWidth#1{(cos(deg(\k*(#1)))-sin(deg(\k*(#1)))/(3.1*\k))}
  \foreach \xoffset/\k/\panel in {0/0/a,5.85/.6/b}{
    \begin{scope}[shift={(\xoffset,0)}]
      \ifdim\k pt=0pt
        \def\firstPatch{\mathscr D_i}
        \def\nextPatch{\mathscr D_j}
        \def\overlap{O_{ij}}
        \let\widthAt\straightWidth
        \def\focus{3.1}
        \def\rim{1.86}
      \else
        \def\firstPatch{\mathscr D_i^{\prime}}
        \def\nextPatch{\mathscr D_j^{\prime}}
        \def\overlap{O_{ij}^{\prime}}
        \let\widthAt\curvedWidth
        \pgfmathsetmacro{\focus}{rad(atan(3.1*\k))/\k}
        \pgfmathsetmacro{\rim}{(rad(acos(.4/sqrt(1+(1/(3.1*\k))^2)))-rad(atan(1/(3.1*\k))))/\k}
      \fi
      \pgfmathsetmacro{\greenBase}{.13*\rim}
      \pgfmathsetmacro{\greenWaist}{\rim+.37}
      \pgfmathsetmacro{\greenTop}{\focus+.82}
      \def\bluePatch{(-.32,-.90)--(.32,-.90)--(1.58,0)
        --plot[domain=0:\rim,samples=30,variable=\u] ({1.58*\widthAt{\u}},\u)
        --(-.632,\rim)
        --plot[domain=\rim:0,samples=30,variable=\u] ({-1.58*\widthAt{\u}},\u)
        --cycle}
      \def\greenPatch{(.05,\greenBase)--(.60,\greenBase)--(2.02,\greenWaist)
        --(.61,\greenTop)--(.12,\greenTop)--(-1.31,\greenWaist)--cycle}
      \path[fill=patchBlue!13] \bluePatch;
      \path[fill=patchGreen!12] \greenPatch;
      \begin{scope}
        \clip \bluePatch;
        \clip \greenPatch;
        \fill[patchBoth] (-2,-1) rectangle (2.1,4.2);
        \foreach \l in {-6,-5.8,...,2.2}
          \draw[patchHatch,line width=.30pt] (\l,-1)--({\l+5},4);
      \end{scope}
      \draw[patchGreen,line width=.85pt] \greenPatch;
      \draw[patchBlue,line width=.85pt] \bluePatch;
      \draw[patchInk,line width=1.05pt] (-.632,\rim)--(.632,\rim);
      \foreach \side in {-1,1}
        \draw[patchBlue,line width=.7pt,dash pattern=on 2.5pt off 1.9pt]
          plot[domain=\rim:\focus,samples=30,variable=\u] ({\side*1.58*\widthAt{\u}},\u);
      \fill[patchBlue] (0,\focus) circle[radius=1.75pt];
      \node[text=patchBlue] at (-.02,-.41) {$\firstPatch$};
      \node[text=patchGreen] at (1.30,\greenWaist) {$\nextPatch$};
      \node[fill=patchBoth,inner sep=.65pt,text=patchInk] at (.24,{.70*\rim}) {$\overlap$};
      \node[anchor=east,inner sep=2.4pt,text=patchBlue] at (-1.58,0) {$S$};
      \draw[patchInk!65,line width=.36pt] (-1.90,\rim)--(-.68,\rim);
      \node[anchor=east,inner sep=2pt,text=patchInk] at (-1.92,\rim) {$\Sigma$};
      \node[anchor=west,text=patchInk] at (-1.94,-.75) {(\panel)};
    \end{scope}
  }
\end{tikzpicture}
\caption{In each spacetime the blue and green causal diamonds overlap in the hatched region. We truncate the blue null cone at $\Sigma$. Its null generators continue along the dashed curves to a caustic (dot), which occurs at a smaller change in affine parameter in (b) because the shear is greater for the same area and expansion at $S$. We match the tetrads and spin connections throughout the overlap. The green null generators then extend our description of the spacetime beyond the caustic.}
\label{fig:bf-local-reconstruction}
\end{figure}

Although~\eqref{bfgeo:integrated-focusing-main:b} ensures $\partial_u\varrho>0$ wherever $v$ is a regular coordinate on the expanding null cone, the Raychaudhuri equation~\eqref{bfgeo:focusing:f} allows shear to bring this derivative to zero while $\varrho>0$. At such a turning point $v=\varrho/\varrho_S$ ceases to be a coordinate, although $u$ still parametrizes a smooth null boundary until neighboring null generators focus at a caustic or distinct null generators intersect while $\varrho>0$~\cite{Reisenberger12113880}. If either happens between $\partial\Sigma_i$ and $S$, the null generators cease to define the embedded null boundary required in~\eqref{bfgeo:diamond-boundary-conditions}. We can still cover the point where these null generators focus or cross with a new causal diamond, since every point of a smooth globally hyperbolic vacuum spacetime admits a causally convex normal neighborhood. Within that neighborhood we choose one vertex to the timelike past of the point and the other to its timelike future, sufficiently close that their null cones intersect at a smooth cut while $\partial_u\varrho$ is still positive on each null generator.\footnote{For example, at a smooth point $x$ in a high energy collision of gravitational waves, take the vertices at $\gamma(\pm L)$ on a unit timelike geodesic through $x=\gamma(0)$. Halving $L$ eventually places the entire causal diamond in a convex normal neighborhood of $x$~\cite{Penrose1972Topology}. With $|\dot\gamma\cdot\ell|=1$ at each vertex and tidal curvature of order $\ell_P^{-2}$, where $\ell_P=\sqrt{G_N\hbar}$ is the Planck length, the geodesic deviation equation predicts a transverse separation $u\,\delta\theta[1+O(u^2/\ell_P^2)]$ for null generators initially an angle $\delta\theta$ apart. Since $u\simeq L$ at $S$, we can take $L\ll\ell_P$ and place the caps at $v=\epsilon$ while preserving the curvature at $x$.} The characteristic data induced on each causal diamond then reconstructs the metric in its domain of dependence~\cite{Rendall1990,ChoquetBruhat10064467,Chrusciel12091971}, including in neighborhoods where a congruence from farther away has already focused (Figure~\ref{fig:bf-local-reconstruction}).

Before the original null generators focus, we can pass from one causal diamond to the next across a spacelike cap $\Sigma$ where their embedded images $\mathscr D_i,\mathscr D_j\subset\mathscr M$ overlap. Writing $X_i,X_j$ for the embeddings of $\Sigma$ as the future cap of $\mathscr M_i\subset\mathscr D_i$ and the past cap of $\mathscr M_j\subset\mathscr D_j$, respectively, we fix its edge by $(X_i^*v)|_{\partial\Sigma}=\epsilon$ so that the affine parameter there can vary with shear as in~\eqref{bfgeo:integrated-focusing-main:d}. To continue the solution from $\mathscr M_i$ into $\mathscr M_j$, we match $X_i^*B_i=X_j^*B_j$ and $X_i^*\omega_i=X_j^*\omega_j$ in one local Lorentz frame. Both portions then induce the same metric and extrinsic curvature with respect to the future directed unit normal to $\Sigma$, so their cap integrals in~\eqref{bfgeo:boundary-action:a} cancel by opposite orientation. Since solutions with matching data on $\Sigma$ have the same local development there, we obtain~\cite{CattaneoMnevReshetikhin2014,CattaneoMnev2023FiberProducts}
\begin{align}
&\mathcal P(\mathscr M_j\circ_\Sigma\mathscr M_i)\simeq
\left.\big(\mathcal P(\mathscr M_i)\times\mathcal P(\mathscr M_j)\big)
\right|_{\mathfrak h_{\Sigma,i}=\mathfrak h_{\Sigma,j}}\big/\!\sim ,
\label{bfgeo:cap-phase-space-gluing}
\end{align}
where we identify simultaneous changes of the embedding and local Lorentz frame on both sides of the internal cap, since these preserve the spacetime obtained by gluing.

On a spatial slice we instead join two subregions at their shared corner $S$. Identifying angular coordinates by $Y$, we match the induced metric and its normal derivatives as in~\eqref{bfgeo:geometric-matching}. The same area element enters the two boost charges with opposite signs. Upon quotienting by simultaneous changes of the local Lorentz frames, $\Omega_++\Omega_-$ descends to the symplectic form on the union, with relative boost angle $\Gamma_R-\Gamma_L$ conjugate to $\mathscr K(y)=\varrho_S(y)^2/(8\pi\GN)$ as in~\eqref{bfgeo:glued-potential-main}~\cite{CF,DF}.

Agreement on $\Sigma$ determines the metric in its domain of dependence~\cite{ChoquetBruhat10064467,ChoquetBruhatGeroch1969}. Wherever the overlap $O_{ij}=\mathscr D_i\cap\mathscr D_j$ extends beyond that domain, we must specify further characteristic data. To keep both causal diamonds in full, we thus match tetrads and spin connections throughout $O_{ij}$ (Figure~\ref{fig:bf-local-reconstruction}) and obtain
\begin{align}
&\mathcal P(\mathscr D_i\cup\mathscr D_j)\simeq
\big(\mathcal P(\mathscr D_i)\times\mathcal P(\mathscr D_j)\big)
\big|_{(e_i,\omega_i)=(e_j,\omega_j)}\big/\!\sim .
\label{bfgeo:overlap-phase-space}
\end{align}
Each pair includes the identification of points and frames on $O_{ij}$. As in~\eqref{bfgeo:cap-phase-space-gluing}, we quotient by changes of the internal embeddings and frames that preserve the glued solution and its boundary data. Moving the surface dividing a spatial slice within $O_{ij}$ adds to one side precisely the integral of the symplectic current that we subtract from the other, since the fields and their variations agree throughout the overlap. Combining the corner terms as in~\eqref{bfgeo:glued-potential-main}, we thus obtain the same reduced symplectic form on the union wherever we divide the slice.

For a globally hyperbolic spacetime $\mathscr M=\bigcup_i\mathscr D_i$, we can now choose a cover separately for each solution.\footnote{A cover by the domains of dependence of intersecting double null sheets was proposed in~\cite{Reisenberger0703134}. Such sheets can be constructed from truncated past null cones~\cite{FuchsReisenberger170406992}.} Writing $L_{ij}$ for the local Lorentz transformation taking $e_j$ to $e_i$ on $O_{ij}$, we impose~\eqref{bfgeo:overlap-phase-space} on all overlaps to obtain
\begin{align}
&\mathcal P(\mathscr M)\simeq
\left(\bigsqcup_{\{\mathscr D_i\}}
\left\{(z_i,L_{ij})\,\middle|\,
z_i\in\mathcal P(\mathscr D_i),\ e_i=L_{ij}e_j,\ L_{ij}L_{jk}=L_{ik}\right\}\right)\big/\!\sim .
\label{bfgeo:patched-phase-space}
\end{align}
Since $\omega_i$ is the torsion free spin connection of $e_i$, the same $L_{ij}$ relates $\omega_i$ and $\omega_j$ as in~\eqref{bfgeo:global-gluing}. Keeping track of the gauge transformations on overlaps likewise allows one to glue Yang Mills fields~\cite{Benini:2017zjv} and to construct global observables in Abelian and linearized gauge theories~\cite{Benini:2015hta,Benini:2019hoc}. We compare two covers on smaller causal diamonds, each contained in one member of each cover. The quotient in~\eqref{bfgeo:patched-phase-space} identifies descriptions whose fields agree there up to diffeomorphism and local Lorentz transformation, with the same data on the external boundary. The matched metrics define a single smooth vacuum spacetime on their union, and every smooth globally hyperbolic vacuum spacetime restricts to such data.

\subsection{The symplectic form with edge modes}
\label{bfgeo:symplectic-edge-modes}

When we join two subregions,~\eqref{bfgeo:cap-phase-space-gluing} identifies simultaneous changes of the embedding and local Lorentz frame on the shared cut that preserve the glued solution. For a single subregion, displacing the cut along its null generators can change its area even on a fixed spacetime. Boosting the local Lorentz frame at the cut can likewise change the event reached at a given affine parameter. The charges for such surface symmetries have been studied at black hole horizons~\cite{Donnay:2015abr,Donnay:2016ejv,Adami:2020amw}, on general null boundaries~\cite{ChandrasekaranFlanaganPrabhu2018,Adami:2021kvx} and on the boundary of a causal diamond~\cite{Chandrasekaran:2019ewn}. To obtain them here, we vary the embedding and local Lorentz frame independently of the bulk fields. A displacement $\xi^a$ of the cut on a fixed solution then defines a variation $\delta_\xi$ with charge variation
\begin{equation}
\delta H_\xi=\Omega(\delta,\delta_\xi).
\label{bfgeo:charge-variation}
\end{equation}
A diffeomorphism acting on the fields and the embedding together preserves pullbacks and amounts to a gauge transformation.

For a radiative observable $\mathcal{O}$ and a charge $H_\xi$ acting on the edge modes $\mathfrak e$, the symplectic form $\Omega=\delta\Theta$ on the reduced phase space determines the Poisson bracket
\begin{equation}
\{\mathcal{O},H_\xi\}=\delta_\xi\mathcal{O}.
\label{bfgeo:charge-action-bracket}
\end{equation}
When radiation crosses the boundary, varying its embedding allows us to define Poisson brackets of the surface charges~\cite{Freidel:2021dxw}. At null infinity, fixing the radiative data yields a Dirac bracket for these charges~\cite{Wieland:2021eth}.

In the symplectic potential $\Theta$ we therefore need to track variations of $\varrho$, $\omega$, $\ell$ and $n$ induced by the null constraints. It is easy to show that the symplectic potential takes the form
\begin{equation}
\bm\theta_{\mathscr M}=\frac{1}{16\pi\GN}B_{\mu\nu}\curlywedge\delta\omega^{\mu\nu},\ \bm\theta=\bm\theta_{\mathscr M}+\dd\bm\alpha.
\label{bfgeo:potential-improvement}
\end{equation}
With $\star_g$ denoting the Hodge dual for the spacetime metric,
\begin{equation}
\bm\alpha=\frac{1}{16\pi\GN}\,\star_g\bigl(e_{\mu}\curlywedge\delta e^{\mu}\bigr).
\label{bfgeo:potential-improvement-form}
\end{equation}
On the constraint surface, $\dd\bm\alpha$ removes the charge associated to a local Lorentz transformation acting on $e^\mu{}_a$ and $\omega^{\mu\nu}$ together~\cite{DePaoliSpeziale2018,Freidel:2020xyx,FreidelGeillerPranzetti2020}. On the other hand displacing a boundary frame changes an observable's location or components, so its charge remains.

For the bulk contribution, we define $\beta_a$ and the one forms $K^\mu$ for $\mu=2,3$ by
\begin{subequations}\label{bfgeo:bulk-connection-forms}
\begin{align}
&\beta_a=-n_\mu\nabla_a\ell^\mu,\label{bfgeo:bulk-connection-forms:a}\\
&K^\mu=K_A{}^\mu\,\dd y^A,\label{bfgeo:bulk-connection-forms:b}\\
&K_A{}^\mu=q^{BC}E_C{}^\mu K_{AB}.\label{bfgeo:bulk-connection-forms:c}
\end{align}
\end{subequations}
Using $\bm\theta_{\mathscr M}$ above, we obtain
\begin{equation}
\int_{\mathscr H}\bm\theta_{\mathscr M}
=\frac1{8\pi\GN}\int_{\mathscr H}
\left[\bm\mu\curlywedge\delta\beta
+\dd u\curlywedge(E^2\curlywedge\delta K^3-E^3\curlywedge\delta K^2)\right].
\label{bfgeo:bf-null-potential-main}
\end{equation}
At fixed tetrad the integrand reduces to
\begin{equation}
\bm\eta\bigl(\delta\kappa+\delta\Theta_{(\ell)}\bigr),
\label{bfgeo:fixed-tetrad-null-integrand}
\end{equation}
where $\kappa=\ell^a\beta_a$. Varying the action on the null boundary cancels this term. For $\nabla_{[a}e^\mu{}_{b]}=0$, we first hold the null normals and angular coordinates fixed, $\delta\ell^a=\delta n^a=0$ and $\delta y^A=0$. Collecting the terms at the cuts in $\Theta_\partial$, we obtain the null symplectic potential~\cite{Reisenberger07122541,CF}
\begin{equation}
\Theta_{\mathscr H}
=\frac1{16\pi\GN}\int_{\mathscr H}\bm\eta
\left[\sigma^{AB}\delta q_{AB}-(2\kappa+\Theta_{(\ell)})\delta\log\varrho^2\right]
+\Theta_\partial.
\label{bfgeo:null-potential-main}
\end{equation}
The shear is conjugate to the conformal metric~\cite{HopfmuellerFreidel2017,CiambelliFreidelLeigh2026PhaseSpace} since $\sigma^{AB}\delta q_{AB}=(1/2)(\partial_u\bar q)^{AB}\,\delta\bar q_{AB}$. In Appendix~\ref{app:bf-geometry} we evaluate~\eqref{bfgeo:bf-null-potential-main} in an adapted tetrad, keeping the corner term $\int_{\partial\mathscr H}\bm\alpha$ and allowing the null normals to vary along $\mathscr H$ as in~\eqref{bfgeo:complete-null-potential}--\eqref{bfgeo:null-potential-flux}. 

To account for displacements of the boundary as well, we must vary its embedding $X$ in $\mathscr M$. Under a displacement $\chi^a=\delta X^a\circ X^{-1}$, the pullback of a spacetime tensor $T$ transforms as~\cite{Speranza},
\begin{equation}
\delta(X^*T)=X^*(\delta T+\Lie_\chi T).
\label{bfgeo:embedding-variation}
\end{equation}
Hence a simultaneous diffeomorphism of $T$ and $X$ preserves $X^*T$, whereas varying $X$ alone produces a surface charge. For variations of the two null normals, the corner symplectic potential is
\begin{equation}
\Theta_S=\frac{1}{8\pi\GN}\int_S\bm\mu
\left(\delta\Gamma+
\frac{\ell_a\delta n^a-n_a\delta\ell^a}{2(-\ell\cdot n)}\right).
\label{bfgeo:normal-corner}
\end{equation}
Here $\Gamma$ parametrizes boosts of the boundary frame, with conjugate momentum $\bm\mu/(8\pi\GN)$~\cite{CarlipTeitelboim1995}. The variations $\delta\ell^a$ and $\delta n^a$ include their dependence on $\mathfrak h$; angular displacements of $X$ couple to the tangential surface momentum~\cite{DF,Speranza}.

For a smooth null cone, substituting~\eqref{bfgeo:integrated-focusing-main} into $\Theta_{\mathscr H}$ and expressing variations at fixed $v$ further cancels out the terms proportional to $\delta u_*$ at $v=1$:
\begin{subequations}
\label{bfgeo:cone-form-main}
\begin{align}
&\Theta
=\int_S\bm\epsilon_0 \mathscr K(y)
(\alpha-2\delta J+\delta\Gamma+\Upsilon),\label{bfgeo:cone-form-main:a}\\
&\begin{aligned}
\Omega=\int_S\bm\epsilon_0\bigl[&\mathscr K(y)\,\delta\alpha+\mathscr K(y)\,\delta\Upsilon
+\delta \mathscr K(y)\wedge(\alpha-2\delta J+\delta\Gamma+\Upsilon)\bigr].
\end{aligned}\label{bfgeo:cone-form-main:b}\\
&\alpha=\frac{1}{4}\int_0^1v^2\,(\partial_v\bar q)^{AB}\,\delta\bar q_{AB}\,\dd v,\label{bfgeo:cone-form-main:c}\\
&J=\frac{1}{16}\int_0^1v(1-v^2)\,(\partial_v\bar q)^{AB}\,\partial_v\bar q_{AB}\,\dd v.\label{bfgeo:cone-form-main:d}
\end{align}
\end{subequations}
The boost generator is the integral over $S$ of the charge aspect $\mathscr K(y)=\varrho_S(y)^2/(8\pi\GN)$, \begin{align}\mathscr K=\int_S\bm\epsilon_0\mathscr K(y)=\frac{\mathscr A}{8\pi\GN}.\end{align} For a spherical corner $\mathscr A$ is also an invariant of the coadjoint orbits of the corner symmetry group~\cite{Donnelly:2020xgu}, and one can construct an area operator by approximating its Lie algebra with matrices~\cite{Donnelly:2022kfs}. Even with the angular embedding held fixed in~\eqref{bfgeo:cone-form-main:b}, $\mathscr{K}(y)$ couples to the conformal metric through $\delta\mathscr K(y)\wedge\alpha$ and to the choice of null normals through the term in~\eqref{bfgeo:normal-corner},
\begin{equation}
\Upsilon=\frac{\ell_a\delta n^a-n_a\delta\ell^a}{2(-\ell\cdot n)}.
\label{bfgeo:normal-variation-upsilon}
\end{equation}
If we also vary the angular embedding, the Lie derivative in~\eqref{bfgeo:embedding-variation} enters the shear term in the symplectic potential and couples the angular edge modes to the conformal metric as in~\eqref{bfgeo:coupled-form}.

Finally, we use $\mathfrak T=0$ for the independent constraints on $\omega_{\mu\nu}$ required to preserve $\mathfrak S=0$ under time evolution, and $\mathfrak C_A=0$ for the constraints solved in the reduction, including $\mathfrak S=0$, $\mathfrak T=0$, the Gauss constraint and the null constraints. We identify configurations related by a gauge transformation of both the fields and the boundary frame. In canonical coordinates $\phi=(q,p)$ on the resulting phase space, the embedding into the constraint surface determines the boundary fields and their symplectic form,
\begin{equation}
 \mathfrak h=\iota(\phi),\ \mathfrak C_A(\iota(\phi))=0,\ \Omega=\iota^*\delta\Theta.\label{bfgeo:constraint-surface-pullback}
\end{equation}
 The inverse of $\Omega$ fixes the commutators of the smeared canonical variables in~\eqref{bfads:rep-weyl}.

\subsection{The reduced Hamiltonian}
\label{bfgeo:reduced-hamiltonian-main}

The pullback in~\eqref{bfgeo:constraint-surface-pullback} fixes the Poisson brackets of the reduced variables $\phi=(q,p)$ on a slice. To find the Hamiltonian that evolves them, we pull the action back to the constraint surface as the slice advances. On the constraints the Hamiltonian of the original fields reduces to a surface charge~\cite{ReggeTeitelboim,HarlowWu2020}. Its value depends on the radiation between the cuts of the null cone, since for a time translation normalized at $S$ the shear can change how far the inner cut advances along the transverse null normal. We choose a foliation and write $\omega^{\mu\nu}=\omega_t^{\mu\nu}\dd t+\omega_\Sigma^{\mu\nu}$ and $B^{\mu\nu}=\dd t\curlywedge B_t^{\mu\nu}+B_\Sigma^{\mu\nu}$. Integrating the term containing $(\nabla_i\omega_t^{\mu\nu})\,\dd x^i$ by parts separates the bulk symplectic potential,
\begin{align}
\label{bfgeo:canonical-pair-main}
\Theta_{\Sigma}=\frac1{16\pi\GN}\int_\Sigma
B_{\Sigma \mu\nu}\curlywedge\delta\omega_\Sigma^{\mu\nu},
\end{align}
from the Hamiltonian
\begin{align}
\label{bfgeo:canonical-hamiltonian}
&H_{\Sigma}=-\frac1{16\pi\GN}\int_\Sigma
B_{t\mu\nu}\curlywedge(F_\Sigma^{\mu\nu}-\Phi^{\mu\nu}{}_{\rho\sigma}B_\Sigma^{\rho\sigma}) \nonumber \\
&-\frac1{16\pi\GN}\int_\Sigma\omega_t^{\mu\nu}\frac12\nabla_{[i}(B_\Sigma)_{|\mu\nu|jk]}\,\dd x^i\curlywedge\dd x^j\curlywedge\dd x^k
+\frac1{16\pi\GN}\int_{\partial\Sigma}
B_{\Sigma \mu\nu}\omega_t^{\mu\nu}.
\end{align}
Here $F_\Sigma^{\mu\nu}$ is the curvature of $\omega_\Sigma^{\mu\nu}$. Varying $B_t^{\mu\nu}$ and $\omega_t^{\mu\nu}$ imposes the constraints
\begin{subequations}\label{bfgeo:hamiltonian-constraints}
\begin{align}
&F_\Sigma^{\mu\nu}=\Phi^{\mu\nu}{}_{\rho\sigma}B_\Sigma^{\rho\sigma},\label{bfgeo:spatial-curvature-constraint}\\
&\nabla_{[i}(B_\Sigma)_{|\mu\nu|jk]}=0,\label{bfgeo:spatial-gauss-constraint}
\end{align}
\end{subequations}
so $H_{\Sigma}$ reduces to a pure boundary term~\cite{ReggeTeitelboim}. Here we have included the boundary action $I_{\partial \mathscr M}$ of~\eqref{bfgeo:boundary-action:a} and varied the edge modes $\mathfrak e$ in $I$ \cite{HarlowWu2020}.\footnote{We note that when varying $I$, $\delta\kappa$ is completely unconstrained around $\kappa = 0$.} Varying the limits of integration evaluates the integrand at the two cuts $u_-$ and $u_+$ with opposite signs. In Appendix~\ref{app:bf-hamiltonian} we evaluate these contributions by writing the action as
\begin{equation}
 I=\int_{t_i}^{t_f}\bigl[\Theta_t(\dot{\mathfrak h})-H\bigr]\,\dd t.
 \label{bfgeo:unreduced-action}
\end{equation}

For a boundary displacement $\xi^a$, the two form $\bm\alpha$ in~\eqref{bfgeo:potential-improvement-form} contributes to the Noether charge and hence to the variation of the Hamiltonian,
\begin{subequations}
\label{bfgeo:charge-main}
\begin{align}
&\bm Q_\xi=\frac1{16\pi\GN}B_{\mu\nu}(i_\xi\omega^{\mu\nu})
+\bm\alpha(\Lie_\xi e),\\
&\delta H_\xi=\int_{\partial\Sigma}
\left(\delta\bm Q_\xi-\bm Q_{\delta\xi}-i_\xi\bm\theta\right).
\end{align}
When $\xi$ is field dependent, its variation enters through $\bm Q_{\delta\xi}$ \cite{FreidelOliveriPranzettiSpeziale2021}. With vanishing torsion~\cite{DePaoliSpeziale2018,IW},
\begin{align}
&(\bm Q_\xi)_{ab}=-\frac{1}{16\pi\GN}(\bm\epsilon_g)_{abcd}\nabla^c\xi^d.\label{bfgeo:torsion-free-noether-charge}
\end{align}
\end{subequations}
For a unit boost at $S$, $\xi|_S=0$ and $\nabla_a\xi^b|_S=\epsilon^b{}_a$, where $\epsilon^a{}_b=n^a\ell_b-\ell^an_b$. Since $\epsilon^a{}_b\epsilon^b{}_a=2$ implies $\epsilon^a{}_b\delta\epsilon^b{}_a=0$, we obtain
\begin{equation}
 \bm Q_\xi=\frac{\bm\mu}{8\pi\GN},\ \bm Q_{\delta\xi}=0.\label{bfgeo:unit-corner-boost-charge}
\end{equation}
 The associated surface charge is thence the boost generator $\mathscr{K}$, as in~\eqref{bfgeo:normal-corner}. When we translate a cap, the map $\mathfrak h=\iota_t(\phi)$ changes even at fixed canonical coordinates $\phi=(q,p)$.

To separate this displacement from the time evolution of $\phi$, we pull the symplectic potential back to the reduced phase space,
\begin{equation}
 \Theta_t(\dot{\mathfrak h})=(\iota_t^*\Theta_t)(\dot\phi)+\Theta_t(\partial_t\iota_t),\ \iota_t^*\Theta_t=p_k\,\delta q^k+\delta\chi_t.\label{bfgeo:moving-constraint-surface}
\end{equation}
Extracting the total derivative $\dd\chi_t/\dd t$ from $\delta\chi_t(\dot\phi)$ puts the action in canonical form,
\begin{subequations}\label{bfq:hcan}
\begin{align}
&I=\int_{t_i}^{t_f}(p_k\dot q^k-H_c)\,\dd t+[\chi_t]_{t_i}^{t_f},\label{bfq:hcan:a}\\
&H_c(\phi,t)=H(\iota_t(\phi),t)-\Theta_t(\partial_t\iota_t)+\partial_t\chi_t .\label{bfq:hcan:b}
\end{align}
\end{subequations}
We take both partial derivatives at fixed $\phi$. For a spacelike cap $\Sigma_t=X_t(\Sigma)$, we read $H$ and $\Theta_t$ from the action including the cap and joint terms in~\eqref{bfgeo:strip-action-cuts}. When $X_t$ moves, the Hamiltonian term $-\Theta_t(\partial_t\iota_t)$ includes the surface charge $-\int_{\partial\Sigma_t}\bm Q_{\partial_t X_t\circ X_t^{-1}}$ of~\eqref{bfgeo:extended-potential}.

\begin{figure}[!htbp]
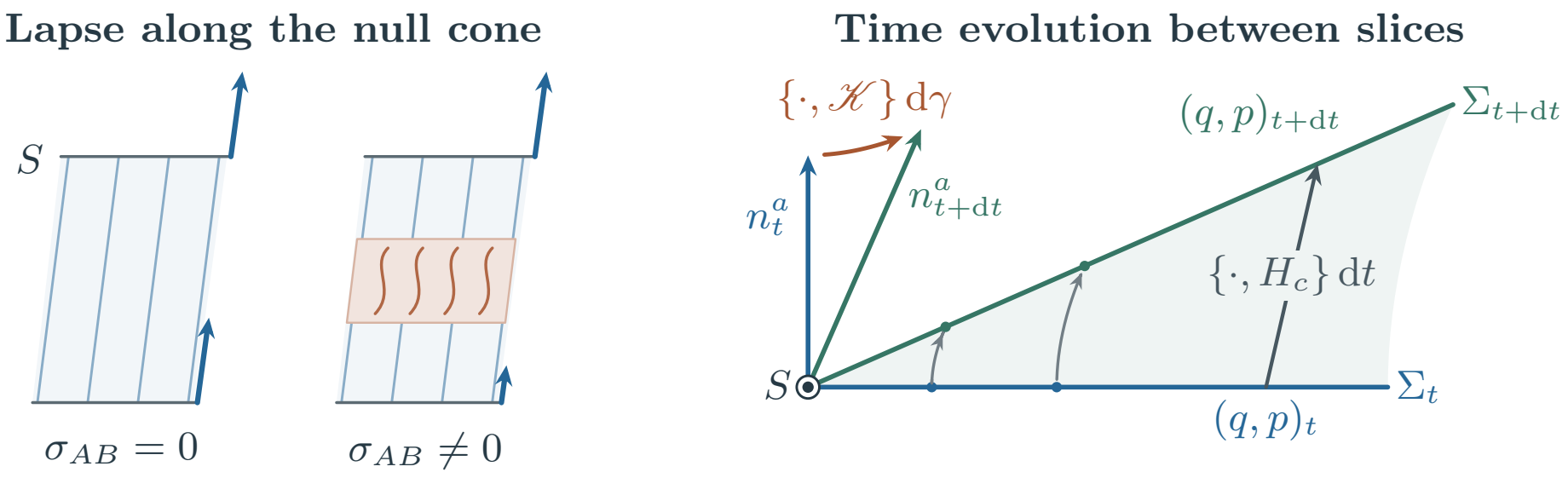

\centering
\SummaryHamiltonian
\caption[Time evolution between slices]{On the left the blue arrows show the lapse at the two cuts for the same displacement parameter. Shear in the shaded strip reduces the lapse at the inner cut even though the shear vanishes there. On the right the gray arrows follow two points as the slice moves. The central arrow shows the time evolution of the canonical pairs generated by $H_c$ in~\eqref{bfq:hcan:b}. At fixed $S$ the future unit normal undergoes a uniform boost generated by $\mathscr K$.}
\label{fig:summary-hamiltonian}
\end{figure}

To see how radiation enters even at fixed cuts, we evaluate the charge on an expanding null cone between closed cuts at $v_-$ and $v_+$. In the coordinates of~\eqref{bfgeo:full-metric}, with $\varrho=v\varrho_S$ and $\bm\epsilon_0=\dd^2y$, the divergence~\eqref{bfgeo:hamiltonian-divergence} reduces~\eqref{bfgeo:canonical-hamiltonian} to
\begin{equation}
 H_\Sigma=\frac{1}{16\pi G_N}\left[
 \int_C\bm\epsilon_0\left\{v^2\varrho_S^2
 \left(\mu_c+\frac{\partial_v F}{2N}\right)-V^A\Pi_A\right\}
 \right]_{v_-}^{v_+}.
 \label{bfgeo:eh-hamiltonian}
\end{equation}
Here we integrate the transverse connection $\mu_c$ against the area element $v^2\varrho_S^2\bm\epsilon_0$ on each cut, while the angular shift $V^A$ couples to the momentum $\Pi_A$. In the area coordinate $v$, the lapse $N$ specifies the displacement along the transverse null normal per unit coordinate time. The null constraint~\eqref{bfgeo:lapse-constraint} relates $N$ at either cut to its value at $S$,
\begin{equation}
 \partial_v\log N=\partial_v\Xi_y,\ N(v,y)=N(1,y)\exp[\Xi_y(v)-\Xi_y(1)].
 \label{bfgeo:lapse-at-the-cuts}
\end{equation}
Thus even at fixed corner normalization $N(1,y)$, the lapse depends on the shear between the cut and $S$ through the same integral~\eqref{bfgeo:integrated-focusing-main:c} that enters the clock (Figure~\ref{fig:summary-hamiltonian}). The angular divergence of shear enters~\eqref{bfgeo:shift-constraint}, which determines $\Pi_A$ and $V^A$ from their values at $S$. With initial data from the intersecting null cone, the transverse constraints~\eqref{bfgeo:complementary-constraints}--\eqref{bfgeo:temporal-reconstruction} also determine $\mu_c$ and the metric coefficient $F$ for the chosen foliation.

\subsection{The metric and dressed observables}
\label{sec:metric-bulk-observables}
\label{sec:bf-reconstruction}

For the same time translation at $S$, the lapse in~\eqref{bfgeo:lapse-at-the-cuts} yields different transverse displacements of the inner cut on solutions with different values of the focusing observable $\Xi_y(1)$. So to measure the same polarization of gravitational radiation on two such solutions, we must specify both the event and the polarization relative to the gravitational boundary data $\mathfrak h$ in each. Integration along the null generators in Section~\ref{bfgeo:null-initial-data} fixes $e^\mu|_\Sigma$, while~\eqref{bfgeo:field-equations} determines how the tetrad varies away from the null boundary. We parallel transport the local Lorentz frame $\mathfrak e$ along a geodesic from the boundary and measure the field in that frame at a specified affine parameter. If gluing places this frame in the interior, we can identify the same observable in both subregions by matching their tetrads and spin connections along the geodesic.

In the $2+2$ formulation, reconstructing the metric off $\mathscr H_p$ also requires its derivative in the transverse null direction. The intersecting null cone $\mathscr H_q$ supplies the transverse expansion and shear at $S$ (Figure~\ref{fig:summary-data}). To transport this data along $\mathscr H_p$ we use the normal connection and the transverse second fundamental form,
\begin{subequations}\label{bfgeo:transverse-null-data}
\begin{align}
&\varpi_a=q_a{}^b\beta_b=-n_\mu q_a{}^b\nabla_b\ell^\mu,\label{bfgeo:transverse-null-data:a}\\
&K^{(n)}_{ab}=q_a{}^cq_b{}^de_{d\mu}\nabla_c n^{\mu}=\sigma^{(n)}_{ab}+\frac{1}{2}\Theta_{(n)}q_{ab}.\label{bfgeo:transverse-null-data:b}
\end{align}
\end{subequations}
The Damour Navier Stokes equation~\cite{HopfmuellerFreidel2018} and cross focusing equation~\cite{Hayward1994Dynamics} are the two curvature constraints $\mathfrak C_{\ell a}=0$ and $\mathfrak C_{\ell n}=0$. In the Carrollian description of a black hole horizon~\cite{Penna:2018gfx,Donnay:2019jiz}, the Raychaudhuri and Damour Navier Stokes equations express local conservation of energy and momentum, respectively~\cite{Ciambelli:2018wre,Ciambelli:2018ojf,Petkou:2022bmz,Freidel:2022bai}.
With $\nabla_{[a}e^\mu{}_{b]}=0$, these equations and the tracefree transverse projection of $F^{\mu\nu}=\Phi^{\mu\nu}{}_{\rho\sigma}B^{\rho\sigma}$ become
\begin{subequations}
\label{bfgeo:transverse-constraints-main}
\begin{align}
&\mathfrak C_{\ell a}=q_a{}^bR_{bc}\ell^c=0,\label{bfgeo:transverse-constraints-main:a}\\
&\mathfrak C_{\ell n}=\frac{1}{2}q^{ab}R_{ab}-\Lambda=(G_{ab}+\Lambda g_{ab})\ell^an^b=0,\label{bfgeo:transverse-constraints-main:b}\\
&(\Lie_\ell+\Theta_{(\ell)})\varpi_a
=\frac12q_a{}^b\nabla_b\Theta_{(\ell)}-q_a{}^bq^{cd}\nabla_c\sigma_{db},\label{bfgeo:transverse-constraints-main:c}\\
&(\Lie_\ell+\Theta_{(\ell)})\Theta_{(n)}
=-\frac12R[q]+q^{ab}\nabla_a\varpi_b+\varpi_a\varpi^a+\Lambda,\label{bfgeo:transverse-constraints-main:d}\\
&(\Lie_\ell\sigma^{(n)}_{ab})^{\rm TF}
=\frac12\Theta_{(\ell)}\sigma^{(n)}_{ab}-\frac12\Theta_{(n)}\sigma_{ab}
+(q_a{}^cq_b{}^d\nabla_{(c}\varpi_{d)}+\varpi_a\varpi_b)^{\rm TF}.\label{bfgeo:transverse-constraints-main:e}
\end{align}
\end{subequations}
Here $\ell^a\nabla_a\ell^b=0$, $R[q]$ is the scalar curvature of $q_{ab}$, and $\mathrm{TF}$ removes the trace with $q_{ab}$. The expansion and shear in $\mathfrak C_{\ell a}=0$ are already determined by $\Lie_\ell q_{ab}$. Multiplying~\eqref{bfgeo:transverse-constraints-main:c} by the area element turns its left hand side into a total derivative along the null generator. With $\varpi_a$ known, the cross focusing equation fixes $\Theta_{(n)}$, which in turn enters the equation for $\sigma^{(n)}_{ab}$. In each case the initial data comes from the intersecting null boundary or from smoothness at a vertex. Here $R_{ab}\ell^an^b=-\Lambda$, and only $\mathfrak C_{\ell n}$ contains $\Lambda$ explicitly.

In Appendix~\ref{app:bf-hamiltonian}, we integrate~\eqref{bfgeo:field-equations} with $\nabla_{[a}e^\mu{}_{b]}=0$ to determine derivatives of $e^\mu{}_a$ and $\omega_a{}^{\mu\nu}$ transverse to $\mathscr H$ and express the lapse, angular shift and $g_{ab}$ in terms of $\mathfrak h$. One can also obtain these equations by taking the null limit of the Einstein equations on a stretched horizon~\cite{Freidel:2022vjq,Freidel:2024emv}. Smooth compatible data determines a unique local solution up to diffeomorphism \cite{Rendall1990,ChoquetBruhat10064467,Chrusciel12091971} in which changing the local Lorentz frame transforms the tetrad and spin connection together while preserving the metric.

We use the edge modes $\mathfrak e$ to dress bulk observables, specifying both the event and the frame in which we measure components of tensors~\cite{CarrozzaEcclesHoehn2024}. More specifically, geodesics launched from the corner embedding $X$ in directions fixed by $\mathfrak e$ define local coordinates, with frames parallel transported by $\omega$~\cite{DonnellyGiddings2016}. Then, under $\delta\mathfrak h$, $\chi^a=(\delta X\circ X^{-1})^a$ is the displacement of a point on a geodesic with tangent $t^a$, holding the affine parameter and its initial location and direction relative to $\mathfrak e$ fixed. This displacement obeys
\begin{equation}
t^b\nabla_b(t^c\nabla_c\chi^a)+R^a{}_{bcd}t^b\chi^ct^d=-\delta\Gamma^a{}_{bc}t^bt^c,
\label{bfgeo:geodesic-displacement}
\end{equation}
where we obtain $\delta\Gamma^a{}_{bc}$ by varying the torsion free spin connection and the tetrad as in~\eqref{bfgeo:linearized-torsion-curvature}, while the edge modes fix how the initial point and tangent to the geodesic vary. For $g_{ab}$, the variations combine as $\delta(X^*g)=X^*(\delta g+\Lie_{\chi} g)$.

For example, we choose $f^{AB}$ with compact support near the event reached at a chosen affine parameter along a null geodesic from $S$. Its components in the local Lorentz frame parallel transported along that geodesic select the polarization of the smeared radiative observable, which we evaluate on the solution with boundary data $\mathfrak h$. Under a diffeomorphism, $\delta T=\Lie_\xi T$ for a spacetime tensor $T$ and $\chi=-\xi$, so~\eqref{bfgeo:embedding-variation} yields
\begin{equation}
 \sigma_f[\mathfrak h]=\int\dd u\,\dd A\,f^{AB}\sigma_{AB},\ \delta \sigma_f[\mathfrak h]=0.\label{bfgeo:dressed-shear-variation}
\end{equation}

\subsection{Matching fields across subregions}
\label{sec:pi-matching}

The smeared shear $\sigma_f[\mathfrak h]$ in~\eqref{bfgeo:dressed-shear-variation} is invariant under a diffeomorphism because the fields and the geodesic specifying its support transform together. When that support lies in the overlap $O_{ij}=\mathscr D_i\cap\mathscr D_j$, the two causal diamonds must agree on the geodesic and on the polarization measured along it. We therefore match the local Lorentz frames parallel transported by $\omega_i$ and $\omega_j$ along the geodesic from $S$. For a vector $v$ of the transported frame and tangent $t^a$ to the geodesic, matching as in~\eqref{bfgeo:overlap-phase-space} requires the Lorentz transformation $L_{ij}$ between the two frames to satisfy
\begin{subequations}
\label{bfgeo:global-gluing}
\begin{align}
&v_i=L_{ij}v_j,\label{bfgeo:global-gluing:a}\\
&(t^a\nabla_a v)_i=L_{ij}(t^a\nabla_a v)_j=0,\label{bfgeo:global-gluing:b}\\
&B_i=L_{ij}B_jL_{ij}^{-1},\label{bfgeo:global-gluing:c}\\
&(\nabla_a L_{ij})^\mu{}_\nu=0,\label{bfgeo:global-gluing:d}
\end{align}
\end{subequations}
where $L_{ij}L_{jk}=L_{ik}$ on triple overlaps. In $\nabla_a L_{ij}$, the spin connection $\omega_i$ acts on the upper index $\mu$ and $\omega_j$ on the lower index $\nu$.
With~\eqref{bfgeo:simplicity}, $e_i=L_{ij}e_j$ implies $g_i=g_j$ on each overlap, and the induced metrics define a metric on the glued cobordism.

\begin{figure}[!htbp]
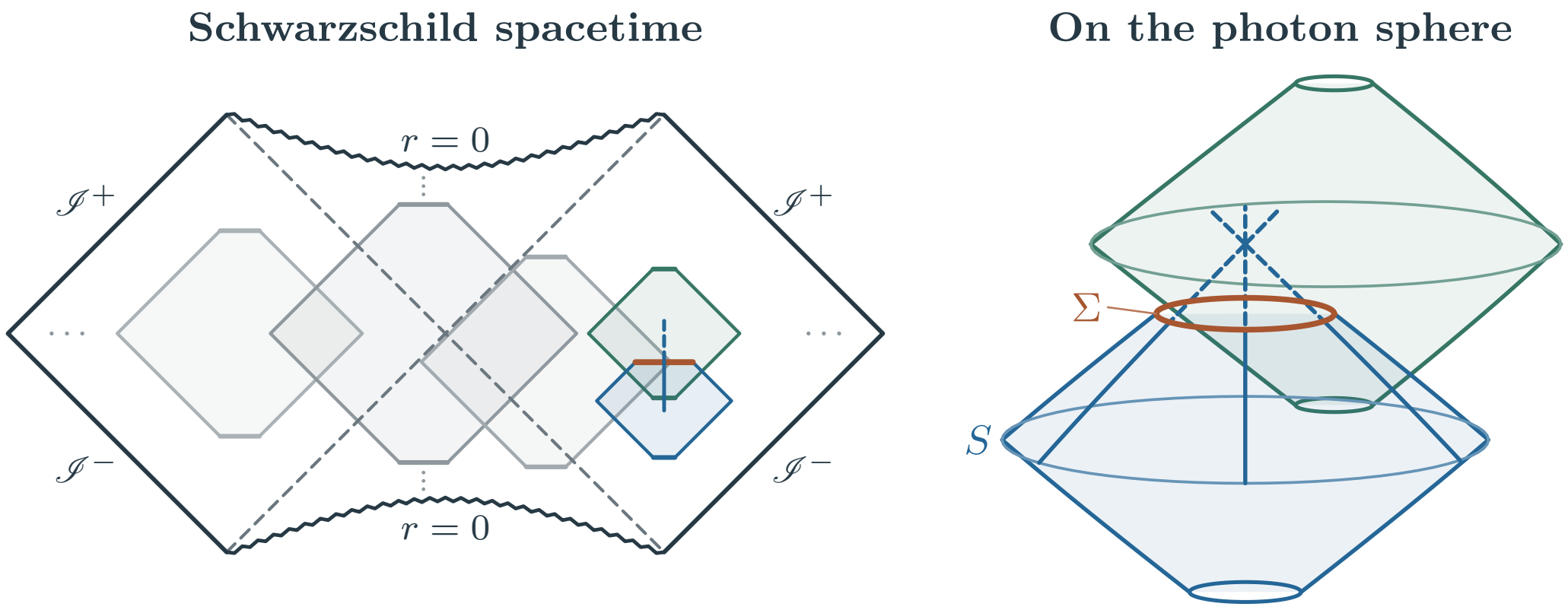

\centering
\resizebox{.82\linewidth}{!}{\SummarySpacetimeCover}
\caption[Overlapping causal diamonds in Schwarzschild spacetime]{The Penrose diagram on the left shows capped causal diamonds in maximally extended Schwarzschild spacetime. The dashed gray lines are the horizons. On the right we draw the blue and green causal diamonds on the photon sphere, with time increasing upwards in both views. Three null generators of the blue null cone reach $\Sigma$ at distinct angles. Their dashed continuations meet inside the green causal diamond, which overlaps the blue one before $\Sigma$.}
\label{fig:summary-cover}
\end{figure}

For example, circular null geodesics that start together on the Schwarzschild photon sphere in different orbital planes meet at its antipode at the same Schwarzschild time~\cite{ZenginogluGalley2012}. We truncate their null cone before they meet and choose a second causal diamond containing the caustic. Matching the tetrad and spin connection on the overlap continues the same Schwarzschild geometry across it (Figure~\ref{fig:summary-cover}).

To match the connection across $S$, we use $Y$ to pull the null frame $(\ell_L^a,n_L^a)$ from the $-$ side back to the $+$ side, where we compare it with $(\ell_R^a,n_R^a)$. For a boost of rapidity $\gamma=\Gamma_R-\Gamma_L$, the normals satisfy
\begin{equation}
\ell_R=\exp(\gamma)Y^*\ell_L,\ n_R=\exp(-\gamma)Y^*n_L.
\label{bfgeo:boosted-null-normals}
\end{equation}
It follows that
\begin{subequations}
\label{bfgeo:geometric-matching}
\begin{align}
&q_+=Y^*q_-,\label{bfgeo:geometric-matching:a}\\
&K^{(\ell)}_+=e^\gamma Y^*K^{(\ell)}_-,\label{bfgeo:geometric-matching:b}\\
&K^{(n)}_+=e^{-\gamma}Y^*K^{(n)}_-,\label{bfgeo:geometric-matching:c}\\
&\varpi_+=Y^*\varpi_-+\dd\gamma.\label{bfgeo:geometric-matching:d}
\end{align}
\end{subequations}
After pullback by $Y$, the derivative of the boost angle drops out of the projected normal curvatures and enters the normal connection. Thus a relative boost affects $\varpi_a$ at fixed $q_{ab}$. Equations~\eqref{bfgeo:geometric-matching} match the metric and both normal derivatives, so the connection agrees on the cut and the constraints in~\eqref{bfgeo:transverse-constraints-main} propagate the same data along the null generators on either side.

Importantly, these two subregions induce opposite orientations on the cut, so their boost terms enter the symplectic potential with opposite signs. After identifying the angular coordinates, let $\varrho_{S,\pm}(y)^{2}$ denote the area per unit solid angle on either side. For the boost charges we write
\begin{subequations}\label{bfgeo:two-sided-boost-charges}
\begin{align}
&\mathscr K_\pm(y)=\frac{\varrho_{S,\pm}(y)^2}{8\pi\GN},\label{bfgeo:two-sided-boost-charges:a}\\
&\mathscr K(y)=\frac{\mathscr K_+(y)+\mathscr K_-(y)}{2},\label{bfgeo:two-sided-boost-charges:b}\\
&\Delta\mathscr K(y)=\mathscr K_+(y)-\mathscr K_-(y).\label{bfgeo:two-sided-boost-charges:c}
\end{align}
\end{subequations}
Matching the induced metrics then yields
\begin{equation}
 \mathscr K_+(y)\delta\Gamma_R-\mathscr K_-(y)\delta\Gamma_L =\mathscr K(y)\,\delta\gamma,\ \Delta \mathscr K(y)=0.\label{bfgeo:two-sided-boost-potential}
\end{equation}
Consequently, a simultaneous shift of $\Gamma_R$ and $\Gamma_L$ then has zero charge and is a gauge transformation. What then remains is the relative boost $\gamma$ conjugate to $\mathscr K(y)$~\cite{CF}. Lastly, when $S$ serves only to subdivide a region, its frame is no longer part of the boundary data after gluing.

For the angular embeddings, we use $g_S$ to map a fixed cut into the first subregion and $Y$ to identify it with the second. Varying their composition yields
\begin{equation}
s_-=Y\circ g_S,\ (s_-^{-1})_*\delta s_- =(g_S^{-1})_*\bigl[(Y^{-1})_*\delta Y\bigr] +(g_S^{-1})_*\delta g_S.
\label{bfgeo:angular-embedding-variation}
\end{equation}
On pulling both angular momentum aspects back to the cut, we find that $J_{-A}$ couples to variations of $Y$. Including this contribution in $\Theta_{+\cup-}$ separates it from the charge $J_{+A}+J_{-A}$ for displacements of $g_S$,
\begin{equation}
\Theta_++\Theta_-
=\Theta_{+\cup-}
+\int_S\dd^2y\,(J_{+A}+J_{-A})[(g_S^{-1})_*\delta g_S]^A+\delta B_S.
\label{bfgeo:glued-potential-main}
\end{equation}
Here $B_S$ comes from the integrals over the corners in $I_{\partial \mathscr M}$. Once the outward charges cancel, changes of $g_S$ are gauge transformations, and the symplectic form is $\Omega_{+\cup-}=\delta\Theta_{+\cup-}$ \cite{DF,Subregion260107915}. In $\Omega_{+\cup-}$, $Y$ and $\gamma$ are coupled to the bulk degrees of freedom, while gluing cancels out the contributions of the two null boundaries to $I_{\partial \mathscr M}$.

On the two copies $S_+$ and $S_-$ of the intermediate cut, our matching conditions identify the integrands in $\delta H_\xi$. Their contributions cancel by opposite orientation, so the charge on the union has support only at its outer boundary. In Appendix~\ref{app:bf-hamiltonian} we check explicitly this cancellation for a translated cut. For adjacent subregions sweeping out $\mathscr M_i$ and $\mathscr M_j$, the internal boundary terms in the action likewise cancel, so $I_{\mathscr M}=I_{\mathscr M_i}+I_{\mathscr M_j}$.

\section{Quantization and dynamics on the null cone}
\label{sec:bf-quantization}

In Section~\ref{sec:pi-matching} we showed that the actions of neighboring subregions add when we match their gravitational boundary data. To glue amplitudes along a spacelike cut $\Sigma$, we also need the Hilbert space on the cut and the inner product between the states on its two sides. We can prepare two such states from the same incoming wavefunction by coupling different sources to the reduced action. Their inner product pairs one source operator with the adjoint of the other, so in Section~\ref{bfq:observable-algebra} we construct the algebra of observables these operators generate~\cite{HaagKastler1964,Haag1996,BuchholzFredenhagen190206062}. In Section~\ref{bfq:bulk-diffeomorphisms} we show that the cocycle under bulk diffeomorphisms vanishes, so changing the foliation in the interior of either preparation preserves the overlap.

At fixed corner area $\mathscr A$ we construct a Fock space from Gaussian wavefunctions of the incoming canonical pairs (Section~\ref{bfq:algebra-states}). To specify where a source turns off, we follow the null generators inward from $S$ by a fixed change in affine parameter. When shear slows their expansion, the cut loses less area over this displacement. In Section~\ref{bfq:fixed-area-uv} we show that this focusing yields finite overlaps for a dense set of radiative states as we include arbitrarily short wavelengths at fixed $G_N$ and $\hbar$. If we first expand one such overlap in $G_N$, we find a UV divergence at two loops.

The sources used for these overlaps preserve the area of $S$. To compute a two point function across $S$, we must also allow the first radiative insertion to change the area of the intermediate state. We therefore choose a pure extension $\omega$ of the Gaussian to the full algebra for which summing over intermediate areas reproduces the overlap we compute from the reduced action with the same sources and incoming wavefunction. In this representation the observables generate a type~I factor. Since the area operator generates relative boosts at $S$, each change in area contributes a distinct phase when we boost one insertion. At fixed incoming area we can thus read off the weight of each intermediate area from the boost dependence of the two point function (Section~\ref{bfq:state-extension}).

We next translate $S$ along its null generators with the reduced Hamiltonian $H_c$ and boost the displaced cut to measure how area fluctuations change. Focusing bounds this change to first order in the displacement, whereas in $g\phi^6$ theory and QED on a fixed background we find UV divergent corrections to fluctuations of the boost generator at fixed bare couplings (Section~\ref{bfq:full-area-uv}). Focusing also bounds the energy flux (Section~\ref{bfq:glued-null-energy}).

To glue amplitudes, we evolve the incoming state forward to $\Sigma$ and the outgoing state back to it. Focusing yields finite norms for the radiative states of Section~\ref{bfq:fixed-area-uv}, and time evolution preserves these norms by unitarity (Section~\ref{bfq:state-extension}). For any two external states of finite norm, Cauchy Schwarz then ensures that their inner product on $\Sigma$ is an absolutely convergent sum over intermediate states (Section~\ref{bfuv:completeness}).

\subsection{Algebra of observables}
\label{bfq:observable-algebra}

In Section~\ref{sec:pi-matching} we matched gravitational boundary data such that the dressed shear $\sigma_f[\mathfrak h]$ in~\eqref{bfgeo:dressed-shear-variation} specifies the same classical measurement in overlapping subregions. To compute its two point function, we need the product of operators in which the later insertion acts on the state the earlier one prepares. We first couple a source $j(t)$ to an observable $\mathcal O[\mathfrak h(t)]$ by adding $F[j]$ to the reduced action $I$. We isolate its action on $\Sigma_0$ by gluing evolution under $I+F$ to the adjoint of evolution under $I$ along $\Sigma_1$ (Figure~\ref{fig:summary-sources}). Identifying the boundary states as in Appendix~\ref{app:bf-quantization}, we obtain the source operator in a Hamiltonian representation,
\begin{equation}
 U_I(t_f,t_i)=\mathrm T\exp\!\left[-(i/\hbar)\int_{t_i}^{t_f}\widehat H(t)\,\dd t\right],\ S_I(F)=U_I^\dagger U_{I+F},\ 
 F[j]=\int j(t)\mathcal{O}[\mathfrak h(t)]\,\dd t.\label{bfq:time-evolution-and-source}
\end{equation}
Here $\mathrm T$ denotes time ordering. For this source, derivatives with respect to $j$ yield operator insertions whose matrix elements determine time ordered correlation functions~\cite{HollandsWald2005}. We calculate these derivatives in Appendix~\ref{bfq:canonical-representation}.

More generally, $F$ can contain nonlinear functions of $\mathfrak h$.
We require $I+F$ to define an action on the same reduced phase space as $I$, including its boundary terms, and with $\mathfrak S=\mathfrak T=0$ as constraints.
In canonical coordinates $\phi(t)$ we write the deformation as
$F[\phi]=\int f_t[\phi(t)]\,\dd t$, with real valued $f_t$ that vanishes outside
a specified range of time. Its dependence on fields across a spatial
boundary may be nonlocal.

With no deformation we have $S_I(0)=\widehat{1}$, while adding a constant $c$ to the action multiplies the unitary by a phase, $S_I(F+c)=\exp(ic/\hbar)S_I(F)$. Changing variables by $\phi(t)\mapsto \phi(t)+s(t)$, with $s$ vanishing
near the boundary, preserves transition amplitudes. Under this change of variables the deformation $F^s+\Delta_s I$ describes the same time evolution as $F$. In the presence of an additional deformation $N$, the time evolution operators must also compose whenever $F$ acts later than $G$. Imposing both requirements on the generators~\cite{BuchholzFredenhagen190206062,Buchholz:2020pmh}, we obtain
\begin{subequations}\label{bfq:relations}
\begin{align}
&F^s[\phi]=F[\phi+s],\label{bfq:relations:a}\\
&\Delta_s I=I^s-I,\label{bfq:relations:b}\\
&I+F^s+\Delta_s I=(I+F)^s,\label{bfq:relations:c}\\
&S_I(F^s+\Delta_s I)=S_I(F),\label{bfq:relations:d}\\
&S_I(F+N)S_I(N)^{-1}S_I(G+N)=S_I(F+G+N).\label{bfq:relations:e}
\end{align}
\end{subequations}
Differentiating~\eqref{bfq:relations:d} with respect to $s$ and $j$ yields the Schwinger Dyson equations for time ordered correlation functions, including terms from varying the insertions~\cite{HollandsWald2005}.

\begin{figure}[!htbp]
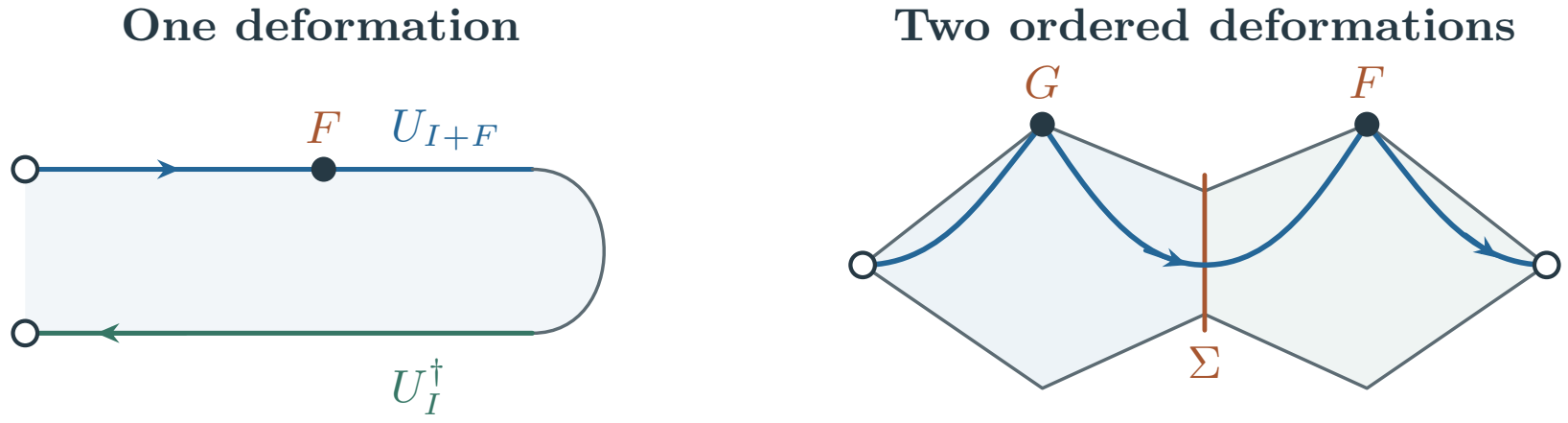

\centering
\SummarySources
\caption[Sources and their time ordering]{On the left both open ends lie on the incoming slice. We obtain $S_I(F)$ by gluing the cobordism with the source to an undeformed copy with opposite orientation. On the right $G$ prepares the state arriving at $\Sigma$. The later source $F$ acts on this state, fixing the ordering of the insertions in~\eqref{bfq:relations:e}.}
\label{fig:summary-sources}
\end{figure}

Shifts that reach the boundary can also act on its edge modes and charges. In particular, Appendix~\ref{app:physical-inner-product} keeps track of the resulting corner phases under gluing. Next, for $F$ to the future of $G$, the causal factorization in~\eqref{bfq:relations:e} reduces to $S_I(F)S_I(G)=S_I(F+G)$ at $N=0$. For a finite sum $x=\sum_j c_jg_j$ of products of these generators and their adjoints, unitarity bounds $\|x\|_{\mathrm u}=\sup_\pi\|\pi(x)\|\le\sum_j|c_j|$, where $\pi$ ranges over unitary representations satisfying~\eqref{bfq:relations:d} and~\eqref{bfq:relations:e} with the specified boundary phases. Quotienting by the kernel of this seminorm and completing defines the $C^*$~algebra $\mathcal A$.

For an additional deformation $F$ of $I+V$, time evolution in the interaction picture is described by the relative $S$ matrix~\cite{DutschFredenhagen2001},
\begin{equation}
 R_V(F)=S_I(V)^{-1}S_I(V+F).\label{bfq:relative-source-unitary}
\end{equation}
For admissible deformations this change of reference action preserves both relations~\eqref{bfq:relations:d} and~\eqref{bfq:relations:e}, with inverse obtained by using $-V$ relative to $I+V$~\cite{BuchholzFredenhagen190206062}. To use the resulting isomorphism for gravitational amplitudes we must also identify the boundary states, including their phases.

When we translate the cap, the boundary term in the reduced action of Section~\ref{bfgeo:reduced-hamiltonian-main} multiplies the matrix element $\langle q_f|U_{\mathscr M}|q_i\rangle$ by $e^{i(\chi_{t_f}-\chi_{t_i})/\hbar}$. Under the same identification of boundary data a wavefunction acquires the phase $e^{i\chi_t/\hbar}$, so the phases of the incoming ket and outgoing bra cancel out this factor. Taking $I_0$ as the reference action identifies $V_{\mathscr M}$ as its deformation,
\begin{equation}
I_0=\int p_k\dot q^k\,\dd t,\ V_{\mathscr M}=-\int H_c\,\dd t,\ U_{\mathscr M}=S_{I_0}(V_{\mathscr M}).
\label{bfq:reduced-path-integral}
\end{equation}
Using the reduced Hamiltonian~\eqref{bfq:hcan:b}, we thus obtain the Lorentzian path integral
\begin{equation}
\langle q_f|U_{\mathscr M}|q_i\rangle
=\int_{q(t_i)=q_i}^{q(t_f)=q_f}\!\mathcal Dq\,\mathcal Dp\,
\exp\!\left[\frac{i}{\hbar}\int_{t_i}^{t_f}\!\bigl(p_k\dot q^k-H_c(q,p,t)\bigr)\,\dd t\right].\label{bfq:reduced-path-integral:d}
\end{equation}
The Liouville measure $\mathcal Dq\,\mathcal Dp$ includes the canonical pairs at the corner (Appendix~\ref{app:semiclassical-determinant}), with the midpoint prescription for $H_c$ implementing Weyl ordering (Appendix~\ref{bfq:canonical-representation}).

\begin{figure}[!htbp]
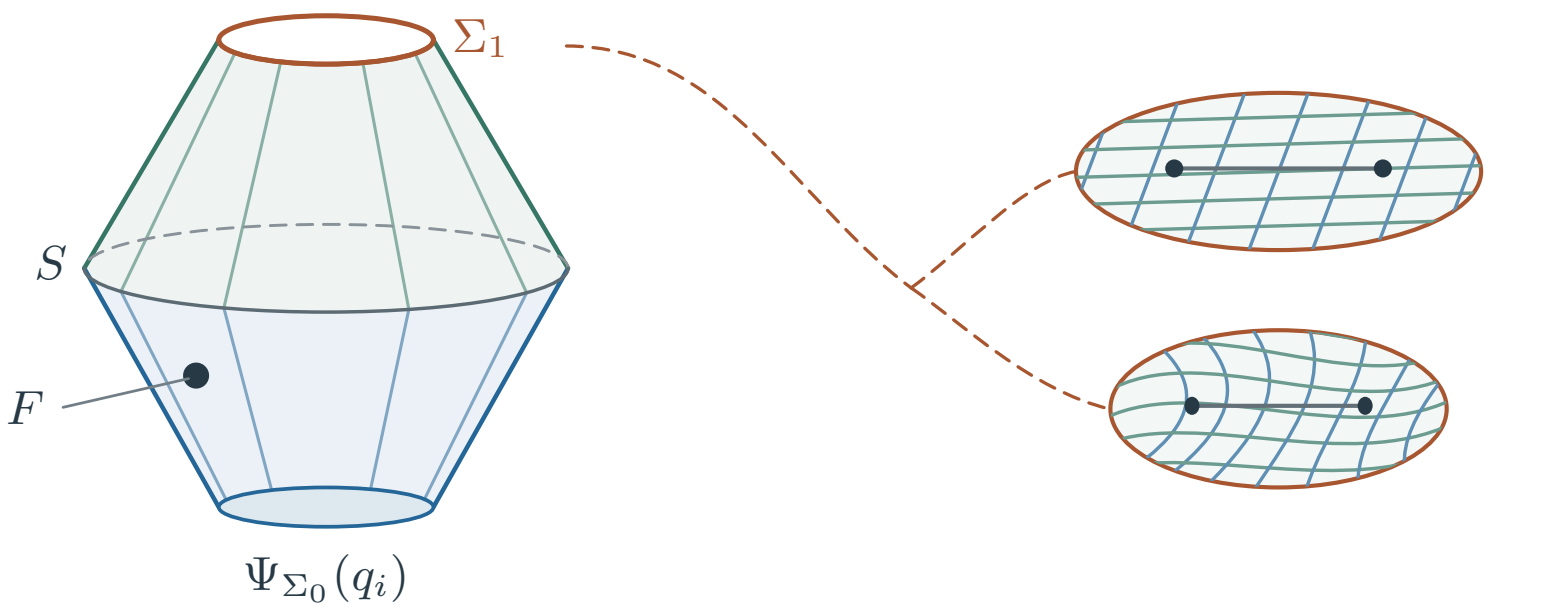

\centering
\SummaryOpenState
\caption[Preparing a wavefunction on the outgoing slice]{The incoming wavefunction on $\Sigma_0$ weights the initial configuration $q_i$, and the filled point marks the source $F$ on the null cone. The same two marked points on $\Sigma_1$ have different proper separations in configurations $q_f$ on the right. Each $q_f$ is an argument of the outgoing wavefunction in~\eqref{bfq:open-state-wavefunction}.}
\label{fig:summary-open-state}
\end{figure}

If the incoming wavefunction is specified on $\Sigma_0$ and we couple a source $F$ to the null cone, we can leave $\Sigma_1$ open. At each outgoing configuration $q_f$, the kernel prepares the wavefunction
\begin{equation}
\Psi_{\Sigma_1}(q_f)=\int\mathcal D q_i\,\langle q_f|U_{I+F}|q_i\rangle\,\Psi_{\Sigma_0}(q_i).
\label{bfq:open-state-wavefunction}
\end{equation}
Two marked points on $\Sigma_1$ can have different proper separations in different configurations $q_f$ (Figure~\ref{fig:summary-open-state}). Pairing this wavefunction with an outgoing bra then yields the transition amplitude.

\subsection{Vanishing cocycle under bulk diffeomorphisms}
\label{bfq:bulk-diffeomorphisms}

In~\eqref{bfq:open-state-wavefunction} we choose a foliation of $\mathscr M$ to evolve the incoming wavefunction from $\Sigma_0$ to $\Sigma_1$. Bulk diffeomorphism invariance requires that we obtain the same wavefunction when we change the foliation at fixed source and boundary data. A priori, even a relative phase could survive when we glue one evolution to the adjoint of the other. We therefore compare two foliations $\Sigma_t$ and $\Sigma'_t$ that agree near the source and the boundary and advance two spacelike separated portions of an intermediate slice in opposite orders.\footnote{Diffeomorphisms that rescale a null normal held fixed at the boundary can lead to an extension of the surface charge algebra~\cite{Chandrasekaran:2020wwn}. We consider local diffeomorphisms connected to the identity and leave the study of large diffeomorphisms to future work.}

For each $\phi=(q,p)$ we obtain the data $\iota_t(\phi)$ and $\iota'_t(\phi)$ on the two slices from the same local solution by the matching in~\eqref{bfgeo:cap-phase-space-gluing}. Write $I_t(\phi)$ for its oriented action between them, including the cap and joint terms in~\eqref{bfgeo:strip-action-cuts}. Varying the solution at fixed $t$ and moving the slices at fixed $\phi$ yield, respectively,
\begin{subequations}\label{bfq:refoliation-action}
\begin{align}
\delta_\phi I_t
&=(\iota'_t)^*\Theta'_t-\iota_t^*\Theta_t
=\delta_\phi(\chi'_t-\chi_t),\label{bfq:refoliation-action:fields}\\
\partial_t I_t
&=\big[\Theta'_t(\partial_t\iota'_t)-H'\big]
 -\big[\Theta_t(\partial_t\iota_t)-H\big].\label{bfq:refoliation-action:slices}
\end{align}
\end{subequations}
The common $p_k\,\delta q^k$ in~\eqref{bfgeo:moving-constraint-surface} cancels in the first line, so $\chi'_t-\chi_t=I_t+c(t)$ with $c$ independent of $\phi$. Using the second line in the reduced Hamiltonian~\eqref{bfq:hcan:b}, and keeping the source $F$ fixed in~\eqref{bfq:reduced-path-integral}, we obtain
\begin{subequations}\label{bfq:refoliation-hamiltonian}
\begin{align}
H'_c-H_c
&=(H'-H)-\Theta'_t(\partial_t\iota'_t)+\Theta_t(\partial_t\iota_t)+\partial_t(\chi'_t-\chi_t)=\partial_t(\chi'_t-\chi_t-I_t)=\dot c(t),\label{bfq:refoliation-hamiltonian:difference}\\
U'_{I+F}
&=S_{I_0}\!\left(V_{\mathscr M}+F-[c]_{t_i}^{t_f}\right)
=\exp\!\left(-\frac{i}{\hbar}[c]_{t_i}^{t_f}\right)U_{I+F}.\label{bfq:refoliation-hamiltonian:kernel}
\end{align}
The foliations coincide on $\Sigma_0$ and $\Sigma_1$, so $I_{t_i}=I_{t_f}=0$. When we glue the two evolutions, the difference $[\chi'_t-\chi_t]_{t_i}^{t_f}$ from~\eqref{bfq:hcan:a} contributes the phase that cancels the one in~\eqref{bfq:refoliation-hamiltonian:kernel},
\begin{equation}
\exp\!\left(\frac{i}{\hbar}[\chi'_t-\chi_t]_{t_i}^{t_f}\right)U_{I+F}^{\dagger}U'_{I+F}
=\exp\!\left(\frac{i}{\hbar}[c]_{t_i}^{t_f}\right)
 \exp\!\left(-\frac{i}{\hbar}[c]_{t_i}^{t_f}\right)\widehat{1}
=\widehat{1}.\label{bfq:refoliation-hamiltonian:cocycle}
\end{equation}
\end{subequations}
The two foliations thus prepare the same wavefunction on $\Sigma_1$, including its phase, at finite $G_N$ and $\hbar$. In deriving~\eqref{bfq:refoliation-hamiltonian:cocycle} we solved the null constraints before quantization. If one quantizes the Raychaudhuri constraint first, the algebra of reparametrizations of each null generator can acquire a Virasoro central charge~\cite{CiambelliFreidelLeigh2024QuantumNull,Klinger:2025hjp}. One can then use the dressing time $V$ as a quantum reference frame~\cite{Freidel:2025ous} to define a covariant normal ordering~\cite{FreidelKirklin2026}.

\subsection{Fock space representation at fixed corner area}
\label{bfq:algebra-states}

The kernel in~\eqref{bfq:open-state-wavefunction} prepares an outgoing state once we supply the incoming wavefunction on $\Sigma_0$. We choose a Gaussian wavefunction for each independent canonical pair at fixed corner area $\mathscr A$, where both radiative polarizations and the distribution of area over $S$ can fluctuate. From this we construct a Fock space representation in which we compute the overlaps of radiative and edge mode excitations at finite $G_N$.

We begin with the unital subalgebra generated by deformations of the incoming data that preserve corner area and by relative boosts at the corner. Recall that for fixed $\mathscr A$ a boost of constant rapidity $\eta$ acts by $e^{i\eta\mathscr A/(8\pi G_N\hbar)}$. On an incoming null cone whose area element increases along each null generator from $\partial\Sigma_0$ to $S$, we label the cuts by $v=\varrho/\varrho_S$ as in Section~\ref{bfgeo:null-initial-data}. In these variables the two radiative polarizations couple to the area density and angular identification at $S$ through the reduced symplectic form~\eqref{bfgeo:coupled-form}. In Appendix~\ref{bfapp:incoming-canonical-pairs} we use the conformal metric along each null generator as initial data for~\eqref{bfapp:characteristic-evolution} to construct a canonical pair $(Q^\alpha,P_\alpha)$ for each polarization. For smooth data with $\epsilon>0$ we can in turn recover the conformal metric from these pairs~\eqref{bfapp:characteristic-pairs}. By shifting the momenta conjugate to the corner variables we absorb the mixed terms in~\eqref{bfgeo:coupled-form}, so the area element on $S$ Poisson commutes with both $Q^\alpha$ and $P_\alpha$. The radiative creation operators in~\eqref{bfq:gaussian-ladder} thus preserve $\mathscr A$, whereas an insertion of boost frequency $k$ shifts it by $8\pi G_N\hbar k$~\eqref{bfq:source-boost-phase:b}.

Expanding each coordinate and its conjugate momentum in dual mode bases yields the canonical pairs $(q_j,p_j)$, $j=1,2,\ldots$.\footnote{The $L^2$ spaces on $S$ and on the interval in $v$ are separable, so products of angular and radial basis functions provide a countable basis for smearings on the incoming null cone.} To realize the canonical commutation relations $[\widehat q_j,\widehat p_j]=i\hbar\widehat{1}$ on wavefunctions $\chi_j(q_j)$, we represent $\widehat q_j$ by multiplication and $\exp(-ia\widehat p_j/\hbar)$ by the shift $\chi_j(q_j)\mapsto\chi_j(q_j-a)$. Unitarity requires translations to preserve overlaps, so intervals of equal length carry equal measure. The measure is therefore proportional to $\dd q_j$, and we set its overall constant to one.

\begin{figure}[!htbp]
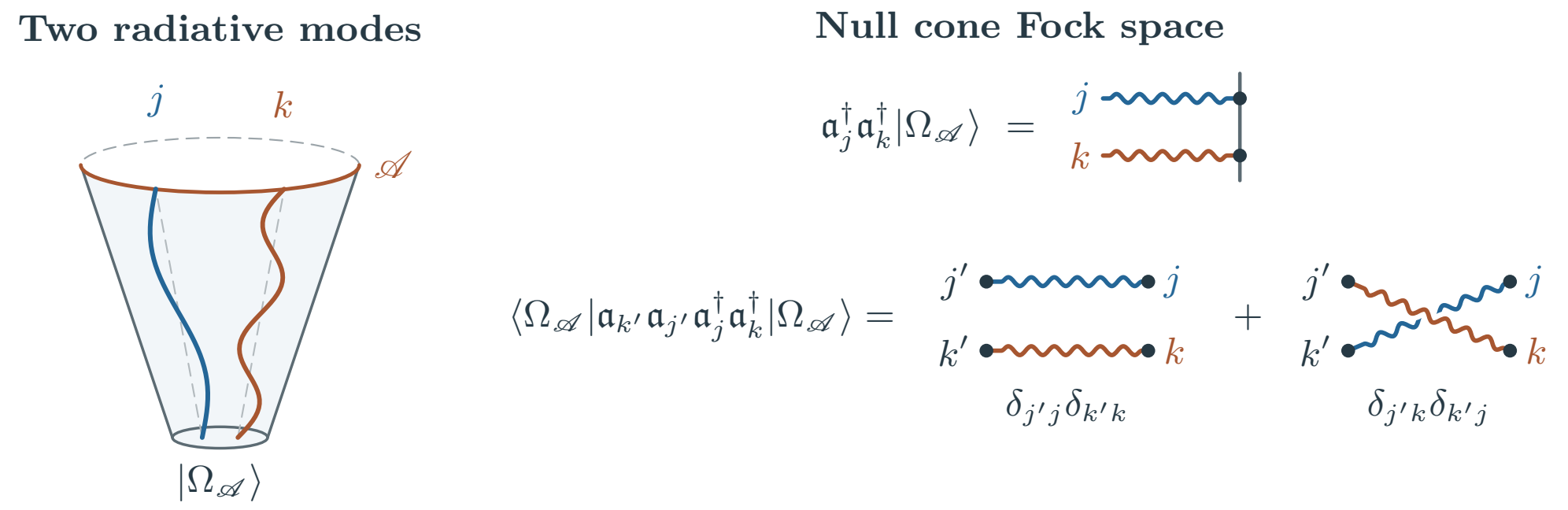

\centering
\SummaryFock
\caption[Occupation number states on the incoming null cone]{The two creation operators excite modes $j$ and $k$ on the incoming null cone. In the overlap with a state prepared in modes $j',k'$, each line carries a Kronecker delta between the two mode indices it joins. When both excitations occupy the same mode the two Wick contractions each contribute one to the squared norm, hence the factor of $1/\sqrt{2!}$ in~\eqref{bfq:fixed-area-fock:a}.}
\label{fig:summary-fock}
\end{figure}

Since the canonical pairs parametrize solutions of the full null constraints, we can work nonperturbatively in $G_N$ with a Gaussian wavefunction. Its centers and widths fix all moments of $(q_j,p_j)$ even though the conformal metric depends nonlinearly on them. The corner factors are Gaussian in the modes of $h$, the logarithm of the area density with its constant part removed, and in the real valued coordinates $\xi_k$ for the angular identification $Y$. Their definitions in~\eqref{bfapp:gaussian-corner-data} fix the total area and keep $Y$ invertible throughout the support of the wavefunction. For centers $(q_j^0,p_j^0)$ and widths $\Delta q_j>0$, normalizing each factor yields
\begin{align}
 &|\Omega_{\mathscr A}\rangle=\bigotimes_{j=1}^\infty|\Omega_j\rangle,\ 
 \langle q_j|\Omega_j\rangle
=\frac{1}{[2\pi(\Delta q_j)^2]^{1/4}}
\exp\!\left[-\frac{(q_j-q_j^0)^2}{4(\Delta q_j)^2}
+\frac{i}{\hbar}p_j^0(q_j-q_j^0)\right].
 \label{bfq:fixed-area-gaussian}
\end{align}
For a finite set $\mathcal L$ of mode indices, take $\phi_j(q_j),\chi_j(q_j)\in L^2(\mathbb R,\dd q_j)$ for $j\in \mathcal L$ and $|\phi_j\rangle=|\chi_j\rangle=|\Omega_j\rangle$ otherwise. Their overlap is
\begin{equation}
\langle\phi|\chi\rangle
 =\int_{\mathbb R^{|\mathcal L|}}\prod_{j\in \mathcal L}\bigl[\phi_j(q_j)^*\chi_j(q_j)\,\dd q_j\bigr]
 =\prod_{j\in \mathcal L}\langle\phi_j|\chi_j\rangle,\ 
 |\chi\rangle=\bigl(\underset{j\in \mathcal L}{\otimes}|\chi_j\rangle\bigr)\otimes
 \bigl(\underset{j\notin \mathcal L}{\otimes}|\Omega_j\rangle\bigr).
\label{bfq:gaussian-product-inner}
\end{equation}
We define the Hilbert space $\mathcal K_{\mathscr A}$ at fixed corner area by completing the span of these product states in the norm induced by~\eqref{bfq:gaussian-product-inner}. To find an orthonormal basis, we use $\widehat p_j=-i\hbar\partial_{q_j}$ in~\eqref{bfq:fixed-area-gaussian} to obtain annihilation and creation operators with $\mathfrak{a}_j|\Omega_j\rangle=0$ and $[\mathfrak{a}_j,\mathfrak{a}_k^\dagger]=\delta_{jk}$,
\begin{equation}
\mathfrak{a}_j=\frac{\widehat q_j-q_j^0}{2\Delta q_j}
 +\frac{i\Delta q_j}{\hbar}(\widehat p_j-p_j^0),\ \mathfrak{a}_j^\dagger=\frac{\widehat q_j-q_j^0}{2\Delta q_j}
 -\frac{i\Delta q_j}{\hbar}(\widehat p_j-p_j^0).
\label{bfq:gaussian-ladder}
\end{equation}
Acting $n$ times with $\mathfrak{a}_j^\dagger$ thus prepares a normalized occupation number state of that mode. These wavefunctions form the Hermite basis of $L^2(\mathbb R,\dd q_j)$, so we can expand each factor in~\eqref{bfq:gaussian-product-inner} in the states $|n\rangle_j$. Multiplying the expansions for $j\in \mathcal L$, we obtain the coefficient $\prod_{j\in \mathcal L}{}_j\langle n_j|\chi_j\rangle$ of the occupation number state $|\boldsymbol n\rangle$, with $n_j=0$ outside $\mathcal L$. The single mode excitations $\mathfrak{a}_j^\dagger|\Omega_{\mathscr A}\rangle$ form an orthonormal basis of the one particle space $\ell^2(\mathbb N)$, so grouping the occupation number states by $m=\sum_j n_j$ identifies $\mathcal K_{\mathscr A}$ with the bosonic Fock space\footnote{One can also construct a Fock space by quantizing the Bondi news at null infinity~\cite{Ashtekar:1981sf} or by mapping the shear along each null generator to an $\mathrm{SU}(1,1)$ current algebra~\cite{WielandPower2025}.} illustrated for two radiative modes in Figure~\ref{fig:summary-fock}
\begin{subequations}\label{bfq:fixed-area-fock}
\begin{align}
&|\boldsymbol n\rangle=\prod_{j\in \mathcal L}\frac{(\mathfrak{a}_j^\dagger)^{n_j}}{\sqrt{n_j!}}\,|\Omega_{\mathscr A}\rangle,\label{bfq:fixed-area-fock:a}\\
&0\le\Big\|\sum_{\boldsymbol n}c_{\boldsymbol n}|\boldsymbol n\rangle\Big\|^2
 =\sum_{\boldsymbol n}|c_{\boldsymbol n}|^2<\infty,\label{bfq:fixed-area-fock:b}\\
&\mathcal K_{\mathscr A}
 \simeq\bigoplus_{m=0}^{\infty}\operatorname{Sym}^m\ell^2(\mathbb N).\label{bfq:fixed-area-fock:c}
\end{align}
\end{subequations}

\begin{figure}[!htb]
\centering
\resizebox{0.96\linewidth}{!}{\begingroup
\definecolor{incoming}{HTML}{386C85}
\definecolor{reflected}{HTML}{4C7866}
\definecolor{corner}{HTML}{216F69}
\definecolor{ink}{HTML}{17333F}
\definecolor{capedge}{HTML}{677A80}
\definecolor{capfill}{HTML}{E4EAEC}
\begin{tikzpicture}[x=1cm,y=1cm,line cap=round,line join=round,>=stealth,
  text=ink,font=\fontsize{12}{14}\selectfont]
\newcommand{\incomingcone}[1]{%
  \path[fill=incoming!11]
    (-1.60,2.23)--(-.30,.27)
    arc[start angle=180,end angle=360,x radius=.30,y radius=.071]
    --(1.60,2.23)
    arc[start angle=0,end angle=-180,x radius=1.60,y radius=.235]--cycle;
  \draw[incoming!47,line width=.56pt]
    (-.26,.234)--(-1.385,2.112) (.26,.234)--(1.385,2.112);
  \draw[incoming,line width=1.06pt]
    (-.30,.27)--(-1.60,2.23) (.30,.27)--(1.60,2.23);
  \draw[corner!68,densely dashed,line width=.72pt]
    (1.60,2.23) arc[start angle=0,end angle=180,x radius=1.60,y radius=.235];
  \draw[corner,line width=1.09pt]
    (-1.60,2.23) arc[start angle=180,end angle=360,x radius=1.60,y radius=.235];
  \filldraw[fill=capfill,draw=capedge,line width=.73pt]
    (0,.27) ellipse[x radius=.30,y radius=.071];
  \node[inner sep=1pt] at (0,1.53) {#1};
}

\begin{scope}[shift={(2.30,0)}]
  \incomingcone{$|\Omega_{\mathscr A}\rangle$}
  \node[corner,anchor=east,inner sep=3pt] at (-1.64,2.23) {$S$};
  \node[incoming,anchor=east,inner sep=3pt] at (-.90,1.00) {$\mathscr H_p$};
  \node[capedge,anchor=north,inner sep=3pt] at (0,.18) {$\Sigma_0$};
\end{scope}

\definecolor{sector}{HTML}{A94A28}
\definecolor{referenceframe}{HTML}{68777C}
\draw[ink!78,->,line width=.88pt](4.02,1.28)--(5.58,1.28);
\node[anchor=south,inner sep=2.4pt] at (4.80,1.33) {$\mathfrak{a}_1^\dagger$};

\begin{scope}[shift={(8.04,0)}]
  \incomingcone{}
  \draw[incoming!43,line width=.57pt]
    (-.705,.880) arc[start angle=180,end angle=360,x radius=.705,y radius=.122]
    (-1.255,1.710) arc[start angle=180,end angle=360,x radius=1.255,y radius=.192];
  \filldraw[fill=incoming!38,draw=incoming,line width=.71pt]
    (-.4985,.7937)
    arc[start angle=-135,end angle=-105,x radius=.705,y radius=.122]
    --(-.1005,.9112)
    arc[start angle=-105,end angle=-135,x radius=.705,y radius=.122]
    --cycle;
  \filldraw[fill=incoming!38,draw=incoming,line width=.71pt]
    (-.8067,1.5629)
    arc[start angle=-130,end angle=-70,x radius=1.255,y radius=.192]
    --(.4972,1.6206)
    arc[start angle=-70,end angle=-130,x radius=1.255,y radius=.192]
    --cycle;
  \draw[incoming!87,line width=.64pt]
    (-.197,.216)
    .. controls (-.29,.40) and (-.43,.63) .. (-.4985,.7937)
    .. controls (-.59,1.02) and (-.73,1.37) .. (-.8067,1.5629)
    .. controls (-.89,1.77) and (-.96,1.92) .. (-1.023,2.049)
    (-.068,.201)
    .. controls (-.12,.40) and (-.21,.62) .. (-.1825,.7622)
    .. controls (-.14,1.00) and (.26,1.33) .. (.4292,1.5296)
    .. controls (.59,1.72) and (.75,1.91) .. (.850,2.031)
    (-.092,.330)
    .. controls (-.15,.47) and (-.36,.77) .. (-.4165,.9427)
    .. controls (-.48,1.14) and (-.65,1.45) .. (-.7387,1.6539)
    .. controls (-.82,1.84) and (-.87,1.95) .. (-.920,2.038)
    (.034,.339)
    .. controls (-.02,.50) and (-.14,.73) .. (-.1005,.9112)
    .. controls (-.04,1.15) and (.33,1.43) .. (.4972,1.6206)
    .. controls (.64,1.80) and (.85,1.97) .. (.957,2.042);
  \filldraw[fill=capfill,draw=capedge,line width=.73pt]
    (0,.27) ellipse[x radius=.30,y radius=.071];
  \begin{scope}[shift={(1.60,2.23)}]
    \draw[referenceframe,line width=.92pt,->] (0,0)--(0,.48);
    \draw[sector,line width=1.00pt,->] (0,0)--(.433,.646);
    \draw[sector!80,line width=.48pt]
      (0,.223) .. controls (.066,.223) and (.137,.244) .. (.201,.300);
    \fill[sector] (0,0) circle[radius=.031];
    \node[sector,anchor=west,inner sep=1.4pt] at (.305,.25) {$\gamma$};
  \end{scope}
  \node[anchor=north,inner sep=2.2pt] at (0,.145) {$|1,0,\ldots\rangle$};
\end{scope}

\draw[ink!78,->,line width=.88pt] (10.53,1.28)
  -- node[midway,above=1.4pt,inner sep=2.4pt] {$\cup_\Sigma$} (12.09,1.28);

\begin{scope}[shift={(14.13,1.24)}]
  \path[fill=incoming!13]
    (-1.43,0)--(-.26,-1.11)
    arc[start angle=180,end angle=360,x radius=.26,y radius=.064]
    --(1.43,0)
    arc[start angle=0,end angle=-180,x radius=1.43,y radius=.19]--cycle;
  \path[fill=reflected!15]
    (-1.43,0)--(-.26,1.11)
    arc[start angle=180,end angle=0,x radius=.26,y radius=.064]
    --(1.43,0)
    arc[start angle=0,end angle=-180,x radius=1.43,y radius=.19]--cycle;
  \draw[incoming,line width=.96pt]
    (-.26,-1.11)--(-1.43,0) (.26,-1.11)--(1.43,0);
  \draw[reflected,line width=.96pt]
    (-1.43,0)--(-.26,1.11) (1.43,0)--(.26,1.11);
  \filldraw[fill=capfill,draw=capedge,line width=.66pt]
    (0,-1.11) ellipse[x radius=.26,y radius=.064];
  \filldraw[fill=capfill,draw=capedge,line width=.66pt]
    (0,1.11) ellipse[x radius=.26,y radius=.064];
  \path[fill=capfill!88] (0,0) ellipse[x radius=1.43,y radius=.19];
  \draw[corner!64,densely dashed,line width=.70pt]
    (1.43,0) arc[start angle=0,end angle=180,x radius=1.43,y radius=.19];
  \draw[corner,line width=.90pt]
    (-1.43,0) arc[start angle=180,end angle=360,x radius=1.43,y radius=.19];
  \node[anchor=south,inner sep=2pt] at (0,1.25) {$\langle1,0,\ldots|$};
  \node[anchor=north,inner sep=2pt] at (0,-1.25) {$|\psi\rangle$};
\end{scope}
\end{tikzpicture}
\endgroup}
\caption{On the left we prepare a Gaussian state at fixed corner area. The middle null cone shows what an excitation by $\mathfrak{a}_1^\dagger$ might look like. Its radiative component shears the blue area element, while its edge mode component is a fluctuation of the relative boost angle $\gamma$ between the null frames at $S$. On the right we glue the reflected excitation to a normalized $|\psi\rangle$ along $\Sigma$. Their squared overlap is the probability $p=|\langle1,0,\ldots|\psi\rangle|^2$.}
\label{fig:gaussian-excitations}
\end{figure}
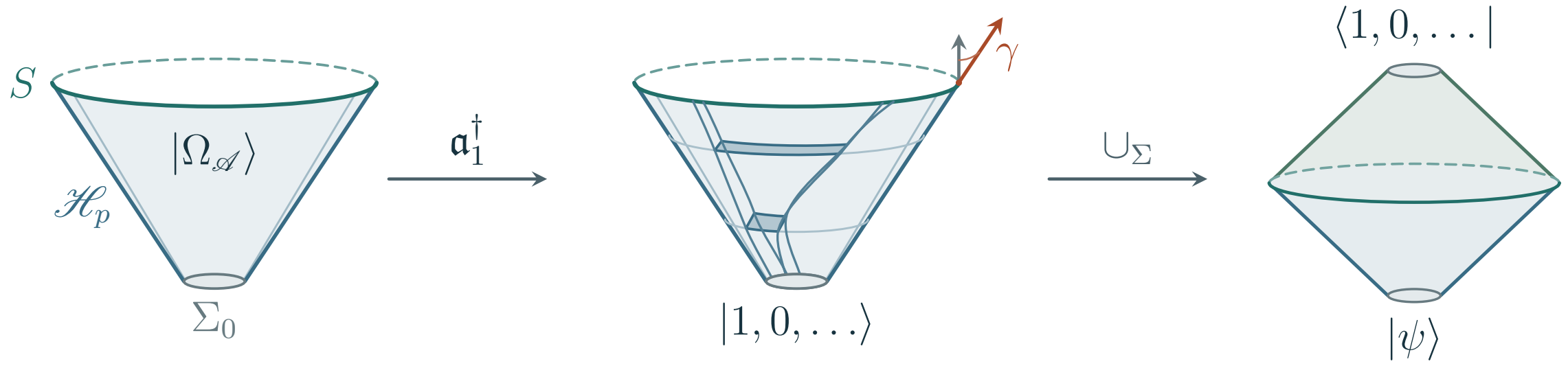

Even at fixed $\mathscr A$, variations of the area element $\dd A$ couple to the conformal metric in~\eqref{bfgeo:coupled-form}. Since $\dd A/(8\pi G_N)$ is conjugate to a local relative boost angle $\gamma=\Gamma_R-\Gamma_L$ between null frames (Section~\ref{sec:pi-matching}), one excitation $|1,0,\ldots\rangle=\mathfrak{a}_1^\dagger|\Omega_{\mathscr A}\rangle$ can change fluctuations in a radiative polarization and in differences of $\gamma$ across $S$ (Figure~\ref{fig:gaussian-excitations}).

To measure $|\langle1,0,\ldots|\psi\rangle|^2$ in a normalized state $|\psi\rangle\in\mathcal K_{\mathscr A}$, we need the projector $|1,0,\ldots\rangle\langle1,0,\ldots|$ to belong to the algebra of incoming observables. For $u,v$ in a dense subspace of $\mathcal K_{\mathscr A}$, coupling boundary sources to functions of the canonical coordinates and momenta yields $|u\rangle\langle v|$ as a strong limit of observables (Appendix~\ref{bfq:fixed-area-source-algebra}). The von Neumann algebra $\mathcal M_{\mathscr A}$ generated by the incoming observables at this area is therefore
\begin{align}
 &\mathcal M_{\mathscr A}=\mathcal B(\mathcal K_{\mathscr A}).
 \label{bfq:fixed-area-type-i-main}
\end{align}
Taking expectation values in $|\Omega_{\mathscr A}\rangle$ thus defines a pure state on the incoming subalgebra.

\subsection{UV finiteness from gravitational focusing in the fixed area Fock space}
\label{bfq:fixed-area-uv}

In Section~\ref{bfq:algebra-states} we constructed the Fock space at fixed corner area by exciting canonical modes of the incoming Gaussian state $|\Omega_{\mathscr A}\rangle$. We now determine whether coupling sources to radiation on the null cone yields finite overlaps in this space as we include arbitrarily short wavelengths. With both cuts fixed, we turn each source off near $S$ over a prescribed change in affine parameter. By the null constraint~\eqref{bfgeo:integrated-focusing-main:d}, stronger focusing along a null generator confines this part of the source to a narrower range of the area coordinate $v$.

On $|\Omega_{\mathscr A}\rangle$ this change in $v$ depends on the focusing operator $\widehat\Xi_y(v)$. To construct it we express $\partial_v\bar q_{AB}$ in canonical variables $\phi^I$ and fix the ordering of the product in~\eqref{bfgeo:integrated-focusing-main:c}. For real valued smearings $\lambda,\lambda'$ with symplectic pairing $\lambda_I J^{IJ}\lambda'_J$, the inverse reduced symplectic form $J^{IJ}$ fixes the Weyl relations,
\begin{align}
 &D(\lambda)=e^{i\lambda_I\widehat\phi^I},\ D(\lambda)D(\lambda')
 =e^{-i\hbar\lambda_IJ^{IJ}\lambda'_J/2}D(\lambda+\lambda').
 \label{bfads:rep-weyl}
\end{align}
Derivatives of $D(\lambda)$ at $\lambda=0$ along smooth smearings yield symmetrized products of the corresponding field operators on their common domain in $\mathcal K_{\mathscr A}$~\cite{Moyal1949}. For a function $f$ of the boundary fields, we express $f\circ\iota_t$ in canonical variables and use Weyl ordering wherever the smeared products are defined. Since $\mathfrak C_A\circ\iota_t=0$, the tetrad, metric and volume form in each insertion refer to the same solution $\mathfrak h=\iota_t(\phi)$.

For the focusing observable the product we need is the norm of $\partial_s\bar q_{AB}$. At each point the conformal metrics of unit determinant form the hyperbolic plane $\mathrm{SL}(2,\mathbb R)/\mathrm{SO}(2)$~\cite{FuchsReisenberger170406992,Wieland:2021vef}. Writing the conformal metric in terms of the two polarizations $z^1,z^2$, we find the metric that contracts its variations,
\begin{subequations}\label{bfq:canonical-conformal-metric}
\begin{align}
&\bar q_{AB}=\begin{pmatrix}e^{z^1}&e^{z^1}z^2\\e^{z^1}z^2&e^{-z^1}+e^{z^1}(z^2)^2\end{pmatrix},\label{bfq:canonical-conformal-metric:a}\\
&\mathsf g_{\alpha\beta}\delta z^\alpha\delta' z^\beta=\frac{1}{2}\bar q^{AC}\bar q^{BD}\delta\bar q_{AB}\,\delta'\bar q_{CD},\label{bfq:canonical-conformal-metric:b}\\
&(\mathsf g_{\alpha\beta})=\operatorname{diag}(1,\exp(2z^1)).\label{bfq:canonical-conformal-metric:c}
\end{align}
\end{subequations}
With $\widehat z^\alpha=\mathcal Q(z^\alpha)$ this metric expresses the focusing operator as a sum of squares,
\begin{equation}
\widehat \Xi_y(v)=\frac14\int_\epsilon^v s\left[
(\partial_s\widehat z^1)^\dagger\partial_s\widehat z^1
+(\partial_s\widehat z^2)^\dagger e^{2\widehat z^1}
\partial_s\widehat z^2\right]\mathrm ds,
\label{bfq:canonical-clock-integral}
\end{equation}
which is nonnegative since $e^{2\widehat z^1}$ is positive.

We first fix the state the source prepares when it acts all the way to $S$. For a smooth wave packet $b$ of either polarization, we couple the source to $\Pi_b$, the radiative momentum smeared with $b$. Appendix~\ref{bfq:fixed-area-radiative-estimate} determines the polarization $H_b^\beta(v,y)$ along each null generator for which
\begin{subequations}\label{bfq:affine-radiative-preparation}
\begin{equation}
 f^{AB}=-\frac{H_b^\beta}{16\pi G_N\varrho^2}\bar q^{AC}\bar q^{BD}
     \frac{\partial\bar q_{CD}}{\partial z^\beta},
 \label{bfq:affine-radiative-preparation:smearing}
\end{equation}
\begin{equation}
 \int\dd u\,\dd A\,f^{AB}\sigma_{AB}=\sum_j b_j p_j=\Pi_b .
 \label{bfq:affine-radiative-preparation:momentum}
\end{equation}
With this smearing fixed, we can turn the source off near $S$ and compute how the state differs from $\widehat{\Pi}_b|\Omega_{\mathscr A}\rangle$. Normalizing the affine parameter $u$ at $\partial\Sigma_0$, where $v=\epsilon>0$, we use~\eqref{bfgeo:integrated-focusing-main:d} to measure its change from $S$,
\begin{equation}
 d_y(v)=\frac{p_{\epsilon,y}}{\varrho_S(y)}[u(1,y)-u(v,y)]
 =\int_v^1 e^{\Xi_y(v')}\,\dd v' .
 \label{bfq:affine-radiative-preparation:parameter}
\end{equation}
We turn off the coupling by multiplying $f^{AB}$ by a smooth real valued $\tau(d_y)$ that vanishes near $d_y=0$ and equals one above a prescribed value. The source now couples to
\begin{equation}
 \Pi_{b,\tau}=-\frac{1}{16\pi G_N}\int_S\dd A_S\int_\epsilon^1 v^2\,\tau(d_y(v))\,
     \mathsf g_{\alpha\beta}(z)\,\partial_v z^\alpha H_b^\beta\,\dd v ,
 \ \Pi_{b,1}=\Pi_b .
 \label{bfq:affine-radiative-preparation:source}
\end{equation}
\end{subequations}
At fixed total area the area element and both polarizations can fluctuate. For each such configuration the integral for $\Pi_{b,\tau}-\Pi_b$ runs only over the part of the null cone where $\tau(d_y)\ne1$. Since $|\partial v/\partial d_y|=e^{-\Xi_y(v)}$ by~\eqref{bfq:affine-radiative-preparation:parameter}, stronger focusing confines that difference to a narrower range of $v$ (Figure~\ref{fig:fixed-area-overlap}a).

To see what this confinement implies for an overlap, we act on the same centered Gaussian wavefunction $|\Omega_{\mathscr A}\rangle$ with the two sources $\Pi_{b,\tau}$ and $\Pi_{c,\tau'}$, where $b,c$ have finite one particle norm. For either polarization we expand the canonical data~\eqref{bfapp:characteristic-pairs} in Fourier modes $\cos(\pi n r/D)$ on $0\le r\le D=(1-\epsilon)/2$ and real spherical harmonics $Y_{\ell m}(y)$ on $S$, with $0\le n\le N$, $0\le\ell\le L$ and $|m|\le\ell$. For each $n$ and polarization we use the same Gaussian width for all $\ell,m$, holding the widths and both sources fixed as we add modes. Since the derivative of the metric entering the shear also enters $\Xi_y$ quadratically, the factor $e^{-\Xi_y}$ suppresses the contribution of short wavelengths to $\Pi_{b,\tau}-\Pi_b$. In~\eqref{bfq:radiative-gaussian-tail} we bound the squared norm of the state this difference prepares by $C_0(C_1/N)^N$ uniformly in $L$, and likewise for $c,\tau'$. Gluing at fixed $G_N$ and $\hbar$ thus yields
\begin{equation}
 \lim_{N,L\to\infty}\sum_{\boldsymbol n}
 \langle\Omega_{\mathscr A}|\mathcal Q(\Pi_{b,\tau,N,L})^\dagger|\boldsymbol n\rangle
 \langle\boldsymbol n|\mathcal Q(\Pi_{c,\tau',N,L})|\Omega_{\mathscr A}\rangle
 =\sum_j\overline{b_j}\,c_j(\Delta p_j)^2 .
 \label{bfq:fixed-area-radiative-sum}
\end{equation}
Since both states have finite norm, the sum over occupation numbers in~\eqref{bfq:fixed-area-radiative-sum} converges absolutely by Cauchy Schwarz.

For the two states in~\eqref{bfq:fixed-area-radiative-sum}, we can independently shift the Gaussian centers $q_k^0$ of finitely many radiative modes. For each fixed choice of centers, the correction from turning off the source near $S$ still tends to zero in norm by~\eqref{bfq:radiative-gaussian-tail}. The overlaps then determine the inner product on the radiative Fock space, since for one fixed $j$ the limiting vectors $\widehat p_j\exp(i\sum_k t_k\widehat p_k/\hbar)|\Omega_{\mathscr A}\rangle$ have dense span. Indeed orthogonality to all of them sets every finite dimensional Fourier transform of $p_j\overline\psi$ against the Gaussian measure to zero, so $\psi=0$ almost everywhere.

In~\eqref{bfq:fixed-area-radiative-sum} we summed over intermediate states at fixed $G_N$. To see what happens if we expand in $G_N$ first, we start with the squared norm, setting $c=b$ and $\tau'=\tau$, and divide by its continuum value $\sum_j|b_j|^2(\Delta p_j)^2$ to obtain $R_{N,L}$. With $s=\mathscr A/(4G_N\hbar)$ we choose the Gaussian variance $(\Delta q_n)^2=1/(2sn)$ for the Fourier modes $n>0$ of either polarization. For the constant Fourier mode $n=0$ of each polarization we hold $s(\Delta q_0)^2$ fixed as we vary $G_N$ and use the same corner wavefunction throughout. For the wave packet $b$ and function $\tau$ of Appendix~\ref{bfq:fixed-area-radiative-estimate}, the linear insertion vanishes and the one loop pair sum is finite at fixed $L$. We expand the overlap in $G_N$ at finite $N$ and then let $N\to\infty$, obtaining
\begin{subequations}\label{bfq:fixed-area-uv-expansion}
\begin{align}
 &R_{N,L}=\frac{a_{1,N,L}}s+\frac{a_{2,N,L}}{s^2}+\cdots,\label{bfq:fixed-area-uv-expansion:series}\\
 &\lim_{N\to\infty}a_{1,N,L}=a_{1,L}>0,\label{bfq:fixed-area-uv-expansion:one-loop}\\
 &a_{2,N,L}=-c_{b,\tau,L}N^2+o_L(N^2),\ c_{b,\tau,L}>0 .\label{bfq:fixed-area-uv-expansion:two-loop}
\end{align}
\end{subequations}
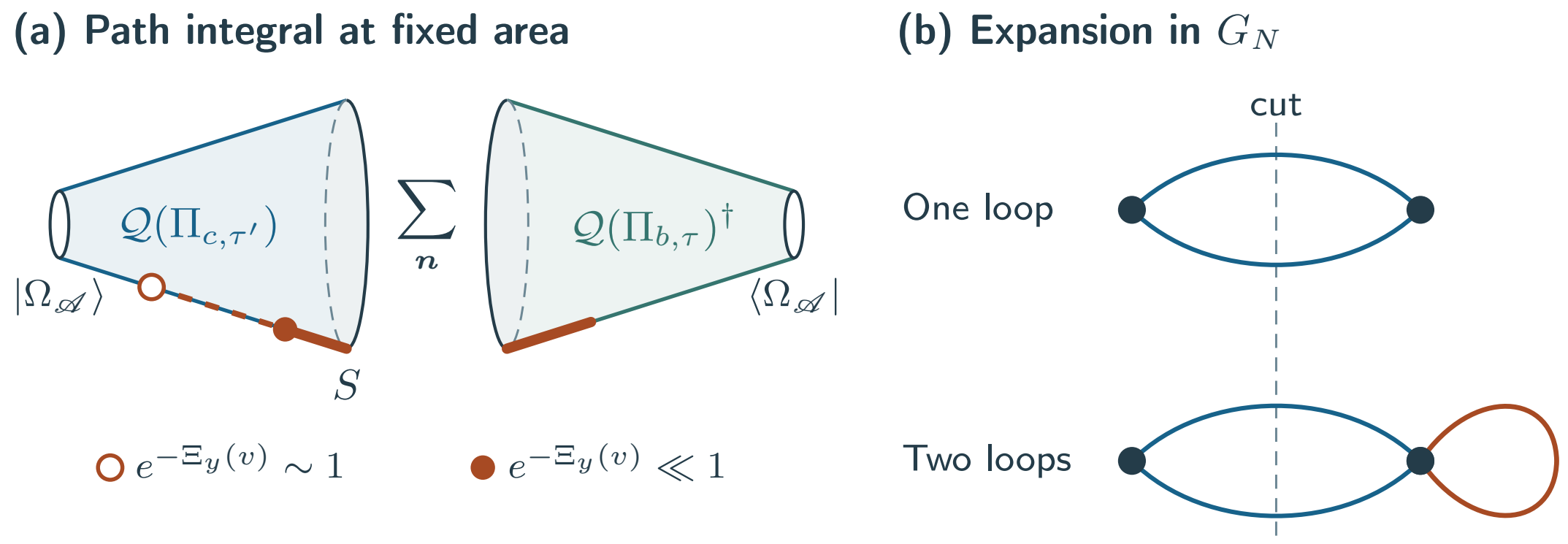
\begin{figure}[!htb]
\centering
\resizebox{.90\linewidth}{!}{%
\begingroup
\definecolor{overlapInk}{HTML}{243C49}
\definecolor{overlapCut}{HTML}{6C8794}
\definecolor{overlapBlue}{HTML}{17628A}
\definecolor{overlapTeal}{HTML}{35756F}
\definecolor{overlapRust}{HTML}{A74A23}
\begin{tikzpicture}[
  x=1cm,y=1cm,font=\sffamily\small,text=overlapInk,
  boundary/.style={draw=overlapInk,line width=.65pt},
  back edge/.style={draw=overlapCut,densely dashed,line width=.46pt},
  pair/.style={draw=overlapBlue,line width=.95pt},
  source/.style={circle,fill=overlapInk,inner sep=2.1pt}
]
\node[anchor=west,inner sep=0pt,font=\sffamily\small\bfseries] at (0,4.40)
  {(a) Path integral at fixed area};
\path[fill=overlapBlue!9!white] (.36,3.22) -- (2.51,3.90) --
  (2.51,2.04) -- (.36,2.72) -- cycle;
\path[fill=overlapTeal!9!white] (3.71,3.90) -- (5.86,3.22) --
  (5.86,2.72) -- (3.71,2.04) -- cycle;
\draw[overlapBlue,line width=.8pt] (.36,3.22) -- (2.51,3.90)
  (.36,2.72) -- (2.51,2.04);
\draw[overlapTeal,line width=.8pt] (3.71,3.90) -- (5.86,3.22)
  (3.71,2.04) -- (5.86,2.72);
\foreach \capx in {.36,5.86}{
  \draw[boundary,fill=white] (\capx,2.97) ellipse[x radius=.07,y radius=.25];
}
\foreach \facex/\front/\back in {2.51/-90/270,3.71/270/-90}{
  \fill[overlapCut!12!white] (\facex,2.97) ellipse[x radius=.16,y radius=.93];
  \draw[back edge] (\facex,3.90) arc[start angle=90,end angle=\back,x radius=.16,y radius=.93];
  \draw[boundary] (\facex,3.90) arc[start angle=90,end angle=\front,x radius=.16,y radius=.93];
}
\node[anchor=north,inner sep=1.5pt] at (2.51,1.94) {$S$};
\node at (3.11,2.97) {$\displaystyle\sum_{\boldsymbol n}$};
\draw[overlapRust,line width=1.25pt,dashed] (1.05,2.502) -- (2.51,2.04);
\draw[overlapRust,line width=2.2pt,line cap=round] (2.05,2.185) -- (2.51,2.04);
\draw[overlapRust,line width=2.2pt,line cap=round] (3.71,2.04) -- (4.34,2.24);
\draw[overlapRust,fill=white,line width=.85pt] (1.05,2.502) circle[radius=2.6pt];
\fill[overlapRust] (2.05,2.185) circle[radius=2.6pt];
\node[inner sep=0pt,font=\footnotesize] at (.36,2.43) {$|\Omega_{\mathscr A}\rangle$};
\node[inner sep=0pt,font=\footnotesize] at (5.86,2.43) {$\langle\Omega_{\mathscr A}|$};
\node[text=overlapBlue] at (1.40,2.95) {$\mathcal Q(\Pi_{c,\tau'})$};
\node[text=overlapTeal] at (4.81,2.95) {$\mathcal Q(\Pi_{b,\tau})^{\dagger}$};
\draw[overlapRust,fill=white,line width=.85pt] (.74,1.15) circle[radius=2.6pt];
\node[anchor=mid west,inner sep=0pt,font=\sffamily\footnotesize] at (.92,1.15)
  {$e^{-\Xi_y(v)}\sim 1$};
\fill[overlapRust] (3.53,1.15) circle[radius=2.6pt];
\node[anchor=mid west,inner sep=0pt,font=\sffamily\footnotesize] at (3.71,1.15)
  {$e^{-\Xi_y(v)}\ll 1$};
\begin{scope}[xshift=6.62cm]
  \node[anchor=west,inner sep=0pt,font=\sffamily\small\bfseries] at (0,4.40)
    {(b) Expansion in $G_N$};
  \draw[back edge] (2.85,.64) -- (2.85,3.75);
  \node[font=\sffamily\footnotesize,anchor=south,inner sep=0pt] at (2.85,3.78) {cut};
  \foreach \height/\description in {3.08/One loop,1.20/Two loops}{
    \begin{scope}[yshift=\height cm]
      \node[anchor=west,inner sep=0pt,font=\sffamily\footnotesize] at (.04,0) {\description};
      \draw[pair] (1.77,0) .. controls (2.32,.55) and (3.38,.55) .. (3.93,0);
      \draw[pair] (1.77,0) .. controls (2.32,-.55) and (3.38,-.55) .. (3.93,0);
      \node[source] at (1.77,0) {};
      \node[source] at (3.93,0) {};
    \end{scope}
  }
  \draw[overlapRust,line width=1.05pt] (3.93,1.20)
    .. controls (4.32,1.77) and (4.94,1.72) .. (4.95,1.20)
    .. controls (4.94,.68) and (4.32,.63) .. (3.93,1.20);
  \node[source] at (3.93,1.20) {};
\end{scope}
\end{tikzpicture}
\endgroup%
}
\caption[Gravitational focusing in the fixed area overlap]{(a) We obtain the overlap in~\eqref{bfq:fixed-area-radiative-sum} by summing over mode occupation states on the cut. Stronger focusing narrows the range of $v$ over which we turn off the source near $S$ (solid orange). (b) Expanding first in $G_N$, we obtain a finite one loop contribution from the pair crossing the cut (blue). At two loops we also contract the two factors of $\partial_v\bar q_{AB}$ in $\Xi_y(v)$ at one source (orange), yielding the $N^2$ divergence in~\eqref{bfq:fixed-area-uv-expansion:two-loop}.}
\label{fig:fixed-area-overlap}
\end{figure}

The divergence in~\eqref{bfq:fixed-area-uv-expansion:two-loop} also enters overlaps between distinct states. To obtain one from the squared norms, we take two linearly independent real valued wave packets $b,\tilde b$ of the form used for the second polarization in Appendix~\ref{bfq:fixed-area-radiative-estimate}, with equal one particle norms. With the same smearing $\tau$ we place $b+\alpha\tilde b$ and $b-\alpha\tilde b$ on opposite sides of the cut. Since~\eqref{bfq:affine-radiative-preparation} is linear in the wave packet, the cross terms cancel in the real part of their overlap. After division by its continuum value, the coefficient of $1/s^2$ is
\begin{equation}
 -\frac{c_{b,\tau,L}-\alpha^2c_{\tilde b,\tau,L}}{1-\alpha^2}\,N^2+o_L(N^2),
 \label{bfq:fixed-area-distinct-source-divergence}
\end{equation}
which diverges for any sufficiently small fixed $\alpha>0$.

In both overlaps the divergence comes from expanding the affine parameter in $\tau(d_y)$. Contracting the two factors of $\partial_v\bar q_{AB}$ in $\Xi_y(v)$ at one source leaves two fields that create the same graviton pair as the quadratic source. Each mode $n$ in the contraction contributes a factor of $n$, so interference between the two amplitudes scales as $N^2$ at two loops in~\eqref{bfq:fixed-area-uv-expansion:two-loop} and~\eqref{bfq:fixed-area-distinct-source-divergence} (Figure~\ref{fig:fixed-area-overlap}b). Keeping $G_N$ finite while summing the same intermediate states instead recovers the overlap in~\eqref{bfq:fixed-area-radiative-sum}, with $R_{N,L}\to1$ on the diagonal.\footnote{Classically, the mass of a charged shell stays finite as we shrink it to a point at fixed $G_N$, even though expanding first in $G_N$ yields divergent self energies~\cite{Arnowitt:1960zz,Arnowitt:1960zza}. Keeping $G_N$ finite likewise yielded finite electron and photon self energies to order $e^2$ in an early calculation in QED coupled to gravity~\cite{Isham:1970aw,Isham:1971es}.}

\subsection{A physical choice of state on the full algebra of observables}
\label{bfq:state-extension}

The preparations of Section~\ref{bfq:fixed-area-uv} preserve the total area $\mathscr A$. In a two point function across $S$, a radiative insertion on one side can connect the incoming state to an intermediate area different from $\mathscr A$. To choose a pure extension $\omega$ of the Gaussian state to the full algebra $\mathcal A$, we require that summing over the intermediate areas reproduce the two point function on the uncut preparation. We use the same incoming wavefunction and couple the sources to the full reduced gravitational action, so the condition also applies when an insertion excites several radiative modes.

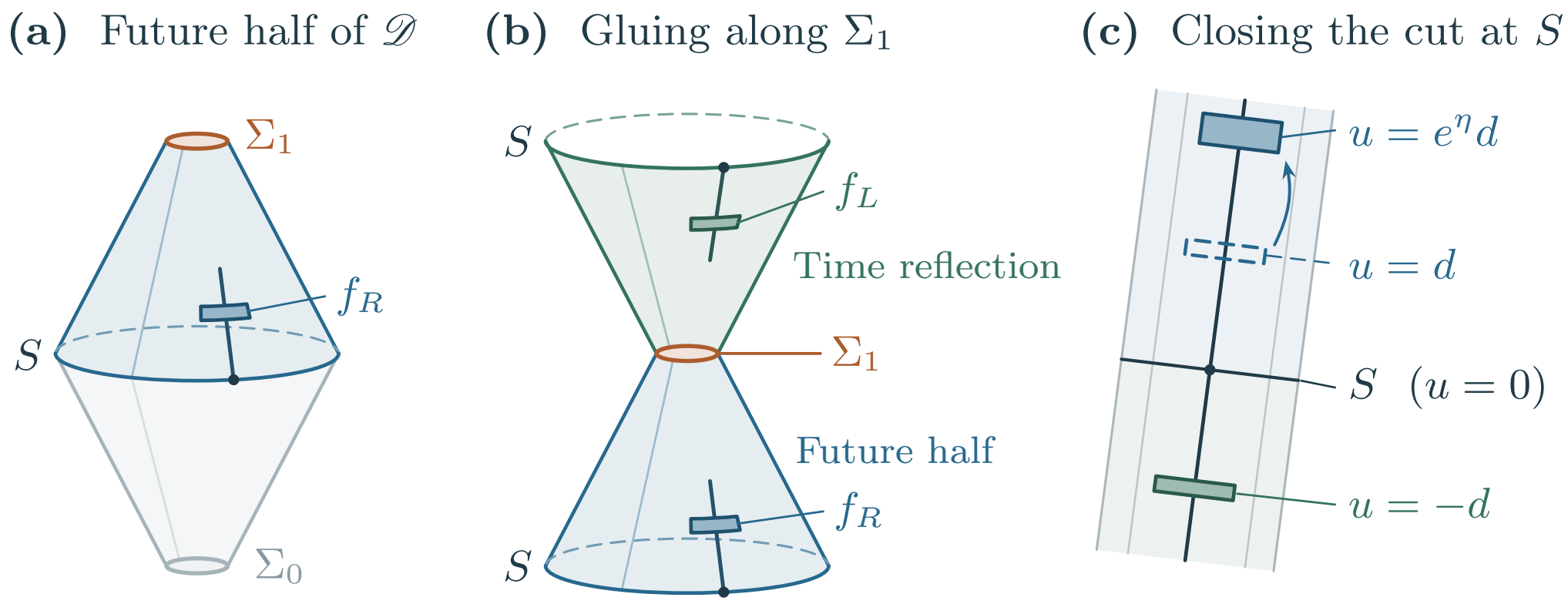
\begin{figure}[!htb]
\centering
\resizebox{.90\linewidth}{!}{%
\begingroup
\definecolor{stateInk}{HTML}{203A47}
\definecolor{stateRight}{HTML}{27688E}
\definecolor{stateLeft}{HTML}{34735E}
\definecolor{stateCap}{HTML}{AC5C2D}
\definecolor{statePast}{HTML}{A6B4BA}

\def\StateHalf#1#2{%
  \begin{scope}
    \pgfmathsetmacro{\rimY}{1.56*(#2)}
    \path[fill=#1!12] (-1.04,\rimY)--(-.225,0)
      arc[start angle=180,end angle=360,x radius=.225cm,y radius=.055cm]
      --(1.04,\rimY)
      arc[start angle=0,end angle=-180,x radius=1.04cm,y radius=.20cm]--cycle;
    \draw[#1,line width=.72pt] (-1.04,\rimY)--(-.225,0) (.225,0)--(1.04,\rimY);
    \draw[#1!46,line width=.44pt] (-.104,-.049)--(-.48,{\rimY-.178});
  \end{scope}%
}
\def\StateCorner#1#2{%
  \draw[#1!66,line width=.5pt,densely dashed] (-1.04,#2)
    arc[start angle=180,end angle=0,x radius=1.04cm,y radius=.20cm];
  \draw[#1,line width=.76pt] (-1.04,#2)
    arc[start angle=180,end angle=360,x radius=1.04cm,y radius=.20cm];
}
\def\StateCap#1{%
  \path[draw=#1,fill=#1!17,line width=.87pt] (0,0) ellipse[x radius=.225cm,y radius=.055cm];
}

\def\StateSource#1#2{%
  \begin{scope}
    \def\sourceNear{.68}
    \def\sourceFar{.74}
    \pgfmathsetmacro{\radNear}{.225+.815*\sourceNear}
    \pgfmathsetmacro{\radFar}{.225+.815*\sourceFar}
    \pgfmathsetmacro{\depthNear}{.055+.145*\sourceNear}
    \pgfmathsetmacro{\depthFar}{.055+.145*\sourceFar}
    \pgfmathsetmacro{\yNear}{(#2)*1.56*\sourceNear}
    \pgfmathsetmacro{\yFar}{(#2)*1.56*\sourceFar}
    \draw[#1!52!stateInk,line width=.80pt]
      ({(.225+.815*.52)*cos(-75)},{(#2)*1.56*.52+(.055+.145*.52)*sin(-75)})
      --({1.04*cos(-75)},{(#2)*1.56+.20*sin(-75)});
    \path[draw=#1!78!black,fill=#1!48,line width=.74pt]
      ({\radNear*cos(-88)},{\yNear+\depthNear*sin(-88)})
      arc[start angle=-88,end angle=-62,x radius=\radNear cm,y radius=\depthNear cm]
      --({\radFar*cos(-62)},{\yFar+\depthFar*sin(-62)})
      arc[start angle=-62,end angle=-88,x radius=\radFar cm,y radius=\depthFar cm]--cycle;
    \fill[stateInk] ({1.04*cos(-75)},{(#2)*1.56+.20*sin(-75)}) circle[radius=1.15pt];
  \end{scope}%
}

\begin{tikzpicture}[x=1cm,y=1cm,font=\small,text=stateInk,line cap=round,line join=round,
                    every node/.style={inner sep=1.4pt}]
  \node[anchor=west] at (0,4.42) {\textbf{(a)}\ \ Future half of $\mathscr D$};
  \node[anchor=west] at (3.50,4.42) {\textbf{(b)}\ \ Gluing along $\Sigma_1$};
  \node[anchor=west] at (7.88,4.42) {\textbf{(c)}\ \ Closing the cut at $S$};

  \begin{scope}[shift={(1.46,.51)}]
    \StateHalf{statePast}{1}
    \StateCap{statePast}
    \node[anchor=west,text=statePast!77!stateInk] at (.36,0) {$\Sigma_0$};
  \end{scope}
  \begin{scope}[shift={(1.46,3.63)}]
    \StateHalf{stateRight}{-1}
    \StateCorner{stateRight}{-1.56}
    \StateCap{stateCap}
    \StateSource{stateRight}{-1}
    \node[anchor=west,text=stateCap] at (.29,.035) {$\Sigma_1$};
    \node[anchor=east] at (-1.08,-1.57) {$S$};
    \draw[stateRight,line width=.45pt] (.37,-1.25)--(.94,-1.14);
    \node[anchor=west,text=stateRight] at (.96,-1.14) {$f_R$};
  \end{scope}

  \begin{scope}[shift={(5.06,2.07)}]
    \StateHalf{stateRight}{-1}
    \StateHalf{stateLeft}{1}
    \StateCorner{stateRight}{-1.56}
    \StateCorner{stateLeft}{1.56}
    \StateSource{stateRight}{-1}
    \StateSource{stateLeft}{1}
    \StateCap{stateCap}
    \node[anchor=east] at (-1.08,1.56) {$S$};
    \node[anchor=east] at (-1.08,-1.56) {$S$};
    \draw[stateCap,line width=.61pt] (.23,0)--(.98,0);
    \node[anchor=west,text=stateCap] at (1.00,0) {$\Sigma_1$};
    \draw[stateLeft,line width=.45pt] (.37,.96)--(1.00,1.19);
    \node[anchor=west,text=stateLeft] at (1.02,1.19) {$f_L$};
    \draw[stateRight,line width=.45pt] (.37,-1.25)--(1.00,-1.16);
    \node[anchor=west,text=stateRight] at (1.02,-1.16) {$f_R$};
    \node[anchor=west,text=stateLeft,font=\footnotesize] at (.72,.65) {Time reflection};
    \node[anchor=west,text=stateRight,font=\footnotesize] at (.74,-.71) {Future half};
  \end{scope}

  \begin{scope}[shift={(8.90,1.95)}]
    \begin{scope}[cm={1,-.12,.13,1,(0,0)}]
      \path[fill=stateLeft!9] (-.65,-1.40) rectangle (.65,0);
      \path[fill=stateRight!9] (-.65,0) rectangle (.65,1.98);
      \draw[stateInk!37,line width=.5pt] (-.65,-1.40)--(-.65,1.98)
                                         (.65,-1.40)--(.65,1.98);
      \draw[stateInk!30,line width=.40pt] (-.42,-1.40)--(-.42,1.98)
                                         (.44,-1.40)--(.44,1.98);
      \draw[stateInk,line width=.76pt] (0,-1.40)--(0,1.98);
      \draw[stateInk,line width=.67pt] (-.65,0)--(.65,0);
      \fill[stateInk] (0,0) circle[radius=1.15pt];
      \path[draw=stateLeft!78!black,fill=stateLeft!48,line width=.75pt]
         (-.29,-.925) rectangle (.29,-.815);
      \path[draw=stateRight,line width=.72pt,densely dashed]
         (-.29,.815) rectangle (.29,.925);
      \path[draw=stateRight!78!black,fill=stateRight!48,line width=.75pt]
         (-.29,1.63) rectangle (.29,1.85);
      \draw[stateRight,-{Stealth[length=3.8pt]},line width=.63pt]
         (.35,.96) to[bend right=19] (.35,1.58);
    \end{scope}
    \draw[stateLeft,line width=.45pt] (.20,-.91)--(.92,-.99);
    \node[anchor=west,text=stateLeft] at (.96,-.99) {$u=-d$};
    \draw[stateInk,line width=.43pt] (.67,-.08)--(.92,-.13);
    \node[anchor=west] at (.96,-.13) {$S$ \ $(u=0)$};
    \draw[stateRight,line width=.43pt,densely dashed] (.42,.84)--(.92,.78);
    \node[anchor=west,text=stateRight] at (.96,.78) {$u=d$};
    \draw[stateRight,line width=.45pt] (.53,1.71)--(.92,1.76);
    \node[anchor=west,text=stateRight] at (.96,1.76) {$u=e^\eta d$};
  \end{scope}
\end{tikzpicture}
\endgroup%
}
\caption[A null cone measurement of the state on the full algebra]{We prepare both states from the same incoming wavefunction $\chi$. (a) The right sided source $f_R$ lies near $S$ on the null boundary of the future half of the causal diamond (blue). (b) Gluing this half to its time reflection (green) along $\Sigma_1$ leaves two copies of $S$, with the reflected source $f_L$ near the second copy. (c) The trace identifies the two copies, so one null generator passes through the marked point on $S$ and both source regions. Boosting the right sided null frame moves $f_R$ from $u=d$ (dashed) to $u=e^\eta d$ while $f_L$ stays at $u=-d$. Our choice of $\omega$ reproduces this overlap as we vary $\eta$ to distinguish transitions between different corner areas.}
\label{fig:sec4-reflected-sources}
\end{figure}

To construct the uncut preparation we place a source near $S$ on the future null cone of the causal diamond and glue its future half to its time reflection along $\Sigma_1$ (Figure~\ref{fig:sec4-reflected-sources}). The trace identifies the two copies of $S$, where matching the fields makes the shear continuous in a freely falling local Lorentz frame~\eqref{bfgeo:geometric-matching}. Setting the affine parameter $u$ to zero there, we center the right sided smearing $f_R^{AB}$ at $u=d>0$ and its reflection $f_L^{AB}$ at $u=-d$. The sources then measure the same polarization on opposite sides of $S$ along one null generator.

To distinguish the areas between the insertions we boost the right sided null frame through rapidity $\eta$ while holding the left sided one fixed. Since the area operator $\widehat{\mathscr A}$ divided by $8\pi G_N$ generates this boost~\eqref{bfgeo:two-sided-boost-potential}, a transition from area $\mathscr A$ to area $\mathscr A'$ acquires the phase $e^{i\eta(\mathscr A'-\mathscr A)/(8\pi G_N\hbar)}$. This action on the corner edge modes is exact at finite $G_N$ and $\hbar$. We set the resolution of the area measurement by the width of the Gaussian $g$ in the conjugate boost angle $\gamma$ of the incoming wavefunction $\chi(\gamma)=g(\gamma)|\Omega_{\mathscr A}\rangle$~\eqref{app:gravarea-restriction}.

For linearized radiative insertions we can evaluate the overlap directly on the uncut preparation. We expand the two external sources to first order in the metric perturbation about a shear free background with a spherical corner of area $4\pi b^2$. For smearings of width $\ell_P\ll\Delta u\ll d\ll b$ and $b$ small compared with the curvature radius, the graviton propagator in the freely falling vacuum yields a two point function proportional to $-(u-u')^{-2}$. The boost moves the right sided source from $d$ to $e^\eta d$ and multiplies its shear by $e^\eta$, while the reflected source stays at $-d$. Their symmetrized two point function $G_{LR}(\eta)$ therefore satisfies
\begin{equation}
 \frac{G_{LR}(\eta)}{G_{LR}(0)}
 \simeq\frac{e^\eta(2d)^2}{(e^\eta d+d)^2}
 =\frac{1}{\cosh^2(\eta/2)} .
\label{bfq:reflected-radiative-overlap}
\end{equation}
We can also resolve the one graviton wave packet emitted on the right into components of boost frequency $\nu$. A component with phase $e^{i\nu\eta}$ changes the intermediate area by $8\pi G_N\hbar\nu$. Continuing it across $S$ multiplies its amplitude by $-e^{-\pi\nu}$ on the reflected half. Summing over the intermediate areas with these emission and reabsorption amplitudes reproduces~\eqref{bfq:reflected-radiative-overlap} (Appendix~\ref{app:dressed-graviton-reflection}).

When an insertion creates several radiative excitations, the boost phase still measures the change in area of the intermediate state. To determine the amplitude for each transition we use the full reduced action $I^{(2)}+V$, where $I^{(2)}$ is its quadratic part and $V$ contains the interactions. With the boundary phases absorbed in the external wavefunctions, we compare evolution with and without the source $F$ through the relative $S$ matrix $R_V(F)=S_{I^{(2)}}(V)^\dagger S_{I^{(2)}}(V+F)$~\eqref{bfq:reference-action-maps}. We require $\omega$ to reproduce the overlap of the states that the two sources prepare on $\chi$,
\begin{equation}
 \omega\!\left(S_{I^{(2)}+V}(F_L)^\dagger S_{I^{(2)}+V}(F_R)\right)
 =\langle\chi|R_V(F_L)^\dagger R_V(F_R)|\chi\rangle .
\label{bfq:interacting-source-overlap}
\end{equation}
If a source that preserves the area acts before $V$, causal factorization~\eqref{bfq:relations:e} cancels the subsequent evolution in $R_V(F)$, so we recover the overlaps of Section~\ref{bfq:algebra-states}.

When the source changes the area, we obtain the interacting two point function across $S$ by differentiating~\eqref{bfq:interacting-source-overlap} once with respect to each source. For each transition $|i\rangle\to|j\rangle$, reflection pairs $\langle j|\widehat{\sigma}_{f_R}|i\rangle$ with its complex conjugate. The transition thus enters with its squared amplitude and the sign and area weight in~\eqref{app:gravarea-reflection-interacting}, even when the insertion excites several radiative modes.

To take the continuum limit of the reflected overlap, we hold the two sources, their strengths, $\chi$ and the relative boost fixed as we include shorter wavelengths. A smooth source suppresses direct creation of two high frequency gravitons because it must supply the sum of their frequencies. A virtual pair can instead contribute through a small difference of two large frequencies. For such a pair we integrate the leading transverse term in the null constraints by parts. The values at the cuts and the integral against the slowly varying expansion stay finite as the frequency difference tends to zero~\eqref{app:gravarea-virtual-transverse-cuts}. At fixed transverse momentum the angular term contributes $O(1/N)$ from modes between $N$ and $2N$, which is summable over successive doublings~\eqref{app:gravarea-virtual-angular}. Appendix~\ref{app:dressed-graviton-reflection} uses these estimates to argue that the reflected overlap converges at fixed $G_N$ and $\hbar$.

The boost frequency likewise fixes the change in area for any product $a$ of relative $S$ matrices and their adjoints. Conjugating $a$ by $e^{i\eta\widehat{\mathscr A}/(8\pi G_N\hbar)}$ produces the boosted insertion $\alpha_\eta(a)$. For eigenstates $|i\rangle,|j\rangle$ of the area operator $\widehat{\mathscr A}$ with eigenvalues $\mathscr A_i,\mathscr A_j$, respectively, the phase of the matrix element fixes the change in area when the insertion on $|i\rangle$ has definite boost frequency $k$,
\begin{subequations}\label{bfq:source-boost-phase}
\begin{align}
&\langle j|\alpha_\eta(a)|i\rangle
=e^{i\eta(\mathscr A_j-\mathscr A_i)/(8\pi G_N\hbar)}\langle j|a|i\rangle,\label{bfq:source-boost-phase:a}\\
&\widehat{\mathscr A}\,a|i\rangle=(\mathscr A_i+8\pi G_N\hbar k)\,a|i\rangle.\label{bfq:source-boost-phase:b}
\end{align}
\end{subequations}
To select an interval $J$ of final areas, choose smooth, compactly supported Fourier transforms bounded between zero and one such that
\begin{equation}
\widetilde g_m(k)=\int_{\mathbb R}g_m(\eta)\exp(ik\eta)\,\dd\eta
\xrightarrow[m\to\infty]{}\mathbf1_J(\mathscr A_i+8\pi G_N\hbar k).
\label{bfq:source-boost-phase:c}
\end{equation}
By the spectral theorem the following limit holds in norm even for a continuous distribution of area,
\begin{equation}
\int_{\mathbb R}g_m(\eta)\alpha_\eta(a)|i\rangle\,\dd\eta
=\widetilde g_m\!\left(\frac{\widehat{\mathscr A}-\mathscr A_i}{8\pi G_N\hbar}\right)a|i\rangle
\xrightarrow[m\to\infty]{}\mathbf1_J(\widehat{\mathscr A})\,a|i\rangle.
\label{bfq:source-area-interval}
\end{equation}

For an incoming occupation number state $|\boldsymbol n\rangle$, equation~\eqref{bfq:source-area-interval} selects the component of $a|\boldsymbol n\rangle$ with final area in $J$. Summing over disjoint intervals that cover the spectrum recovers the Fock space inner product since $a^\dagger a=\widehat 1$,
\begin{equation}
 \sum_J\langle\boldsymbol m|a^\dagger\mathbf1_J(\widehat{\mathscr A})a|\boldsymbol n\rangle=\delta_{\boldsymbol m\boldsymbol n}.\label{bfq:sum-over-area-intervals}
\end{equation}
 The states $a|\boldsymbol n\rangle$ thus form an orthonormal basis of $a\mathcal K_{\mathscr A}$, with creation operators $a \mathfrak{a}_j^\dagger a^\dagger$ acting on $a|\Omega_{\mathscr A}\rangle$. A state with definite occupation numbers in this basis can have components across several areas.

To superpose preparations made with different unitary insertions $a,b$, we also need their mutual overlap, which enters the norm of the superposition. The Gaussian state fixes this overlap whenever $a^\dagger b$ belongs to the incoming subalgebra generated by relative boosts at $S$ and deformations of the incoming data that preserve corner area. For general insertions, $b$ can take the preparation to a different area and $a^\dagger$ can bring it back. By~\eqref{bfq:fixed-area-type-i-main} the Gaussian wavefunction defines a pure state on the incoming subalgebra, so we can choose a pure state $\omega$ on $\mathcal A$ with $\omega(c)=\langle\Omega_{\mathscr A}|c|\Omega_{\mathscr A}\rangle$ for every $c$ in that subalgebra~\cite{KadisonSinger1959}. Our choice must also reproduce the overlap of the reflected radiative insertions to the area resolution of the preparation. Two boundary preparations $a,b\in\mathcal A$ then have GNS inner product~\cite{Haag1996}
\begin{equation}
\langle a|b\rangle_\omega=\omega(a^\dagger b).
\label{bfq:pairing}
\end{equation}

We obtain the GNS Hilbert space $\mathcal H_\omega$ by identifying preparations whose difference has zero norm in~\eqref{bfq:pairing} and completing in this inner product. Each $a\in\mathcal A$ acts by left multiplication, $\pi_\omega(a)|b\rangle=|ab\rangle$. For incoming $a$, the map $a|\Omega_{\mathscr A}\rangle\mapsto|a\rangle$ preserves inner products by~\eqref{bfq:pairing} and embeds $\mathcal K_{\mathscr A}$ in $\mathcal H_\omega$. Taking $a=\mathbf1$ identifies $|\Omega_{\mathscr A}\rangle$ with the vector $|\mathbf1\rangle$ prepared with no insertion. Since $|a\rangle=\pi_\omega(a)|\Omega_{\mathscr A}\rangle$ and products of the source unitaries and their adjoints span a dense subalgebra, $\mathcal H_\omega$ is the closed span of the spaces $\pi_\omega(a)\mathcal K_{\mathscr A}$ as $a$ ranges over these products. In particular, $\mathcal H_\omega$ is infinite dimensional before we restrict the clock readings. Moreover, $\omega$ is pure, so the algebra acts irreducibly and
\begin{align}
 &\pi_\omega(\mathcal A)''=\mathcal B(\mathcal H_\omega).
 \label{bfq:gns-factor}
\end{align}
The von Neumann algebra generated in this representation is therefore a type~I factor. For states prepared by cobordisms, gluing one to the reflection of the other computes their overlap in~\eqref{bfq:pairing}, with reflection acting as the adjoint.

\begin{figure}[!htb]
\centering
\includegraphics[width=.77\linewidth]{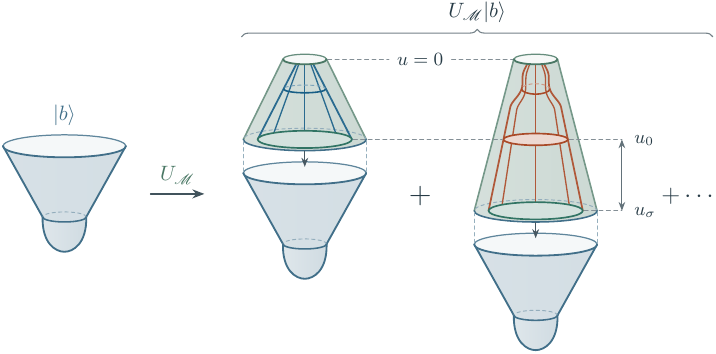}
\caption[Time evolution and the gravitational clock]{After gluing $\mathscr M$ to $|b\rangle$, the outgoing state $U_{\mathscr M}|b\rangle$ can be a superposition of different gravitational clock readings. With the initial area element and expansion held fixed, the shear free null generator (blue) reaches $S$ at $u_0$ and the sheared null generator (red) reaches the same corner area element at $u_\sigma>u_0$ by~\eqref{bfgeo:capped-clock}.}
\label{fig:outgoing-clock-readings}
\end{figure}

To time evolve $|b\rangle$ between two caps, we insert the real valued deformation $V_{\mathscr M}=-\int_{t_i}^{t_f}H_c\,\dd t$ of $I_0$. When $I_0+V_{\mathscr M}$ defines an action on the same reduced phase space as $I_0$, with associated boundary terms and constraints $\mathfrak S=\mathfrak T=0$, $U_{\mathscr M}=S_{I_0}(V_{\mathscr M})$ is a unitary in $\mathcal A$. Using~\eqref{bfq:pairing} to evaluate the norm of $U_{\mathscr M}b$, we find
\begin{align}
 &\big\|\,|U_{\mathscr M}b\rangle\big\|_\omega^2
 =\omega(b^\dagger U_{\mathscr M}^\dagger U_{\mathscr M}b)
 =\omega(b^\dagger b).
 \label{bfq:cap-evolution-norm}
\end{align}
Multiplication by $U_{\mathscr M}$ therefore preserves null states and extends by continuity to $\mathcal H_\omega$, with inverse multiplication by $U_{\mathscr M}^\dagger$. Time evolution for this displacement is therefore unitary in each GNS representation (Figure~\ref{fig:outgoing-clock-readings}). In particular, the states $U_{\mathscr M}|\boldsymbol n\rangle$ form an orthonormal basis of $U_{\mathscr M}\mathcal K_{\mathscr A}$ on the final cap.

In any GNS representation we can therefore use~\eqref{bfq:pairing} to evaluate the overlap of $|U_{\mathscr M}b\rangle$ with an outgoing state $|a\rangle$ and, by the Born rule, the probability of an outcome $0\le\widehat E\le\widehat 1$,
\begin{subequations}
\label{bfq:predictions}
\begin{align}
&\langle a|U_{\mathscr M}|b\rangle=\omega(a^\dagger U_{\mathscr M}b),
\label{bfq:amplitude}\\
&p(\widehat E\mid \mathscr M,b)=\omega(b^\dagger U_{\mathscr M}^\dagger \widehat E U_{\mathscr M}b),
\label{bfq:probability}
\end{align}
\end{subequations}
where $\omega(b^\dagger b)=1$.
Interference between states and the sum over intermediate states therefore use one and the same inner product. Moreover, inserting a complete basis of states at each intermediate cap yields~\eqref{bfq:reduced-path-integral:d}, as in Section~\ref{bfuv:completeness}.

To see how our choice of $\omega$ determines the outgoing state, we couple a source $F$ to the metric on $\Sigma_1$. For $F_L=0$ and $F_R=F$ to the future of $V$, causal factorization~\eqref{bfq:relations:e} reduces~\eqref{bfq:interacting-source-overlap} to the expectation value of $S_{I^{(2)}}(F)$ in $S_{I^{(2)}}(V)|\chi\rangle$. Although we represent the incoming state using $I^{(2)}$, a shift $s$ of the interior fields varies the action by $\Delta_s I^{(2)}+V^s-V$~\eqref{bfq:reference-action-causal-relations}. The continuum correlation functions obtained by varying the sources thus obey the Schwinger Dyson equations of the full reduced Einstein Hilbert action.

For boundary states sharply peaked about smooth classical data, $\langle\widehat g_{ab}\rangle\to g_{0\,ab}$ as $\hbar\to0$, with $G_{ab}[g_0]+\Lambda g_{0\,ab}=0$ (Section~\ref{sec:bf-classical}). Gluing joins these solutions across successive caps (Figure~\ref{fig:bf-composition}), so in this limit a clock measures proper time along its worldline in the resulting spacetime.

\subsection{UV finiteness from gravitational focusing in the full Hilbert space}
\label{bfq:full-area-uv}
\label{bfq:pair-null-translation}

In Minkowski spacetime one can quantize the gravitational boundary data of a spherical causal diamond~\cite{Bub:2024nan} and bound the variance of its corner area from below~\cite{Ciambelli:2025flo}. The extension of Section~\ref{bfq:state-extension} lets us compare areas on different cuts using the same incoming wavefunction $\chi$. If we move $S$ inward along its null generators, focusing reduces how much area they lose for a prescribed displacement. To measure the change in area fluctuations we boost the displaced cut and compare the overlap with the boost about $S$. We keep $\chi$, $G_N$ and $\hbar$ fixed throughout.

For the overlap at fixed $S$, we boost the future directed unit normal to the right sided spacelike slice through rapidity $\eta$ relative to the left sided normal. On the constraints the generator reduces to the uniform corner charge $\mathscr K=\mathscr A/(8\pi G_N)$~\eqref{bfgeo:normal-corner}. The boost $B_\eta=\exp[i\eta\widehat{\mathscr A}/(8\pi G_N\hbar)]$ therefore acts by a phase on each eigenspace of the area operator. On the wavefunction $\chi(\gamma)=g(\gamma)|\Omega_{\mathscr A}\rangle$ of Section~\ref{bfq:state-extension}, this boost translates $g$ through $\eta$. Taking the overlap of the two Gaussians yields the return probability for a finite displacement at fixed $S$,
\begin{equation}
 \big|\langle\chi|B_\eta|\chi\rangle\big|^2
 =\left|\int_{\mathbb R}\overline{g(\gamma)}g(\gamma+\eta)\,\dd\gamma\right|^2
 = e^{-\eta^2/[4(\Delta\gamma)^2]} .
\label{bfq:fixed-corner-return}
\end{equation}
Here $\Delta\gamma$ is the width of the incoming Gaussian wavefunction in the relative boost angle. The probability holds at finite $G_N$ and $\hbar$ for any $\eta$ in this representation~\eqref{app:gravarea-boost-action},~\eqref{app:gravarea-restriction}.

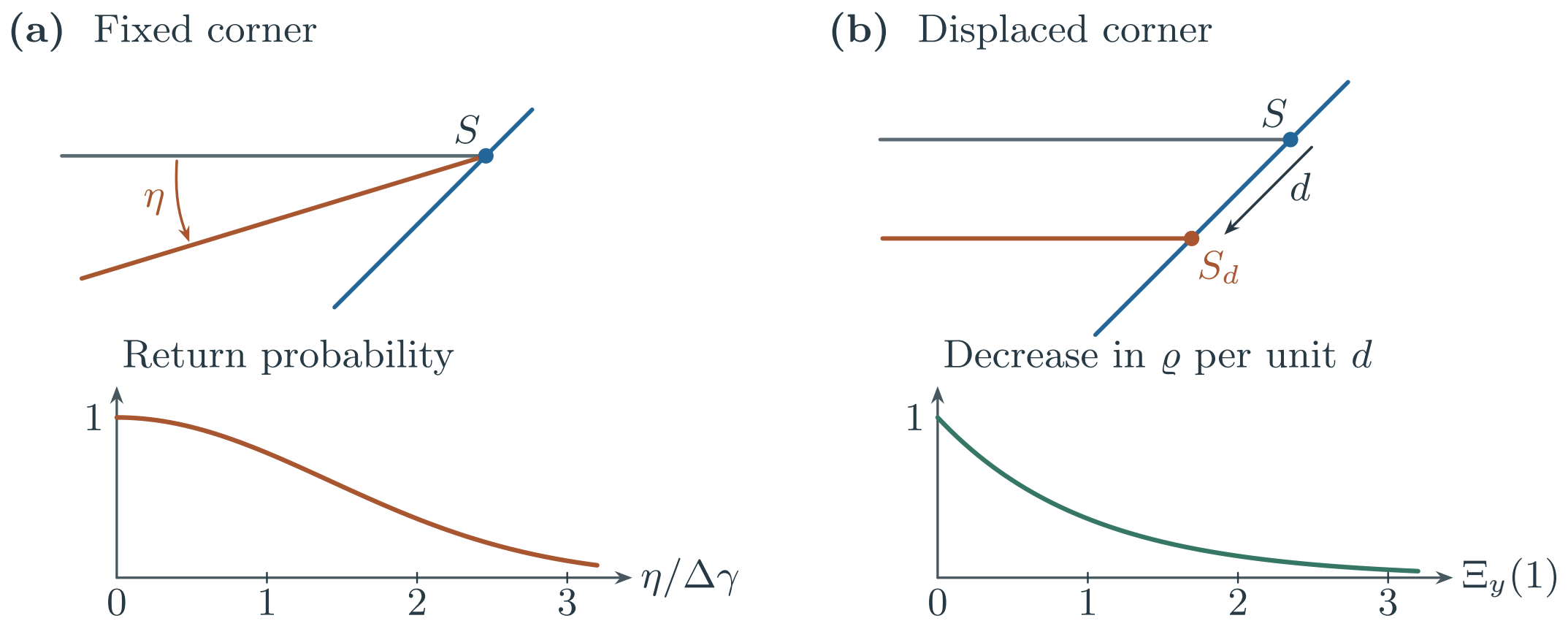
\begin{figure}[!htb]
\centering
\resizebox{.90\linewidth}{!}{%
\begingroup
\definecolor{motionInk}{HTML}{263944}
\definecolor{motionBlue}{HTML}{246697}
\definecolor{motionGreen}{HTML}{367665}
\definecolor{motionRust}{HTML}{A75630}
\begin{tikzpicture}[x=1cm,y=1cm,font=\small,text=motionInk,
  line cap=round,line join=round,every node/.style={inner sep=1.6pt},
  axis/.style={draw=motionInk!85,line width=.55pt,-{Stealth[length=4pt]}},
  old cap/.style={draw=motionInk!75,line width=.8pt},
  new cap/.style={draw=motionRust,line width=1pt},
  null generator/.style={draw=motionBlue,line width=.95pt}]
 \node[anchor=west] at (0,4.9) {\textbf{(a)}\ \ Fixed corner};
 \draw[null generator] (2.68,2.70)--(4.26,4.28);
 \draw[old cap] (.5,3.91)--(3.89,3.91);
 \draw[new cap] (.66,2.93)--(3.89,3.91);
 \draw[motionRust,-{Stealth[length=3.6pt]},line width=.6pt]
   (1.42,3.87) to[bend right=12] (1.52,3.22);
 \node[text=motionRust,anchor=east] at (1.39,3.54) {$\eta$};
 \fill[motionBlue] (3.89,3.91) circle (1.75pt);
 \node[above left] at (3.91,3.95) {$S$};
 \begin{scope}[shift={(.94,.54)}]
  \node[anchor=west] at (-.02,1.76) {Return probability};
  \draw[axis] (0,0)--(4.12,0) node[right] {$\eta/\Delta\gamma$};
  \draw[axis] (0,0)--(0,1.53);
  \draw[motionRust,line width=1.05pt] plot[domain=0:3.20,samples=72]
       ({1.20*\x},{1.28*exp(-\x*\x/4)});
  \node[left] at (-.04,1.28) {$1$}; \node[below] at (0,-.03) {$0$};
  \foreach \n in {1,2,3} {
    \draw[motionInk] ({1.20*\n},-.04)--({1.20*\n},.04);
    \node[below] at ({1.20*\n},-.03) {$\n$};
  }
 \end{scope}

 \begin{scope}[shift={(6.56,0)}]
  \node[anchor=west] at (0,4.9) {\textbf{(b)}\ \ Displaced corner};
  \draw[null generator] (2.20,2.48)--(4.22,4.50);
  \draw[old cap] (.48,4.04)--(3.76,4.04);
  \draw[new cap] (.50,3.25)--(2.97,3.25);
  \fill[motionBlue] (3.76,4.04) circle (1.75pt);
  \fill[motionRust] (2.97,3.25) circle (1.75pt);
  \node[above left] at (3.80,4.08) {$S$};
  \node[below right,text=motionRust] at (2.95,3.21) {$S_d$};
  \draw[motionInk,-{Stealth[length=3.6pt]},line width=.6pt]
    (3.93,3.98)--(3.23,3.28) node[pos=.45,right=2pt] {$d$};
  \begin{scope}[shift={(.94,.54)}]
   \node[anchor=west] at (-.02,1.76) {Decrease in $\varrho$ per unit $d$};
   \draw[axis] (0,0)--(4.12,0) node[right] {$\Xi_y(1)$};
   \draw[axis] (0,0)--(0,1.53);
   \draw[motionGreen,line width=1.05pt] plot[domain=0:3.20,samples=72]
       ({1.20*\x},{1.28*exp(-\x)});
   \node[left] at (-.04,1.28) {$1$}; \node[below] at (0,-.03) {$0$};
   \foreach \n in {1,2,3} {
     \draw[motionInk] ({1.20*\n},-.04)--({1.20*\n},.04);
     \node[below] at ({1.20*\n},-.03) {$\n$};
   }
  \end{scope}
 \end{scope}
\end{tikzpicture}
\endgroup%
}
\caption[A finite boost at fixed corner and the area lost along a null generator]{(a) The two spacelike slices meet at the same cut $S$. Their future directed unit normals differ there by a Lorentz boost of rapidity $\eta$ in the plane normal to $S$. The return probability on $\chi$ depends only on $\eta/\Delta\gamma$. (b) Moving the cut inward by $d$ along the same null generator takes $S$ to $S_d$. With the affine parameter $u$ normalized at $\partial\Sigma_0$, focusing reduces the area lost to first order in $d$ by the factor $e^{-\Xi_y(1)}$ along that null generator.}
\label{fig:sec4-corner-motions}
\end{figure}

We now move $S$ inward to $S_d$ along each incoming null generator by the same clock reading $d=p_{\epsilon,y}[u(S,y)-u(S_d,y)]$, with the affine parameter $u$ normalized at $\partial\Sigma_0$. By~\eqref{bfgeo:integrated-focusing-main:b}, $\partial_d\varrho|_{d=0}=-e^{-\Xi_y(1)}$. Since $\dd A=\varrho_S^2\bm\epsilon_0$, the change in corner area under this null translation satisfies
\begin{equation}
 -\left.\partial_d\mathscr A(S_d)\right|_{d=0}
 =2\int_S \varrho_S(y)e^{-\Xi_y(1)}\bm\epsilon_0
 \le 4\sqrt{\pi\mathscr A} .
\label{bfq:null-translation-area}
\end{equation}
The inequality follows from $\Xi_y(1)\ge0$ and $\int_S\bm\epsilon_0=4\pi$. A round shear free null cone saturates it. Shear reduces the area lost along each null generator by the factor $e^{-\Xi_y(1)}$ to first order in $d$ (Figure~\ref{fig:sec4-corner-motions}).

To measure the area operator $\widehat{\mathscr A}(S_d)$ on the same $\chi$, we insert $B_\eta$ on the displaced cut. Let $U_{d,N}$ implement this null translation with the spacelike caps and their joints carried along. With $N$ incoming canonical modes, write $J_N=-\partial_d\mathscr A(S_d)|_{d=0}/(8\pi G_N)$. We hold the incoming Gaussian wavefunction fixed as we include shorter wavelengths. Since $\mathscr K$ is conjugate to the uniform relative boost angle $\gamma$, Hamilton's equation for~\eqref{bfq:hcan:b} yields $\partial_\gamma H_c=J_N$. As $J_N$ is independent of $\gamma$, $B_\eta\mathcal Q(H_c)B_{-\eta}=\mathcal Q(H_c+\eta J_N)$. Differentiating the overlap at $d=0$ thus yields
\begin{equation}
 C_{\eta,N}(d)=\langle\chi|U_{d,N}^\dagger B_\eta U_{d,N}|\chi\rangle,\ 
 \left.\partial_d C_{\eta,N}(d)\right|_{d=0}
 =-\frac{i\eta}{\hbar}\left\langle e^{i\eta\mathscr K/\hbar}J_N\right\rangle_{W_\chi}.
\label{bfq:displaced-corner-overlap}
\end{equation}
Here we integrate $e^{i\eta\mathscr K/\hbar}J_N$ over the canonical pairs $(q,p)$ with the Wigner function $W_\chi$ of the incoming Gaussian wavefunction. Since $W_\chi\ge0$,~\eqref{bfq:null-translation-area} bounds $|\partial_d C_{\eta,N}(0)|$ by $|\eta|\sqrt{\langle\widehat{\mathscr A}\rangle_\chi/(4\pi)}/(G_N\hbar)$ for any $N$. Along null generators with $\Xi_y(1)\ge X$, the area lost is further suppressed by $e^{-X}$.\footnote{By~\eqref{bfgeo:integrated-focusing-main:c}, if $\Xi_y(1)\le X$, the conformal metric can change by at most $O(\sqrt{X/N})$ between cuts separated by $1/N$ in $v$. On a shear free background the Fourier modes in the affine parameter $u$ with $N/2<k\le N$ vary on this scale. With $(\Delta q_k)^2=1/(2s_b k)$ as in~\eqref{bfq:translated-pair-state}, their variances sum to $(\log 2)/(2s_b)+O(N^{-1})$. This suggests that the probability of $\Xi_y(1)\le X$ decreases with $N$. If it tends to zero for the full conformal metric at each fixed $X$ and almost every $y$, the derivative in~\eqref{bfq:displaced-corner-overlap} converges.}

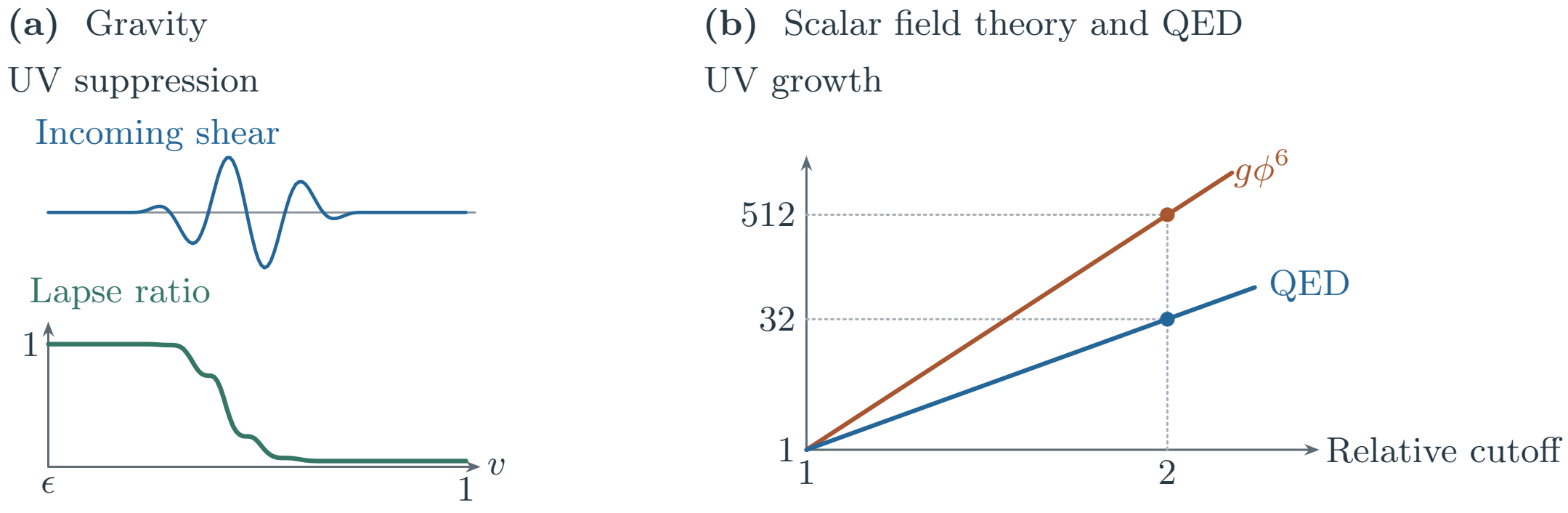
\begin{figure}[!htb]
\centering
\resizebox{.90\linewidth}{!}{%
\begingroup
\definecolor{varianceInk}{HTML}{263944}
\definecolor{varianceBlue}{HTML}{246697}
\definecolor{varianceGreen}{HTML}{367665}
\definecolor{varianceRust}{HTML}{A75630}
\begin{tikzpicture}[x=1cm,y=1cm,font=\small,text=varianceInk,
  line cap=round,line join=round,every node/.style={inner sep=1.35pt},
  axis/.style={draw=varianceInk!75,line width=.55pt,-{Stealth[length=3.8pt]}},
  momentum/.style={line width=1pt,-{Stealth[length=4.2pt]}},
  net/.style={draw=varianceRust,line width=1.1pt,-{Stealth[length=3.7pt]}}]
 \node[anchor=west] at (-.08,4.40) {\textbf{(a)}\ \ Gravity};
 \node[anchor=west] at (-.08,3.90) {UV suppression};
 \node[anchor=west,text=varianceBlue] at (.17,3.44) {Incoming shear};
 \draw[varianceInk!55,line width=.5pt] (.34,2.76)--(4.10,2.76);
 \draw[varianceBlue,line width=.85pt] plot coordinates {(0.340,2.760) (0.368,2.760) (0.395,2.760) (0.423,2.760) (0.450,2.760) (0.478,2.760) (0.506,2.760) (0.533,2.760) (0.561,2.760) (0.588,2.760) (0.616,2.760) (0.644,2.760) (0.671,2.760) (0.699,2.760) (0.726,2.760) (0.754,2.760) (0.782,2.760) (0.809,2.760) (0.837,2.760) (0.864,2.760) (0.892,2.760) (0.920,2.760) (0.947,2.760) (0.975,2.760) (1.002,2.760) (1.030,2.760) (1.058,2.760) (1.085,2.760) (1.113,2.761) (1.140,2.763) (1.168,2.768) (1.196,2.776) (1.223,2.786) (1.251,2.797) (1.278,2.807) (1.306,2.813) (1.334,2.813) (1.361,2.805) (1.389,2.786) (1.416,2.756) (1.444,2.717) (1.472,2.671) (1.499,2.621) (1.527,2.572) (1.554,2.530) (1.582,2.500) (1.610,2.487) (1.637,2.495) (1.665,2.526) (1.692,2.581) (1.720,2.656) (1.748,2.749) (1.775,2.852) (1.803,2.958) (1.830,3.058) (1.858,3.144) (1.886,3.207) (1.913,3.242) (1.941,3.243) (1.968,3.211) (1.996,3.145) (2.024,3.051) (2.051,2.934) (2.079,2.804) (2.106,2.671) (2.134,2.545) (2.162,2.435) (2.189,2.350) (2.217,2.295) (2.244,2.273) (2.272,2.286) (2.300,2.331) (2.327,2.402) (2.355,2.494) (2.382,2.597) (2.410,2.703) (2.438,2.803) (2.465,2.891) (2.493,2.960) (2.520,3.007) (2.548,3.030) (2.576,3.031) (2.603,3.012) (2.631,2.977) (2.658,2.932) (2.686,2.882) (2.714,2.833) (2.741,2.789) (2.769,2.753) (2.796,2.727) (2.824,2.711) (2.852,2.706) (2.879,2.708) (2.907,2.716) (2.934,2.726) (2.962,2.737) (2.990,2.747) (3.017,2.754) (3.045,2.758) (3.072,2.760) (3.100,2.760) (3.128,2.760) (3.155,2.760) (3.183,2.760) (3.210,2.760) (3.238,2.760) (3.266,2.760) (3.293,2.760) (3.321,2.760) (3.348,2.760) (3.376,2.760) (3.404,2.760) (3.431,2.760) (3.459,2.760) (3.486,2.760) (3.514,2.760) (3.542,2.760) (3.569,2.760) (3.597,2.760) (3.624,2.760) (3.652,2.760) (3.680,2.760) (3.707,2.760) (3.735,2.760) (3.762,2.760) (3.790,2.760) (3.818,2.760) (3.845,2.760) (3.873,2.760) (3.900,2.760) (3.928,2.760) (3.956,2.760) (3.983,2.760) (4.011,2.760) (4.020,2.760)};
 \node[anchor=west,text=varianceGreen] at (.12,2.05) {Lapse ratio};
 \draw[axis] (.34,.52)--(4.15,.52) node[right] {$v$};
 \draw[axis] (.34,.52)--(.34,1.80);
 \node[left] at (.32,1.60) {$1$};
 \draw[varianceGreen,line width=1.15pt] plot coordinates {(0.340,1.600) (0.368,1.600) (0.395,1.600) (0.423,1.600) (0.450,1.600) (0.478,1.600) (0.506,1.600) (0.533,1.600) (0.561,1.600) (0.588,1.600) (0.616,1.600) (0.644,1.600) (0.671,1.600) (0.699,1.600) (0.726,1.600) (0.754,1.600) (0.782,1.600) (0.809,1.600) (0.837,1.600) (0.864,1.600) (0.892,1.600) (0.920,1.600) (0.947,1.600) (0.975,1.600) (1.002,1.600) (1.030,1.600) (1.058,1.600) (1.085,1.600) (1.113,1.600) (1.140,1.600) (1.168,1.600) (1.196,1.600) (1.223,1.600) (1.251,1.599) (1.278,1.598) (1.306,1.596) (1.334,1.594) (1.361,1.593) (1.389,1.592) (1.416,1.592) (1.444,1.591) (1.472,1.588) (1.499,1.579) (1.527,1.561) (1.554,1.531) (1.582,1.492) (1.610,1.446) (1.637,1.401) (1.665,1.363) (1.692,1.338) (1.720,1.326) (1.748,1.323) (1.775,1.322) (1.803,1.309) (1.830,1.273) (1.858,1.210) (1.886,1.124) (1.913,1.031) (1.941,0.946) (1.968,0.878) (1.996,0.831) (2.024,0.804) (2.051,0.791) (2.079,0.788) (2.106,0.788) (2.134,0.782) (2.162,0.766) (2.189,0.739) (2.217,0.705) (2.244,0.672) (2.272,0.644) (2.300,0.623) (2.327,0.610) (2.355,0.602) (2.382,0.599) (2.410,0.598) (2.438,0.598) (2.465,0.597) (2.493,0.595) (2.520,0.591) (2.548,0.587) (2.576,0.582) (2.603,0.578) (2.631,0.575) (2.658,0.573) (2.686,0.572) (2.714,0.571) (2.741,0.571) (2.769,0.571) (2.796,0.571) (2.824,0.571) (2.852,0.571) (2.879,0.571) (2.907,0.571) (2.934,0.571) (2.962,0.571) (2.990,0.571) (3.017,0.571) (3.045,0.571) (3.072,0.571) (3.100,0.571) (3.128,0.571) (3.155,0.571) (3.183,0.571) (3.210,0.571) (3.238,0.571) (3.266,0.571) (3.293,0.571) (3.321,0.571) (3.348,0.571) (3.376,0.571) (3.404,0.571) (3.431,0.571) (3.459,0.571) (3.486,0.571) (3.514,0.571) (3.542,0.571) (3.569,0.571) (3.597,0.571) (3.624,0.571) (3.652,0.571) (3.680,0.571) (3.707,0.571) (3.735,0.571) (3.762,0.571) (3.790,0.571) (3.818,0.571) (3.845,0.571) (3.873,0.571) (3.900,0.571) (3.928,0.571) (3.956,0.571) (3.983,0.571) (4.011,0.571) (4.020,0.571)};
 \node[below] at (.34,.48) {$\epsilon$}; \node[below] at (4.02,.48) {$1$};

 \begin{scope}[shift={(7.02,.67)}]
  \node[anchor=west] at (-.97,3.73) {\textbf{(b)}\ \ Scalar field theory and QED};
  \node[anchor=west] at (-.96,3.23) {UV growth};
  \draw[axis] (0,0)--(4.52,0) node[right] {Relative cutoff};
  \draw[axis] (0,0)--(0,2.59);
  \draw[varianceInk!40,densely dotted,line width=.55pt] (0,2.07)--(3.18,2.07)--(3.18,0);
  \draw[varianceInk!40,densely dotted,line width=.55pt] (0,1.15)--(3.18,1.15);
  \draw[varianceRust,line width=1.1pt] (0,0)--(3.75,2.44);
  \draw[varianceBlue,line width=1.1pt] (0,0)--(3.95,1.43);
  \fill[varianceRust] (3.18,2.07) circle (1.85pt);
  \fill[varianceBlue] (3.18,1.15) circle (1.85pt);
  \node[anchor=west,text=varianceRust] at (3.71,2.48) {$g\phi^6$};
  \node[anchor=west,text=varianceBlue] at (4.02,1.44) {QED};
  \node[left] at (-.04,0) {$1$}; \node[left] at (-.04,1.15) {$32$};
  \node[left] at (-.04,2.07) {$512$};
  \node[below] at (0,-.04) {$1$}; \node[below] at (3.18,-.04) {$2$};
 \end{scope}
\end{tikzpicture}
\endgroup%
}
\caption[Gravitational focusing and UV growth]{We move the cut inward and measure the change in fluctuations of the boost generator. (a) In gravity $N(\epsilon,y)/N(v,y)=e^{-\Xi_y(v)}$~\eqref{bfgeo:lapse-at-the-cuts}. The same factor at $v=1$ bounds the change in area fluctuations~\eqref{bfq:null-translation-area-variance}. (b) For a spherical cut of radius $b$ on a fixed background, the contributions to $-\partial_d\operatorname{Var}_\chi K_{b-d}(d)|_{d=0}$ grow as $g^2b^4\Lambda_{\rm UV}^9$ in scalar field theory and $e^2b^4\Lambda_{\rm UV}^5$ in QED~\eqref{app:matter-boost-powers}. We normalize each curve at the reference momentum cutoff and both axes are logarithmic.}
\label{fig:sec4-boost-variance}
\end{figure}

The expectation value and variance of $\widehat{\mathscr A}(S_d)$ enter $\log C_{\eta,N}(d)$ in~\eqref{bfq:displaced-corner-overlap} at first and second order in $\eta$, respectively, so
\begin{equation}
 \frac{1}{(8\pi G_N)^2}\left.\partial_d\operatorname{Var}_\chi[\widehat{\mathscr A}(S_d)]\right|_{d=0}
 =-\hbar^2\left.\partial_\eta^2\partial_d\log|C_{\eta,N}(d)|\right|_{\eta=d=0}
 =-2\operatorname{Cov}_{W_\chi}(\mathscr K,J_N).
\label{bfq:null-translation-area-variance}
\end{equation}
For an incoming wavefunction with finite area variance,~\eqref{bfq:null-translation-area} and the Cauchy Schwarz inequality thus bound~\eqref{bfq:null-translation-area-variance} uniformly in $N$. In Appendix~\ref{app:dressed-graviton-scalar} we compute the same change in the connected two point function of the boost generator for a spherical cut on a fixed Minkowski background. Even on the free vacuum, the contributions from six scalar quanta in $g\phi^6$ theory and from a photon and an electron positron pair in QED diverge at fixed bare couplings~\eqref{app:matter-boost-powers} (Figure~\ref{fig:sec4-boost-variance}). In gravity the short wavelength shear itself reduces the area lost under the null translation.

To see how the shear of individual gravitons enters~\eqref{bfq:null-translation-area}, we emit two gravitons on the incoming null cone and reabsorb them after boosting the displaced cut. We select one polarization and one real valued spherical harmonic of the conformal metric and denote the Fourier modes along the null generators by $k$. For the incoming Gaussian wavefunction $\chi$ in~\eqref{app:gravarea-restriction}, we choose $q_k^0=p_k^0=0$ and $(\Delta q_k)^2=1/(2s_b k)$, with $s_b=\langle\widehat{\mathscr A}\rangle_\chi/(4G_N\hbar)$. Writing $|2_k\rangle=(\mathfrak a_k^\dagger)^2|\chi\rangle/\sqrt{2}$ for the state with two gravitons in mode $k$, we obtain
\begin{equation}
 |\psi_{2,N}\rangle =\sum_{k=1}^N\big[\widehat q_k^{\,2}-(\Delta q_k)^2\big]|\chi\rangle
 =\sum_{k=1}^N c_k|2_k\rangle,\ 
 c_k=\sqrt{2}(\Delta q_k)^2=\frac{1}{\sqrt{2}\,s_b k}.
 \label{bfq:translated-pair-state}
\end{equation}
We compare the boosts about $S_d$ and $S$ by computing the connected overlap
\begin{equation}
 \Delta C_N(d)=\langle U_{d,N}^\dagger B_\eta U_{d,N}B_{-\eta}\rangle_c,\ 
 \langle X\rangle_c=\langle\psi_{2,N}|X|\psi_{2,N}\rangle
 -\|\psi_{2,N}\|^2\langle\chi|X|\chi\rangle .
 \label{bfq:translated-pair-overlap}
\end{equation}
Since the area divided by $8\pi G_N$ is conjugate to $\gamma$, its commutator with $\mathcal Q(H_c)$ differentiates $H_c$ with respect to $\gamma$ exactly. For the null translation~\eqref{bfq:null-translation-area} we thus have
\begin{equation}
 \left.\partial_d\widehat{\mathscr A}(S_d)\right|_{d=0}=\frac{[\widehat{\mathscr A},\mathcal Q(H_c)]}{i\hbar}
 =-8\pi G_N\mathcal Q(J_N),\ 
 J_N=\frac{1}{4\pi G_N}\int_S\varrho_S(y)e^{-\Xi_y(1)}\bm\epsilon_0 .
 \label{bfq:translated-pair-generator}
\end{equation}
As $J_N$ is independent of $\gamma$,~\eqref{bfq:translated-pair-overlap} yields $\left.\partial_d\Delta C_N(d)\right|_{d=0}=-i\eta\langle\mathcal Q(J_N)\rangle_c/\hbar$. To evaluate $\langle\mathcal Q(J_N)\rangle_c$, we integrate $J_N$ against the difference of Wigner functions
\begin{equation}
 W_{\psi_{2,N}}-\|\psi_{2,N}\|^2W_\chi
 =8W_\chi\left(\left|\sum_{k=1}^N c_k\alpha_k^2\right|^2
                 -\sum_{k=1}^N c_k^2|\alpha_k|^2\right).
 \label{bfq:translated-pair-wigner}
\end{equation}
Here the real and imaginary parts of the oscillator coordinate $\alpha_k$ each have variance $1/4$ in $\chi$. At fixed corner area $\mathscr A$, integrating either term in parentheses against $W_\chi$ yields $\|\psi_{2,N}\|^2/2$. Since $0\le J_N\le\sqrt{\mathscr A/(4\pi)}/G_N$ by~\eqref{bfq:null-translation-area}, we can bound the two nonnegative integrals separately. Averaging over $\mathscr A$ with $\langle\widehat{\mathscr A}\rangle_\chi=4\pi b^2$ then yields
\begin{equation}
 \frac{1}{8\pi b\hbar}\left|\left\langle[\widehat{\mathscr A},\mathcal Q(H_c)]\right\rangle_c\right|
 \le 4\|\psi_{2,N}\|^2
 =\frac{2}{s_b^2}\sum_{k=1}^N\frac1{k^2}
 \le\frac{\pi^2}{3s_b^2}.
 \label{bfq:translated-pair-focusing-bound}
\end{equation}
We can thus include arbitrarily short wavelengths at fixed $G_N$ and $\hbar$ while the connected matrix element stays bounded.\footnote{We can keep all $N$ modes in $\Xi_y(1)$ and emit the two gravitons only in modes $k\le M$ of~\eqref{bfq:translated-pair-state}. By~\eqref{bfq:translated-pair-wigner} this changes the matrix element by $O(M^{-1/2})$ uniformly in $N\ge M$, so it suffices to establish convergence as $N\to\infty$ for each fixed $M$.}

To expand first in $G_N$, we hold $N$ fixed and perturb the metric of a round shear free null cone. Since both states in~\eqref{bfq:translated-pair-overlap} have the same corner wavefunction, the constant term in $e^{-\Xi_y(1)}=1-\Xi_y(1)+\cdots$ drops out. By~\eqref{bfgeo:integrated-focusing-main:c}, $\Xi_y(1)$ is quadratic in the derivative of the conformal metric along the null generator. The leading UV contribution to $\langle[\widehat{\mathscr A},\mathcal Q(H_c)]\rangle_c/(8\pi b\hbar)$ is thus proportional to
\begin{equation}
 i\sum_{k=1}^N k^2\langle\widehat q_k^{\,2}\rangle_c
 =4i\sum_{k=1}^N c_k^2 k^2(\Delta q_k)^2
 =\frac{i}{s_b^3}\sum_{k=1}^N\frac{1}{k}
 =\frac{i}{s_b^3}\big(\log N+O(1)\big).
 \label{bfq:translated-pair-divergence}
\end{equation}
The shear responsible for the logarithmic UV divergence at order $G_N^3$ thus suppresses the change in area at finite $G_N$~\eqref{bfq:translated-pair-focusing-bound}.

\subsection{Finite energy under gluing of null cones}
\label{bfq:glued-null-energy}

In quantum field theory on a fixed background, a state can agree with the vacuum on either side of $S$ and yet have arbitrarily large energy near $S$. For a free scalar field $\phi$ on a spatial lattice of spacing $a$, the term $|\nabla\phi|^2/2$ in the Hamiltonian density depends on the squared difference of $\phi$ at neighboring sites. Taking the product of the reduced vacuum density matrices on either side of $S$ removes the correlation across each link through $S$. In four dimensions this costs an energy of order $\hbar/a$ per link.

In gravity the shear measures how rapidly the conformal metric varies along a null generator~\eqref{bfgeo:focusing:d}. We might therefore expect a similar divergence in the energy flux near $S$ when we glue two independently prepared copies of $\chi$. As we saw in Section~\ref{bfq:full-area-uv}, however, the square of the shear that enters the flux also slows the growth of the area element. Requiring each cone to expand from the cap to $S$ thus limits how much energy can cross it.

If the expansion were discontinuous along a null generator through $S$, the Raychaudhuri equation~\eqref{bfgeo:focusing:e} would require a source localized there. To compare the expansions we use $Y$ to identify the generators and $\gamma$ to relate their null frames. Taking the trace of each null extrinsic curvature in~\eqref{bfgeo:geometric-matching} yields
\begin{equation}
 \Theta_{(\ell_R)}=e^\gamma Y^*\Theta_{(\ell_L)},\
 \Theta_{(n_R)}=e^{-\gamma}Y^*\Theta_{(n_L)}.
 \label{bfq:matched-null-expansions}
\end{equation}
Expressed in the frame $(\ell_R,n_R)$, both expansions are continuous across $S$, so the boundary terms at $S$ cancel when we integrate the null constraint along the generator.

To determine the flux through either expanding null cone $\mathscr H$, we take $u$ to increase from the cap to $S$, with $\partial_u\varrho=p_{\epsilon,y}$ at the cap. Integrating~\eqref{bfgeo:focusing:f} and using~\eqref{bfgeo:integrated-focusing-main:b}, we find
\begin{equation}
 \frac{1}{2p_{\epsilon,y}}\int_{u(\epsilon,y)}^{u(1,y)}
 \varrho\,\sigma_{AB}\sigma^{AB}\,\dd u
 =1-\frac{\left.\partial_u\varrho\right|_S}{p_{\epsilon,y}}
 =1-e^{-\Xi_y(1)}.
 \label{bfq:integrated-shear-slope}
\end{equation}
However short the wavelengths of the radiation, the decrease in $\partial_u\varrho$ cannot exceed its initial value as long as the cone is expanding.

To measure the energy with respect to the affine parameter $p_{\epsilon,y}u$ of the clock~\eqref{bfgeo:capped-clock}, we evaluate the shear term in the symplectic potential~\eqref{bfgeo:null-potential-main} on a translation along $\ell^a/p_{\epsilon,y}$. Only the trace free part $2\sigma_{AB}$ of $\Lie_\ell q_{AB}$ enters this term~\eqref{bfgeo:focusing:a}. The radiative energy crossing $\mathscr H$ between the cut at $v$ and $S$ is thus
\begin{equation}
 E_{\mathscr H}(v)
 =\frac{1}{8\pi G_N}\int_S\bm\epsilon_0\int_{u(v,y)}^{u(1,y)}
 \frac{\varrho^2\,\sigma_{AB}\sigma^{AB}}{p_{\epsilon,y}}\,\dd u
 =-\frac{1}{4\pi G_N}\int_S\varrho_S(y)\bm\epsilon_0
 \int_v^1 w\,\partial_w e^{-\Xi_y(w)}\,\dd w .
 \label{bfq:null-shear-energy}
\end{equation}

Both the area element at $S$ and the shear fluctuate in the state we prepare by gluing. Integrating~\eqref{bfq:null-shear-energy} by parts, we quantize the flux in terms of the focusing operator~\eqref{bfq:canonical-clock-integral} with the ordering
\begin{equation}
 \widehat E_{\mathscr H}(v)
 =\frac{1}{4\pi G_N}\int_S\bm\epsilon_0\,\widehat\varrho_S(y)^{1/2}
 \left[v e^{-\widehat\Xi_y(v)}+\int_v^1 e^{-\widehat\Xi_y(w)}\,\dd w
 -e^{-\widehat\Xi_y(1)}\right]\widehat\varrho_S(y)^{1/2}.
 \label{bfq:quantum-null-shear-energy}
\end{equation}
By positivity of $\widehat\Xi_y(v)$, we have
\begin{equation}
 0\le v e^{-\widehat\Xi_y(v)}+\int_v^1 e^{-\widehat\Xi_y(w)}\,\dd w\le\widehat{1}.
 \label{bfq:bounded-focusing-exponential}
\end{equation}
Since the same bound holds for $\exp(-\widehat\Xi_y(1))$, the difference in brackets in~\eqref{bfq:quantum-null-shear-energy} has norm at most one. For the normalized state $|\Psi\rangle$ of the union, Cauchy Schwarz and~\eqref{bfq:canonical-area-density:e} thus yield
\begin{equation}
 G_N\big|\langle\widehat E_{\mathscr H}(v)\rangle_\Psi\big|
 \le\frac{1}{4\pi}\int_S\langle\widehat\varrho_S(y)\rangle_\Psi\,\bm\epsilon_0
 \le\left(\frac{\langle\widehat{\mathscr A}\rangle_\Psi}{4\pi}\right)^{1/2}.
 \label{bfq:glued-null-energy-bound}
\end{equation}

It remains to determine the area of $S$ in the state of the union. This is simplest when each incoming state has definite total area. Indeed, matching the induced metric~\eqref{bfgeo:geometric-matching:a} requires
\begin{equation}
 \mathscr A_+=\mathscr A_-=\mathscr A,\
 \frac{\dd A_+}{\mathscr A}=Y^*\!\left(\frac{\dd A_-}{\mathscr A}\right).
 \label{bfq:matched-area-profile}
\end{equation}
To measure $\mathscr A$ we can boost the right sided frame through a constant rapidity $\eta$ before gluing. Both null extrinsic curvatures on the right transform as in~\eqref{bfgeo:geometric-matching} with $\gamma$ replaced by $\gamma+\eta$, while $\dd\gamma$ is unchanged. By~\eqref{bfgeo:two-sided-boost-potential} this amounts to acting with $B_\eta$ after gluing.

In particular, we can glue two independently prepared copies of $\omega$ with no source insertions. Since $\omega$ agrees with $|\Omega_{\mathscr A}\rangle$ on the incoming subalgebra, the boost acts by a single phase. With $N$ modes on either side, a nonzero state $|\Psi_{\mathscr A,N}\rangle$ of the union therefore satisfies
\begin{equation}
 \frac{\langle\Psi_{\mathscr A,N}|B_\eta|\Psi_{\mathscr A,N}\rangle}
      {\langle\Psi_{\mathscr A,N}|\Psi_{\mathscr A,N}\rangle}
 =\omega(B_\eta)=e^{i\eta\mathscr A/(8\pi G_N\hbar)}
 \ \Longrightarrow\ \langle\widehat{\mathscr A}\rangle_{\Psi_{\mathscr A,N}}=\mathscr A.
 \label{bfq:glued-area-boost}
\end{equation}
At fixed $G_N$ and $\hbar$ we therefore find for either null cone
\begin{equation}
 \sup_N \left|\langle\widehat E_{\mathscr H}(v)\rangle_{\Psi_{\mathscr A,N}}\right|
 \le \frac{1}{G_N}\left(\frac{\mathscr A}{4\pi}\right)^{1/2}
 =\frac{b}{G_N},\ \epsilon\le v\le1 .
 \label{bfq:glued-uniform-flux-bound}
\end{equation}

For two independent copies of $\chi$ as in Section~\ref{bfq:full-area-uv}, the area fluctuates. Matching the other fields at $S$ can now change the probability of each area. To estimate the expectation value in~\eqref{bfq:glued-null-energy-bound} we compare two nearby choices of the incoming wavefunction. For $s=\mathscr A/(4G_N\hbar)>0$ we take the Gaussian of width $\Delta s=\pi/\Delta\gamma$ in~\eqref{app:gravarea-restriction}. Shifting its center from $s_b$ to $s_b+\delta$ with $0<\delta\ll\Delta s$ yields
\begin{equation}
 \chi_\delta(s)\propto e^{\delta s/[2(\Delta s)^2]}\chi(s),\
 1-|\langle\chi|\chi_\delta\rangle|^2
 =O\!\left(\frac{\delta^2}{(\Delta s)^2}\right).
 \label{bfq:shifted-area-packet}
\end{equation}

Let $|\Psi_N\rangle$ and $|\Psi_{N,\delta}\rangle$ be the normalized states obtained by gluing two copies of $\chi$ and two copies of $\chi_\delta$, respectively. At each $s$ we keep the wavefunction of all other modes fixed. Since the integral over boosts~\eqref{eq:collar-boost-match} pairs equal areas, the factor in~\eqref{bfq:shifted-area-packet} enters the amplitude twice, once from either side of $S$. For any $s_0>0$ we thus find
\begin{equation}
 \langle\hat s\rangle_{\Psi_N}
 =\frac{\langle \hat s\,e^{-\lambda\hat s}\rangle_{\Psi_{N,\delta}}}
        {\langle e^{-\lambda\hat s}\rangle_{\Psi_{N,\delta}}}
 \le \frac{e^{\lambda s_0-1}}{\lambda\,\Pr_{\Psi_{N,\delta}}(s\le s_0)},\
 \lambda=\frac{2\delta}{(\Delta s)^2}.
 \label{bfq:glued-gaussian-area-bound}
\end{equation}
For the expected area in $|\Psi_N\rangle$ to diverge, the probability of every finite interval of areas in $|\Psi_{N,\delta}\rangle$ would thus have to tend to zero. Yet we can make the change in either incoming state arbitrarily small, independently of $N$, without exciting any of the short wavelength modes we add. We therefore expect that for some finite $s_0$ the probability in~\eqref{bfq:glued-gaussian-area-bound} stays bounded away from zero as we increase $N$. This suffices to bound the expected energy flux in~\eqref{bfq:glued-null-energy-bound} under spatial gluing.

\subsection{Transition amplitudes from gluing}
\label{bfuv:completeness}
\label{sec:bf-uv}

We now compose the amplitudes for two cobordisms along a spacelike cut $\Sigma$. Focusing yields finite norms for the states we prepare with the radiative sources of Section~\ref{bfq:fixed-area-uv}, and unitary time evolution~\eqref{bfq:cap-evolution-norm} preserves these norms on $\Sigma$. For any two fixed external states of finite norm in $\mathcal H_\omega$, Cauchy Schwarz bounds the sum of absolute values over intermediate states on $\Sigma$ by the product of the two norms.

\begin{figure}[!htb]
\centering
\includegraphics[width=.84\linewidth]{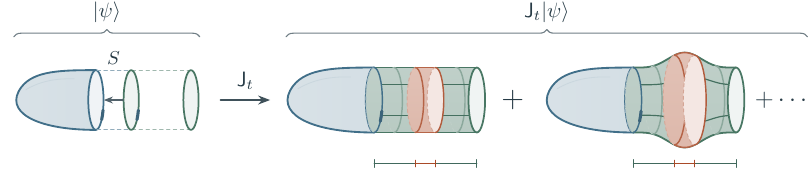}
\caption[Spatial gluing]{Different spatial metrics in the interior can satisfy the same matching conditions at $S$. We form $\mathsf J_t|\psi\rangle$ by joining each pair with its amplitude in $|\psi\rangle$. At fixed proper width $w$ and induced metrics on both faces, larger intermediate areas make the second collar enclose more volume.}
\label{fig:prologue-space-time}
\end{figure}

To join two neighboring subregions in space, the entangling product matches their fields at $S$ and identifies states related by a simultaneous change of frame~\cite{DF,CiambelliKlinger2026}.\footnote{In electromagnetism one can likewise glue the algebras of two subregions by treating edge modes as quantum reference frames for gauge transformations at their common boundary~\cite{Fewster:2025ijg}. If one reduces by these transformations on each side separately, the electric flux density at the corner labels superselection sectors~\cite{Riello:2022din}.} Let $\mathsf J_t$ map the matched states at time $t$ to states on the union. For two subregions prepared at fixed corner areas $\mathscr A_+$ and $\mathscr A_-$, the charges for a simultaneous uniform boost at $S$ cancel only when $\mathscr A_+=\mathscr A_-$. In the occupation number bases of Section~\ref{bfq:algebra-states}, $\mathsf J_t$ joins superpositions of $|\boldsymbol n\rangle_+\otimes|\boldsymbol m\rangle_-$ that match the fields in~\eqref{bfgeo:geometric-matching} and are invariant under simultaneous surface symmetries at $S$ (Figure~\ref{fig:prologue-space-time}).

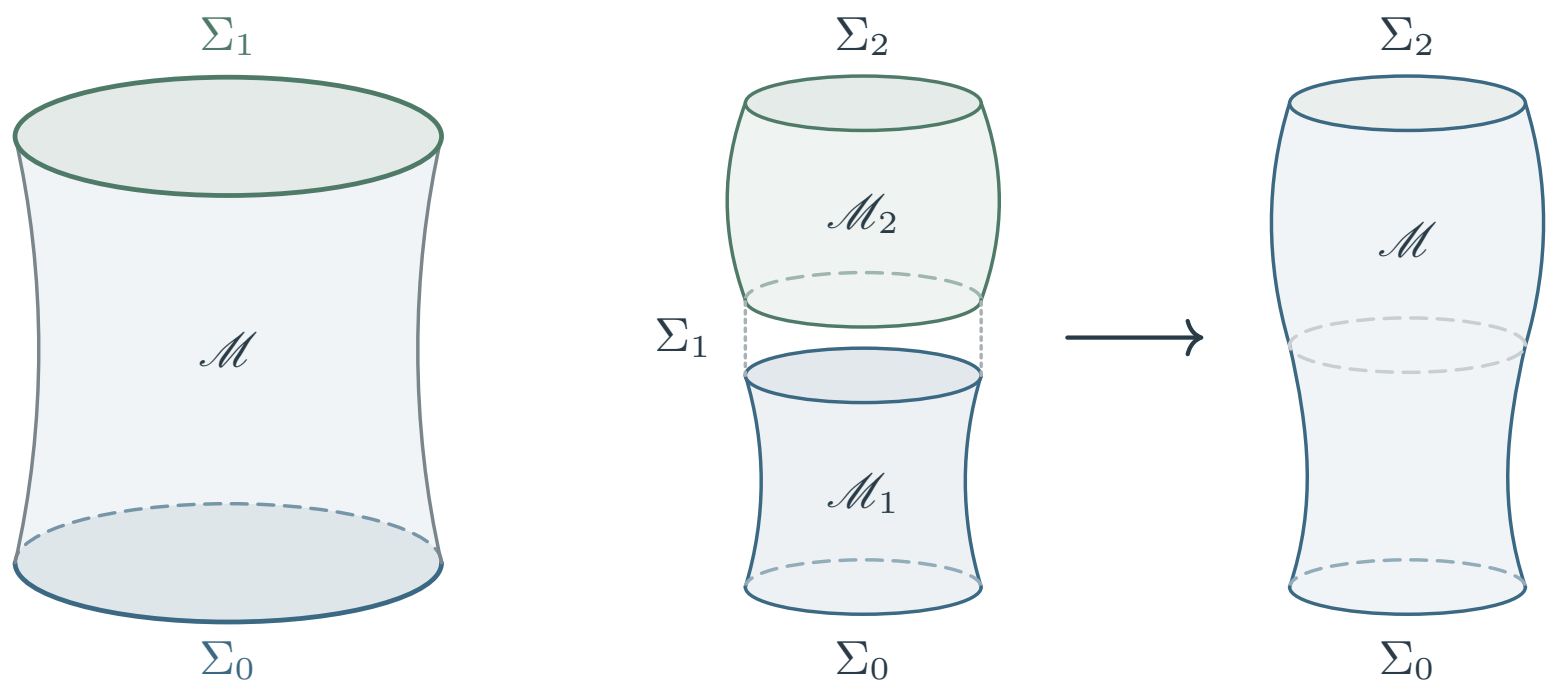
\begin{figure}[htbp]
\centering
\begingroup
\definecolor{bfFigBlue}{HTML}{3B6983}
\definecolor{bfFigGreen}{HTML}{4E7A67}
\definecolor{bfFigInk}{HTML}{283B46}
\begin{tikzpicture}[x=1cm,y=1cm,line width=.65pt,
  line cap=round,line join=round,font=\small,text=bfFigInk]
\begin{scope}[xshift=-4.20cm]
\begin{scope}[yshift=.16cm,scale=.91]
\fill[bfFigBlue!7]
  (-1.55,0) arc[start angle=180,end angle=360,x radius=1.55,y radius=.43]
  .. controls (1.32,1.00) and (1.32,2.10) .. (1.55,3.10)
  arc[start angle=0,end angle=180,x radius=1.55,y radius=.43]
  .. controls (-1.32,2.10) and (-1.32,1.00) .. cycle;
\fill[bfFigBlue!17] (0,0) ellipse[x radius=1.55,y radius=.43];
\draw[bfFigBlue!70,densely dashed] (1.55,0)
  arc[start angle=0,end angle=180,x radius=1.55,y radius=.43];
\draw[bfFigBlue,line width=.9pt] (-1.55,0)
  arc[start angle=180,end angle=360,x radius=1.55,y radius=.43];
\draw[bfFigInk!62] (-1.55,0)
  .. controls (-1.32,1.00) and (-1.32,2.10) .. (-1.55,3.10);
\draw[bfFigInk!62] (1.55,0)
  .. controls (1.32,1.00) and (1.32,2.10) .. (1.55,3.10);
\filldraw[fill=bfFigGreen!16,draw=bfFigGreen,line width=.9pt]
  (0,3.10) ellipse[x radius=1.55,y radius=.43];
\node at (0,1.55) {$\mathscr M$};
\end{scope}
\node[below,text=bfFigBlue] at (0,-.23) {$\Sigma_0$};
\node[above,text=bfFigGreen] at (0,3.39) {$\Sigma_1$};
\end{scope}
\fill[bfFigBlue!9] (-.78,0)
  arc[start angle=180,end angle=360,x radius=.78,y radius=.18]
  .. controls (.64,.46) and (.64,.94) .. (.78,1.40)
  arc[start angle=0,end angle=180,x radius=.78,y radius=.18]
  .. controls (-.64,.94) and (-.64,.46) .. cycle;
\draw[bfFigBlue] (-.78,0) .. controls (-.64,.46) and (-.64,.94) .. (-.78,1.40);
\draw[bfFigBlue] (.78,0) .. controls (.64,.46) and (.64,.94) .. (.78,1.40);
\draw[bfFigBlue!55,densely dashed] (.78,0)
  arc[start angle=0,end angle=180,x radius=.78,y radius=.18];
\draw[bfFigBlue] (-.78,0)
  arc[start angle=180,end angle=360,x radius=.78,y radius=.18];
\filldraw[fill=bfFigBlue!15,draw=bfFigBlue] (0,1.40) ellipse[x radius=.78,y radius=.18];
\fill[bfFigGreen!9] (-.78,1.90)
  arc[start angle=180,end angle=360,x radius=.78,y radius=.18]
  .. controls (.94,2.34) and (.94,2.77) .. (.78,3.20)
  arc[start angle=0,end angle=180,x radius=.78,y radius=.18]
  .. controls (-.94,2.77) and (-.94,2.34) .. cycle;
\draw[bfFigGreen] (-.78,1.90) .. controls (-.94,2.34) and (-.94,2.77) .. (-.78,3.20);
\draw[bfFigGreen] (.78,1.90) .. controls (.94,2.34) and (.94,2.77) .. (.78,3.20);
\draw[bfFigGreen!55,densely dashed] (.78,1.90)
  arc[start angle=0,end angle=180,x radius=.78,y radius=.18];
\draw[bfFigGreen] (-.78,1.90)
  arc[start angle=180,end angle=360,x radius=.78,y radius=.18];
\filldraw[fill=bfFigGreen!15,draw=bfFigGreen] (0,3.20) ellipse[x radius=.78,y radius=.18];
\draw[bfFigInk!40,densely dotted] (-.78,1.40) -- (-.78,1.90);
\draw[bfFigInk!40,densely dotted] (.78,1.40) -- (.78,1.90);
\node at (0,.62) {$\mathscr M_1$};
\node at (0,2.47) {$\mathscr M_2$};
\node[below] at (0,-.23) {$\Sigma_0$};
\node[anchor=east,inner sep=2.5pt] at (-.90,1.65) {$\Sigma_1$};
\node[above] at (0,3.39) {$\Sigma_2$};
\draw[->,bfFigInk,line width=.8pt] (1.35,1.65) -- (2.25,1.65);
\begin{scope}[xshift=3.6cm]
\fill[bfFigBlue!7] (-.78,0)
  arc[start angle=180,end angle=360,x radius=.78,y radius=.18]
  .. controls (.60,.54) and (.65,1.03) .. (.78,1.60)
  .. controls (.94,2.14) and (.94,2.69) .. (.78,3.20)
  arc[start angle=0,end angle=180,x radius=.78,y radius=.18]
  .. controls (-.94,2.69) and (-.94,2.14) .. (-.78,1.60)
  .. controls (-.65,1.03) and (-.60,.54) .. cycle;
\draw[bfFigBlue] (-.78,0)
  .. controls (-.60,.54) and (-.65,1.03) .. (-.78,1.60)
  .. controls (-.94,2.14) and (-.94,2.69) .. (-.78,3.20);
\draw[bfFigBlue] (.78,0)
  .. controls (.60,.54) and (.65,1.03) .. (.78,1.60)
  .. controls (.94,2.14) and (.94,2.69) .. (.78,3.20);
\draw[bfFigBlue!55,densely dashed] (.78,0)
  arc[start angle=0,end angle=180,x radius=.78,y radius=.18];
\draw[bfFigBlue] (-.78,0)
  arc[start angle=180,end angle=360,x radius=.78,y radius=.18];
\filldraw[fill=bfFigGreen!12,draw=bfFigBlue] (0,3.20) ellipse[x radius=.78,y radius=.18];
\draw[bfFigInk!25,densely dashed] (0,1.60) ellipse[x radius=.78,y radius=.18];
\node at (0,2.30) {$\mathscr M$};
\node[below] at (0,-.23) {$\Sigma_0$};
\node[above] at (0,3.39) {$\Sigma_2$};
\end{scope}
\end{tikzpicture}
\endgroup
\caption[Composition across an intermediate cut]{The outgoing cap $\Sigma_1$ on the left becomes an intermediate cut when we glue $\mathscr M_2$ to $\mathscr M_1$ by matching $\mathfrak h_{\Sigma_1}$ on their oppositely oriented caps (center). To obtain the amplitude for $\mathscr M_2\circ_{\Sigma_1}\mathscr M_1$ on the right, we therefore sum over states on $\Sigma_1$ (dashed ellipse) in the composition law~\eqref{bfgeo:sewing-main}, holding fixed the incoming state on $\Sigma_0$, the outgoing state on $\Sigma_2$ and the state on the rest of the boundary.}
\label{fig:bf-composition}
\end{figure}

To compose amplitudes in time, we now fix a spacelike cut $\Sigma$ and match the fields induced on the future cap of $\mathscr M_1$ to those on the past cap of $\mathscr M_2$ as in~\eqref{bfgeo:cap-phase-space-gluing} (Figure~\ref{fig:bf-composition}).\footnote{For two covers of the same solution~\eqref{bfgeo:patched-phase-space}, the change of variables on $\Sigma$ preserves the amplitude (Appendix~\ref{app:physical-inner-product}). Pairing boundary states also implements gluing in perturbative quantizations of gauge theories on manifolds with boundary~\cite{Cattaneo:2015vsa,Cattaneo:2016hqk}.} Because the two caps have opposite orientations, their terms in the action cancel. The action deformations thus add, and causal factorization~\eqref{bfq:relations:e} yields $U_{\mathscr M_2\circ_\Sigma\mathscr M_1}=U_{\mathscr M_2}U_{\mathscr M_1}$. By unitarity~\eqref{bfq:cap-evolution-norm}, the adjoints cancel the two time evolutions upon closing the cut, inserting the identity (Figure~\ref{fig:auxiliary-cap-orderings}a). Resolving this identity in an orthonormal basis $|e_\alpha\rangle$ on $\Sigma$ yields
\begin{equation}
\langle\phi|U_{\mathscr M_2}U_{\mathscr M_1}|\psi\rangle
=\sum_\alpha
\langle\phi|U_{\mathscr M_2}|e_\alpha\rangle
\langle e_\alpha|U_{\mathscr M_1}|\psi\rangle.
\label{bfgeo:sewing-main}
\end{equation}

The two factors in each summand of~\eqref{bfgeo:sewing-main} are components of $U_{\mathscr M_2}^\dagger|\phi\rangle$ and $U_{\mathscr M_1}|\psi\rangle$ in the same basis on $\Sigma$. Since time evolution preserves their norms, the Cauchy Schwarz inequality implies
\begin{equation}
\sum_\alpha
\big|\langle\phi|U_{\mathscr M_2}|e_\alpha\rangle
\langle e_\alpha|U_{\mathscr M_1}|\psi\rangle\big|
\leq\sqrt{\langle\phi|\phi\rangle\langle\psi|\psi\rangle}.
\label{bfq:sewing-bound}
\end{equation}
Thus the resolution of identity in~\eqref{bfgeo:sewing-main} converges absolutely in the amplitude, even when $\Sigma$ admits infinitely many independent states. Changing variables or boundary frame on $\Sigma$ acts by $V_{ji}(t)$ on the incoming factor and by $V_{ji}(t)^{-1}$ on the outgoing factor, including the phases from the boundary action~\eqref{bfq:hcan:a}. These unitaries cancel in the sum, as shown in Appendix~\ref{app:physical-inner-product}.

For several cuts, let $P_j$ project onto the states included on cap $j$, and let $U_j$ denote time evolution between consecutive caps. Restoring the omitted states from the incoming cap onwards, we find that the state reaching cut $j$ is $U_j\cdots U_1|\psi\rangle$. Only its component orthogonal to the range of $P_j$ is missing there, and none of the remaining unitary operators or projections can increase the norm of this component. Thus
\begin{equation}
\big|\langle\phi|(U_m\cdots U_1-U_mP_{m-1}U_{m-1}\cdots P_1U_1)|\psi\rangle\big|\le\sqrt{\langle\phi|\phi\rangle}\sum_{j=1}^{m-1}\|(\widehat{1}-P_j)U_j\cdots U_1|\psi\rangle\|.
\label{bfq:intermediate-state-error}
\end{equation} On each cut, $\|(\widehat{1}-P_j)U_j\cdots U_1|\psi\rangle\|\to0$ as we include the remaining states, independently of the sums on the other cuts. Restoring all intermediate states therefore recovers the amplitude between $|\psi\rangle$ on the initial cap and $|\phi\rangle$ on the final cap for any finite subdivision.

\section{Hilbert space of a causal diamond and black hole entropy}
\label{sec:diamond-hilbert-entropy}

In Section~\ref{bfuv:completeness} we glued amplitudes by summing over a complete basis of $\mathcal H_\omega$ on a spacelike cut. To determine how many orthogonal states we can prepare in a causal diamond $\mathscr D$ of given maximum size, we now construct the gravitational clock operator and restrict its readings as we did in JT gravity (Section~\ref{jt:prelude-quantum}). There we bounded the proper width by projecting onto all readings up to $\ell(\mathscr D)$ at once, preserving coherence among them. Even with the dilaton fixed at both corners and the matter field in the same state, varying only the phase of the clock wavefunction yielded infinitely many orthogonal states with identical clock probabilities.

In four dimensions shear changes the affine parameter at which a null generator reaches a given area element at $S$~\eqref{bfgeo:capped-clock}, so an upper limit on the clock reading also restricts the radiative states we can prepare at fixed corner area. Indeed, semiclassically the covariant entropy bound~\cite{Bousso1999CovariantEntropy} suggests that the corner area limits the number of orthogonal states this projection admits. A finite dimensional Hilbert space for each causal diamond is a starting point of holographic spacetime~\cite{Banks:2020zcr} and has been argued for on more general grounds~\cite{Bao:2017rnv}. One can also estimate its dimension at fixed spatial volume~\cite{Jacobson:2022jir} and its entropy at fixed temperature~\cite{Jacobson:2022gmo} using the Euclidean gravitational path integral.

To show that this number is finite, we measure a relative boost at $S$ by inserting the smeared radiative operator on each side of the corner. Closing the cut at $S$ inserts the identity as in~\eqref{bfgeo:sewing-main}, so the sum over intermediate states reproduces the unnormalized amplitude for the same preparation and insertions on the uncut causal diamond. By comparing the change under the boost in the cut and uncut amplitudes, we show that the Hilbert space of the causal diamond is finite dimensional and hence that its area operator $\widehat{\mathscr A}$ has a discrete spectrum. We use the distinct outcomes of an area measurement in this causal diamond to then also bound the number of states in a finite interval of energy at a reflecting AdS boundary. When the corner is a marginally trapped surface, the trace over the allowed interior states yields the Bekenstein Hawking entropy and, for a large Schwarzschild black hole, its logarithmic correction.

\subsection{Gravitational clock operator on the null cone}
\label{bfq:canonical-quantization}

To measure the size of $\mathscr D$ we quantize the average clock reading on its incoming null cone~\eqref{bfgeo:capped-clock}. Even at fixed total area $\mathscr A$ this average depends on the focusing operator~\eqref{bfq:canonical-clock-integral} along each null generator and on how the area of $S$ is distributed among them. A real valued function $f_S$ on $S$ specifies $\mathrm dA=\varrho_S^2\bm\epsilon_0$ through
\begin{equation}
\varrho_S(y)^2=\frac{\mathscr A\exp(f_S(y))}{\int_S\exp(f_S(y'))\bm\epsilon_0(y')}.
\label{bfq:canonical-area-density}
\end{equation}
With
$\widehat{\mathscr A}=\mathscr A\widehat{1}$ and $\widehat f_S(y)=\mathcal Q(f_S(y))$, we use the normalization
\begin{subequations}\label{bfq:canonical-area-density:b}
\begin{equation}
\widehat w_y=
\left[\int_Se^{\widehat f_S(y')}\bm\epsilon_0(y')\right]^{-1/2}
e^{\widehat f_S(y)}
\left[\int_Se^{\widehat f_S(y')}\bm\epsilon_0(y')\right]^{-1/2},\label{bfq:canonical-area-weight:a}
\end{equation}
\begin{equation}
\widehat{\varrho}_S(y)=\sqrt{\mathscr A}\,\widehat w_y^{1/2},\ \int_S\widehat w_y\bm\epsilon_0=\widehat{1}.\label{bfq:canonical-area-weight:b}
\end{equation}
\end{subequations}
Thus the density integrates to the fixed corner area,
\begin{equation}
\int_S\widehat{\varrho}_S(y)^2\bm\epsilon_0=\mathscr A\widehat{1}.
\label{bfq:canonical-area-density:e}
\end{equation}
Using the same angular average as in the classical expression, the clock operator takes the form
\begin{equation}
 \widehat{\mathsf a}=\int_S\widehat{\varrho}_S(y)\bm\epsilon_0,\ 
 \widehat{\bar u}_*=\widehat{\mathsf a}^{-1/2}
\left[\int_S\bm\epsilon_0(y)\int_\epsilon^1\mathrm dv\,
\widehat{\varrho}_S(y) e^{\widehat \Xi_y(v)}\widehat{\varrho}_S(y)\right]
\widehat{\mathsf a}^{-1/2}.\label{bfq:canonical-clock}
\end{equation}
Each factor in~\eqref{bfq:canonical-clock} acts within $\mathcal K_{\mathscr A}$, so the clock preserves the corner area on incoming wavefunctions.
Inverse powers of $\widehat{\mathsf a}$ act on its positive spectral support, so with the ordering in~\eqref{bfq:canonical-clock}, the clock is defined by the closed quadratic form~\cite[Chapter VI, Section 2]{Kato1995}\footnote{In Appendix~\ref{bfq:canonical-representation} we argue that~\eqref{bfent:clock-form} is finite on a dense set of wavefunctions, as required for it to define a self adjoint clock operator on $\mathcal K_{\mathscr A}$.}
\begin{equation}
\label{bfent:clock-form}\vphantom{\int\limits_S}
\left\|\widehat{\bar u}_*^{1/2}|\psi\rangle\right\|^2
=\int_S\boldsymbol\epsilon_0(y)\int_\epsilon^1\mathrm dv\,
\left\|
e^{\widehat \Xi_y(v)/2}\widehat{\varrho}_S(y)
\widehat{\mathsf a}^{-1/2}|\psi\rangle
\right\|^2.
\end{equation}

\subsection{Existence of states in a causal diamond}
\label{bfq:diamond-state-existence}

We can use the clock operator~\eqref{bfq:canonical-clock} to select a state of a macroscopic causal diamond if the incoming wavefunction assigns nonzero probability to readings at most $\ell(\mathscr D)$. We estimate this probability at finite $G_N$ in the Gaussian wavefunction~\eqref{bfq:fixed-area-gaussian} centered on a spherically symmetric incoming null cone of corner area $\mathscr A=4\pi b^2$, with $b\gg\ell_P$. On this null cone the shear vanishes and $\varrho_S(y)=b$ on every null generator, so the classical clock reads $(1-\epsilon)b$. Although both radiative polarizations and the distribution of area over $S$ fluctuate in $|\Omega_{\mathscr A}\rangle$, the clock preserves total area and we can project this state within $\mathcal K_{\mathscr A}$.

By~\eqref{bfq:pairing} we can regard the Fock space $\mathcal K_{\mathscr A}$ as a subspace of $\mathcal H_\omega$. In either space we select the states of the causal diamond by restricting the readings of the clock operator~\eqref{bfent:clock-form} to $[0,\ell(\mathscr D)]$,
\begin{align}
&\mathcal H_{\mathscr D}=P_{\mathscr D}\mathcal H_\omega,\ P_{\mathscr D}\mathcal K_{\mathscr A}=\mathcal K_{\mathscr A}\cap\mathcal H_{\mathscr D},\label{bfq:clock-selected-space}\\
&P_{\mathscr D}
 =\mathbf1_{[0,\ell(\mathscr D)]}(\widehat{\bar u}_*) .\label{bfq:clock-selected-states:b}
\end{align}
Since $\widehat{\bar u}_*$ commutes with $\widehat{\mathscr A}$, selecting this range of clock readings preserves any specified interval of corner areas. If an ideal observation of the clock in a normalized state $|\psi\rangle\in\mathcal H_{\mathscr D}$ yields $\bar u_*\in I$, the conditional probability that a subsequent observation finds the corner area $\mathscr A$ in $J$ is
\begin{align}
&\Pr_\psi(\mathscr A\in J\mid\bar u_*\in I)
=\frac{\|\mathbf1_J(\widehat{\mathscr A})\mathbf1_I(\widehat{\bar u}_*)|\psi\rangle\|^2}
       {\|\mathbf1_I(\widehat{\bar u}_*)|\psi\rangle\|^2} ,
\label{prologue:conditional-area}
\end{align}
whenever the denominator is nonzero. Thus a reading of the gravitational clock $\widehat{\bar u}_*$ can condition a measurement of the area operator $\widehat{\mathscr A}$ even when neither observable has a sharp value in $|\psi\rangle$.

\begingroup
But before we can condition a measurement of the area operator $\widehat{\mathscr A}$ on the clock operator as in~\eqref{prologue:conditional-area}, we need an allowed reading of $\widehat{\bar u}_*$ to occur with nonzero probability in $|\Omega_{\mathscr A}\rangle$. Radiative fluctuations enter the clock operator through $\exp(\widehat\Xi_y(v))$ in~\eqref{bfq:canonical-clock}, so $\langle\Omega_{\mathscr A}|\widehat{\bar u}_*|\Omega_{\mathscr A}\rangle$ can diverge even when very long time delays carry little probability. To bound the probability of a reading above $\ell(\mathscr D)$, we seek a state $|\Psi\rangle$ close to $|\Omega_{\mathscr A}\rangle$ in norm for which $\langle\Psi|\widehat{\bar u}_*|\Psi\rangle<\infty$. To that aim, by the triangle inequality, we have as a starting point
\begin{equation}
 \langle\Omega_{\mathscr A}|(\widehat 1-P_{\mathscr D})|\Omega_{\mathscr A}\rangle
 \le \bigl[\|(\widehat 1-P_{\mathscr D})|\Psi\rangle\|
 +\|\,|\Omega_{\mathscr A}\rangle-|\Psi\rangle\,\|\bigr]^2 .
 \label{bfq:incoming-clock-probability-bound}
\end{equation}

To construct this state we first bound the exponential in~\eqref{bfq:canonical-clock} by replacing $\widehat\Xi_y(v)$ with $\min(\widehat\Xi_y(v),\Xi_*)$. The resulting clock operator $\widehat{\bar u}_*(\Xi_*)$ is positive and self adjoint~\eqref{bfq:capped-clock-form}. We then solve for an unnormalized wavefunction $|\Psi\rangle$ that minimizes the functional
\begin{equation}
 \|\,|\Psi\rangle-|\Omega_{\mathscr A}\rangle\,\|^2
 +\frac{\langle\Psi|\widehat{\bar u}_*(\Xi_*)|\Psi\rangle}{L} ,
 \label{bfq:incoming-clock-functional}
\end{equation}
where $1/L>0$ is the Lagrange multiplier for $\langle \widehat{\bar u}_*\rangle$. At each reading $u$ the second term weights $|\langle u|\Psi\rangle|^2$ by $u/L$, so varying $\langle\Psi|u\rangle$ yields $(1+u/L)\langle u|\Psi\rangle=\langle u|\Omega_{\mathscr A}\rangle$~\eqref{bfq:clock-variational-equation}. At fixed $L$ this wavefunction converges in norm as we remove the focusing cutoff~\eqref{bfq:clock-resolvent-limit},
\begin{equation}
 |\Psi_L\rangle=\lim_{\Xi_*\to\infty}
 \frac{L}{L+\widehat{\bar u}_*(\Xi_*)}|\Omega_{\mathscr A}\rangle .
 \label{bfq:clock-filter-definition}
\end{equation}
Moreover, at fixed $\Xi_*$ and finite $u$, $\langle u|\Psi\rangle$ tends to $\langle u|\Omega_{\mathscr A}\rangle$ as $L$ grows. To bound the distance from $|\Psi_L\rangle$ to $|\Omega_{\mathscr A}\rangle$, we then integrate $[u/(L+u)]^2\le u/(L+u)$ over the clock readings in $|\Omega_{\mathscr A}\rangle$ before taking $\Xi_*\to\infty$~\eqref{bfq:clock-form-approximation},
\begin{equation}
 \|\,|\Omega_{\mathscr A}\rangle-|\Psi_L\rangle\,\|^2
 \le 1-\langle\Omega_{\mathscr A}|\Psi_L\rangle
 =\lim_{\Xi_*\to\infty}\left\langle
 \frac{\widehat{\bar u}_*(\Xi_*)}{L+\widehat{\bar u}_*(\Xi_*)}
 \right\rangle_{\Omega_{\mathscr A}} .
 \label{bfq:clock-overlap-limit}
\end{equation}

At zeroth order in the Weyl product~\eqref{bfq:canonical-products}, we evaluate the fraction in~\eqref{bfq:clock-overlap-limit} with the full nonlinear classical clock $\bar u_*(q,p)$ at fixed $G_N$, where $(q,p)$ are the reduced canonical pairs on the incoming null cone. As before, we use $v=\varrho/\varrho_S$ as a coordinate in~\eqref{bfgeo:integrated-focusing-main} when the area element increases along each incoming null generator from $\partial\Sigma_0$ to $S$, so that $\partial_u\varrho>0$ throughout. Then, at fixed angular and radial resolution, the focusing observable in~\eqref{bfgeo:integrated-focusing-main:c} is finite for a conformal metric smooth in $v$, hence averaging the clock readings over $S$ as in~\eqref{bfgeo:area-and-clock-average} yields
\endgroup
\begin{equation}
\bar u_*\le\max_{y\in S}u_*(y)
\le(1-\epsilon)\max_{y\in S}\bigl[\varrho_S(y)e^{\Xi_y(1)}\bigr]<\infty .
\label{bfq:finite-classical-clock}
\end{equation}
\begingroup

In order to obtain $1-\langle\Omega_{\mathscr A}|\Psi_L\rangle$ in~\eqref{bfq:clock-overlap-limit}, we now average over the incoming state. The Weyl ordering in~\eqref{bfads:rep-weyl} represents each Fourier mode of $f(q,p)$ by the displacement operator $D(\lambda)$, so the measure with which we perform the average is the inverse Fourier transform of $\langle\Omega_{\mathscr A}|D(\lambda)|\Omega_{\mathscr A}\rangle$, namely the Wigner function~\cite{Moyal1949}. At zeroth order in $\hbar$ the Weyl product is ordinary multiplication, so the fraction removed at each clock reading in~\eqref{bfq:clock-filter-definition} fixes the leading term $f_0$,
\begin{equation}
 \langle\mathcal Q(f)\rangle_{\Omega_{\mathscr A}}
 =\int\dd q\,\dd p\,W_{\Omega_{\mathscr A}}(q,p)\,f(q,p),\
 f_0=\frac{\bar u_*}{L+\bar u_*}.
 \label{bfq:wigner-expectation-value}
\end{equation}
For each finite clock reading in~\eqref{bfq:finite-classical-clock}, $f_0$ tends to zero as $L$ grows. Since $0\le f_0\le1$, its integral against the positive, normalized Wigner function~\eqref{bfq:incoming-wigner-density} of the Gaussian state~\eqref{bfq:fixed-area-gaussian} also tends to zero, even if $\langle\bar u_*\rangle=\infty$~\eqref{bfq:classical-clock-expectation-limit}.

Using the full nonlinear classical clock at fixed $G_N$, we have thus shown at zeroth order in $\hbar$ that $\langle\Omega_{\mathscr A}|\Psi_L\rangle\to1$ as $L\to\infty$. To see whether each higher order correction vanishes in the same limit, we determine $f$ from the inverse in~\eqref{bfq:clock-overlap-limit} at finite $\Xi_*$,
\begin{equation}
 \mathcal Q(f)=\frac{\widehat{\bar u}_*(\Xi_*)}{L+\widehat{\bar u}_*(\Xi_*)},\ (L+\widehat{\bar u}_*(\Xi_*))\,\mathcal Q(f)=\widehat{\bar u}_*(\Xi_*).
 \label{bfq:overlap-weyl-identity}
\end{equation}
With the clock operator at its classical order in~\eqref{bfq:overlap-weyl-identity}, $f$ still depends on $\hbar$ through the Weyl product~\eqref{bfq:canonical-products}. Since $f_0$ in~\eqref{bfq:wigner-expectation-value} depends only on $\bar u_*$, we have
\begin{subequations}\label{bfq:clock-classical-weyl-product}
\begin{align}
 &(L+\bar u_*)\star f=\bar u_*,\label{bfq:clock-classical-weyl-product:a}\\
 &\{L+\bar u_*,f_0\}=0.\label{bfq:clock-classical-weyl-product:b}
\end{align}
\end{subequations}
With $\phi=(q,p)$ and $J^{IJ}=\{\phi^I,\phi^J\}$ as in~\eqref{bfads:rep-weyl}, expanding~\eqref{bfq:clock-classical-weyl-product:a} to first nonvanishing order thus yields
\begin{equation}
 f-f_0=-\frac{L\hbar^2}{8(L+\bar u_*)}J^{IJ}J^{KM}\frac{\partial^2\bar u_*}{\partial\phi^I\partial\phi^K}\frac{\partial^2}{\partial\phi^J\partial\phi^M}\left(\frac{1}{L+\bar u_*}\right)+O(\hbar^3).
 \label{bfq:clock-weyl-derivatives}
\end{equation}

Thus far we have computed the first correction to $f$ from the Weyl product at fixed $(q,p)$ in~\eqref{bfq:clock-weyl-derivatives}. To now obtain the overlap in~\eqref{bfq:clock-overlap-limit}, we use~\eqref{bfq:wigner-expectation-value} to average $f$ over fluctuations of the incoming Gaussian state about the spherical data $\phi_0=(q^0,p^0)$. For $\Delta\widehat\phi^I=\widehat\phi^I-\phi_0^I$, holding $G_N$, $b$ and $(\Delta q_j)^2/\hbar$ fixed in~\eqref{bfq:fixed-area-gaussian} yields the propagator
\begin{equation}
 \Sigma^{IJ}=\frac{1}{2}\langle\{\Delta\widehat\phi^I,\Delta\widehat\phi^J\}\rangle_{\Omega_{\mathscr A}}=O(\hbar).
 \label{bfq:clock-canonical-propagator}
\end{equation}
The classical term $f_0$ therefore already contributes to the overlap at order $\hbar$ through the single Wick contraction in Figure~\ref{fig:clock-overlap-diagrams},
\begin{equation}
 \left.\frac{\Sigma^{IJ}}{2}
 \frac{\partial^2}{\partial\phi^I\partial\phi^J}
 \left(\frac{\bar u_*}{L+\bar u_*}\right)\right|_{\phi_0}
 =\left.\frac{L\Sigma^{IJ}}{2(L+\bar u_*)^2}
 \left(\frac{\partial^2\bar u_*}{\partial\phi^I\partial\phi^J}
 -\frac{2}{L+\bar u_*}
 \frac{\partial\bar u_*}{\partial\phi^I}\frac{\partial\bar u_*}{\partial\phi^J}\right)
 \right|_{\phi_0}.
 \label{bfq:semiclassical-clock-overlap}
\end{equation}

\noindent\begin{minipage}{\linewidth}
\captionsetup{type=figure}
\centering
\begingroup
\definecolor{clockInk}{HTML}{263944}
\definecolor{contractionBlue}{HTML}{246697}
\begin{tikzpicture}[x=1cm,y=1cm,font=\small,text=clockInk,line cap=round,line join=round,
    every node/.style={inner sep=1.6pt},
    contraction/.style={draw=contractionBlue,line width=1.15pt},
    vertex/.style={circle,draw=clockInk,fill=clockInk,inner sep=0pt,minimum size=4.7pt}]
 \node[anchor=base west,inner sep=0pt] (lhs) at (0,0)
    {$\displaystyle
      \tfrac12\Sigma^{IJ}\partial_I\partial_J
      \!\left(\frac{\bar u_*}{L+\bar u_*}\right)={}$};
 \begin{scope}[shift={(lhs.base east)}]
   \draw[contraction] (.62,.09)
      ..controls(.56,.32) and(.19,.44)..(.23,.72)
      ..controls(.28,1.04) and(.96,1.04)..(1.01,.72)
      ..controls(1.05,.44) and(.68,.32)..(.62,.09);
   \node[vertex] at (.62,.09) {};
   \node[below,font=\footnotesize,inner sep=1pt] at (.62,-.13)
       {$\displaystyle\frac{L}{2(L+\bar u_*)^2}$};
   \node[anchor=base west,inner sep=0pt] (relativeSign) at (1.24,0)
       {$\displaystyle -$};
 \end{scope}
 \begin{scope}[shift={(relativeSign.base east)}]
   \draw[contraction] (.34,.09)..controls(.34,1.10) and(1.34,1.10)..(1.34,.09);
   \node[vertex] at (.34,.09) {};
   \node[vertex] at (1.34,.09) {};
   \node[below,font=\footnotesize,inner sep=1pt] at (.84,-.13)
       {$\displaystyle\frac{L}{(L+\bar u_*)^3}$};
 \end{scope}
 \begin{scope}[shift={(current bounding box.east)}]
   \draw[contraction] (.75,0)--(1.37,0);
   \node[anchor=west] at (1.46,0) {$\Sigma^{IJ}$};
 \end{scope}
\end{tikzpicture}%
\endgroup\par
\caption[Wick contraction in the clock overlap]{Each blue line represents the propagator $\Sigma^{IJ}$ of canonical fluctuations in the incoming state.}
\label{fig:clock-overlap-diagrams}
\end{minipage}\par

A simple calculation shows that the contraction in~\eqref{bfq:semiclassical-clock-overlap} is $O(L^{-1})$. For the full clock, expanding~\eqref{bfq:overlap-weyl-identity} with $\widehat{\varrho}_S$ and $e^{\widehat\Xi_y}$ ordered as in~\eqref{bfq:canonical-clock} yields the same suppression for every fixed derivative of each coefficient of $f$ at $\phi_0$~\eqref{bfq:clock-local-weyl-limit}. Since each Wick contraction supplies one power of $\hbar$, only finitely many of these derivatives enter the coefficient of $\hbar^n$ in $1-\langle\Omega_{\mathscr A}|\Psi_L\rangle$. This coefficient is therefore $O(L^{-1})$ for every $n$~\eqref{bfq:clock-gaussian-coefficients}.

Of course, derivatives at $\phi_0$ suffice order by order, but at finite $\hbar$ the desired overlap also depends on the probability of time delays in the incoming state itself. Given that a reading $u\ge L$ removes at least half its weight $|\langle u|\Omega_{\mathscr A}\rangle|^2$ from the overlap, since $u/(L+u)\ge1/2$ in~\eqref{bfq:clock-overlap-limit}, we must estimate the probability of clock readings of order $L$ or larger as $L\to\infty$, after first taking $\Xi_*\to\infty$ at fixed $L$.

To estimate this probability we begin with clock readings that can be large even when $\Xi_y(v)=0$ along every null generator. At fixed $\mathscr A$, concentrating the corner area density $\varrho_S(y)^2$ on a small part of $S$ decreases the denominator of the angular average in~\eqref{bfgeo:area-and-clock-average} and increases the clock reading. In the clock operator~\eqref{bfq:canonical-clock} this denominator is the positive operator $\widehat{\mathsf a}$. At $\Xi_*=0$ we have $e^{\min(\widehat\Xi_y(v),\Xi_*)}=\widehat 1$, so the area normalization~\eqref{bfq:canonical-area-density:e} fixes the clock at finite $\hbar$. Bounding the same exponential at finite $\Xi_*$ then yields
\begin{equation}
 \widehat{\bar u}_*(0)=(1-\epsilon)\mathscr A\,\widehat{\mathsf a}^{-1},\
 \widehat{\bar u}_*(\Xi_*)\le e^{\Xi_*}\widehat{\bar u}_*(0).
 \label{bfq:exact-corner-clock}
\end{equation}
In~\eqref{bfq:exact-corner-clock} we choose $\Xi_*=\tfrac12\log(L/b)$ so that the cutoff grows with $L$ while $Le^{-\Xi_*}=\sqrt{bL}$ suppresses the corner contribution~\eqref{bfq:clock-corner-resolvent}.

Taking $\Xi_*\to\infty$ at this same $L$ can increase a classical reading only if the focusing along some null generator exceeds the chosen cutoff. Since $u/(L+u)\le1$, we estimate the further loss of overlap by the probability of this event, using both polarizations of the full conformal metric to evaluate $\Xi_y(1)$~\eqref{bfq:clock-full-focusing},
\begin{equation}
 1-\langle\Omega_{\mathscr A}|\Psi_L\rangle
 \ \lesssim\
 \left\langle\frac{(1-\epsilon)\mathscr A}
 {(1-\epsilon)\mathscr A+\sqrt{bL}\,\widehat{\mathsf a}}\right\rangle_{\Omega_{\mathscr A}}
 +\int\dd q\,\dd p\,W_{\Omega_{\mathscr A}}(q,p),\ L>b,
 \label{bfq:clock-finite-hbar-estimate}
\end{equation}
where we integrate over $(q,p)$ with $\Xi_y(1)>\tfrac12\log(L/b)$ for some $y\in S$. At fixed $\mathscr A>0$, the corner term in~\eqref{bfq:clock-finite-hbar-estimate} tends to zero because its denominator grows with $L$ for every $a>0$ and $a=0$ has zero probability~\eqref{bfq:clock-corner-resolvent}. On each smooth incoming null cone the focusing is bounded over $S$, so the probability of exceeding the growing cutoff also tends to zero~\eqref{bfq:clock-radiation-probability}. Together with~\eqref{bfq:clock-overlap-limit}, this provides strong evidence at finite $G_N$ and $\hbar$ that $\|\,|\Psi_L\rangle-|\Omega_{\mathscr A}\rangle\,\|\to0$ as $L\to\infty$.\footnote{Integrating~\eqref{bfq:clock-classical-focusing-difference} against $W_{\Omega_{\mathscr A}}$ bounds the change in the classical expectation value of $\bar u_* /(L+\bar u_*)$ as $\Xi_*$ increases from $\tfrac12\log(L/b)$ to infinity at fixed $L$. Using it to estimate the increase of the exact matrix element~\eqref{bfq:clock-overlap-limit} is a heuristic for the noncommuting operators in~\eqref{bfq:canonical-clock}.}

We can now combine this estimate of the distance to $|\Omega_{\mathscr A}\rangle$ with a bound on readings above $\ell(\mathscr D)$ in $|\Psi_L\rangle$, as in~\eqref{bfq:incoming-clock-probability-bound}. The weight $u[L/(L+u)]^2$ in Figure~\ref{fig:clock-expectation-value} is at most $L/4$~\eqref{bfq:clock-form-approximation}, which bounds this probability at finite $G_N$ and $\hbar$,
\endgroup
\begin{equation}
\big\|(\widehat 1-P_{\mathscr D})|\Psi_L\rangle\big\|^2
\le\frac{\langle\Psi_L|\widehat{\bar u}_*|\Psi_L\rangle}{\ell(\mathscr D)}
\le\frac{L}{4\ell(\mathscr D)} .
\label{bfq:filtered-clock-tail}
\end{equation}

\noindent\begin{minipage}{\linewidth}
\captionsetup{type=figure}
\centering
\begingroup
\definecolor{clockInk}{HTML}{263944}
\definecolor{clockGreen}{HTML}{367665}
\definecolor{clockRust}{HTML}{A75630}
\begin{tikzpicture}[x=1cm,y=1cm,font=\small,text=clockInk,line cap=round,line join=round,
    every node/.style={inner sep=1.6pt},
    vertex/.style={circle,draw=clockInk,fill=clockInk,inner sep=0pt,minimum size=4.7pt},
    spectral/.style={draw=clockGreen,line width=1.0pt},
    weight/.style={rectangle,draw=clockGreen,fill=white,inner sep=0pt,
                   minimum width=5.3pt,minimum height=5.3pt,line width=.85pt}]
 \node[anchor=base west,inner sep=0pt] (lhs) at (0,0)
    {$\displaystyle\langle\Psi_L|\widehat{\bar u}_*|\Psi_L\rangle={}$};
 \begin{scope}[shift={(lhs.base east)}]
   \draw[spectral] (.28,.09)--(3.98,.09);
   \node[weight] at (1.04,.09) {};
   \node[above] at (1.04,.24) {$\displaystyle\frac{L}{L+u}$};
   \node[vertex,draw=clockRust,fill=clockRust] at (2.14,.09) {};
   \node[above] at (2.14,.24) {$u$};
   \node[weight] at (3.24,.09) {};
   \node[above] at (3.24,.24) {$\displaystyle\frac{L}{L+u}$};
   \node[below] at (.49,-.06) {$|\Omega_{\mathscr A}\rangle$};
   \node[below] at (3.79,-.06) {$\langle\Omega_{\mathscr A}|$};
 \end{scope}
 \begin{scope}[shift={(current bounding box.east)}]
   \draw[spectral] (.75,0)--(1.37,0);
   \node[anchor=west] at (1.46,0) {$\langle u'|u\rangle$};
 \end{scope}
\end{tikzpicture}%
\endgroup\par
\caption[Expectation value of the gravitational clock]{The green line represents the propagator $\langle u'|u\rangle$ in the clock basis, so the same reading $u$ passes through all three insertions.}
\label{fig:clock-expectation-value}
\end{minipage}\par

Combining~\eqref{bfq:incoming-clock-probability-bound} and~\eqref{bfq:filtered-clock-tail} for $|\Psi\rangle=|\Psi_L\rangle$, we thus obtain
\begin{equation}
\|(\widehat 1-P_{\mathscr D})|\Omega_{\mathscr A}\rangle\|
\le \|\,|\Omega_{\mathscr A}\rangle-|\Psi_L\rangle\,\|+\frac12\sqrt{\frac{L}{\ell(\mathscr D)}} .
\label{bfq:clock-projection-estimate}
\end{equation}
We first choose $L$ large to suppress the first term in~\eqref{bfq:clock-projection-estimate}, then raise $\ell(\mathscr D)$ at that same $L$ to suppress the second, and thus argue that at fixed $b$
\begin{equation}
 \lim_{\ell(\mathscr D)\to\infty}\big\|P_{\mathscr D}|\Omega_{\mathscr A}\rangle\big\|^2=1.
 \label{bfq:state-clock-weight}
\end{equation}
The limit in~\eqref{bfq:state-clock-weight} implies that the incoming wavefunction has nonzero projection for some finite maximum reading $\ell(\mathscr D)$. Normalizing this projection prepares $|\psi\rangle$. Since $P_{\mathscr D}$ commutes with $\widehat{\mathscr A}$, $|\psi\rangle$ has the same corner area $\mathscr A$ as the incoming Gaussian state.\footnote{Small displacements of the Gaussian state centered on spherically symmetric data vary continuously in norm, so nearby Gaussian states centered on data with shear in either polarization or angular variation in $\varrho_S$ also have nonzero projection. Time evolution to $\Sigma_1$ preserves the norm of $|\psi\rangle$ by~\eqref{bfq:cap-evolution-norm}. The probability that the clock on $\Sigma_1$ reads in an interval $I$ is $\|\mathbf1_I(\widehat{\bar u}_*)U_{\mathscr M}|\psi\rangle\|^2$.}

\begin{figure}[!htbp]
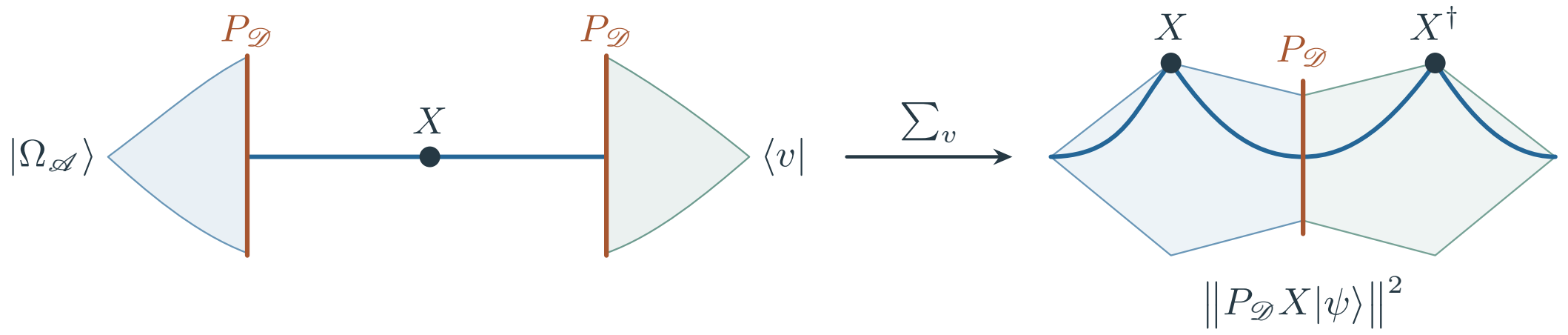

\centering
\resizebox{.85\linewidth}{!}{\SummaryPreparation}
\caption[A clock projection on a state and its adjoint]{On the left the first clock projection prepares $|\psi\rangle$ from the Gaussian state, and the second selects allowed readings after the insertion $X$. Summing the squared amplitudes over outgoing states $|v\rangle$ glues the preparation to its adjoint on the right. The clock projections from the two sides of the shared cut select the same subspace.}
\label{fig:summary-preparation}
\end{figure}

At fixed $\mathscr A$, acting on $|\psi\rangle$ can excite either radiative polarization or change the area element on $S$. Since $|\psi\rangle\ne0$, insertions from the full algebra generate a dense subspace of $\mathcal H_\omega$ by the argument leading to~\eqref{bfq:gns-factor}. Projecting these preparations onto the allowed clock readings yields
\begin{equation}
 |\psi\rangle=\frac{P_{\mathscr D}|\Omega_{\mathscr A}\rangle}{\|P_{\mathscr D}|\Omega_{\mathscr A}\rangle\|},\ 
 \overline{\{P_{\mathscr D}\pi_\omega(X)|\psi\rangle:X\in\mathcal A\}}
 =P_{\mathscr D}\mathcal H_\omega=\mathcal H_{\mathscr D}.\label{bfq:diamond-cyclic-packet}
\end{equation}
The first projection prepares $|\psi\rangle$. We apply $P_{\mathscr D}$ again only after the last insertion in $X$ (Figure~\ref{fig:summary-preparation}), so between successive insertions the state can range over $\mathcal H_\omega$ as in~\eqref{bfq:compressed-source-products}. By~\eqref{bfq:gns-factor}, $P_{\mathscr D}\pi_\omega(\mathcal A)^{\prime\prime}P_{\mathscr D}=\mathcal B(\mathcal H_{\mathscr D})$ on the allowed states. In particular, every interval of area with nonzero spectral projection in this space can occur in a preparation from $|\psi\rangle$. In Section~\ref{bfent:subregion-boost-main} we identify a radiative insertion that connects two distinct allowed areas.\footnote{We defer to future work whether different choices of $\omega$ in Section~\ref{bfq:state-extension} yield unitarily equivalent representations for the same range of clock readings.}

\Needspace*{6\baselineskip}
\subsection{The discrete spectrum of the area operator}
\label{bfent:subregion-boost-main}
\label{bfent:trace-density-main}
\label{bfent:clock-area-commute-main}

\begin{figure}[!htbp]
\centering
\includegraphics[width=.79\linewidth]{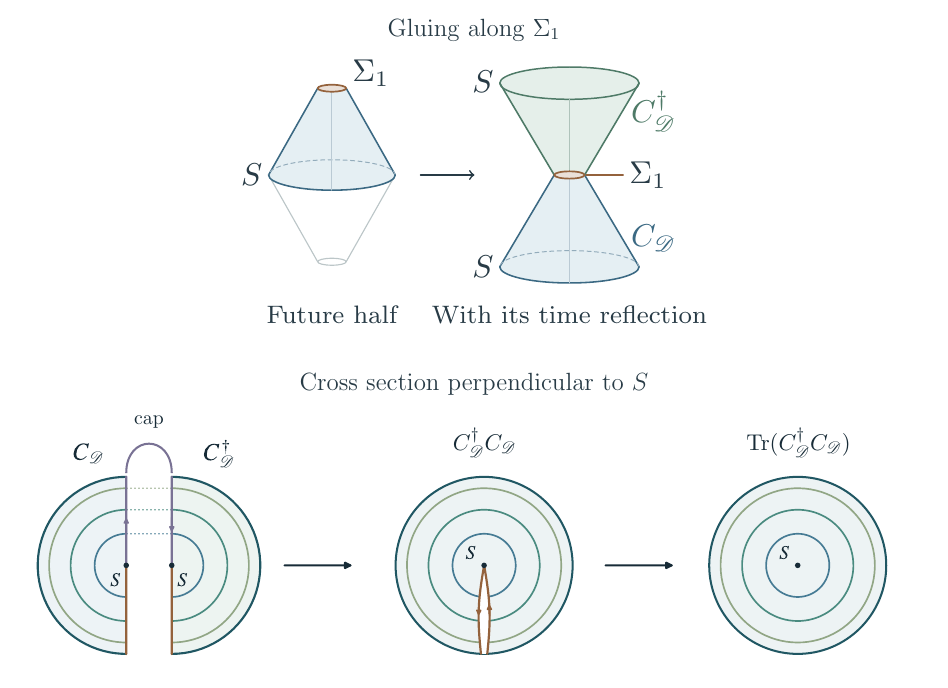}
\caption{Gluing the future half of $\mathscr D$ (blue) to its time reflection (green) along $\Sigma_1$ forms $C_{\mathscr D}^\dagger C_{\mathscr D}$. In cross section perpendicular to $S$, the cut remains open until we take $\operatorname{Tr}(C_{\mathscr D}^\dagger C_{\mathscr D})$, which identifies the two copies of $S$.}
\label{fig:bf-cap-and-trace}
\end{figure}

In~\eqref{bfq:diamond-cyclic-packet} we restricted every preparation from $|\psi\rangle$ to clock readings at most $\ell(\mathscr D)$, which also limits the corner area through the focusing in~\eqref{bfent:clock-form}. If $\mathcal H_{\mathscr D}$ were infinite dimensional, we could therefore find infinitely many orthogonal states supported in an arbitrarily narrow interval of corner area. A relative boost at $S$ acts by nearly the same phase throughout such an interval. On a smooth null cone, the same boost changes the separation of two radiative insertions across $S$ and hence their two point function. To compare these two descriptions, we glue the future half of $\mathscr D$ to its time reflection along $\Sigma_1$ (Figure~\ref{fig:bf-cap-and-trace}). One copy of $S$ carries a bra and the other a ket, so closing the cut at $S$ sums over the intermediate states in $\mathcal H_{\mathscr D}$.

We can read the clock along either null boundary, from $\partial\Sigma_0$ to $S$ on $\mathscr H_p$ or from $\partial\Sigma_1$ to $S$ on $\mathscr H_q$. On each null generator the affine parameter is normalized at $\partial\Sigma_i$ as in~\eqref{bfgeo:capped-clock}, and $\widehat{\bar u}_*$ averages its change from $v=\epsilon$ there to $v=1$ at $S$.\footnote{Closely related work uses null constraints to construct a gravitational clock~\cite{CiambelliFreidelLeigh2024DressingTime} and quantize fields along null generators~\cite{CiambelliFreidelLeigh2024QuantumNull}. Their classical solution reproduces the nonlinear dependence of our clock on shear after changing to our affine parameter (Appendix~\ref{app:bf-geometry}), but their treatment of gravitons is perturbative. In~\cite{CiambelliKlinger2026} one quantizes the bulk fields about a classical background and matches area and tangential momentum where two null boundaries intersect. The type of the resulting algebra and the existence of a trace remain open in that construction. In our quantization of the full nonlinear reduced phase space, the pure state we choose in Section~\ref{bfq:state-extension} yields an irreducible representation in which the observables generate a type~I factor.} At fixed area, shear can only increase the clock reading. Indeed, $\widehat\Xi_y(v)\ge0$ in~\eqref{bfent:clock-form} bounds the clock from below, and Cauchy Schwarz on $S$ bounds the angular integral of the area density. Since $[\widehat{\bar u}_*,\widehat{\mathscr A}]=0$, an upper limit on $\widehat{\bar u}_*$ also bounds $\widehat{\mathscr A}$ on $\mathcal H_{\mathscr D}$,
\begin{subequations}\label{bfent:clock-area-main}
\begin{align}
&\widehat{\bar u}_*\ge(1-\epsilon)\widehat{\mathscr A}\widehat{\mathsf a}^{-1},\label{bfent:clock-area-main:a}\\
&\widehat{\mathsf a}^{\,2}\le4\pi\int_S\widehat{\varrho}_S(y)^2\bm\epsilon_0=4\pi\widehat{\mathscr A},\label{bfent:clock-area-main:b}\\
&\widehat{\bar u}_*\ge(1-\epsilon)(\widehat{\mathscr A}/4\pi)^{1/2},\label{bfent:clock-area-main:c}\\
&0\le P_{\mathscr D}\widehat{\mathscr A}P_{\mathscr D}\le\frac{4\pi\ell(\mathscr D)^2}{(1-\epsilon)^2}P_{\mathscr D}.\label{bfent:clock-area-main:d}
\end{align}
\end{subequations}

To calculate the weight of a state of given area, we follow the normal frame around the corner $S$. Recall that real valued boosts are generated by $\widehat{\mathscr K}=\widehat{\mathscr A}/(8\pi G_N)$,
\begin{equation}
\alpha_\eta(\widehat{\mathcal O})=e^{i\eta\widehat{\mathscr K}/\hbar}\widehat{\mathcal O}e^{-i\eta\widehat{\mathscr K}/\hbar}.
\end{equation} 
To prepare the future half of $\mathscr D$, we analytically continue the boost to $i\pi$ and use $U$ for time evolution from $S$ to $\Sigma_1$ (Appendix~\ref{app:bf-quantization}). Gluing to its time reflection along $\Sigma_1$ pairs $U$ with $U^\dagger$ (Figure~\ref{fig:bf-cap-and-trace}). By unitarity, only the area dependence of the imaginary boost and the restriction on the clock remain in $C_{\mathscr D}^\dagger C_{\mathscr D}$,
\begin{subequations}
\label{bfent:full-gluing-main}
\begin{align}
&C_{\mathscr D}
=U e^{-\widehat{\mathscr A}/(8G_N\hbar)}P_{\mathscr D},\\
&C_{\mathscr D}^{\dagger}C_{\mathscr D}
=P_{\mathscr D}e^{-\widehat{\mathscr A}/(4G_N\hbar)}P_{\mathscr D}.
\end{align}
\end{subequations}
For the state of area $\mathscr A$ prepared in Section~\ref{bfq:diamond-state-existence}, $P_{\mathscr D}|\psi\rangle=|\psi\rangle$, so $\|C_{\mathscr D}|\psi\rangle\|^2=\exp[-\mathscr A/(4G_N\hbar)]>0$. Closing the cut at $S$ inserts the identity on $\mathcal H_{\mathscr D}$, so by~\eqref{bfq:diamond-cyclic-packet} we can use a basis prepared from $|\psi\rangle$ before we know the dimension. Each basis state enters with the weight fixed by~\eqref{bfent:full-gluing-main}.

To compare the amplitude on a smooth geometry with the sum over all allowed states, we choose a preparation centered on the corner area of that geometry. We write the dimensionless area operator as
\begin{equation}
\hat s=\frac{\widehat{\mathscr A}}{4G_N\hbar}.
\label{bfent:dimensionless-area-operator}
\end{equation}
On the bounded range of $\hat s$ in~\eqref{bfent:clock-area-main:d}, $e^{-s}f_b(s)$ is bounded away from zero for the positive continuous weight $f_b$ centered at any $s_b=\pi b^2/(G_N\hbar)$, as in~\eqref{bfent:collar-comparison}. We insert $f_b(\hat s)$ in the trace by replacing $C_{\mathscr D}$ with
\begin{equation}
C_b=C_{\mathscr D}f_b(\hat s)^{1/2},\ f_b(s)=\exp[-(s-s_b)^2/(2\sigma_s^2)].
\label{bfent:shear-preparation}
\end{equation}
To probe the shear in $C_b$, we separate the two factors of $\sigma_{AB}\sigma^{AB}$ in the Raychaudhuri equation~\eqref{bfgeo:focusing:e} with the reflected insertions of Section~\ref{bfq:state-extension}. We hold $\widehat\sigma_{f_L}$ fixed on $\Sigma_1$ as we boost the right sided null frame defining $\widehat\sigma_{f_R}$.

To evaluate this two point function, we expand about a smooth solution $g_0$ of the Einstein equations with the chosen data on the spacelike cap and $\bar u_*[g_0]<\ell(\mathscr D)$ (Section~\ref{bfq:classical-solution}). We can now choose $4\pi b^2=\mathscr A[g_0]$ so that $f_b$ peaks at its corner area. In the local Lorentz frame parallel transported from the boundary (Section~\ref{sec:bf-reconstruction}), the radius of curvature of $g_0$ is set by $R_c^{-2}=\sup_{x\in\mathscr D}\max_{\mu,\nu,\rho,\sigma}|R_{\mu\nu\rho\sigma}[g_0](x)|$. We therefore take the induced metric on $S$ to be spherically symmetric in the flat space approximation $b \ll R_c$, with fractional corrections of order $b^2/R_c^2$ from the curvature of $g_0$.

For a boost of rapidity $\eta$ of order one, we can localize both insertions near $S$ by separating their widths from their distance to the corner. For a smearing of width $\Delta u$ in $u$ we take
\begin{equation}
 \ell_P\ll\Delta u\ll d\ll b\ll R_c.\label{bfent:shear-scale-hierarchy}
\end{equation}
 We use the same transverse smearing on both sides, of size at least $\Delta u$ in a basis parallel transported along the congruence. To see why perturbation theory applies at these distances, we write $g_{ab}=g_{0\,ab}+h_{ab}$, with $h_{AB}$ the trace free transverse part of $h_{ab}$ in this basis. For a graviton wave packet of size $\Delta u$, the quadratic part of the reduced action in~\eqref{bfq:reduced-path-integral:d} fixes the size of fluctuations with $I^{(2)}[h]/\hbar=O(1)$,
\begin{equation}
 \frac{I^{(2)}[h]}{\hbar}\sim(\Delta u/\ell_P)^2h_{AB}h^{AB},\ h_{AB}=O(\ell_P/\Delta u).\label{bfent:shear-fluctuation-estimate}
\end{equation}
Each additional graviton loop then carries a factor $G_N\hbar/(\Delta u)^2$.

\begin{figure}[!htbp]
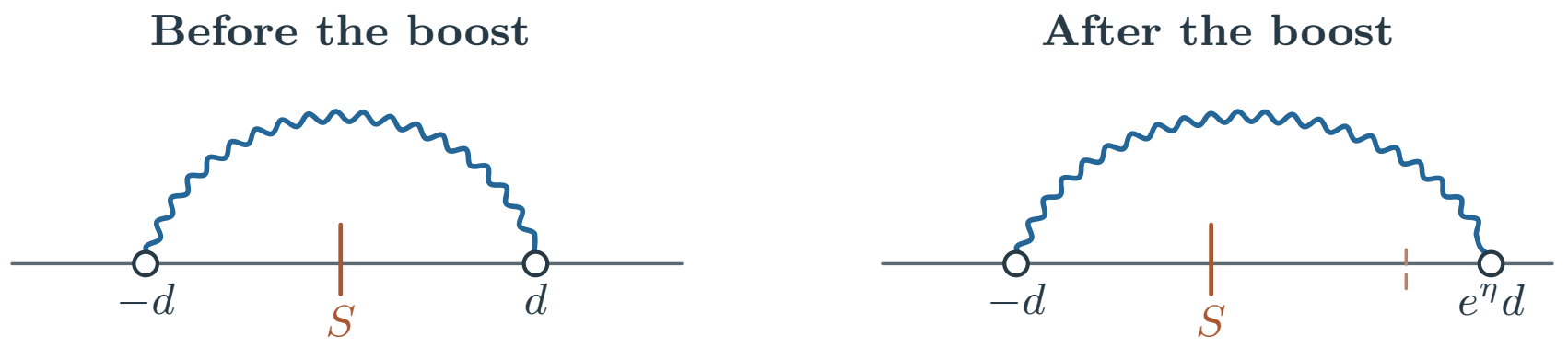

\centering
\SummaryBoost
\caption[A relative boost of two radiative insertions]{The wavy line contracts the two radiative insertions across $S$ in~\eqref{bfent:shear-boost-ratio}. The left sided smearing stays fixed while the right sided smearing moves along the null generator. The dashed tick marks its position before the boost.}
\label{fig:summary-boost}
\end{figure}

To find how a relative boost changes the correlation across $S$, we first glue the two halves at fixed $C_b,j_L,j_R$ to obtain the unnormalized amplitude $Z(j_L,j_R)$. On the shear free background the sources couple linearly to the fluctuations at one loop. Completing the square with the Hessian $K_0$ in~\eqref{bfapp:shear-source-integral:a} yields
\begin{subequations}\label{bfent:graviton-source-calculation}
\begin{equation}
\frac{Z(j_L,j_R)}{Z(0,0)}
=\frac{(\det_{\mathcal C}K_0)^{-1/2}}{(\det_{\mathcal C}K_0)^{-1/2}}\,
 \exp\!\left[-\frac{i\hbar}{2}\sum_{a,c=L,R}j_a j_c
 (\delta_I\sigma_{f_a})(K_0^{-1})^{IJ}(\delta_J\sigma_{f_c})\right],
\label{bfent:graviton-source-ratio}
\end{equation}
\begin{equation}
\frac{1}{2}\big\langle\{\widehat\sigma_{f_L},\widehat\sigma_{f_R}\}\big\rangle
=-\left.\operatorname{Re}\frac{\partial^2\log Z(j_L,j_R)}{\partial j_L\,\partial j_R}\right|_{j_L=j_R=0} .
\label{bfent:graviton-mixed-source}
\end{equation}
For $u<0<u'$, we use $\sigma_{AB}^{(1)}=\partial_u h_{AB}/2$ to obtain
\begin{equation}
\mathcal G_\sigma(u,u')=\frac{1}{2}\langle\{\widehat\sigma_{AB}(u),\widehat\sigma^{AB}(u')\}\rangle
\simeq\frac{1}{8}\big\langle\{\partial_u h_{AB}(u),\partial_{u'}h^{AB}(u')\}\big\rangle,
\label{bfent:graviton-correlator}
\end{equation}
and in the freely falling vacuum, before smearing in $u,u'$,
\begin{equation}
\mathcal G_{\sigma,0}(u,u')=-\frac{c_\perp}{(u-u')^2},
\label{bfent:freely-falling-graviton-correlator}
\end{equation}
\end{subequations}
where $c_\perp>0$ is fixed by the transverse smearing. Applying the relative boost to the right sided insertion at $u'=d$, with the left sided insertion fixed at $u=-d$ (Figure~\ref{fig:summary-boost}), yields
\begin{equation}
\frac{e^\eta\mathcal G_{\sigma,0}(-d,e^\eta d)}{\mathcal G_{\sigma,0}(-d,d)}=\frac{4e^\eta}{(1+e^\eta)^2}.
\label{bfent:shear-boost-ratio}
\end{equation}

For smearing functions reflection symmetric about $-d$ and $d$, respectively, the boost thus reduces the magnitude of the two point function by a fraction $\tanh^2(\eta/2)+O((\Delta u/d)^2)$. The supports of the two insertions are disjoint, and their transverse normalization $c_\perp$ cancels in~\eqref{bfent:shear-boost-ratio}.

The vacuum subtracted correlators in~\eqref{bfapp:shear-focusing} have finite coincidence limits at $S$ in the point split Raychaudhuri equation. With data on the cap and transverse smearing fixed in units of $b$, $\mathcal G_\sigma-\mathcal G_{\sigma,0}=O(c_\perp/b^2)$ near $S$ on the smooth background. Its contribution to~\eqref{bfent:shear-boost-ratio} is therefore of order $(d/b)^2$.

A boost acting by a single phase would preserve this two point function. If there were infinitely many intermediate states, however, arbitrarily many orthogonal states could lie in a narrow area interval where the boost phases are nearly equal. To bound the effect of the states outside that interval, we use a bounded function of the two radiative insertions whose expectation value also changes under the boost (Appendix~\ref{app:collar-return}).
On the intermediate states the boost in~\eqref{bfent:shear-boost-ratio} acts by $e^{i\eta\hat s/(2\pi)}$. Over the range of areas allowed by~\eqref{bfent:clock-area-main:d}, the Gaussian weight $f_b$ in~\eqref{bfent:shear-preparation} has a positive minimum, so
\begin{subequations}
\begin{align}
&0\le\hat s\le s_*,\ s_*=\frac{\pi\ell(\mathscr D)^2}{G_N\hbar(1-\epsilon)^2},
\label{bfent:collar-comparison:a}\\*
&\widetilde\rho_{\mathscr D}=C_{\mathscr D}^\dagger C_{\mathscr D},\ m_{b,\mathscr D}=\min_{0\le s\le s_*} f_b(s)>0,
\label{bfent:collar-comparison:b}\\*
&e^{-s_*}m_{b,\mathscr D}\widehat{1}\le m_{b,\mathscr D}\widetilde\rho_{\mathscr D}\le C_b^\dagger C_b=f_b(\hat s)\widetilde\rho_{\mathscr D}\le \widetilde\rho_{\mathscr D}\le\widehat{1} .
\label{bfent:collar-comparison}
\end{align}
\end{subequations}
For a finite set of orthonormal intermediate states $|i\rangle$, write $P=\sum_i|i\rangle\langle i|$. Pairing each $|i\rangle$ with its time reflection $|i^*\rangle$ across $S$, we obtain the expectation value of a bounded observable $\widehat{\mathcal O}$ before the boost,
\begin{equation}
\langle\widehat{\mathcal O}\rangle_{0,P}
=\frac{\sum_{i,j}(\langle i|C_b^\dagger\otimes\langle i^*|)\,
 \widehat{\mathcal O}\,(C_b|j\rangle\otimes|j^*\rangle)}
 {\sum_{i}\langle i|C_b^\dagger C_b|i\rangle}.
\label{bfent:shear-finite-observable}
\end{equation}
If nearly all the weight of the normalized state $\rho_P$ lies near $s_0$, the boost acts nearly by the same phase on those states. In Appendix~\ref{app:collar-return} we bound the resulting change in $\langle\widehat{\mathcal O}\rangle$ by
\begin{equation}
 \rho_P=\frac{PC_b^\dagger C_bP}{\Tr(PC_b^\dagger C_b)},\ 
 \big|\langle\widehat{\mathcal O}\rangle_{\eta,P}
       -\langle\widehat{\mathcal O}\rangle_{0,P}\big|
\le\frac{|\eta|}{\pi}\|\widehat{\mathcal O}\|\,
       \sqrt{\Tr[\rho_P(\hat s-s_0)^2]} .\label{bfent:boost-spread}
\end{equation}

\begin{figure}[!htbp]
\centering
\resizebox{.98\linewidth}{!}{\begingroup
\definecolor{cutInk}{HTML}{243D48}
\definecolor{cutBlue}{HTML}{326B83}
\definecolor{cutGreen}{HTML}{477967}
\definecolor{cutCopper}{HTML}{A85B31}
\begin{tikzpicture}[x=1cm,y=1cm,line cap=round,line join=round,
  font=\small,text=cutInk,
  outline/.style={draw=cutInk!88,line width=.78pt},
  slice/.style={draw=cutCopper,line width=.9pt,dash pattern=on 2.9pt off 2.2pt},
  exposed/.style={draw=cutCopper,line width=.91pt}]
\begin{scope}[shift={(.15,0)},scale=1.23]
\node at (.04,1.11) {(a)};
\foreach \cx/\bow in {2.21/0,7.02/.56,11.82/0}{
  \begin{scope}[shift={(\cx,0)}]
  \path[fill=cutBlue!5] (-1.40,0)--(-.35,1.05)--(.35,1.05)--
    (1.40,0)--(.35,-1.05)--(-.35,-1.05)--cycle;
  \ifdim\cx pt<10pt
    \path[fill=cutBlue!11] (-1.4,0).. controls (-.45,\bow) and (.45,\bow)..(1.4,0)
      --(.35,-1.05)--(-.35,-1.05)--cycle;
    \path[fill=cutGreen!10] (-1.4,0).. controls (-.45,\bow) and (.45,\bow)..(1.4,0)
      --(.35,1.05)--(-.35,1.05)--cycle;
    \draw[slice] (-1.4,0).. controls (-.45,\bow) and (.45,\bow)..(1.4,0);
  \fi
  \draw[outline] (-1.4,0)--(-.35,1.05)--(.35,1.05)--
    (1.4,0)--(.35,-1.05)--(-.35,-1.05)--cycle;
  \draw[cutGreen,line width=1.18pt] (-.35,1.05)--(.35,1.05);
  \draw[cutBlue,line width=1.18pt] (-.35,-1.05)--(.35,-1.05);
  \fill[cutInk] (.30,.64) circle[radius=1.8pt];
  \fill[cutInk] (-.31,-.48) circle[radius=1.8pt];
  \end{scope}
}
\node[left,inner sep=2pt] at (.75,0) {$S$};
\node[above,text=cutCopper,inner sep=2pt] at (2.21,.11) {$\Sigma$};
\node[below,text=cutCopper,inner sep=2pt] at (7.02,.31) {$\Sigma'$};
\node at (4.62,0) {$=$};
\node at (9.42,0) {$=$};
\node[above,text=cutGreen,inner sep=3.5pt] at (11.82,1.05) {$\Sigma_1$};
\node[below,text=cutBlue,inner sep=3.5pt] at (11.82,-1.05) {$\Sigma_0$};
\node[left,inner sep=5.8pt] at (12.12,.64) {$j_R$};
\node[right,inner sep=5.8pt] at (11.51,-.48) {$j_L$};

\end{scope}

\node at (.19,-1.89) {(b)};
\newcommand{\cutprojector}[1]{\lvert e_{#1}\rangle\langle e_{#1}\rvert}
\begin{scope}[shift={(1.77,-3.03)}]
  \filldraw[outline,fill=cutGreen!10] (-1.08,.255)--(-.33,1.005)--(.33,1.005)--(1.08,.255)--cycle;
  \filldraw[outline,fill=cutBlue!11] (-1.08,-.255)--(-.33,-1.005)--(.33,-1.005)--(1.08,-.255)--cycle;
  \draw[exposed] (-1.08,.255)--(1.08,.255) (-1.08,-.255)--(1.08,-.255);
  \draw[cutGreen,line width=1.12pt] (-.33,1.005)--(.33,1.005);
  \draw[cutBlue,line width=1.12pt] (-.33,-1.005)--(.33,-1.005);
  \fill[cutInk] (.28,.655) circle[radius=1.55pt];
  \fill[cutInk] (-.28,-.655) circle[radius=1.55pt];
  \node at (-.32,.585) {$\mathscr M_2$};
  \node at (.32,-.585) {$\mathscr M_1$};
  \node[anchor=base,inner sep=0pt] at (0,-.080) {$\cutprojector{\alpha}$};
  \node[above,inner sep=2pt] at (0,1.005) {$\langle\phi|$};
  \node[below,inner sep=2pt] at (0,-1.005) {$|\psi\rangle$};
  \node at (-1.49,0) {$\displaystyle\sum_{\alpha}$};
\end{scope}
\newcommand{\projecteddiamond}[3]{%
  \begin{scope}[shift={(#1,#2)}]
    \filldraw[outline,fill=cutGreen!10,line width=.62pt] (-.55,.22)--(-.16,.61)--(.16,.61)--(.55,.22)--cycle;
    \filldraw[outline,fill=cutBlue!11,line width=.62pt] (-.55,-.24)--(-.16,-.63)--(.16,-.63)--(.55,-.24)--cycle;
    \draw[exposed,line width=.68pt] (-.55,.22)--(.55,.22) (-.55,-.24)--(.55,-.24);
    \draw[cutGreen,line width=.88pt] (-.16,.61)--(.16,.61);
    \draw[cutBlue,line width=.88pt] (-.16,-.63)--(.16,-.63);
    \fill[cutInk] (.115,.42) circle[radius=1.0pt];
    \fill[cutInk] (-.115,-.44) circle[radius=1.0pt];
    \node[anchor=base,inner sep=0pt,font=\footnotesize] at (0,-.080) {$\cutprojector{#3}$};
  \end{scope}}
\foreach \cx/\idx in {3.97/1,5.38/2,6.79/3,8.20/4}
  {\projecteddiamond{\cx}{-3.03}{\idx}}
\foreach \cx/\idx in {10.88/1,12.32/3,14.29/2,15.73/4}
  {\projecteddiamond{\cx}{-3.03}{\idx}}
\foreach \cx in {3.11,9.79} \node at (\cx,-3.03) {$=$};
\foreach \cx in {4.675,6.085,7.495,8.94,11.60,13.30,15.01,16.75}
  \node at (\cx,-3.03) {$+$};
\foreach \cx in {9.31,17.13} \node at (\cx,-3.03) {$\cdots$};
\foreach \cx in {10.10,13.51}
  \draw[cutInk!72,line width=.65pt] (\cx+.13,-2.33).. controls (\cx-.05,-2.58)
     and (\cx-.05,-3.50)..(\cx+.13,-3.75);
\foreach \cx in {13.09,16.50}
  \draw[cutInk!72,line width=.65pt] (\cx-.13,-2.33).. controls (\cx+.05,-2.58)
     and (\cx+.05,-3.50)..(\cx-.13,-3.75);
\end{tikzpicture}
\endgroup}
\caption[Gluing on an intermediate cut]{(a) The cuts $\Sigma$ and $\Sigma'$ share a boundary $S$, where we match the gravitational boundary data. At fixed external states and sources $j_L,j_R$, diffeomorphism invariance of~\eqref{bfq:reduced-path-integral:d} requires the amplitude be the same regardless of where we make the cut or whether we make one at all. (b) Closing the cut inserts the identity. Grouping the terms of~\eqref{bfgeo:sewing-main} into cycles of $V$ includes each $|e_\alpha\rangle\langle e_\alpha|$ once.}
\label{fig:auxiliary-cap-orderings}
\end{figure}
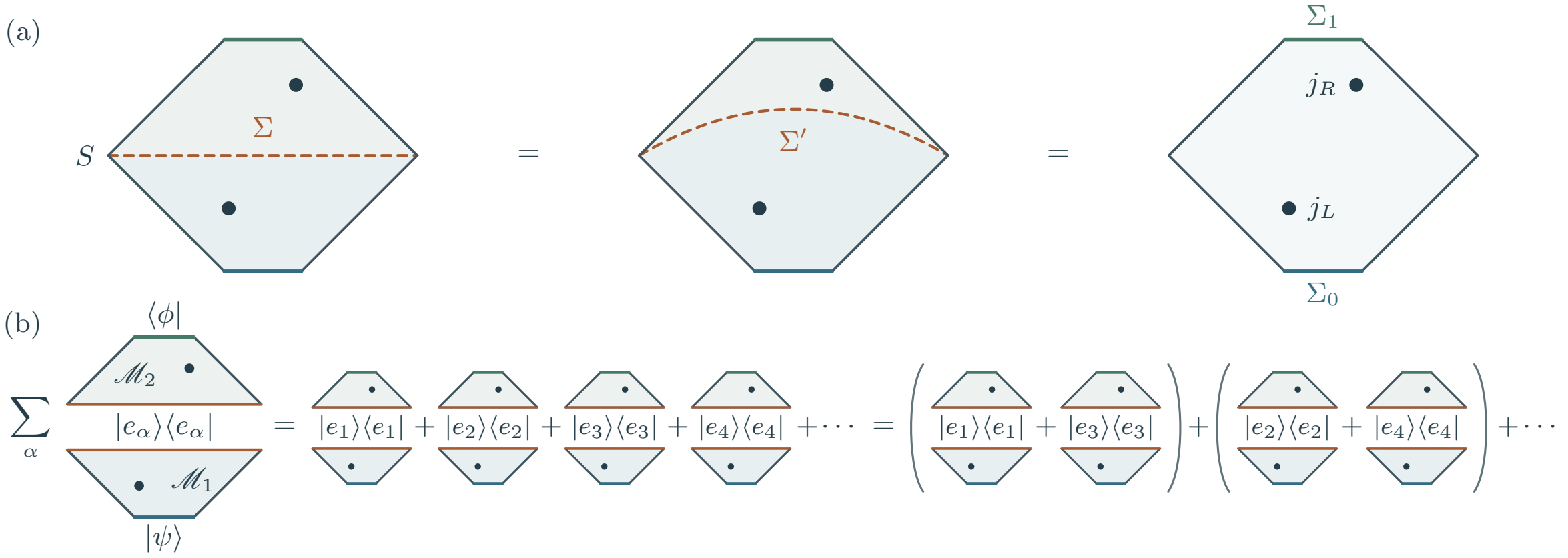

We evaluate~\eqref{bfent:boost-spread} by coupling real valued sources $j_L$ and $j_R$ to the two radiative insertions and choosing an observable with $\|\widehat{\mathcal O}\|\le1$,
\begin{equation}
\widehat{\mathcal O}=\sin(j_L\widehat\sigma_{f_L})\sin(j_R\widehat\sigma_{f_R}),\ \langle\widehat\sigma_{f_L}\widehat\sigma_{f_R}\rangle=\left.\partial_{j_L}\partial_{j_R}\langle\widehat{\mathcal O}\rangle\right|_{j_L=j_R=0}.
\label{bfent:bounded-graviton-observable}
\end{equation}
The relative boost in~\eqref{bfent:shear-boost-ratio} therefore deforms $\langle\widehat{\mathcal O}\rangle$ when the right sided insertion moves from $d$ to $e^\eta d$. For the freely falling vacuum $|0\rangle$, we may fix $j_L,j_R$ so that
\begin{equation}
\Delta_\eta\widehat{\mathcal O}=\widehat{\mathcal O}_\eta-\widehat{\mathcal O}_0,\ \delta_\eta=\langle0|\Delta_\eta\widehat{\mathcal O}|0\rangle\ne0.
\label{bfent:graviton-boost-deformation}
\end{equation}
How closely $C_b$ reproduces this change at the two separations $2d$ and $(1+e^\eta)d$ depends on the excitation of the polarization common to both insertions. For this polarization we use the smeared radiative operator
\begin{equation}
\widehat\sigma_\perp(u)=\int_S f_\perp^{AB}\widehat\sigma^{(1)}_{AB}(u,y)\,\dd A.
\label{bfent:transversely-smeared-graviton}
\end{equation}
Suppressing the boost deformation requires an excitation of this polarization on an interval containing both insertions. Appendix~\ref{app:collar-return} bounds the integral of its normal ordered square from below for states that suppress the deformation and from above for the uncut preparation $C_b$,
\begin{subequations}\label{bfent:shear-interval-cost}
\begin{align}
&\big|\langle\Delta_\eta\widehat{\mathcal O}\rangle\big|\le \frac{|\delta_\eta|}{2}
\ \Longrightarrow\ \int g^2\langle\normalorder{\widehat\sigma_\perp^2}\rangle\,\dd u\gtrsim \frac{c_f}{d},\label{bfent:shear-interval-cost:a}\\
&\int g^2\langle\normalorder{\widehat\sigma_\perp^2}\rangle_{C_b}\,\dd u=O(c_f B/b^2),\ d\ll B\ll b .\label{bfent:shear-interval-cost:b}
\end{align}
\end{subequations}
Here the real valued smearing function $g$ equals one on both insertions before and after the boost~\eqref{bfapp:shear-interval-window} and $c_f>0$ fixes the vacuum two point function of $\widehat\sigma_\perp$.

\begin{figure}[!htbp]
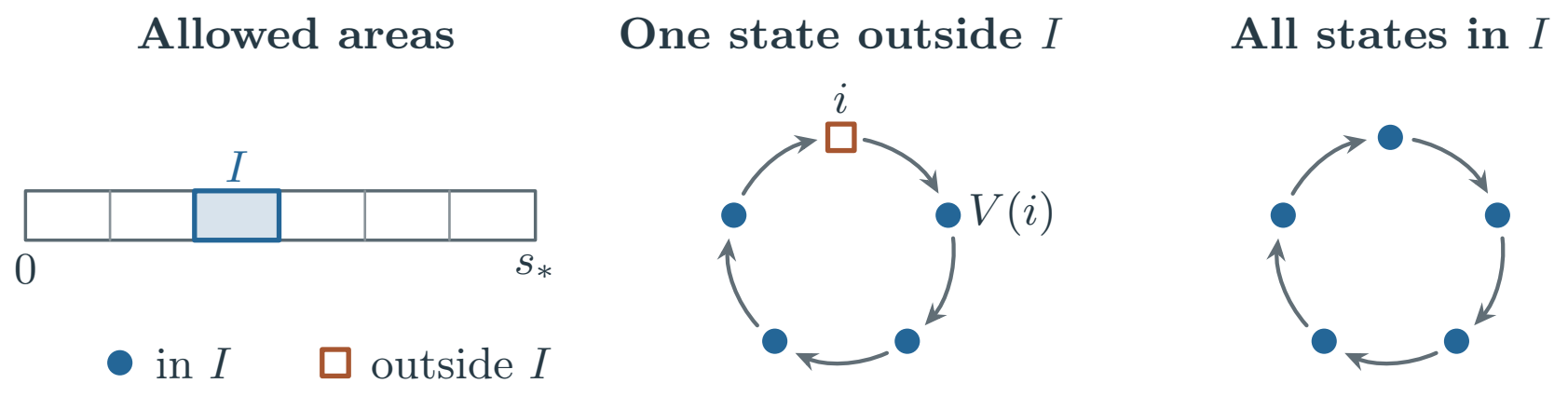

\centering
\SummaryCycles
\caption[Cycles of intermediate states of nearly equal area]{The left panel partitions the allowed areas up to $s_*$. Dots are basis vectors in the narrow interval $I$, and the square is a vector outside it. The arrows follow $V$ around two cycles of length five. The first contains one vector outside $I$, and the second lies wholly within it.}
\label{fig:summary-cycles}
\end{figure}

If $\mathcal H_{\mathscr D}$ were infinite dimensional, one of finitely many narrow intervals partitioning the allowed areas would have a spectral subspace of the same dimension as $\mathcal H_{\mathscr D}$. In a basis adapted to that interval, we place each vector outside it in a cycle with $r-1$ vectors inside, then group the remaining vectors inside into cycles of the same length (Figure~\ref{fig:summary-cycles}). The resulting permutation $V$ satisfies $V^r(i)=i$, and every cycle contains at most one vector outside the interval. The bounds on the weight of each state in~\eqref{bfent:collar-comparison} make the fraction of the weight outside this interval small in every finite union of cycles as $r$ increases. We can choose the interval sufficiently narrow and $r$ sufficiently large that~\eqref{bfent:boost-spread} bounds the change in $\langle\widehat{\mathcal O}\rangle_P$ by $|\delta_\eta|/2$ for every such projection $P$ (Appendix~\ref{app:collar-return}). Every state enters with its cycle, as illustrated for $r=2$ in Figure~\ref{fig:auxiliary-cap-orderings}b, so $P\to\widehat{1}$ strongly as we include the cycles, ordered by inclusion for an uncountable basis. By diffeomorphism invariance, closing this internal cut with $C_b$ and both insertions fixed must reproduce the unnormalized amplitude on the uncut preparation (Section~\ref{sec:pi-matching}).

\begin{figure}[!htbp]
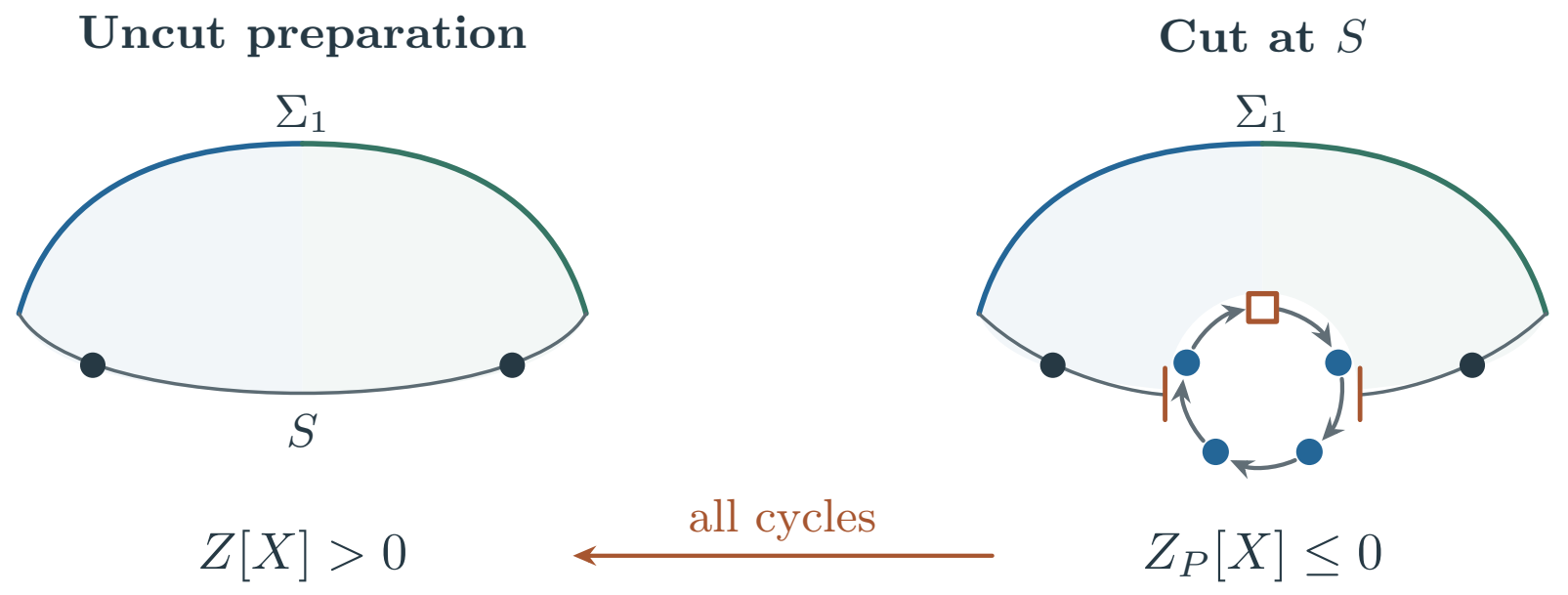

\centering
\SummaryContradiction
\caption[Cut and uncut amplitudes for the same preparation]{In both amplitudes we glue $C_b$ to its time reflection on $\Sigma_1$ with the same radiative insertions. The ring at $S$ shows a cycle of the basis labels in Figure~\ref{fig:summary-cycles}. Each ket is sewn to its own bra. Including all the cycles inserts the identity on the cut, so the partial sums in~\eqref{bfent:shear-cycle-contradiction} must reproduce the uncut amplitude.}
\label{fig:summary-contradiction}
\end{figure}

To compare the two amplitudes, we subtract half the vacuum expectation value $\delta_\eta$ in~\eqref{bfent:graviton-boost-deformation} from $\Delta_\eta\widehat{\mathcal O}$ and insert the resulting operator $X$. Since the observable in~\eqref{bfent:bounded-graviton-observable} is odd in $j_L$, we choose the sign of $j_L$ so that $\delta_\eta>0$. The remainder estimate~\eqref{bfapp:shear-interval-contrast} then lets us choose $B$ and $b$ so that $X$ has a positive expectation value in the uncut preparation $C_b$,
\begin{subequations}
\begin{equation}
X=\Delta_\eta\widehat{\mathcal O}-\frac{\delta_\eta}{2}\widehat{1},
\label{bfent:boost-deformation-insertion}
\end{equation}
\begin{equation}
\bigl|\langle\Delta_\eta\widehat{\mathcal O}\rangle_P\bigr|
\le\frac{\delta_\eta}{2}
<\langle\Delta_\eta\widehat{\mathcal O}\rangle_{C_b}.
\label{bfent:boost-deformation-bounds}
\end{equation}
\end{subequations}
Gluing $C_b$ to its time reflection multiplies $\langle X\rangle_{C_b}$ by its squared norm, so the unnormalized amplitude $Z[X]$ is strictly positive (Figure~\ref{fig:summary-contradiction}). On the cut, the same subtraction makes the unnormalized amplitude nonpositive for every nonempty finite union of cycles,
\begin{equation}
Z_P[X]=\Tr(PC_b^\dagger C_b)\left(\langle\Delta_\eta\widehat{\mathcal O}\rangle_P-\frac{\delta_\eta}{2}\right)\le 0 < Z[X].
\label{bfent:shear-cycle-contradiction}
\end{equation}

Every basis vector belongs to a cycle, so gluing at $S$ requires these partial sums to approach the uncut amplitude $Z[X]$ by~\eqref{bfgeo:sewing-main}. Their sign in~\eqref{bfent:shear-cycle-contradiction} makes this impossible. Hence $\mathcal H_{\mathscr D}$ is finite dimensional.\footnote{Restricting $\widehat{\bar u}_*$ to a smaller interval selects a subspace of $\mathcal H_{\mathscr D}$, so the conclusion also holds when the interval is too short to contain the smooth causal diamond used to evaluate $\mathcal G_\sigma$.}

Since $0<\widetilde\rho_{\mathscr D}\le\widehat{1}$, finiteness of the dimension also bounds the thermal trace. For a boundary insertion $\widehat{\mathcal O}$ the normalized amplitude is $\Tr(\rho_{\mathscr D}\widehat{\mathcal O})$, where
\begin{subequations}
\label{bfent:finite-trace-main}
\begin{align}
&\rho_{\mathscr D}
=\frac{e^{-\widehat{\mathscr A}/(4G_N\hbar)}}{Z_{\mathscr D,1}},
\label{bfent:density-main}\\*
&Z_{\mathscr D,1}
=\operatorname{Tr}_{\mathcal H_{\mathscr D}}
 e^{-\widehat{\mathscr A}/(4G_N\hbar)}<\infty,\label{bfent:interior-partition-function}\\*
&d_{\mathscr D}=\dim\mathcal H_{\mathscr D}
=Z_{\mathscr D,1}\operatorname{Tr}\!\left(\rho_{\mathscr D}e^{\widehat{\mathscr A}/(4G_N\hbar)}\right)<\infty.
\label{bfent:trace-bound-main}
\end{align}
\end{subequations}
One also obtains a discrete area spectrum by quantizing the gravitational boundary data in terms of spinors~\cite{Wieland:2017zkf}, in which case the area is proportional to the difference of two number operators~\cite{Wieland:2017cmf}. For an impulsive gravitational wave the areas of both cuts have discrete spectra~\cite{Wieland:2024dop}, and its luminosity at null infinity takes discrete values below a threshold set by $G_N$ and the Immirzi parameter~\cite{Wieland:2024kzw}. Here~\eqref{bfent:trace-bound-main} bounds both the number of distinct values of the corner area and the multiplicity of each by $d_{\mathscr D}$. Denoting its eigenvalues by $\mathscr A_1,\ldots,\mathscr A_m$, we can resolve the physical Hilbert space by area,
\begin{equation}
\mathcal H_{\mathscr D}=\bigoplus_{j=1}^m \mathcal H_{\mathscr D,\mathscr A_j}.
\label{bfent:finite-area-decomposition-main}
\end{equation}

As an independent check, in Appendix~\ref{app:focusing-graviton-states} we evaluate the amplitudes for two copies of $C_b^\dagger C_b$ joined cyclically or closed separately through one loop. The calculation also illustrates how focusing restricts the radiative modes. For a graviton of given angular momentum, a higher radial frequency reduces the effect of angular variation on the inward expansion at $S$ but strengthens focusing along the past null boundary. The resulting bounds on occupation numbers for both polarizations yield a positive ratio in~\eqref{eq:collar-regular-return}.

We can now identify an insertion that changes the area of a state prepared from $|\psi\rangle$. Within the chosen range of clock readings, $P_{\mathscr A_j}=P_{\mathscr D}\mathbf1_{\{\mathscr A_j\}}(\widehat{\mathscr A})$ projects onto states of area $\mathscr A_j$. If $\sin(j_R\widehat\sigma_{f_R})$ acted only within each area eigenspace, the boost phases of its bra and ket would cancel. By~\eqref{bfapp:shear-interval-contrast} we have $\langle\Delta_\eta\widehat{\mathcal O}\rangle_{C_b}\ne0$, so the right sided insertion has a matrix element between different eigenspaces of $\widehat{\mathscr A}$. By~\eqref{bfq:diamond-cyclic-packet}, the vectors $P_{\mathscr A_j}\pi_\omega(X)|\psi\rangle$ span the eigenspace at $\mathscr A_j$. Hence for some $X\in\mathcal A$ and $k\ne j$ we have
\begin{equation}
P_{\mathscr A_k}\sin(j_R\widehat\sigma_{f_R})P_{\mathscr A_j}\pi_\omega(X)|\psi\rangle\ne0.
\label{bfent:area-changing-insertion}
\end{equation}

The two distinct area eigenvalues connected by~\eqref{bfent:area-changing-insertion} also let us constrain the spectrum of energy at a reflecting AdS boundary. We close a thermal cylinder disjoint from $\mathscr D$ and use a measurement of area on $\mathscr D$ to determine whether the sum over states on the joint cut converges. We take the conformal boundary metric to be
\begin{equation}
\dd s^2=-\dd t^2+L_{\rm AdS}^2\,\dd\Omega_2^2.
\label{bfent:ads-boundary-metric}
\end{equation}
Taking $E_0$ to be the lower bound of the boundary Hamiltonian $\widehat H$, we measure energies from $E_0$ and write the operator for time evolution through half the period $\hbar\beta$ as
\begin{equation}
C=\exp\!\left[-\frac{\beta}{2}(\widehat H-E_0)\right],\ C^\dagger C=e^{-\beta(\widehat H-E_0)}\le\widehat{1}.
\label{bfent:ads-half-period}
\end{equation}
We choose an area band $I$ containing one of the eigenvalues distinguished by the relative boost~\eqref{bfent:shear-boost-ratio} and excluding the other. Its probability in the thermal state on $\mathscr D$ then satisfies $0<p<1$, with $p=\Tr[\rho_{\mathscr D}\mathbf1_I(\widehat{\mathscr A})]$. On the causal diamond we use an orthonormal area eigenbasis $|a\rangle$ and write $v_a=\langle a|\rho_{\mathscr D}|a\rangle$. On the AdS boundary we can choose any orthonormal basis $|i\rangle$ and write $w_i=\langle i|C^\dagger C|i\rangle$. We insert $\mathbf1_I(\widehat{\mathscr A})-p\widehat{1}$ on the causal diamond before closing the joint cut. Time evolution acts independently on the two components, so summing over a finite set $F$ of pairs $|i\rangle\otimes|a\rangle$ yields the unnormalized amplitude
\begin{equation}
 \sum_{(i,a)\in F} w_i v_a\bigl[\mathbf1_I(\mathscr A_a)-p\bigr].
 \label{bfent:ads-joint-probability}
\end{equation}

For each basis vector $|i\rangle$ on the reflecting AdS boundary, summing over all $|a\rangle$ in~\eqref{bfent:ads-joint-probability} yields zero since $\sum_a v_a=1$ and $\sum_{\mathscr A_a\in I}v_a=p$. Since the causal diamond has a finite basis, every pair on the joint cut enters as we complete the sum for each $|i\rangle$. Closing the joint cut with the insertion held fixed must therefore yield zero. Now choose $|a_0\rangle$ with $\mathscr A_{a_0}\notin I$. After completing any finite set of pairs, we can add $|i\rangle\otimes|a_0\rangle$ for a finite set $J$ of unused indices $i$, changing the sum by $-p\,v_{a_0}\sum_{i\in J}w_i$. Since $p\,v_{a_0}>0$, a divergent $\sum_i w_i$ would allow arbitrarily negative sums after any such completion, contradicting convergence to zero. Hence for each admitted $\beta>0$,
\begin{equation}
 Z_\beta=\Tr(C^\dagger C)=\sum_i w_i<\infty .
 \label{bfent:ads-joint-bound}
\end{equation}
For states in $[E_0,E_0+\Lambda]$, $C^\dagger C\ge e^{-\beta\Lambda}\widehat{1}$, so
\begin{equation}
 \dim\!\left(\mathbf1_{[E_0,E_0+\Lambda]}(\widehat H)\mathcal H\right)
 \le e^{\beta\Lambda} Z_\beta <\infty .
 \label{bfent:ads-energy-count}
\end{equation}
Since $C^\dagger C$ is compact, $\widehat H$ has a lowest eigenvalue with finite multiplicity.

To identify the observables on the sector generated from a ground state, we let $\mathcal M_\partial$ denote the von Neumann algebra generated by bounded functions of the smeared boundary stress tensor, observables of the boundary frame and bounded functions of $\widehat H$. In particular the ground state projection $Q_0$ belongs to $\mathcal M_\partial$~\cite{Marolf08082842}. Its range is finite dimensional, so $Q_0\mathcal M_\partial Q_0$ contains a minimal nonzero projection $P_0$. For a normalized ground state $|0\rangle$ in its range and for every $\widehat{\mathcal{O}}\in\mathcal M_\partial$, minimality implies
\begin{equation}
 P_0\mathcal M_\partial P_0=\mathbb C P_0,\ P_0 \widehat{\mathcal{O}}|0\rangle=\langle0|\widehat{\mathcal{O}}|0\rangle|0\rangle.\label{bfent:vacuum-projection}
\end{equation}
 Thus $P_0$ acts as $|0\rangle\langle0|$ on the sector $\mathcal H_\partial=\overline{\mathcal M_\partial|0\rangle}$. For $|v_i\rangle=\widehat{\mathcal{O}}_i|0\rangle$, we therefore have on this sector
\begin{equation}
 \widehat{\mathcal{O}}_i P_0 \widehat{\mathcal{O}}_j^\dagger=|v_i\rangle\langle v_j|.\label{bfent:boundary-matrix-units}
\end{equation}
 Since these vectors span a dense subspace, the boundary algebra on $\mathcal H_\partial$ is $\mathcal B(\mathcal H_\partial)$.

Time translations preserve $\mathcal H_\partial$, so closing the thermal circle with an insertion \(\widehat{\mathcal O}\) computes its expectation value \(\operatorname{Tr}_{\mathcal H_\partial}(\rho\widehat{\mathcal O})\) in a normal Hartle Hawking state. This state satisfies the KMS condition on \(\mathcal B(\mathcal H_\partial)\), including for operators that change area. Since \(C^\dagger C=e^{-\beta(\widehat H-E_0)}\), it follows that
\begin{equation}
\rho=\frac{e^{-\beta\widehat H}}{Z},\ Z=\operatorname{Tr}_{\mathcal H_\partial}e^{-\beta\widehat H}\le e^{-\beta E_0}Z_\beta<\infty.
\rule[-9pt]{0pt}{9pt}
\label{bfent:canonical-ensemble-main}
\end{equation}

\subsection{Entropy of marginally trapped surfaces}
\label{sec:predictions}
\label{sec:bf-entropy}
\begin{figure}[!hb]
\centering
\smallskip
\resizebox{.51\textwidth}{!}{\input{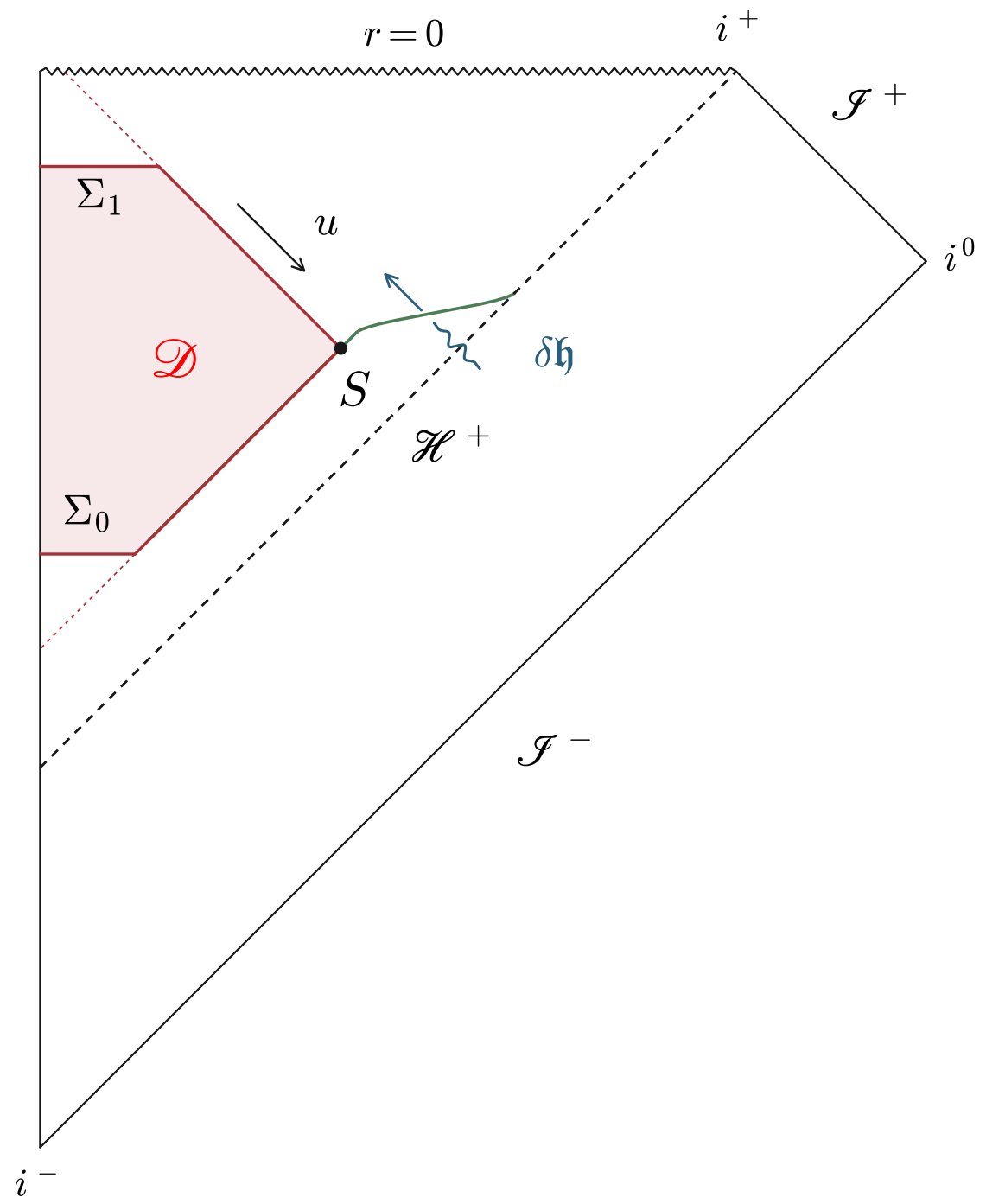}}
\caption{Penrose diagram of a black hole formed from collapse. The shaded region is a radial section of $\mathscr D$ bounded on the left by the regular centre. The green curve tracks marginally trapped surfaces and joins the event horizon $\mathscr H^+$ after an incoming perturbation $\delta\mathfrak h$.}
\label{fig:bf-black-hole-interior}
\end{figure}

In~\eqref{bfent:finite-trace-main} we obtained a density operator on the finite dimensional Hilbert space of a causal diamond, with thermal weights determined by the area operator $\widehat{\mathscr A}$. At fixed $\mathscr A$ these weights are equal, much as the entanglement spectrum of a holographic state of fixed area is flat to leading order in $G_N$~\cite{DongHarlowMarolf181105382,AkersRath181105171}. Taking the trace over all of $\mathcal H_{\mathscr D}$, we obtain
\begin{equation}
S(\rho_{\mathscr D})=-\Tr(\rho_{\mathscr D}\log\rho_{\mathscr D})
=\frac{\langle\widehat{\mathscr A}\rangle}{4G_N\hbar}+\log Z_{\mathscr D,1}.
\label{bfent:canonical-entropy-main}
\end{equation}
To identify the area term in~\eqref{bfent:canonical-entropy-main} with the Bekenstein Hawking entropy, we need a condition on $S$ that we can impose on the interior states in the trace before gluing on an exterior. But whether $S$ lies on the event horizon depends on which exterior we glue to the same interior data, since later infall can trap outgoing null generators that would otherwise reach future null infinity~\cite{Hayward1994Dynamics} (Figure~\ref{fig:bf-black-hole-interior}).

For this reason we use a marginally trapped surface as our local notion of a black hole. More precisely, on a spherically symmetric cut of the Schwarzschild horizon, $\ell^a$ is tangent to the stationary horizon, so the area element is constant along it, whereas it decreases along $n^a$. With $\ell\cdot n=-1$, we vary $\mathscr A$ subject to
\begin{equation}
\left.\Theta_{(\ell)}\right|_S=0,\ \left.\Theta_{(n)}\right|_S<0.
\label{bfent:marginal-surface}
\end{equation}
With $S$ specified, the correction to the area term in~\eqref{bfent:canonical-entropy-main} depends on how $Z_{\mathscr D,1}$ grows with $\mathscr A$. To define the gravitational clock that selects the states in this trace, we truncate the future inward null cone of $S$ at $\Sigma_1$ before its null generators reach a caustic and normalize their affine parameter there~\eqref{bfgeo:capped-clock}. Fixing the maximum reading $\ell(\mathscr D)$ then specifies which interior states enter~\eqref{bfent:interior-partition-function}.

\begin{figure}[!htb]
\centering
\begingroup
\input{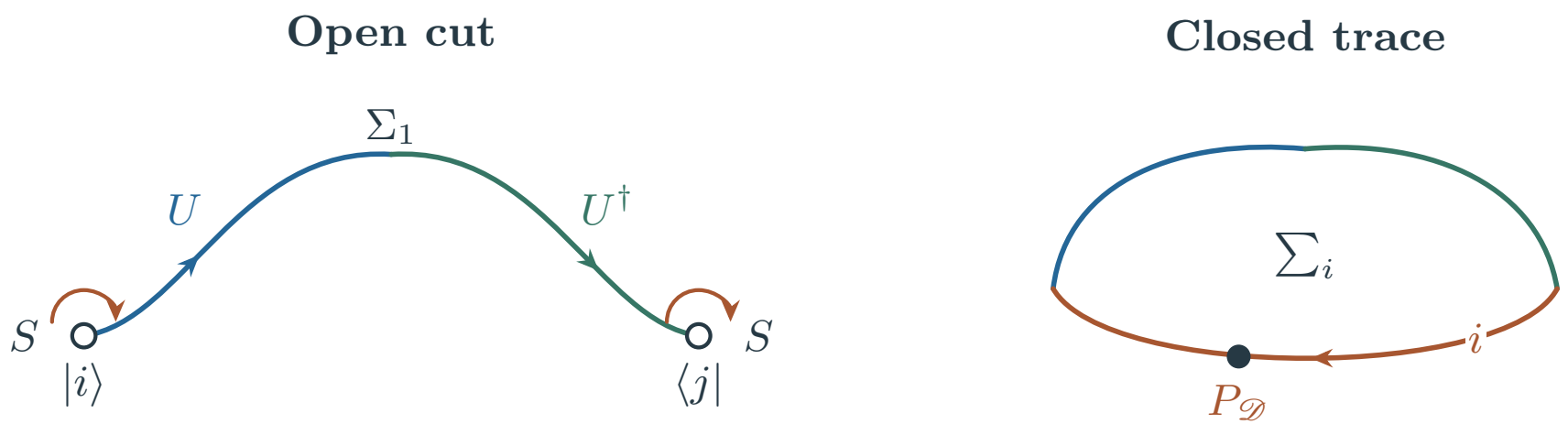}
\begin{tikzpicture}[amplitude]
 \begin{scope}
  \node[head] at (0,2.49) {Open cut};
  \draw[leg,direction mark=.39] (-2.22,.32)..controls(-1.50,.42)and(-1.10,1.70)..(0,1.63);
  \draw[notegreen,line width=1.05pt,direction mark=.67] (0,1.63)..controls(1.10,1.70)and(1.50,.42)..(2.22,.32);
  \node[above left,text=noteblue,inner sep=1.4pt] at (-1.31,1.07) {$U$};
  \node[above right,text=notegreen,inner sep=1.4pt] at (1.31,1.07) {$U^\dagger$};
  \node[external,label={[inner sep=6.2pt]left:$S$}] at (-2.22,.32) {};
  \node[external,label={[inner sep=6.2pt]right:$S$}] at (2.22,.32) {};
  \node[below] at (-2.22,.19) {$|i\rangle$}; \node[below] at (2.22,.19) {$\langle j|$};
  \node[above,inner sep=2pt] at (0,1.63) {$\Sigma_1$};
  \draw[oriented,noterust] (-2.45,.42) arc[start angle=180,end angle=0,radius=.23cm];
  \draw[oriented,noterust] (1.99,.42) arc[start angle=180,end angle=0,radius=.23cm];
 \end{scope}
 \begin{scope}[shift={(6.60cm,0)}]
  \node[head] at (0,2.49) {Closed trace};
  \draw[leg] (-1.82,.66)..controls(-1.72,1.46)and(-.88,1.75)..(0,1.67);
  \draw[notegreen,line width=1.05pt] (0,1.67)..controls(.95,1.75)and(1.70,1.38)..(1.82,.66);
  \draw[noterust,line width=1.05pt,direction mark=.49] (1.82,.66)..controls(1.46,-.01)and(-1.43,-.01)..(-1.82,.66)
    node[pos=.23,fill=white,inner xsep=.5pt,inner ysep=0pt,text=noterust] {$i$};
  \node[vertex] at (-.48,.17) {};
  \node[below,text=noterust,inner sep=2pt] at (-.48,.045) {$P_{\mathscr D}$};
  \node at (0,.90) {$\sum_i$};
 \end{scope}
\end{tikzpicture}%
\endgroup
\caption[Trace over the interior states]{The blue arrow follows the time evolution from one copy of $S$ to $\Sigma_1$, and the green arrow follows its adjoint back to the other copy. Gluing at $\Sigma_1$ identifies the outgoing label $j$ with the incoming label $i$. The orange semicircles show the two corner boosts analytically continued to half turns. Closing at $S$ sums the weights $\exp(-s_i)$ over the states admitted by the clock projection $P_{\mathscr D}$.}
\label{fig:summary-thermal}
\end{figure}

To determine the weight of each state admitted by the clock, we glue the future half of $\mathscr D$ to its time reflection along $\Sigma_1$ (Figure~\ref{fig:summary-thermal}). The time evolution cancels against its adjoint in the sum over states on $\Sigma_1$~\eqref{bfent:full-gluing-main}, so the weight is fixed by the relative boost at $S$. Analytically continuing only this boost through a half turn on each copy yields the full $2\pi$ rotation. As at any corner of a causal diamond, a state of dimensionless area $s_i$~\eqref{bfent:dimensionless-area-operator} therefore enters the trace with weight $\exp(-s_i)$~\eqref{bfent:density-main}. This boost weight yields the Bekenstein Hawking term~\cite{Bekenstein1973,Hawking1975} in~\eqref{bfent:canonical-entropy-main} at finite $G_N$~\cite{CLPW220610780,CPW220910454,Subregion260107915}.

For a large Schwarzschild black hole in equilibrium, we compare the average clock reading on the inward null cone with its maximum $\ell(\mathscr D)$ to show that $\log Z_{\mathscr D,1}$ contributes only a logarithmic correction to the entropy. We will evaluate this correction at one loop from the Lorentzian trace and, at fixed mass with angular momenta summed, recover the result Sen obtained from the Euclidean path integral~\cite{SenSchwarzschild2013}.

Recall that we end each null generator on $\Sigma_1$ at $\varrho=\epsilon\varrho_S(y)$, with fixed $\epsilon\ll1$. Choosing $u=0$ and $\partial_u\varrho=1$ there makes $u_*(y)$ in~\eqref{bfgeo:capped-clock} just the value of $u$ at $S$.

On the Schwarzschild background we choose $\bm\epsilon_0$ invariant under rotations. Then $\varrho$ is constant on each spherically symmetric cut by~\eqref{bfgeo:null-cut-geometry:d}, and its two null derivatives determine the expansions in~\eqref{bfent:marginal-surface}. We thus obtain
\begin{subequations}\label{bfent:marginal-horizon-clock}
\begin{equation}
\left.\nabla_a\varrho\,\nabla^a\varrho\right|_S
=-\frac{\varrho_S^2}{2}\left.\Theta_{(\ell)}\Theta_{(n)}\right|_S=0.
\label{bfent:marginal-horizon-location}
\end{equation}
Thus on the Schwarzschild background the area $\mathscr A$ entering the clock is the horizon area. On the future inward null cone of $S$, spherical symmetry and~\eqref{bfgeo:capped-clock} yield
\begin{equation}
\bar u_*=\varrho_S\int_\epsilon^1\dd v
=(1-\epsilon)\sqrt{\frac{\mathscr A}{4\pi}}.
\label{bfent:marginal-horizon-classical-clock}
\end{equation}
\end{subequations}
Although focusing can increase this reading in a microstate at the same area, the classical contribution to the average time delay vanishes at each horizon area in the expansion about Schwarzschild. Thus, as $r_h/\ell_P\to\infty$ at fixed $G_N$,
\begin{equation}
 \langle\widehat{\bar u}_*-(1-\epsilon)(\widehat{\mathscr A}/4\pi)^{1/2}\rangle=o(r_h),\ 4\pi r_h^2=\langle\widehat{\mathscr A}\rangle.\label{bfent:spherical-clock-expectation}
\end{equation}

To compare these interiors as the black hole grows, we must specify how the maximum clock reading scales with horizon area. In~\eqref{bfgeo:capped-clock} the integral multiplying $\varrho_S(y)$ depends only on the conformal metric along null generator $y$ as a function of $v$. Rescaling lengths in a given interior therefore rescales its clock reading by the same factor. At fixed $\ell(\mathscr D)/r_h>1-\epsilon$ the maximum exceeds $(1-\epsilon)r_h$ by an amount of order $r_h$. The $o(r_h)$ average time delay in~\eqref{bfent:spherical-clock-expectation} therefore keeps the average reading a finite fraction below the maximum as the black hole grows. For the limiting ratio $q$ we obtain an upper bound from concavity of the square root and a lower bound from focusing along the null generators,
\begin{subequations}\label{bfent:canonical-clock-depth}
\begin{align}
&\langle\widehat{\mathscr A}^{1/2}\rangle
\le \langle\widehat{\mathscr A}\rangle^{1/2}=\sqrt{4\pi}\,r_h,
\label{bfent:canonical-clock-depth:a}\\*
&q=\lim_{r_h/\ell_P\to\infty}\frac{\langle\widehat{\bar u}_*\rangle}{\ell(\mathscr D)}
\le (1-\epsilon)\frac{r_h}{\ell(\mathscr D)}<1,
\label{bfent:canonical-clock-depth:b}\\*
&q\ge\left[(1-\epsilon)\frac{r_h}{\ell(\mathscr D)}\right]^2>0 .
\label{bfent:canonical-clock-depth:c}
\end{align}
\end{subequations}
Combining the bound~\eqref{bfent:clock-area-main:c} from the Raychaudhuri equation with $\bar u_*\le\ell(\mathscr D)$, we obtain the lower bound from the second moment,
\begin{equation}
 (1-\epsilon)^2r_h^2\le\langle(\widehat{\bar u}_*)^2\rangle\le\ell(\mathscr D)\langle\widehat{\bar u}_*\rangle.\label{bfent:clock-second-moment-bounds}
\end{equation}

To relate $\langle\widehat{\bar u}_*\rangle$ to $Z_{\mathscr D,1}$, we vary only the upper limit $\ell$ in the clock projector. We write $Z(\ell)$ for the thermal trace and $\langle\cdot\rangle_\ell$ for expectation values with $\mathbf1_{[0,\ell]}(\widehat{\bar u}_*)$ in place of $P_{\mathscr D}$. Since $\widehat s$ and $\widehat{\bar u}_*$ commute by~\eqref{bfent:clock-area-main}, a joint eigenstate starts to contribute weight $e^{-s_i}$ as soon as the upper limit reaches its reading $\bar u_{*,i}$. In the integral over $\ell'$ this state contributes from $\bar u_{*,i}$ to $\ell$, so
\begin{equation}
 Z(\ell)=\sum_{\bar u_{*,i}\le \ell}e^{-s_i},\ 
 \int_0^\ell Z(\ell')\,\dd \ell'
=\sum_{\bar u_{*,i}\le \ell}(\ell-\bar u_{*,i})e^{-s_i}
=Z(\ell)\bigl(\ell-\langle\widehat{\bar u}_*\rangle_\ell\bigr),\label{bfent:integrated-clock-trace}
\end{equation}
where the sum counts independent states with the same $(s_i,\bar u_{*,i})$ separately. This integral is continuous across eigenvalues of $\widehat{\bar u}_*$, and between them \eqref{bfent:integrated-clock-trace} implies
\begin{subequations}\label{bfent:clock-trace-slope}
\begin{align}
&\ell\,\frac{\dd}{\dd \ell}\log\!\left[\frac{1}{\ell}\int_0^\ell Z(\ell')\,\dd \ell'\right]
=\frac{\langle\widehat{\bar u}_*\rangle_\ell}{\ell-\langle\widehat{\bar u}_*\rangle_\ell}.\label{bfent:clock-trace-slope:a}\\
&\frac{1}{\ell}\int_0^\ell Z(\ell')\,\dd \ell'=[1-q+o(1)]Z(\ell).\label{bfent:clock-trace-slope:b}
\end{align}
\end{subequations}
Since $q<1$, the average in~\eqref{bfent:clock-trace-slope:b} is asymptotically a positive multiple of $Z(\ell)$ by~\eqref{bfent:canonical-clock-depth:b} and~\eqref{bfent:integrated-clock-trace}. The right hand side of~\eqref{bfent:clock-trace-slope:a} tends to $q/(1-q)$, so integrating from any fixed $\ell_0$ with $\int_0^{\ell_0}Z(\ell')\,\dd\ell'>0$ up to $\ell(\mathscr D)$ yields
\begin{equation}
\log Z_{\mathscr D,1}=
\int_{\ell_0}^{\ell(\mathscr D)}\frac{\sum_{\bar u_{*,i}\le \ell'}\bar u_{*,i} e^{-s_i}}
{\sum_{\bar u_{*,i}\le \ell'}(\ell'-\bar u_{*,i})e^{-s_i}}\,\frac{\dd \ell'}{\ell'}
+O(1)
=\left[\frac{q}{1-q}+o(1)\right]\log\frac{\ell(\mathscr D)}{\ell_0}.
\label{bfent:entropy-variation-main}
\end{equation}
To express~\eqref{bfent:entropy-variation-main} in terms of horizon area at fixed $\ell(\mathscr D)/r_h$,\footnotemark\ we use
\begingroup
\postdisplaypenalty=10000\relax
\begin{equation}
\langle\widehat s\rangle=\frac{\pi r_h^2}{G_N\hbar},\ \log\frac{\ell(\mathscr D)}{\ell_0}
=\frac12\log\langle\widehat s\rangle+\log\!\left(\frac{\ell(\mathscr D)}{r_h}\frac{\ell_P}{\sqrt{\pi}\,\ell_0}\right).
\label{bfent:clock-area-logarithm}
\end{equation}
\endgroup
\footnotetext{We can also evaluate the coefficient from \(r_h^2=\langle\widehat{\mathscr A}\rangle_\ell/(4\pi)\). With \(x=(1-\epsilon)r_h/\ell\), the focusing inequality~\eqref{bfent:clock-area-main:c} and the Schwarzschild expectation value in~\eqref{bfent:spherical-clock-expectation} fix the second moment of the clock. To recover the first moment from the second, we use the following identity for each reading $u$,
\begin{equation}
 \langle(\widehat{\bar u}_*/\ell)^2\rangle_\ell=x^2+o(1),\ u=\frac{u^2}{\ell}+\int_u^\ell (u/t)^2\,\dd t.\label{bfent:clock-moment-integral}
\end{equation}
Summing this identity over clock readings as in~\eqref{bfent:integrated-clock-trace} yields
\begin{equation}
 q=x^2(2-q),\ q=\frac{2x^2}{1+x^2},\ \frac{q}{2(1-q)}=\frac{x^2}{1-x^2}.\label{bfent:clock-logarithmic-coefficient}
\end{equation}
}

For $q<1$, $\log Z_{\mathscr D,1}$ grows only logarithmically with the maximum clock reading~\eqref{bfent:entropy-variation-main}. Using~\eqref{bfent:clock-area-logarithm} to compare the correction with the Bekenstein Hawking term at large horizon area, we obtain
\begin{equation}
\frac{\log Z_{\mathscr D,1}}{\langle\widehat s\rangle}
=O\!\left(\frac{\log\langle\widehat s\rangle}{\langle\widehat s\rangle}\right).
\label{bfent:subleading-trace}
\end{equation}
The entropy~\eqref{bfent:canonical-entropy-main} of a large Schwarzschild black hole is therefore\footnote{We can also infer how many orthogonal states lie below a given area from their contribution to \(Z(\ell)\). To see this, write \(s_*=\pi\ell^2/[(1-\epsilon)^2\ell_P^2]\) for the upper bound on \(s_i\) from focusing. Using~\eqref{bfent:integrated-clock-trace} and the relation between clock reading and area given just above~\eqref{bfent:canonical-clock-depth}, we find for any fixed \(0<a\le1\) as the black hole grows
\begin{equation}
 \Pr_{\rho_{\mathscr D}}(s_i\le a s_*)\to a^{q/[2(1-q)]}.\label{bfent:area-cumulative-probability}
\end{equation}
 For any fixed \(0<\delta<a\), the band \(((a-\delta)s_*,a s_*]\) has positive limiting probability in \(\rho_{\mathscr D}\). Since each state has weight \(e^{-s_i}\) in the trace and \(\log Z(\ell)=o(s_*)\), the number of orthogonal states below the given area satisfies
\begin{equation}
 \dim\!\left(\mathbf1_{[0,a s_*]}(\widehat s)\mathcal H_{\mathscr D}\right)=\exp[a s_*+o(s_*)].\label{bfent:integrated-state-multiplicity}
\end{equation}
}
\begin{equation}
S(\rho_{\mathscr D})=\frac{\langle\widehat{\mathscr A}\rangle}{4G_N\hbar}
+\frac{q}{2(1-q)}\log\!\left(\frac{\langle\widehat{\mathscr A}\rangle}{4G_N\hbar}\right)
+o\!\left(\log\langle\widehat s\rangle\right).
\label{bfent:entropy-leading-main}
\end{equation}

\subsection{Computing logarithmic correction explicitly at one loop}
\label{bfent:one-loop-clock}

At a marginally trapped surface the corner boost fixes the Boltzmann weight $\exp(-s_i)$ of an interior state of dimensionless area $s_i=\mathscr A_i/(4G_N\hbar)$. Summing over states with clock readings at most $\ell$ yields the causal diamond partition function $Z(\ell)$ and hence the entropy $\langle\hat s\rangle+\log Z(\ell)$ in~\eqref{bfent:canonical-entropy-main}. The area term follows from the exact boost generator. Its subleading correction however depends on how $Z(\ell)$ grows with the maximum clock reading $\ell$. For a large Schwarzschild black hole in equilibrium we showed in~\eqref{bfent:entropy-leading-main} that this correction is logarithmic in the horizon area, with coefficient $q/[2(1-q)]$, where $q$ is the expectation value of the clock relative to its maximum allowed value. We now compute this coefficient explicitly in the canonical ensemble of gravitational fluctuations about the classical Schwarzschild interior at one loop, resolving the trace by area eigenvalues.

\begin{figure}[!htb]
\centering
\begingroup
\definecolor{clockMapInk}{HTML}{243C49}
\definecolor{clockMapTeal}{HTML}{28675F}
\definecolor{clockMapBlue}{HTML}{315F87}
\definecolor{clockMapRust}{HTML}{945234}
\begin{tikzpicture}[
  x=1cm,y=1cm,font=\sffamily\small,text=clockMapInk,
  every node/.style={align=left},
  box/.style={draw,line width=.65pt,inner xsep=5pt,inner ysep=4pt,
              minimum height=1.32cm,text width=6.78cm},
  input/.style={text width=4.65cm,minimum height=1.88cm},
  interior/.style={box,draw=clockMapTeal,fill=clockMapTeal!5!white},
  exterior/.style={box,draw=clockMapBlue,fill=clockMapBlue!5!white},
  common/.style={box,draw=clockMapInk,fill=clockMapInk!4!white},
  compare/.style={box,draw=clockMapRust,fill=clockMapRust!5!white},
  density/.style={minimum height=2.10cm},
  result/.style={minimum height=2.30cm},
  flow/.style={-{Stealth[length=4.5pt,width=4pt]},draw=clockMapInk,line width=.72pt}
]
\node[interior,input] (nullcone) at (-5.15,0)
  {\textbf{Black hole interior}\\
   Marginally trapped surface $S$\\
   Backreaction from sources};
\node[common,input] (shared) at (0,0)
  {\textbf{Shared input at one loop}\\
   Schwarzschild black hole\\
   Graviton and ghost heat kernel};
\node[exterior,input] (boundary) at (5.15,0)
  {\textbf{Black hole exterior}\\
   Stationary, asymptotically flat\\
   Bifurcate Killing horizon};
\node[interior] (corner) at (-4.02,-2.32)
  {\textbf{Lorentzian replica trick}\\
   Cyclic gluing of interiors at $S$\\
   using the area operator $\widehat{\mathscr A}$\\
   \mbox{$8\pi\int_S R[q]\,\dd A=64\pi^2$} for \mbox{$S\simeq S^2$}};
\node[exterior] (outside) at (4.02,-2.32)
  {\textbf{Euclidean cigar}\\
   Thermal circle contracts smoothly\\
   at the bifurcation surface\\
   \mbox{$\int R_{abcd}R^{abcd}\,\dd V=64\pi^2$}};
\node[interior,density] (area) at (-4.02,-4.61)
  {\textbf{Weighted density of states per unit area}\\
   \mbox{$W_\ell(s)=\exp(-s)\,\dd N_\ell/\dd s$}\\[2pt]
   \mbox{$\log W_\ell(s)=\tfrac12\bigl(\tfrac{212}{45}-1\bigr)\log s+O(1)$}\\[2pt]
   \mbox{$\log(Z_{\rm band}/\Delta s)=\tfrac{167}{90}\log s_*+O(1)$}};
\node[exterior,density] (mass) at (4.02,-4.61)
  {\textbf{Sen's ensemble}\\
   Mass $M$ fixed at infinity\\
   Angular momenta summed\\
   \mbox{$s_M=4\pi G_N M^2/\hbar$}\\
   \mbox{$S(M)-s_M\sim\tfrac{106}{45}\log s_M$}};
\node[interior,result] (clock) at (-4.02,-7.19)
  {\textbf{Interior ensemble}\\
   \mbox{$s\leq s_*\propto\ell^2$}, maximum clock reading $\ell$\\
   \mbox{$Z(\ell)=\Tr[P_\ell\exp(-\hat s)]$}\\
   \mbox{$\Pr(\widehat{\bar u}_*\le x\ell)=Z(x\ell)/Z(\ell)$}\\[2pt]
   \mbox{$c_{\mathscr D}=\tfrac{167}{90}+1=\tfrac{257}{90}$}};
\node[compare,result] (conversion) at (4.02,-7.19)
  {\textbf{Conversion to density per unit mass}\\
   \mbox{$\dd s_M/\dd M=8\pi G_NM/\hbar\propto\sqrt{s_M}$}\\[3pt]
   \mbox{$\log\!\left(\frac{Z_{\rm band}}{\Delta s}\frac{\dd s_M}{\dd M}\right)
           =c_M\log s_M+O(1)$}\\[3pt]
   \mbox{$c_M=c_{\mathscr D}-1+\tfrac12=\tfrac{106}{45}$}};
\draw[flow,draw=clockMapTeal] (nullcone.south) -- ([xshift=-1.13cm]corner.north);
\draw[flow] (shared.south) -- ++(0,-.22) -| ([xshift=1.22cm]corner.north);
\draw[flow] (shared.south) -- ++(0,-.22) -| ([xshift=-1.22cm]outside.north);
\draw[flow,draw=clockMapBlue] (boundary.south) -- ([xshift=1.13cm]outside.north);
\draw[flow,draw=clockMapTeal] (corner.south) -- (area.north);
\draw[flow,draw=clockMapBlue] (outside.south) -- (mass.north);
\draw[flow,draw=clockMapTeal] (area.south) -- (clock.north);
\draw[flow,draw=clockMapRust] (clock.east) -- (conversion.west);
\draw[flow,draw=clockMapRust] (conversion.north) -- (mass.south);
\end{tikzpicture}
\endgroup
\caption[Interior clock readings and the fixed mass logarithm]{The Lorentzian trace determines $c_{\mathscr D}$ and the distribution of clock readings~\eqref{bfent:clock-law:b}. To compare with Sen's ensemble we divide $Z_{\rm band}$ by the width $\Delta s$ of the area band and convert to a density per unit mass~\eqref{bfent:schwarzschild-fixed-mass-coefficient}.}
\label{fig:one-loop-clock-map}
\end{figure}
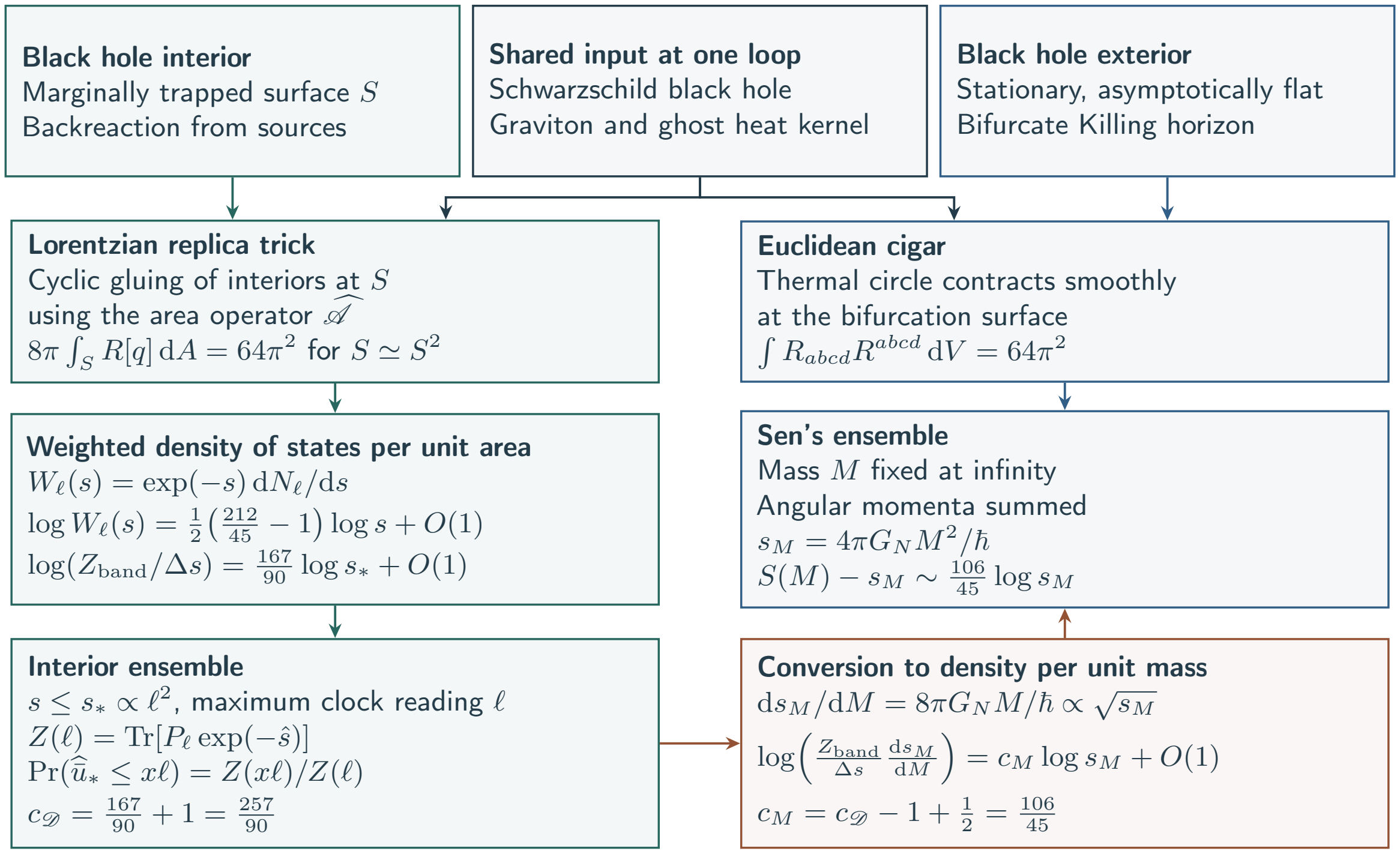

The dependence of the trace on $\ell$ also determines the distribution of clock readings. On each null generator we set $u=0$ and $\partial_u\varrho=1$ at $\partial\Sigma_1$ as in~\eqref{bfgeo:capped-clock}. The clock reads the value of $u$ at the marginally trapped surface $S$, averaged over the generators. We vary only the upper limit $t$ in the clock projector $P_t=\mathbf 1_{[0,t]}(\widehat{\bar u}_*)$. Lowering the maximum from $\ell=\ell(\mathscr D)$ to $x\ell$ selects the states in the original trace with $\widehat{\bar u}_*\le x\ell$. For $0<x<1$ their probability is
\begin{equation}
 \Pr_{\rho_{\mathscr D}}\!\left(\frac{\widehat{\bar u}_*}{\ell}\le x\right)
 =\frac{\Tr\!\left(P_{x\ell}\exp(-\hat s)\right)}
        {\Tr\!\left(P_\ell\exp(-\hat s)\right)}
 =\frac{Z(x\ell)}{Z(\ell)} .
 \label{bfent:clock-cdf-exact}
\end{equation}
Equation~\eqref{bfent:clock-cdf-exact} follows from quantizing the gravitational data on the null cone and holds for the discrete joint spectrum of the clock and area at finite $G_N$ and $\hbar$. We can form the interior trace at any smooth marginally trapped surface $S$ bounding a capped null cone, including a cut inside the event horizon during collapse (Figure~\ref{fig:bf-black-hole-interior}). Sources coupled to the shear prepare gravitational radiation on the cone. Its backreaction changes the expansion through the Raychaudhuri equation~\eqref{bfgeo:focusing:e} and hence the clock reading at a given area~\eqref{bfgeo:capped-clock}. We take the sources to zero for the equilibrium distribution. Figure~\ref{fig:one-loop-clock-map} relates this Lorentzian replica calculation to Sen's entropy calculation on the Euclidean cigar.

For each state admitted by $P_\ell$, null focusing~\eqref{bfent:clock-area-main:c} bounds the clock reading from below by $(1-\epsilon)(\mathscr A/4\pi)^{1/2}$. The upper limit $\ell$ on the reading therefore bounds the dimensionless area $s$ by
\begin{equation}
 s_*=\frac{\pi\ell^2}{(1-\epsilon)^2G_N\hbar}.
 \label{bfent:one-loop-area-bound}
\end{equation}
We denote by $Z_{\rm band}$ the contribution to $Z(\ell)$ from states with $a_1s_*<s\le a_2s_*$, for fixed $0<a_1<a_2<1$. The width of this area band is $\Delta s=(a_2-a_1)s_*$. Taking the difference of the cumulative probabilities~\eqref{bfent:area-cumulative-probability} at its two ends yields
\begin{equation}
 \frac{Z_{\rm band}}{Z(\ell)}
 =\Pr_{\rho_{\mathscr D}}\!\left(a_1<\frac{\hat s}{s_*}\le a_2\right)
 \longrightarrow a_2^{q/[2(1-q)]}-a_1^{q/[2(1-q)]}>0 .
 \label{bfent:one-loop-band-probability}
\end{equation}
The ratio tends to a nonzero constant as the black hole grows. We can therefore determine the growth of the full trace by computing the weight per unit $s$ in this band and integrating it over the band.

\begin{figure}[!htb]
\centering
\begingroup
\input{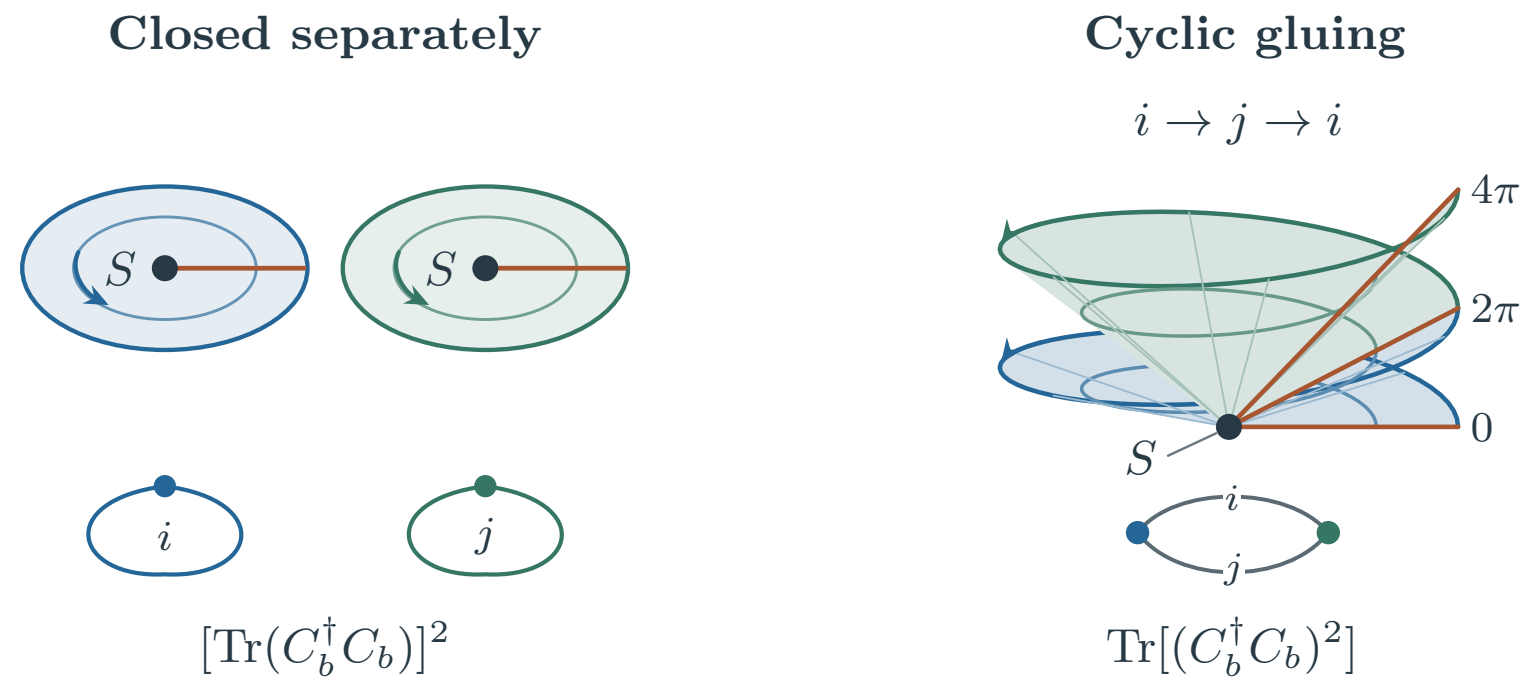}
\begin{tikzpicture}[amplitude,scale=1.06]
 \begin{scope}
  \node[head] at (0,2.99) {Closed separately};
  \foreach \x/\shade/\name in {-1.00/noteblue/i,1.00/notegreen/j}{
   \begin{scope}[shift={(\x cm,0)}]
    \filldraw[fill=\shade!12,draw=\shade,line width=.82pt] (0,1.55) ellipse[x radius=.89cm,y radius=.51cm];
    \draw[\shade!70,line width=.55pt] (0,1.55) ellipse[x radius=.57cm,y radius=.32cm];
    \draw[oriented,\shade] (-.54,1.65)..controls(-.60,1.52)and(-.49,1.39)..(-.35,1.32);
    \draw[noterust,line width=.9pt] (0,1.55)--(.87,1.55);
    \node[vertex] at (0,1.55) {};
    \node[anchor=east,inner sep=1.2pt] at (-.13,1.55) {$S$};
    \begin{scope}[yshift=-.23cm]
     \draw[\shade,line width=.79pt] (0,.42)..controls(-.70,.33)and(-.57,-.16)..(0,-.13)
        ..controls(.58,-.16)and(.69,.33)..(0,.42);
     \node[circle,fill=\shade,inner sep=0pt,minimum size=4.2pt] at (0,.42) {};
     \node[font=\footnotesize] at (0,.11) {$\name$};
    \end{scope}
   \end{scope}
  }
  \node at (0,-.81) {$[\Tr(C_b^\dagger C_b)]^2$};
 \end{scope}
 \begin{scope}[shift={(5.64cm,0)}]
  \node[head] at (.10,2.99) {Cyclic gluing};
  \foreach \lo/\hi/\shade in {0/360/noteblue,360/720/notegreen}{
   \foreach \t in {0,6,...,354}{
    \pgfmathsetmacro{\aang}{\lo+\t}
    \pgfmathsetmacro{\bang}{\aang+6}
    \path[fill=\shade!20,draw=\shade!20,line width=.18pt]
      (0,.56)--({1.43*cos(\aang)},{.56+.40*sin(\aang)+.74*\aang/360})
      --({1.43*cos(\bang)},{.56+.40*sin(\bang)+.74*\bang/360})--cycle;
   }
   \draw[\shade,line width=.85pt] plot[domain=\lo:\hi,samples=65,variable=\t]
    ({1.43*cos(\t)},{.56+.40*sin(\t)+.74*\t/360});
   \draw[\shade!75,line width=.65pt] plot[domain=\lo:\hi,samples=65,variable=\t]
    ({.92*cos(\t)},{.56+.255*sin(\t)+.476*\t/360});
   \foreach \t in {40,100,160,220,280,340}{
    \pgfmathsetmacro{\aang}{\lo+\t}
    \draw[\shade!45,line width=.35pt] (0,.56)--({1.43*cos(\aang)},{.56+.40*sin(\aang)+.74*\aang/360});}
   \pgfmathsetmacro{\astar}{\lo+145}
   \pgfmathsetmacro{\bstar}{\lo+174}
   \draw[oriented,\shade] plot[domain=\astar:\bstar,samples=8,variable=\t]
     ({1.43*cos(\t)},{.56+.40*sin(\t)+.74*\t/360});
  }
  \foreach \yy/\mark in {.56/0,1.30/{2\pi},2.04/{4\pi}}{
   \draw[noterust,line width=.87pt] (0,.56)--(1.43,\yy);
   \node[anchor=west,font=\footnotesize,inner sep=1pt] at (1.47,\yy) {$\mark$};}
  \node[vertex] at (0,.56) {};
  \draw[noteink!68,line width=.46pt] (-.38,.38)--(-.08,.52);
  \node[anchor=east,inner sep=1.5pt] at (-.39,.37) {$S$};
  \node at (.06,2.45) {$i\to j\to i$};
  \begin{scope}[yshift=-.23cm]
   \draw[noteink!75,line width=.75pt] (-.57,.13)..controls(-.34,.43)and(.37,.43)..(.62,.13)
     ..controls(.36,-.20)and(-.32,-.20)..(-.57,.13);
   \node[circle,fill=noteblue,inner sep=0pt,minimum size=4.4pt] at (-.57,.13) {};
   \node[circle,fill=notegreen,inner sep=0pt,minimum size=4.4pt] at (.62,.13) {};
   \node[font=\scriptsize,fill=white,inner sep=.2pt] at (.02,.35) {$i$};
   \node[font=\scriptsize,fill=white,inner sep=.2pt] at (.03,-.10) {$j$};
   \node at (.02,-.58) {$\Tr[(C_b^\dagger C_b)^2]$};
  \end{scope}
 \end{scope}
\end{tikzpicture}
\endgroup
\caption[Cyclic gluing at the marginally trapped surface]{The disks show the plane normal to $S$ on two copies of the preparation. Closing the copies separately returns to the same copy in one turn and sums the state indices $i$ and $j$ independently. Cyclic gluing joins the ket of each copy to the bra of the other, so the two copies form a single trace with opening angle $4\pi$ at $S$.}
\label{fig:summary-replica}
\end{figure}

In~\eqref{bfent:shear-preparation} we prepared states near a chosen $s_b$ by multiplying the amplitude by $f_b(\hat s)^{1/2}$. A state at the center comes equipped with its full boost weight, while states farther from the center contribute less. With $\sigma_s^2=\lambda s_b$ for any fixed $\lambda>0$, we can recover $Z(\ell)$ exactly by integrating over centers,
\begin{equation}
 \begin{aligned}
 &\int_0^\infty\frac{\dd s_b}{\sqrt{2\pi\lambda s_b}}\,
   \exp\!\left[-\frac{(s-s_b)^2}{2\lambda s_b}\right]=1,\ s\ge0,\\
 &Z(\ell)=\int_0^\infty\frac{\dd s_b}{\sqrt{2\pi\lambda s_b}}\,\Tr(C_b^\dagger C_b).
 \end{aligned}
 \label{bfent:area-resolved-trace}
\end{equation}
The first integral holds the area $s$ of each state fixed.\footnote{For $s>0$ we set $s_b=t^2$. The exponent is then $-(t-s/t)^2/(2\lambda)$. Averaging the integral with its image under $t\mapsto s/t$ replaces $2\,\dd t$ by $\dd(t-s/t)$. For $s=0$ we integrate a Gaussian over $t>0$.} To compute the weight near $s_b$, we now hold the center fixed and sum over states. Denoting their counting measure by $\dd N_\ell(s)$, we have
\begin{equation}
 \begin{aligned}
 &\dd N_\ell(s)=\sum_{\bar u_{*,i}\le\ell}\delta(s-s_i)\,\dd s,\\
 &\dd\mu_\ell(s)=\exp(-s)\,\dd N_\ell(s).
 \end{aligned}
 \label{bfent:clock-counting-measures}
\end{equation}
Cyclic gluing carries the same state through all $n$ copies (Figure~\ref{fig:summary-replica}). Each further copy contributes a factor $\exp(-s)f_b(s)$, so the spectrum determines the continuation to $n\sim1$,
\begin{equation}
 Z_{b,n}=\Tr\!\left[(C_b^\dagger C_b)^n\right]
 =\exp\!\left[-(n-1)s_b+\frac{(n-1)^2}{2n}\sigma_s^2\right]
 \int_{[0,s_*]}\exp\!\left[-\frac{n(s-s_n)^2}{2\sigma_s^2}\right]\dd\mu_\ell(s).
\label{bfent:clock-shared-area}
\end{equation}
Here we completed the square and wrote $s_n=s_b-(n-1)\sigma_s^2/n$.

Closing the copies separately yields $Z_{b,1}^n$. The ratio of the two traces determines the von Neumann entropy of $\rho_b=C_b^\dagger C_b/Z_{b,1}$,
\begin{equation}
 S(\rho_b)=-\Tr(\rho_b\log\rho_b)
 =-\left.\partial_n\log\!\left(\frac{Z_{b,n}}{Z_{b,1}^n}\right)\right|_{n=1}.
 \label{bfent:clock-replica-entropy}
\end{equation}

We can write the Gaussian in~\eqref{bfent:clock-shared-area} as a normalized Gaussian times $\sqrt{2\pi\sigma_s^2/n}$. There is one such factor for the common area in the cyclic trace and one for each separately closed copy. Thus the normalization contributes the following term to~\eqref{bfent:clock-replica-entropy},
\begin{equation}
 -\left.\partial_n\log\!
 \left[\frac{(2\pi\sigma_s^2/n)^{1/2}}{(2\pi\sigma_s^2)^{n/2}}\right]\right|_{n=1}
 =\frac12\log(2\pi e\,\sigma_s^2)
 =\frac12\log s_b+O(1).
 \label{bfent:shared-area-entropy}
\end{equation}
We can see how this term enters the entropy directly from the probability of each state admitted by the clock,
\begin{equation}
 \rho_{b,i}=\langle i|\rho_b|i\rangle
 =\frac{\exp(-s_i)f_b(s_i)}{Z_{b,1}}.
 \label{bfent:preparation-probability}
\end{equation}
Taking the entropy of these probabilities and substituting the Gaussian weight yields
\begin{equation}
 S(\rho_b)=-\sum_i\rho_{b,i}\log\rho_{b,i}
 =\langle\hat s\rangle_b+
 \frac{\langle(\hat s-s_b)^2\rangle_b}{2\sigma_s^2}+\log Z_{b,1}.
 \label{bfent:preparation-area-entropy}
\end{equation}
For one copy centered at $s_b$ the corresponding average of the weight per unit $s$ is
\begin{equation}
 \frac{Z_{b,1}}{\sqrt{2\pi}\,\sigma_s}
 =\int_{[0,s_*]}\frac{1}{\sqrt{2\pi}\,\sigma_s}
 \exp\!\left[-\frac{(s-s_b)^2}{2\sigma_s^2}\right]\dd\mu_\ell(s).
 \label{bfent:preparation-log-coefficient}
\end{equation}

We now compute the weight per unit $s$ directly by holding the total area fixed in the Lorentzian replica trick. We denote the one loop density of $\dd\mu_\ell(s)$ by $W_\ell(s)$. Each additional copy in the cyclic trace carries the boost weight of the same state, so the corresponding density is $W_{\ell,n}(s)=\exp[-(n-1)s]W_\ell(s)$. Comparing the two gluings at this area, we obtain
\begin{equation}
 -\left.\partial_n\log\!\left(\frac{W_{\ell,n}(s)}{W_\ell(s)^n}\right)\right|_{n=1}
 =s+\log W_\ell(s).
 \label{bfent:replica-area-density}
\end{equation}

The first term on the right follows from the corner action. We write the classical weight as $\exp(-I_n)$ and denote the contribution of $S$ to $I_n$ by $I_{S,n}$. We continue only the relative boost at $S$ to a rotation~\cite{ColinEllerinEtAl201200828}. Going around $S$ through all $n$ copies, we turn through $2\pi n$. The disk that closes the trace in the normal plane contributes the term $-s$ to $I_n$, canceling the action of one full turn~\eqref{eq:collar-regular-integrals}. At fixed opening angle the boost charge~\eqref{bfgeo:unit-corner-boost-charge} thus yields the boundary variation
\begin{equation}
 \delta I_{S,n}
 =\frac{2\pi n-2\pi}{8\pi G_N\hbar}\,\delta\mathscr A
 =\frac{n-1}{8G_N\hbar}\int_S q^{AB}\delta q_{AB}\,\dd A
 =(n-1)\,\delta s .
 \label{bfent:replica-corner-variation}
\end{equation}
Here $\delta q_{AB}$ includes the displacement of the cut. In computing $W_{\ell,n}(s)$ we sum over geometries $\phi$ for which the area of $S$ is $4G_N\hbar s$. Writing $s[\phi]=\mathscr A[\phi]/(4G_N\hbar)$, we enforce $s[\phi]=s$ by adding the Lagrange multiplier term $\nu(s[\phi]-s)$ to $I_n$. At the saddle point of this path integral, stationarity under variations of the area fixes $\nu=1-n$ by~\eqref{bfent:replica-corner-variation}, so the second variation at $S$ reduces to
\begin{equation}
 \delta^2\!\left[I_{S,n}+\nu(s[\phi]-s)\right]
 =(n-1+\nu)\,\delta^2 s[\phi]+2\,\delta\nu\,\delta s[\phi]
 =2\,\delta\nu\,\delta s[\phi].
 \label{bfent:replica-corner-hessian}
\end{equation}
Integrating over $\delta\nu$ fixes the total area. At each $s$ we still sum over the two polarizations of the graviton and the distribution of area over $S$~\eqref{bfq:fixed-area-gaussian}.

At one loop we evaluate this sum using the second variation of the action~\eqref{semiapp:metric-hessian}. Written as a quadratic form $\langle h,Hh\rangle$ in the metric perturbation $h$, it defines the fluctuation operator $H$. For eigenvalues $\lambda_j$ of $H$ the trace of its heat kernel is $\Tr\exp(-\tau H)=\sum_j\exp(-\tau\lambda_j)$. We denote the combined trace of the graviton and ghost heat kernels on the joined interiors $\mathscr D_n$ by $k_n(\tau)$.

On the Schwarzschild background $k_n(\tau)$ contains the term $53\,\mathcal I_n/[45(4\pi)^2]$ independent of the proper time $\tau$~\cite{Christensen:1979iy,BastianelliBonezziMelis2022}, with $\mathcal I_n$ as in~\eqref{bfent:schwarzschild-trace-one-loop} below. Since the eigenvalues scale as $b^{-2}$ with the Schwarzschild radius $b$, integration with $\dd\tau/(2\tau)$ yields the factor $\log(b/\ell_P)$.

Since the smooth interior and cap occur $n$ times in either gluing, their contributions cancel in $\mathcal I_n-n\mathcal I_1$. Smoothing the cone at $S$ and expanding to first order in the angular deficit yields~\cite{FursaevPatrushevSolodukhin2013}
\begin{equation}
 \begin{aligned}
 \mathcal I_n-n\mathcal I_1
 &=\int_{\mathscr D_n} R_{abcd}R^{abcd}\,\boldsymbol\epsilon_g
   -n\int_{\mathscr D_1}R_{abcd}R^{abcd}\,\boldsymbol\epsilon_g\\
 &=8\pi(1-n)\int_S\!\left(2K^{ab}K^{(n)}_{ab}
            -2R_{abcd}\ell^a n^b\ell^c n^d\right)\dd A+O\!\left((n-1)^2\right).
 \end{aligned}
 \label{bfent:schwarzschild-trace-one-loop}
\end{equation}
Here $K_{ab}$ and $K^{(n)}_{ab}$ are the extrinsic curvatures of $S$ along its two null normals $\ell^a$ and $n^a$, respectively~\eqref{bfgeo:optical-tensor},~\eqref{bfgeo:transverse-null-data:b}. Because the Schwarzschild metric is Ricci flat, the Gauss equation expresses this integrand in terms of the intrinsic scalar curvature $R[q]$ and the null expansions. The outgoing expansion vanishes on our marginally trapped surface~\eqref{bfent:marginal-surface}, so
\begin{equation}
 2K^{ab}K^{(n)}_{ab}-2R_{abcd}\ell^a n^b\ell^c n^d
 =R[q]+2\Theta_{(\ell)}\Theta_{(n)}=R[q].
 \label{bfent:schwarzschild-trace-one-loop:b}
\end{equation}
For a cut of two sphere topology the Gauss Bonnet theorem then yields
\begin{equation}
 8\pi\int_S R[q]\,\dd A=32\pi^2\chi(S)=64\pi^2 .
 \label{bfent:schwarzschild-trace-one-loop:c}
\end{equation}

To complete the determinant calculation we need to determine which fluctuations the copies share at $S$. We first hold the gravitational boundary data on each null cone fixed. In gluing two copies we identify points on $S$ by the map $Y$ and relate their null normals by a boost of rapidity $\gamma$~\eqref{bfgeo:boosted-null-normals}. Writing $X=(Y^{-1})_*\delta Y$, we obtain the zero modes of the matching conditions~\eqref{bfgeo:geometric-matching} on a spherically symmetric cut of the Schwarzschild horizon with $\Theta_{(n)}<0$,
\begin{equation}
 \Lie_X q_{AB}=0,\ \dd\delta\gamma=0,\ \Theta_{(n)}\,\delta\gamma=0
 \implies X\in\mathfrak{so}(3),\ \delta\gamma=0 .
 \label{bfent:replica-gluing-modes}
\end{equation}
Since a simultaneous change of angular frame on both sides of $S$ is gauge~\eqref{bfgeo:glued-potential-main}, we can fix the frame on the first copy. Let $R_i$ rotate the frame on copy $i$ into that on copy $i+1$. In taking the trace we identify the frame on copy $n+1$ with that on copy $1$. Thus $R_n\cdots R_1=\mathbf 1$, and with normalized Haar measure $\dd R$ we obtain
\begin{equation}
 \int_{\mathrm{SO}(3)^n}\!\prod_{i=1}^n\dd R_i\,\delta_{\mathrm{SO}(3)}(R_n\cdots R_1)
 =\left(\int_{\mathrm{SO}(3)}\dd R\right)^{n-1}=1.
 \label{bfent:replica-rotation-integral}
\end{equation}
For each spin $j$, $\Tr_j(R_n\cdots R_1)=2j+1$. All the spin states therefore enter the trace over $\Sigma_1$. At linear order the null constraint equates their angular momentum on $\partial\Sigma_1$ with that on $S$~\eqref{bfent:clock-spin-conservation}.

Matching the induced metrics also equates the areas on all $n$ copies~\eqref{eq:collar-boost-match}. We therefore integrate over one Schwarzschild radius $b$ in the cyclic trace and over $b_1,\ldots,b_n$ when we close the copies separately. Varying $b$ along the family of Schwarzschild solutions $\phi_{n,s}$, we have
\begin{equation}
 \delta\phi=\partial_b\phi_{n,s}\,\delta b=v_n\,\delta s,\ 
 v_n=\partial_s\phi_{n,s}.
 \label{bfent:replica-radius-tangent}
\end{equation}
Denoting the amplitude along the normalized mode by $z$, we find~\eqref{app:replica-area-tangent},~\eqref{app:replica-area-norm}
\begin{equation}
 H_n v_n=0,\ 
 \dd z=\mathcal J_n\,\dd s,\ 
 \mathcal J_n=\frac{\sqrt{\operatorname{Vol}(\mathscr D_n)}}{\ell_P^2s},\ 
 \partial_b \mathcal J_n=0 .
 \label{bfent:replica-radius-measure}
\end{equation}
Thus the integral over $z$ is the area integral in the trace. To compute the density per unit $s$ we take the Jacobian $\mathcal J_n$ in~\eqref{bfent:replica-radius-measure} times the Gaussian integral over fluctuations orthogonal to $v_n$~\eqref{app:replica-area-determinant}. Denoting the trace of the heat kernel on this subspace by $k_{n,\perp}$ and using $\exp(-\tau H_n)v_n=v_n$, we obtain
\begin{equation}
 k_{n,\perp}-n k_{1,\perp}=(k_n-1)-n(k_1-1)=k_n-nk_1+n-1.
 \label{bfent:replica-heat-kernels}
\end{equation}
Using~\eqref{bfent:schwarzschild-trace-one-loop:c} and $s=\pi b^2/\ell_P^2$, we obtain from~\eqref{bfent:replica-area-density}
\begin{equation}
 \log W_\ell(s)
 =-\left.\partial_n\!\left[\frac{53(\mathcal I_n-n\mathcal I_1)}{45(4\pi)^2}+n-1\right]\right|_{n=1}
   \log\!\left(\frac{b}{\ell_P}\right)+O(1)
 =\frac{167}{90}\log s+O(1).
 \label{bfent:clock-smoothed-density}
\end{equation}

To recover $Z_{\rm band}$ in~\eqref{bfent:one-loop-band-probability} we integrate $W_\ell(s)$ over the area band. Its width $\Delta s$ grows in proportion to $s_*$. Setting $s=y s_*$, we obtain
\begin{equation}
 \log Z_{\rm band}=\log\!\left(s_*\int_{a_1}^{a_2} W_\ell(ys_*)\,\dd y\right)
 =\left(\frac{167}{90}+1\right)\log s_*+O(1).
 \label{bfent:clock-band-width}
\end{equation}
By~\eqref{bfent:one-loop-band-probability} this is also the coefficient $c_{\mathscr D}$ of the logarithm in the entropy~\eqref{bfent:entropy-leading-main}. Hence
\begin{equation}
 c_{\mathscr D}=\frac{q}{2(1-q)}=\frac{167}{90}+1=\frac{257}{90}
 \implies q=\frac{257}{302}.
 \label{bfent:schwarzschild-clock-one-loop}
\end{equation}

With $q$ fixed,~\eqref{bfent:area-cumulative-probability} determines the distribution over areas. To infer the distribution of clock readings~\eqref{bfent:clock-cdf-exact}, we recall that classically $\bar u_*/\ell=\sqrt{s/s_*}$~\eqref{bfent:marginal-horizon-classical-clock}. Focusing delays the clock at fixed area. By~\eqref{bfent:clock-area-main:c} and~\eqref{bfent:spherical-clock-expectation},
\begin{equation}
 \lim_{r_h/\ell_P\to\infty}
 \left\langle\left|\frac{\widehat{\bar u}_*}{\ell}-\sqrt{\frac{\hat s}{s_*}}\right|\right\rangle_\ell=0 .
 \label{bfent:clock-radius-limit}
\end{equation}
Thus for $0<x<1$,
\begin{equation}
 \lim_{r_h/\ell_P\to\infty}
 \Pr_{\rho_{\mathscr D}}\!\left(\frac{\widehat{\bar u}_*}{\ell}\le x\right)
 =\lim_{r_h/\ell_P\to\infty}
 \Pr_{\rho_{\mathscr D}}\!\left(\frac{\hat s}{s_*}\le x^2\right)
 =x^{2c_{\mathscr D}}=x^{257/45}.
 \label{bfent:clock-law:b}
\end{equation}
The first moment of this distribution is $q$. Since we can measure the area and clock simultaneously, the same argument yields their joint distribution for $0<x,a<1$,
\begin{equation}
 \lim_{r_h/\ell_P\to\infty}\Pr_{\rho_{\mathscr D}}\!
 \left(\frac{\widehat{\bar u}_*}{\ell}\le x,\ \frac{\hat s}{s_*}\le a\right)
 =\bigl[\min(x,\sqrt a)\bigr]^{257/45}.
 \label{bfent:clock-area-joint-law}
\end{equation}

In~\eqref{bfent:clock-area-joint-law} both the clock reading and the horizon area fluctuate at fixed $\ell$. In equilibrium we can also specify the mass $M$ of the black hole as measured at infinity. To compare~\eqref{bfent:schwarzschild-clock-one-loop} with Sen's logarithmic correction to the entropy~\cite{SenSchwarzschild2013}, we need the density of states per unit $M$, summed over angular momenta.

Both calculations use the same curvature term in the graviton and ghost heat kernel (Figure~\ref{fig:one-loop-clock-map}). Sen integrates it over the Euclidean black hole exterior. Requiring the thermal circle to contract smoothly at the bifurcation surface fixes its period at infinity to $4\pi r_h$. The integral over the exterior then equals the surface integral~\eqref{bfent:schwarzschild-trace-one-loop:c} in our Lorentzian calculation,
\begin{equation}
 (4\pi r_h)(4\pi)\int_{r_h}^{\infty}\frac{12r_h^2}{r^4}\,\dd r
 =64\pi^2=8\pi\int_S R[q]\,\dd A .
 \label{bfent:exterior-interior-curvature}
\end{equation}
To convert our result to a density we first divide $Z_{\rm band}$ in~\eqref{bfent:clock-band-width} by the width $\Delta s$ of the area band,
\begin{equation}
 \log\!\left(\frac{Z_{\rm band}}{\Delta s}\right)
 =(c_{\mathscr D}-1)\log s_*+O(1).
 \label{bfent:band-density-per-area}
\end{equation}

Even at fixed $M$ the horizon area varies with angular momentum $J$. We denote the dimensionless area of the Schwarzschild black hole by $s_M=4\pi G_N M^2/\hbar$. For a Kerr black hole of the same mass, Christodoulou's mass formula~\cite{Christodoulou1970} reads
\begin{equation}
 M^2=\frac{\mathscr A}{16\pi G_N^2}+\frac{4\pi J^2}{\mathscr A}
 \implies s_M-s=\frac{4\pi^2}{s}\left(\frac{J}{\hbar}\right)^2 .
 \label{bfent:clock-kerr-reweight}
\end{equation}
We can estimate the size of this area fluctuation from frame dragging. In units of the Schwarzschild radius $b$, the metric perturbation due to rotation is of order $G_N J/b^2$. For a given perturbation the quadratic term in the Einstein Hilbert action in units of $\hbar$ scales as $b^2/\ell_P^2$~\eqref{app:replica-spin-scaling}. The root mean square amplitude of the perturbation is therefore of order $\ell_P/b$. Averaging at fixed $s$, we find at one loop~\eqref{app:replica-spin-covariance}
\begin{equation}
 \frac{G_N}{b^2}\sqrt{\langle J^2\rangle_{\ell,s}}=O(\ell_P/b)
 \implies \frac{\langle J^2\rangle_{\ell,s}}{\hbar^2}=O(s),\ 
 \langle s_M-s\rangle_{\ell,s}=O(1).
 \label{bfent:clock-spin-scale}
\end{equation}
To extract the correction to $s_M$ in the entropy we divide the density of states $\dd N_\ell/\dd M$ by $\exp(s_M)$. Relative to the Boltzmann weight $\exp(-s)$ in the causal diamond partition function $Z(\ell)$ this introduces the factor $\exp(s-s_M)$. Integrating over the three components of $J$ in~\eqref{app:replica-spin-determinant}, we find that this factor changes the entropy by $O(1)$. Since $\dd s_M/\dd M=8\pi G_N M/\hbar\propto\sqrt{s_M}$, for $s_M$ in the area band we obtain
\begin{equation}
 \log\!\left[\exp(-s_M)\frac{\dd N_\ell}{\dd M}\right]
 =\log\!\left(\frac{Z_{\rm band}}{\Delta s}\frac{\dd s_M}{\dd M}\right)+O(1)
 =\left(c_{\mathscr D}-1+\frac12\right)\log s_M+O(1).
 \label{bfent:band-density-per-mass}
\end{equation}
We thus recover Sen's result for the entropy at fixed mass~\cite{SenSchwarzschild2013},\footnote{In loop quantum gravity the entropy of the $\mathrm{SU}(2)$ Chern Simons states on a nonrotating isolated horizon of fixed area has a logarithmic correction with coefficient $-3/2$~\cite{Kaul:2000kf}.}
\begin{equation}
 S(M)=s_M+c_M\log s_M+O(1),\ 
 c_M=c_{\mathscr D}-1+\frac12=\frac{106}{45}.
 \label{bfent:schwarzschild-fixed-mass-coefficient}
\end{equation}
\FloatBarrier

\subsection{Spatial topology and null states}
\label{sec:spatial-topology}
\label{topology:relation}

When we build a spatial slice from causal diamonds, we can hold the data at its boundaries fixed and still vary how we join them in the interior. For example, a slice can connect two boundaries, or it can have two components, one ending on each boundary. To determine which fillings prepare independent quantum states, we compute their overlaps by gluing~\eqref{bfgeo:sewing-main}.

This issue arises even for a single boundary \(\Sigma\) in the thermal trace~\eqref{bfent:interior-partition-function}, since we can add arbitrarily many handles to the interior. For the states \(|v_i\rangle=|\mathscr M_i\rangle\) prepared by these fillings, gluing \(\mathscr M_j\) to the time reflection of \(\mathscr M_i\) along \(\Sigma\) yields
\begin{equation}
\mathsf G_{ij}=\langle v_i|v_j\rangle,\ \bigl\|\sum_i c_i|v_i\rangle\bigr\|^2=c^\dagger\mathsf Gc,\ \operatorname{rank}\mathsf G\le d_{\mathscr D}.
\label{topology:overlaps}
\end{equation}
Any set of more than \(d_{\mathscr D}\) such fillings therefore admits a null superposition~\cite{MarolfMaxfield2020}. Since the boundary preparations in~\eqref{bfq:diamond-cyclic-packet} already span \(\mathcal H_{\mathscr D}\), adding a handle can change an amplitude without adding an independent state to the trace.

With two boundaries we can prepare product states by filling each boundary separately. Gluing the connected filling $\mathscr M$ to its time reflection yields the squared norm of the state it prepares. A natural, as well as important, question is whether the Hilbert space of the two boundaries factorizes~\cite{HarlowJafferis2020,MarolfMaxfield2020}, in which case the squared overlaps with an orthonormal basis of product states must sum to this norm.

To this aim, we compute each overlap by gluing one filling to each of the two boundaries of $\mathscr M$. These overlaps are the matrix elements of an operator $N_{\mathscr M}$ on $\mathcal H_\Sigma$. Time reflection conjugates amplitudes, so gluing \(\mathscr M_2\) to the reflection of \(\mathscr M_1\) sums their products over states at both boundaries,
\begin{equation}
\label{topology:full-pairing}
\langle\mathscr M_1|\mathscr M_2\rangle
=\operatorname{Tr}_{\mathcal H_{\Sigma}}
\bigl(N_{\mathscr M_1}^\dagger N_{\mathscr M_2}\bigr).
\end{equation}
We can prepare a complete orthonormal basis \(|u_i\rangle\) independently at each boundary when boundary sources generate \(\mathcal B(\mathcal H_\Sigma)\). For the cap Hilbert space constructed in Section~\ref{bfq:diamond-state-existence}, boundary sources generate \(\mathcal B(\mathcal H_{\mathscr D})\), so we can choose the basis states to have definite area or to be superpositions of different areas. The disjoint union $q_{ij}$ corresponds to \(|u_i\rangle\langle u_j|\),
\begin{equation}
q_{ij}=u_i\sqcup u_j^\dagger,\ \langle q_{ij}|\mathscr M\rangle=(N_{\mathscr M})_{ij}.
\label{topology:product-state-overlaps}
\end{equation}
Subtracting the superposition over these product states with their respective overlaps yields
\begin{equation}
\bigl\||\mathscr M\rangle-\sum_{i,j}(N_{\mathscr M})_{ij}|q_{ij}\rangle\bigr\|^2
=\operatorname{Tr}(N_{\mathscr M}^\dagger N_{\mathscr M})
-\sum_{i,j}|(N_{\mathscr M})_{ij}|^2=0.
\label{topology:full-residual}
\end{equation}
At an AdS boundary the sum in~\eqref{topology:full-residual} can be infinite. With the boundary energies restricted to bounded intervals, $\operatorname{Tr}(N_{\mathscr M}^\dagger N_{\mathscr M})<\infty$ (Appendix~\ref{sec:lorentzian-construction}), so the superposition of product states converges in the physical norm. Since \(|\mathscr M\rangle\) and this superposition differ by a null state, they represent the same physical state. Interactions in \(H_c\) enter the coefficients \((N_{\mathscr M})_{ij}\), while~\eqref{topology:full-residual} follows directly from gluing.

The reflected states \(|u_j^*\rangle\) form a basis of the conjugate Hilbert space \(\mathcal H_\Sigma^*\), so for the two boundaries we have
\begin{equation}
\label{topology:complete-two-boundary}
\mathcal K_{\Sigma}\simeq
\mathcal H_{\Sigma}\otimes\mathcal H_{\Sigma}^*.
\end{equation}
In the model of~\cite{MarolfMaxfield2020}, a state of connected topology can have a nonzero component orthogonal to all product states. Here that component is null by~\eqref{topology:full-residual}, so spatial topology does not correspond to a superselection sector within \(\mathcal B(\mathcal H_\Sigma)\), whose center contains only multiples of the identity. At fixed area, we use \(C_{\mathscr D}=c_{\mathscr A}U\) from Section~\ref{bfent:subregion-boost-main}, where \(c_{\mathscr A}=\exp[-\mathscr A/(8G_N\hbar)]\). In the basis \(U|u_i\rangle\) at the cap, we obtain coefficients \(c_{\mathscr A}\delta_{ij}\); each term on the right of~\eqref{topology:full-residual} is then \(d_{\mathscr A}|c_{\mathscr A}|^2\), where \(d_{\mathscr A}=\dim\mathcal H_{\Sigma,\mathscr A}\). After normalization, all $d_{\mathscr A}$ states on either side have equal probability, since the physical trace assigns unit weight to each rank one projection, so the entropy on either side is $\log d_{\mathscr A}$. 

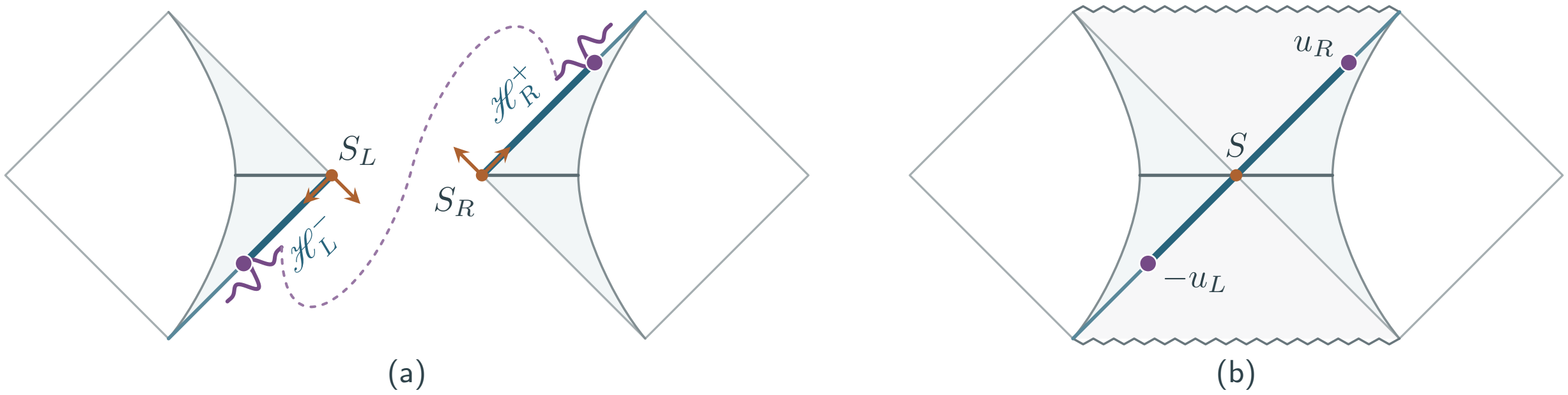
\begin{figure}[!htb]
\centering
\begingroup
\definecolor{pcInk}{HTML}{2E424A}
\definecolor{pcNull}{HTML}{28647C}
\definecolor{pcMode}{HTML}{754A86}
\definecolor{pcFrame}{HTML}{AD6230}
\begin{tikzpicture}[x=1.69cm,y=1.69cm,>=stealth,line join=round,line cap=round,
    pc outline/.style={draw=pcInk!43,line width=.60pt},
    pc wall/.style={draw=pcInk!60,line width=.65pt},
    pc ray/.style={draw=pcNull!77,line width=1.05pt},
    pc interval/.style={draw=pcNull,line width=2.0pt},
    pc label/.style={font=\normalsize,text=pcInk,inner sep=1pt},
    pc panel/.style={font=\small\sffamily,text=pcInk,inner sep=0pt}]
  \def\pcExterior{%
    \fill[pcNull!6] (0,0)--(1,1)
      .. controls (.80,.76) and (.59,.31) .. (.59,0)
      .. controls (.59,-.31) and (.80,-.76) .. (1,-1)--cycle;
    \draw[pc outline] (0,0)--(1,1)--(2,0)--(1,-1)--cycle;
    \draw[pc wall] (1,1) .. controls (.80,.76) and (.59,.31) .. (.59,0)
      .. controls (.59,-.31) and (.80,-.76) .. (1,-1);
    \draw[pcInk!78,line width=.95pt] (0,0)--(.59,0);}
  \def\pcFrameAt#1#2{%
    \begin{scope}[shift={#1},yscale=#2]
      \draw[pcFrame,->,line width=1.0pt] (0,0)--(-.175,.175);
      \draw[pcFrame,->,line width=1.0pt] (0,0)--(.175,.175);
      \fill[pcFrame] (0,0) circle[radius=1.8pt];
    \end{scope}}
  \def\pcCut#1{\filldraw[fill=pcMode,draw=white,line width=.5pt] #1 circle[radius=2.55pt];}
  \def\pcInsertion#1#2{%
    \begin{scope}[shift={#1},rotate=#2]
      \draw[pcMode!65,line width=.6pt] (0,0)--(0,.094);
      \draw[pcMode,line width=1.16pt] plot[domain=-1:1,samples=61]
        ({.235*\x},{.094+.086*sin(450*\x)*cos(90*\x)^2});
    \end{scope}
    \pcCut{#1}}
  \begin{scope}[shift={(-.46,0)}]
    \begin{scope}[xscale=-1]\pcExterior\end{scope}
    \draw[pc ray] (-1,-1)--node[pos=.73,sloped,below=3pt,pc label,text=pcNull]
      {$\mathscr H_L^-$} (0,0);
    \draw[pc interval] (-.54,-.54)--(0,0);
    \pcInsertion{(-.54,-.54)}{225}
    \pcFrameAt{(0,0)}{-1}
    \node[pc label,anchor=south west] at (.01,.04) {$S_L$};
  \end{scope}
  \begin{scope}[shift={(.46,0)}]
    \pcExterior
    \draw[pc ray] (0,0)--node[pos=.34,sloped,above=3pt,pc label,text=pcNull]
      {$\mathscr H_R^+$} (1,1);
    \draw[pc interval] (0,0)--(.69,.69);
    \pcInsertion{(.69,.69)}{45}
    \pcFrameAt{(0,0)}{1}
    \node[pc label,anchor=north east] at (-.01,-.05) {$S_R$};
  \end{scope}
  \draw[overlay,pcMode!75,line width=.78pt,dash pattern=on 1.25pt off 2.35pt]
    (-.767,-.440) .. controls (-.77,-1.16) and (-.17,-.72) .. (.03,.04)
    .. controls (.205,.705) and (.76,1.30) .. (.917,.590);
  \node[pc panel] at (0,-1.22) {(a)};
  \begin{scope}[shift={(5.08,0)}]
    \fill[pcInk!4] (0,0)--(-1,1)--(1,1)--cycle
                    (0,0)--(-1,-1)--(1,-1)--cycle;
    \begin{scope}[xscale=-1]\pcExterior\end{scope}
    \pcExterior
    \foreach \side in {-1,1} {
      \draw[pcInk!72,line width=.65pt] (-1,\side)
        \foreach \i in {1,...,32} {--({-1+\i/16},{\side*(1+.035*mod(\i,2))})};}
    \draw[pc ray] (-1,-1)--(1,1);
    \draw[pc interval] (-.54,-.54)--(.69,.69);
    \pcCut{(-.54,-.54)} \pcCut{(.69,.69)}
    \fill[pcFrame] (0,0) circle[radius=1.8pt];
    \node[pc label,anchor=south] at (0,.09) {$S$};
    \node[pc label,anchor=west] at (-.48,-.64) {$-u_L$};
    \node[pc label,anchor=east] at (.63,.79) {$u_R$};
    \node[pc panel] at (0,-1.22) {(b)};
  \end{scope}
\end{tikzpicture}
\endgroup
\caption{On the disconnected spacetime we dress each radiative insertion (violet) to the null frame at its own corner (orange), where the area operator $\widehat{\mathscr A}$ divided by $8\pi G_N$ generates a uniform boost. Violet dashes denote entanglement between the modes in~\eqref{topology:entangled-mode}. In this state the Raychaudhuri equation along each horizon yields the same two point function~\eqref{topology:horizon-area-correlator:b} of second null derivatives of area as at the two marked cuts on one generator through $S$.}
\label{fig:topology-horizon-cuts}
\end{figure}

The Hartle Hawking state of a two sided Schwarzschild black hole illustrates the geometric content of~\eqref{topology:full-residual}~\cite{MaldacenaSusskind2013}. In the connected spacetime, a generator of the left past horizon continues through the bifurcation surface onto the right future horizon. On the disconnected caps, each horizon has its own corner (Figure~\ref{fig:topology-horizon-cuts}). To compare the two descriptions, we place a reflecting wall in each exterior. At inverse temperature $\beta_R$ measured at either wall, the path integral over half the thermal circle assigns the product state $|q_{ij}\rangle$ the amplitude $\langle u_i|e^{-b\widehat H/\hbar}|u_j\rangle$ up to normalization, where $b=\hbar\beta_R/2$ and $\widehat H=\mathcal Q(H_c)$ generates proper time translations at the wall~\cite{HartleHawking1976,LoukoWhiting1995}. For a horizon of radius $r_h\gg\ell_P$, we expand $H_c$ to second order and retain one odd parity quadrupole~\cite{Moncrief1974}. At wall frequency $\omega$, the thermal amplitude pairs the state of $n$ gravitons on the left with its time reflection on the right,
\begin{equation}
|\Psi_\omega\rangle\propto\sum_{n=0}^\infty e^{-nb\omega}|n\rangle_L|n^*\rangle_R,
\langle\mathfrak{a}_L\mathfrak{a}_R\rangle=\frac{1}{2\sinh(b\omega)},
\label{topology:entangled-mode}
\end{equation}
where $\mathfrak{a}_L$ and $\mathfrak{a}_R$ are the annihilation operators for the respective graviton modes.

Choose future directed affine parameters $-u_L<0$ and $u_R>0$ on the two horizons, measured from their respective corners with the same normalization under time reflection. A translation in proper time at the right wall rescales $u_R$ by $e^{\pi\delta t_R/b}$, so a pair at frequency $\nu=b\omega/\pi$ enters the two point function of the radiative insertions with phase $(u_L/u_R)^{i\nu}$. For $\ell_P\ll u_L,u_R\ll r_h$, with $c_\perp$ as in~\eqref{bfapp:shear-singularity}, summing the paired modes yields
\begin{equation}
\mathcal G_\sigma(-u_L,u_R)=-\frac{c_\perp}{u_Lu_R}\int_0^\infty\frac{\nu\cos[\nu\log(u_L/u_R)]}{\sinh(\pi\nu)}\,\dd\nu
=-\frac{c_\perp}{(u_L+u_R)^2}.
\label{topology:horizon-correlator}
\end{equation}

The dependence on $u_L+u_R$ in~\eqref{topology:horizon-correlator} follows from entangling the left and right graviton modes as in~\eqref{topology:entangled-mode}, even on two fixed Schwarzschild backgrounds. When the horizons themselves fluctuate, the same entanglement correlates the areas of the two cuts, since the shear on each horizon focuses its null generators. For the odd parity quadrupole $Y_{AB}$ with $\int\dd A_0\,Y_{AB}Y^{AB}=1$, the expansion vanishes through first order, so the Raychaudhuri equation~\eqref{bfgeo:focusing:e} yields
\begin{subequations}\label{topology:horizon-area-correlator}
\begin{align}
&\partial_u^2\widehat{\mathscr A}^{(2)}(u)=-\int\dd A_0\,\normalorder{\widehat\sigma^{(1)}_{AB}(u)\widehat\sigma_{(1)}^{AB}(u)},\label{topology:horizon-area-correlator:a}\\
&\big\langle\partial_{u_L}^2\widehat{\mathscr A}^{(2)}_L(u_L)\,\partial_{u_R}^2\widehat{\mathscr A}^{(2)}_R(u_R)\big\rangle_c
=2\mathcal G_\sigma(-u_L,u_R)^2=\frac{2c_\perp^2}{(u_L+u_R)^4}.\label{topology:horizon-area-correlator:b}
\end{align}
\end{subequations}
The factor of two counts the Wick contractions between cuts after normal ordering in the freely falling vacuum. Thus the area fluctuations on the disconnected caps reproduce the dependence on affine parameter along one Schwarzschild horizon generator from $-u_L$ to $u_R$.

\section{The semiclassical limit}
\label{sec:bf-classical}

A state sharply peaked about smooth initial data $\mathfrak h_{\Sigma_0}$ specifies a classical geometry on the incoming null cone. For its time evolution to describe a single spacetime, the metrics we reconstruct in neighboring causal diamonds must agree on their overlap. Comparing them requires us to identify both the point in the overlap and the directions in which we measure distances there. As in Section~\ref{sec:metric-bulk-observables}, we locate the point along a geodesic from the boundary and parallel transport the local Lorentz frame along it. Matching the tetrad and spin connection~\eqref{bfgeo:global-gluing} then ensures that both diamonds assign the same metric to the overlap. We now derive the Einstein equations for this metric from invariance of the amplitude under field shifts in the interior.

At the next order the center of the incoming state no longer suffices to determine the metric. Its fluctuations propagate into both the bulk fields and the boundary frame relative to which they are dressed. For an incoming coherent state, the connected two point function of the metric through $O(\hbar)$ determines the correction $\langle\widehat g_{ab}\rangle-g_{0\,ab}$ to the classical metric $g_0$.

\subsection{The classical solution}
\label{bfq:classical-solution}

In an incoming coherent state the expectation values of the canonical variables specify smooth boundary data $\mathfrak h_{\Sigma_i}$. We can use their time evolution to determine which classical spacetime the state describes. In canonical variables $\phi=(q,p)$, with $\mathfrak h=\iota_t(\phi)$, we take $\hbar\to0$ at fixed initial data $\phi_i$. The variance of each smeared canonical variable is $O(\hbar)$, so over a time interval on which the state remains localized in the reduced phase space, the expectation value of a smoothly smeared observable follows the classical trajectory through $\phi_i$ generated by $H_c$ in~\eqref{bfq:hcan:b}. 

With the quantization of Appendix~\ref{bfq:canonical-representation}, the Heisenberg time evolution of $\mathcal Q(\mathcal O)$ reduces at leading order to Hamiltonian time evolution. For $\iota_{t_i}(\phi_i)=\mathfrak h_{\Sigma_i}$, the classical trajectory $\phi(t)$ satisfies Hamilton's equations with incoming data $\phi(t_i)=\phi_i$,
\begin{samepage}
\begin{subequations}\label{bfq:semiclassical-hamiltonian-equations}
\begin{equation}
\frac{i}{\hbar}[\mathcal Q(H_c),\mathcal Q(\mathcal{O})] =\mathcal Q(\{\mathcal{O},H_c\})+O(\hbar^2),\label{bfq:semiclassical-hamiltonian-equations:a}
\end{equation}
\begin{equation}
\dot q^k=\{q^k,H_c\},\ \dot p_k=\{p_k,H_c\}.\label{bfq:semiclassical-hamiltonian-equations:b}
\end{equation}
\end{subequations}
\end{samepage}
Given a dimensionless functional $f[\phi]$, we deform the action to $I_{\mathscr M}+\hbar f$ and calculate
\begin{equation}
\langle R_{V_{\mathscr M}}(\hbar f)\rangle_{\phi_i}
=e^{if[\phi]}\bigl[1+O(\hbar)\bigr],
\label{bfq:semiclassical-functional}
\end{equation}
where $V_{\mathscr M}=-\int H_c\,\dd t$ and $\langle\cdot\rangle_{\phi_i}$ denotes expectation values in the incoming state. For smooth $f$ the correction starts at $O(\hbar)$ because $\langle\widehat\phi^I-\phi_i^I\rangle_{\phi_i}=0$ and $\langle(\widehat\phi^I-\phi_i^I)^2\rangle_{\phi_i}=O(\hbar)$ for each smeared component of the incoming data.

Hamiltonian time evolution specifies a trajectory for the incoming data, so to establish that the metric along it solves the Einstein equations, we use the invariance of the amplitude under changes of the fields in the interior. For a shift $\phi\mapsto\phi+rs$ supported there,~\eqref{bfq:relations:d} implies invariance for every $r$. Differentiating at $r=0$ and evaluating the variation on the classical trajectory at leading order using~\eqref{bfq:semiclassical-functional} yields
\begin{equation}
 R_{V_{\mathscr M}}(\Delta_{rs}I_{\mathscr M})=\widehat{1},\ \langle\delta_s I_{\mathscr M}\rangle_{\phi_i}=0,\ \delta_s I_{\mathscr M}[\phi]=0.\label{bfq:stationarity-of-the-reduced-action}
\end{equation}
On $\mathfrak S=\mathfrak T=0$, stationarity imposes the field equations~\eqref{bfgeo:field-equations}. Using $B^{\mu\nu}=(1/2)\epsilon^{\mu\nu}{}_{\rho\sigma}e^\rho\curlywedge e^\sigma$ and $\epsilon^{\mu\nu\rho\sigma}\Phi_{\mu\nu\rho\sigma}=\nobreak4\Lambda$, we then obtain
\begin{equation}
 R_{ab}-\frac{1}{2}Rg_{ab}+\Lambda g_{ab}=0.\label{bfq:classical-vacuum-einstein-equation}
\end{equation}
Thus $\phi(t)$ determines a solution of the nonlinear Einstein equations with incoming data $\mathfrak h_{\Sigma_i}$.

To calculate the outgoing expectation values, we write $E_{\mathscr M}(\mathfrak h_{\Sigma_i})$ for the classical data on $\Sigma_f$. For $f_+$ supported after $V_{\mathscr M}$, composition in time implies
$S_{I_0}(V_{\mathscr M}+\hbar f_+) =S_{I_0}(\hbar f_+)S_{I_0}(V_{\mathscr M})$,
so
$R_{V_{\mathscr M}}(\hbar f_+) =U_{\mathscr M}^{\dagger}S_{I_0}(\hbar f_+)U_{\mathscr M}$.
Consequently,
\begin{equation}
\left\langle U_{\mathscr M}^{\dagger}
S_{I_0}(\hbar f_+)U_{\mathscr M}\right\rangle_{\phi_i}
=e^{if_+[E_{\mathscr M}(\mathfrak h_{\Sigma_i})]}\bigl[1+O(\hbar)\bigr].
\label{bfq:classical-outgoing-state}
\end{equation}
Matching the classical solutions across a glued boundary implies $E_{\mathscr M_2\circ \mathscr M_1}=E_{\mathscr M_2}\circ E_{\mathscr M_1}$. 

\subsection{Graviton fluctuations and backreaction}
\label{bfq:graviton-backreaction}

A graviton fluctuation with vanishing expectation value can still source the metric through the quadratic term in the Einstein tensor. The relevant quantity at one loop is the connected graviton two point function. For an incoming Gaussian state $\rho_G$, fluctuations of the initial data have variance $O(\hbar)$, whereas the first correction to Hamiltonian time evolution in the commutator of Section~\ref{bfq:classical-solution} is $O(\hbar^2)$~\cite{Moyal1949}. Through one loop, $\langle\widehat g_{ab}\rangle$ therefore depends on classical time evolution of fluctuations in incoming data. 

For fixed smearings, let $\phi^I$ denote the components of the incoming canonical pairs $(q,p)$. The two point function of their fluctuations is
\begin{equation}
 \Sigma^{IJ}=\frac{1}{2}\langle\{\Delta\widehat{\phi}^I,\Delta\widehat{\phi}^J\}\rangle_{\rho_G}=O(\hbar),\label{bfq:canonical-fluctuation-propagator}
\end{equation}
where $\Delta\widehat{\phi}^I=\widehat{\phi}^I-\phi_0^I$ and $\phi_0^I=\langle\widehat{\phi}^I\rangle_{\rho_G}$. To propagate these fluctuations to the interior, we vary the initial data and solve for the resulting perturbation of the classical fields. At a fixed spacetime point, varying~\eqref{bfgeo:field-equations} yields
\begin{subequations}
\label{semi:linearized-bf}
\begin{align}
&\nabla_{[a}(\delta B)^{\mu\nu}{}_{bc]}+\delta\omega_{[a}{}^\mu{}_\rho B^{\rho\nu}{}_{bc]}+\delta\omega_{[a}{}^\nu{}_\rho B^{\mu\rho}{}_{bc]}=0,\\
&2\nabla_{[a}\delta\omega_{b]}{}^{\mu\nu}=(\delta\Phi)^{\mu\nu}{}_{\rho\sigma}B^{\rho\sigma}{}_{ab}+\Phi^{\mu\nu}{}_{\rho\sigma}(\delta B)^{\rho\sigma}{}_{ab},
\end{align}
\end{subequations}
with $\delta\mathfrak S=0$. Varying $\nabla_{[a}e^\mu{}_{b]}=0$ fixes $\delta\omega$ in terms of $\delta e^\mu{}_a$. Together with the linearized simplicity constraint, the second equation then implies $\delta R_{ab}=\Lambda\,\delta g_{ab}$. We use the reduced symplectic form $\Omega$ to normalize the graviton modes that solve these linearized equations.

When we dress the metric to the boundary frame, the same variation can displace
the event at which we measure the field. We therefore vary $X^*g$ using~\eqref{bfgeo:embedding-variation}, with $\chi^a=(\delta X\circ X^{-1})^a$ for the map $X$ of Section~\ref{sec:bf-reconstruction}~\cite{Speranza,Subregion260107915}.
Under a second variation $\delta'$, the event at which we evaluate $\delta g$ moves by $\chi'{}^a=(\delta' X\circ X^{-1})^a$, and $\chi^a$ itself varies. For commuting variations we therefore have
\begin{equation}
 \delta'\delta(X^*g)=X^*\bigl[\delta'\delta g+\Lie_{\chi'}\delta g+\Lie_\chi\delta' g
 +\Lie_{\delta'\chi}g+\Lie_{\chi'}\Lie_\chi g\bigr].
 \label{semiapp:pullback-second}
\end{equation}
At a fixed reference point $p$, expanding the pulled back metric $g_{ab}(p)$ about $\phi_0=\langle\widehat\phi\rangle$ yields
\begin{subequations}
\label{semi:metric-moments}
\begin{align}
&\bar g_{ab}(p)
=g_{ab}(p)+\frac12(\delta_I\delta_J g_{ab})(p)
                 \Sigma^{IJ}+O(\hbar^2),
\label{semi:metric-moments:a}\\
&\mathcal G_{ab c'd'}(p,p')
=(\delta_I g_{ab})(p)\Sigma^{IJ}
 (\delta_J g_{c'd'})(p')+O(\hbar^2).
\label{semi:metric-moments:b}
\end{align}
\end{subequations}
Here $\bar g_{ab}=\langle\widehat g_{ab}\rangle$ and $\mathcal G$ is the two point function
of metric fluctuations, symmetrized in its arguments. Primed indices refer to
the second point. Moreover, with $J^{IJ}=\{\phi^I,\phi^J\}$, the connected Wightman function is
\begin{equation}
\langle\widehat g_{ab}(p)\widehat g_{c'd'}(p')\rangle_c
=(\delta_I g_{ab})(p)
\left(\Sigma^{IJ}+\frac{i\hbar}{2}J^{IJ}\right)
(\delta_J g_{c'd'})(p')+O(\hbar^2).
\label{semi:metric-wightman-main}
\end{equation}

To determine $\bar g_{ab}-g_{0\,ab}$, we use $E_{ab}[g(\phi)]=0$ for every solution in the family containing $g_0=g(\phi_0)$, where $E_{ab}[g]=G_{ab}[g]+\Lambda g_{ab}$ by~\eqref{bfgeo:field-equations}. With the convention
\begin{equation}
 E_{ab}[g_0+\varepsilon h]=\varepsilon E^{(1)}_{ab}[h]+\varepsilon^2 E^{(2)}_{ab}[h,h]+O(\varepsilon^3),
 \label{semi:einstein-variation-convention}
\end{equation}
commuting variations $\delta,\delta'$ of the incoming data satisfy
$E^{(1)}_{ab}[\delta g]=0$ and
\begin{equation}
E^{(1)}_{ab}[\delta'\delta g]=-2E^{(2)}_{ab}[\delta g,\delta' g].
\label{semi:equation-variations}
\end{equation}
To obtain backreaction from~\eqref{semi:equation-variations}, we use the two point function $\Sigma^{IJ}$ of fluctuations in the incoming data. By~\eqref{semi:metric-moments}, $\tfrac12\Sigma^{IJ}\delta_I\delta_Jg$ is the shift $\bar g-g_0$ to order $\hbar$, while $\Sigma^{IJ}$ pairs $\delta_I g$ with $\delta_J g$ to form the graviton two point function $\mathcal G$. Keeping terms through first order in $\hbar$ in~\eqref{semi:einstein-variation-convention} and~\eqref{semi:equation-variations}, the equation for the expectation value becomes
\begin{equation}
 E_{ab}[\bar g]+\langle E^{(2)}_{ab}[h,h]\rangle_{\mathcal G}
=O(\hbar^2).
\label{semi:semiclassical-einstein}
\end{equation}
Here $\langle h_{ab}(p)h_{c'd'}(p')\rangle_{\mathcal G}=\mathcal G_{ab c'd'}(p,p')$ is the associated symmetrized two point function. With smearing held fixed, we differentiate at separated points before taking $p'\to p$, for example
\begin{equation}
 \langle\nabla_eh_{ab}\nabla_fh_{cd}\rangle_{\mathcal G}
 =\left[\nabla_e\nabla_{f'}\mathcal G_{ab c'd'}(p,p')\right]_{p'=p}.
 \label{semiapp:propagator-derivatives}
\end{equation}

We expand the contracted Bianchi identity for $g_0+\varepsilon h$ with $h_{ab}=\delta g_{ab}$. At order $\varepsilon^2$ the terms from varying the divergence operator vanish by the background field equations and their linearization, so the quadratic source in~\eqref{semi:semiclassical-einstein} is conserved,
\begin{align}
&E[g_0]=E^{(1)}[h]=0,\ \nabla^a E^{(2)}_{ab}[h,h]=0.\label{semi:backreaction}\\
\intertext{With the initial and boundary data fixed, the retarded solution for $\bar g=g_0+\Delta g$ obeys}
&\Delta g=\Delta g_{0}-\mathcal G_R
 \langle E^{(2)}[h,h]\rangle_{\mathcal G}+O(\hbar^2),\label{semi:backreaction:c}
\end{align}
where $\mathcal G_R$ is the retarded Green's function of the linearized Einstein equations on the reduced phase space. The homogeneous solution $\Delta g_0$ propagates perturbations of initial data from $\Sigma_0$, with only graviton and edge mode fluctuations in the causal past of the event at which we evaluate $\Delta g$ entering the integral against $\mathcal G_R$.

Finally, joining the forward time evolution to its adjoint at the final cap sums over final states while keeping the incoming state fixed~\cite{CalzettaHu1987}. Appendix~\ref{app:semiclassical-determinant} shows that, with the reduced measure and the same boundary conditions, the resulting one loop generating functional agrees with that of constrained general relativity~\cite{DeWitt1967II}. For coherent incoming data, its source derivatives yield the graviton two point function in~\eqref{semi:metric-moments} that drives the backreaction in~\eqref{semi:semiclassical-einstein}.

\section{Discussion of open questions}
\label{sec:open-questions}

In Section~\ref{sec:bf-classical} we showed how the semiclassical Einstein equations follow from our proposed quantization of gravity. We now return to the converse problem that motivated us in Section~\ref{sec:bf-introduction}, namely how much of the quantum theory follows from general relativity. With our choice of state $\omega$ on the algebra of observables, we obtained the transition amplitudes~\eqref{bfq:amplitude} from the reduced Einstein Hilbert action without requiring either the initial or the final state to be peaked about a classical geometry.

In quantizing gravitational data on a null cone we found that backreaction of radiative modes can render amplitudes UV finite at fixed $G_N$ and $\hbar$. Two quantizations with this property and the same semiclassical limit could nevertheless predict different probabilities for the same measurement. We would like to know whether diffeomorphism invariance and gluing suffice to fix these probabilities, irrespective of how we impose the gravitational constraints, construct the physical inner product and assign operators to classical observables.

\subsection{Uniqueness of the state \texorpdfstring{$\omega$}{omega}}
\label{sec:discussion-state-extension}

In Section~\ref{bfq:algebra-states} we specified the state $|\Omega_{\mathscr A}\rangle$ on the incoming subalgebra at fixed area $\mathscr A$. For source operators $a,b$ in this subalgebra we can therefore compute $\omega(a^\dagger b)$~\eqref{bfq:pairing}. More generally, $b$ can change the area of $S$ and $a^\dagger$ can restore it. To compute such matrix elements we chose a pure extension $\omega$ to the full algebra $\mathcal A$ in Section~\ref{bfq:state-extension}.

Once we specify a state on the full algebra for the quadratic action $I^{(2)}$, the change of reference action~\eqref{bfq:reference-null-transport} fixes $\omega$ uniquely. Indeed, gluing extends~\eqref{bfq:interacting-source-overlap} to successive source insertions (Appendix~\ref{app:dressed-graviton-reflection}), and time reflection implements the adjoint. Products of the source unitaries and their adjoints span a norm dense subspace of $\mathcal A$.

We would like to know whether, for the same reduced action, matching and gluing already fix these matrix elements from $|\Omega_{\mathscr A}\rangle$ alone. To compare two pure extensions we can boost $b$ relative to $a$. With $P$ the spectral measure of the area operator,~\eqref{bfq:source-boost-phase:a} yields
\begin{equation}
 \omega\big(a^\dagger\alpha_\eta(b)\big)
 =\int\exp\!\left[\frac{i\eta(\mathscr A'-\mathscr A)}{8\pi G_N\hbar}\right]
 \langle a|P(\dd\mathscr A')|b\rangle_\omega .
 \label{disc:interference}
\end{equation}
Fourier transformation in the rapidity $\eta$ would reveal any difference between the two extensions in the overlap for a given interval of intermediate areas.

\subsection{The spectrum of the gravitational clock}
\label{sec:discussion-clock}

In Section~\ref{bfent:subregion-boost-main} we argued that a causal diamond of bounded size admits only finitely many independent states. Since the gravitational clock~\eqref{bfq:canonical-clock} preserves the area $\mathscr A$, we can diagonalize it on $P_\ell\mathcal K_{\mathscr A}$~\eqref{bfq:clock-selected-space} for any finite upper bound $\ell$ on its reading.

If the clock operator $\widehat{\bar u}_*$ is self adjoint on the incoming Fock space $\mathcal K_{\mathscr A}$ of Section~\ref{bfq:algebra-states}, the spectral theorem implies for each fixed normalized $|\psi\rangle\in\mathcal K_{\mathscr A}$ that
\begin{equation}
 \lim_{\ell\to\infty}\langle\psi|P_\ell|\psi\rangle=1 .
 \label{disc:clock-domain}
\end{equation}
The size of the diamond needed to approximate $|\psi\rangle$ to a given accuracy depends on the state. By~\eqref{disc:clock-domain} the eigenstates we obtain as we increase $\ell$ form a complete basis of $\mathcal K_{\mathscr A}$. Since $\mathcal K_{\mathscr A}$ is infinite dimensional and $P_\ell$ has finite rank, the clock then has a discrete spectrum unbounded above.

In Appendix~\ref{bfq:canonical-representation} we argued that wavefunctions $|\psi\rangle$ with $\langle\widehat{\bar u}_*\rangle_\psi<\infty$ are dense in $\mathcal K_{\mathscr A}$, as required for self adjointness on this space. When we bound the focusing along each null generator by $\Xi_*$, density follows from~\eqref{bfq:capped-clock-form}. Upon taking $\Xi_*\to\infty$, the closed quadratic form~\eqref{bfent:clock-form} defines a self adjoint clock on the closure $M$ of its domain. With $P_M$ the projection onto $M$, the limit on the left of~\eqref{disc:clock-domain} equals $\|P_M\psi\|^2$.

To show that a given $|\psi\rangle$ belongs to $M$, we used the wavefunctions $|\Psi_L\rangle$ of~\eqref{bfq:clock-resolvent-limit}, which converge to $P_M|\psi\rangle$ as $L\to\infty$. For $|\psi\rangle=|\Omega_{\mathscr A}\rangle$ we found that $1-\langle\psi|\Psi_L\rangle$ tends to zero order by order in $\hbar$~\eqref{bfq:clock-gaussian-coefficients}. At fixed $G_N$ and $\hbar$ we estimated this loss of overlap from the probability of strong focusing on classical null cones~\eqref{bfq:clock-finite-hbar-estimate}.

To prove $M=\mathcal K_{\mathscr A}$ fully rigorously it suffices to show directly for the quantum clock~\eqref{bfq:canonical-clock} that this loss tends to zero in each occupation number state~\eqref{bfq:fixed-area-fock:a}, or in the states $|\Omega_\alpha\rangle$ of~\eqref{bfapp:displaced-occupation-state} in any finite set of modes, which remains to be done.

\subsection{Diffeomorphism invariance and the choice of quantization}
\label{sec:discussion-diffeomorphisms}

We chose the ordering in~\eqref{bfq:canonical-clock} so that the clock $\widehat{\bar u}_*$ is a positive operator. An ordering ambiguity in the clock could affect which states we count in a causal diamond, just as one in the Hamiltonian could affect their time evolution. One might expect diffeomorphism invariance of the amplitudes to restrict this freedom.

Recall, however, that in Section~\ref{bfq:bulk-diffeomorphisms} we held the reduced canonical coordinates fixed when comparing two foliations. Since the two Hamiltonians then differ by a function of time alone~\eqref{bfq:refoliation-hamiltonian:difference}, the cancellation of the boundary phases holds for many orderings. Even Weyl ordering in two sets of coordinates related by a nonlinear canonical transformation need not yield equivalent operators.

Two quantizations would predict identical correlation functions if on each cut $\Sigma$ a unitary $W_\Sigma$ identifies their states and, for every source $F$ coupled to the same dressed observable,
\begin{equation}
 \widetilde U_{I+F}(\Sigma_f,\Sigma_i)
 =W_{\Sigma_f}U_{I+F}(\Sigma_f,\Sigma_i)W_{\Sigma_i}^\dagger .
 \label{disc:ordering}
\end{equation}
Indeed, $W_\Sigma$ would cancel against its adjoint on each intermediate cut. It remains to determine whether gluing and diffeomorphism invariance imply~\eqref{disc:ordering}.

\subsection{Relation to refined algebraic quantization}
\label{sec:discussion-refined-quantization}

In comparing different operator orderings we have assumed that we first solve the gravitational constraints. We would also like to know whether we obtain the same transition amplitudes if we quantize the original fields and then impose the constraints on states. To compare the two prescriptions we need a physical inner product on solutions of the quantum constraints.

In refined algebraic quantization a rigging map assigns to a wavefunction of the original fields a distributional solution of the constraints~\cite{GiuliniMarolfRAQ1999}. Evaluating this solution on a second wavefunction yields the physical inner product. Since we compute~\eqref{bfq:pairing} by gluing one preparation to the time reflection of the other, we can try to use gluing itself to define the rigging map.

For a uniform boost at $S$ we can see how this works. Instead of identifying the areas $s_L,s_R$ on the two sides before quantizing, we take wavefunctions $a,b$ of both areas with measure $\dd s_L\,\dd s_R$. Since the two sides induce opposite orientations on $S$, averaging over a simultaneous boost of the two frames yields~\eqref{eq:collar-boost-match}
\begin{equation}
 \int_{\mathbb R}\frac{\dd\eta}{4\pi^2}
 \left\langle a\left|\exp\!\left[\frac{i\eta}{2\pi}(\hat s_L-\hat s_R)\right]\right|b\right\rangle
 =\int_0^\infty\dd s\,\overline{a(s,s)}\,b(s,s) .
 \label{disc:boost-average}
\end{equation}
If we act on $a$ or $b$ with $\hat s_L-\hat s_R$, the overlap~\eqref{disc:boost-average} vanishes. Gluing thus imposes the boost constraint and yields the same inner product as matching the areas first.

In Section~\ref{bfq:bulk-diffeomorphisms} we varied a spacelike cut $\Sigma$ after solving the constraints that generate its deformations. To impose these constraints on wavefunctions of the original fields we might try to average over deformations of $\Sigma$ as we did over boosts in~\eqref{disc:boost-average}. Boost rapidities add independently of the metric. The bracket of two normal deformations of $\Sigma$, however, yields a tangential displacement that depends on the metric on $\Sigma$. Thus the uniqueness theorem for averaging over a Lie group~\cite{GiuliniMarolfUniqueness1999} leaves open whether gluing imposes the bulk constraints and yields~\eqref{bfq:pairing}.

\subsection{The graviton \texorpdfstring{$S$}{S} matrix}
\label{sec:discussion-scattering}

We began in Section~\ref{sec:bf-introduction} by recalling the difficulty of computing graviton scattering amplitudes from the Einstein Hilbert action owing to the UV divergences of perturbation theory. The overlap we computed in Section~\ref{bfq:fixed-area-uv} is finite at fixed $G_N$ and $\hbar$ even though its expansion in $G_N$ diverges at two loops. We can likewise compute a scattering amplitude by evolving the initial state with the full reduced action $I$, using the quadratic action $I^{(2)}$ about Minkowski spacetime only to identify and freely propagate asymptotic graviton states.

Consider scattering by $\pi/2$ in the center of mass frame. We take four orthogonal wave packets of equal energy and helicity, with the incoming pair $1,2$ moving along $\pm\bm e_z$ and the outgoing pair $3,4$ along $\pm\bm e_x$. We propagate the outgoing wave packets freely from $T$ back to $-T$. On this initial slice let $|ij\rangle$ have one graviton in each of the wave packets $i,j$ and none in the other two. We denote the state of the remaining modes, including the edge modes, by $|\chi_\perp\rangle$ and take it to have no further gravitons. By~\eqref{bfq:time-evolution-and-source} the $2\to2$ amplitude is
\begin{equation}
 \mathcal S_{34,12}(T)=\langle34|S_T|12\rangle,\ 
 S_T=U_{I^{(2)}}(T,-T)^\dagger U_I(T,-T).
 \label{disc:finite-scattering}
\end{equation}
We choose $|\chi_\perp\rangle$ and the wave packets so that a rotation through $\pi/2$ about $\bm e_y$ interchanges $|12\rangle$ and $|34\rangle$. If the boundary data and the quantization respect this rotation, the symmetric and antisymmetric combinations cannot mix. The scattering amplitude is then a difference of two return amplitudes,
\begin{equation}
 \mathcal S_{34,12}(T)=\tfrac12\big(\langle+|S_T|+\rangle-\langle-|S_T|-\rangle\big),\
 |\pm\rangle=(|12\rangle\pm|34\rangle)/\sqrt2 .
 \label{disc:rotation-scattering}
\end{equation}

In gluing across a cut through the collision we can sum analytically over all occupation numbers of the four wave packets. To see this, we Fourier transform $\exp(-i H_c\Delta t/\hbar)$ in~\eqref{bfq:reduced-path-integral:d} with respect to the four canonical pairs $q_j,p_j$ of the wave packets. Each Fourier component acts on these oscillators by a displacement, so for one time step
\begin{equation}
 U=\int_{\mathbb C^4}\dd^8z\,D(z)\otimes\widehat F_U(z),\
 D(z)=\exp\!\left[\sum_{j=1}^4(z_j\mathfrak a_j^\dagger-\overline z_j\mathfrak a_j)\right],
 \label{disc:scattering-step}
\end{equation}
where $\widehat F_U(z)$ is the Weyl quantization of the Fourier coefficient in the other modes. A product of two displacements is itself a displacement up to a phase~\eqref{bfads:rep-weyl}. Hence gluing two steps yields
\begin{equation}
 \widehat F_{U_2U_1}(z)=\int_{\mathbb C^4}\dd^8w\,
 \exp\!\big[i\operatorname{Im}(z\cdot\overline w)\big]\,
 \widehat F_{U_2}(z-w)\widehat F_{U_1}(w).
 \label{disc:scattering-gluing}
\end{equation}
After iterating~\eqref{disc:scattering-gluing} for $S_T$, we need the matrix element of $D(z)$ that annihilates one graviton in each incoming wave packet and creates one in each outgoing wave packet. Normal ordering $D(z)$ in~\eqref{disc:finite-scattering} yields
\begin{equation}
 \mathcal S_{34,12}(T)=\int_{\mathbb C^4}\dd^8z\,\overline z_1\overline z_2z_3z_4\,
 \exp(-|z|^2/2)\,\langle\chi_\perp|\widehat F_{S_T}(z)|\chi_\perp\rangle .
 \label{disc:scattering-four-modes}
\end{equation}

The difficulty in evaluating~\eqref{disc:scattering-four-modes} analytically lies in gravitational backreaction. Consider for instance a cross section $C_v$ of the incoming null cone at $v<1$, with the corner $S$ at $v=1$. At fixed lapse at $S$, shear between $C_v$ and $S$ reduces the lapse at $C_v$~\eqref{bfgeo:lapse-at-the-cuts} and hence the displacement of $C_v$ along the transverse null normal in a time step $\Delta t$ (Figure~\ref{fig:summary-hamiltonian}).

As we vary the four pairs $q_j,p_j$ in the Fourier transform defining $\widehat F_U(z)$, the shear changes, so we have to recompute the metric entering $H_c$. With finitely many modes we can solve the null constraints of Appendix~\ref{app:bf-hamiltonian} numerically on a grid in $q_j,p_j$, holding the other modes fixed in this transform. In gluing successive steps~\eqref{disc:scattering-gluing} we must still sum over the excitations of these other modes, even those unexcited in both asymptotic states.

To obtain the $S$ matrix we take $T\to\infty$ in~\eqref{disc:finite-scattering}. In four dimensions the radiation can permanently change the shear at null infinity, so the asymptotic states must allow for gravitational memory~\cite{PS}. At finite energy resolution we can instead sum probabilities over undetected soft radiation~\cite{DonoghueTorma1999}.

An explicit evaluation of the $2\to2$ amplitude~\eqref{disc:scattering-four-modes} at finite $G_N$ and $\hbar$ would be an important test of our quantization. We leave this calculation to future work.

\Needspace*{10\baselineskip}

\section*{Acknowledgements}
For helpful discussions and comments on the draft, I thank Ahmed Abdalla, Raphael Bousso, Xi Dong, Éanna Flanagan, Laurent Freidel, Tom Hartman, Luca Iliesiu, Adam Levine, Guanda Lin, Juan Maldacena, Geoff Penington, Pratik Rath, Eva Silverstein, Andy Strominger, Gabriele Di Ubaldo, Misha Usatyuk, Aron Wall, and especially Wayne Weng. I also thank Sam Altman, Sebastien Bubeck, and my OpenAI colleagues for encouragement throughout the project. Several of the calculations and proofs in this paper were developed through detailed, back and forth collaboration with GPT-6 Astra over the course of many months, amounting to $\sim \ $400M total tokens spent; I thank OpenAI for providing this compute.

\appendix
\allowdisplaybreaks[2]

\section{Null constraints and symplectic reduction}
\label{app:bf-geometry}

To quantize the gravitational boundary data of Section~\ref{sec:bf-action}, we need the Poisson brackets of the radiative modes on intersecting null cones, both with one another and with the degrees of freedom at the corner. We derive these brackets here from the first order action, allowing the cuts to move as we vary the fields. After imposing the null constraints, we invert the reduced symplectic form including the terms that couple the corner area and boundary frame to the radiative modes.

\subsection{The tetrad and spin connection on a null boundary}

To identify the momentum conjugate to the conformal metric on a null cone, we separate the terms supported on the two cuts bounding it from the radiative part of the variation of the first order action. In~\eqref{bfgeo:bf-null-potential-main} the area form $\bm\mu$ pairs with $\delta\beta$. For $\ell\cdot n=-1$, the components of $\beta_a=-n_\mu\nabla_a\ell^\mu$ in~\eqref{bfgeo:transverse-data:h} are
\begin{equation}
\beta_v=\partial_v\log N,\ \beta_A=w_A-\frac{1}{2}\mathring{\nabla}_A f,\ f=-\log N.
\label{bfapp:null-normal-connection-components}
\end{equation}

Holding the tetrad fixed in~\eqref{bfgeo:bf-null-potential-main}, we obtain
\begin{equation}
\left.\int_{\mathscr H}\bm\theta_{\mathscr M}\right|_{\delta e=0}
=-\frac{\epsilon_{\mathscr H}}{8\pi G_N}\int_{\mathscr H}\bm\eta\bigl(\delta\kappa+\delta\Theta_{(\ell)}\bigr).
\label{bfapp:fixed-tetrad-null-potential}
\end{equation}
The variation of the null boundary action cancels this term for the orientation in~\eqref{bfgeo:boundary-action:b}~\cite{Lehner,JubbSamuelSorkinSurya161200149}. To see how this cancellation works before we impose vanishing torsion, we use the relation between the expansion and the area two form,
\begin{equation}
\Lie_\ell\bm\mu =(\Theta_{(\ell)}+E_{\mu}{}^AT^{\mu}_{\ell A})\bm\mu,\ I_{\mathscr H}=\frac{\epsilon_{\mathscr H}}{8\pi G_N}\int_{\mathscr H}\bm\eta\bigl(\kappa-E_{\mu}{}^AT^{\mu}_{\ell A}\bigr).
\label{bfgeo:null-boundary-torsion}
\end{equation}
For $T^\mu=\delta T^\mu=0$ we thus have
\begin{equation}
\delta I_{\mathscr H}=\frac{\epsilon_{\mathscr H}}{8\pi G_N}\,\delta\int_{\mathscr H}\bm\eta\,\kappa .
\label{bfapp:null-boundary-action-variation}
\end{equation}
On $\Sigma_0$ and $\Sigma_1$ the extrinsic curvature terms in~\eqref{bfgeo:boundary-action:a} cancel the terms proportional to $\delta K_i$. We can now read off the boundary momenta from the remaining coefficients of $\delta e^\mu$. In particular, separating the trace of $K_{AB}$ from the shear identifies the momenta conjugate to $\varrho^2$ and the conformal metric in Section~\ref{sec:bf-action}. The improvement $\dd\bm\alpha$ in~\eqref{bfgeo:potential-improvement} makes a simultaneous local Lorentz transformation of the tetrad and spin connection a degeneracy of the symplectic form on torsion free configurations~\cite{DePaoliSpeziale2018}.

With both null normals allowed to vary, the variations of $\nabla_{[a}e^\mu{}_{b]}=0$ and $F^{\mu\nu}{}_{ab}$ are
\begin{equation}
\nabla_{[a}\delta e^\mu{}_{b]}+\delta\omega^\mu{}_{\nu[a}e^\nu{}_{b]}=0,\ \delta F^{\mu\nu}{}_{ab}=2\nabla_{[a}\delta\omega_{b]}{}^{\mu\nu}.
\label{bfgeo:linearized-torsion-curvature}
\end{equation}
Writing $h_{ab}=2e_{\mu(a}\delta e^\mu{}_{b)}$ and $h=g^{ab}h_{ab}$, we use the torsion free connection in~\eqref{bfgeo:linearized-torsion-curvature} to express the potential $\bm\theta_{\mathscr M}+\dd\bm\alpha$ in metric variables~\cite{DePaoliSpeziale2018},
\begin{equation}
(\bm\theta)_{abc}=\frac{\epsilon_{dabc}}{16\pi G_N}(\nabla_e h^{de}-\nabla^d h).
\label{bfgeo:metric-variation}
\end{equation}
Writing the null normal as
\begin{equation}
\ell_a=-\exp(\psi)\nabla_a t,
\label{bfapp:null-normal-covector}
\end{equation}
we separate a change in its normalization from a change in its direction by setting $\upsilon^a=\delta\ell^a-\ell^a\delta\psi=-h^{ab}\ell_b$. Indeed, $\ell_a\upsilon^a=0$, and it is the angular part of $\upsilon^a$ that changes the direction. With $q^{ab}h_{ab}=h-2n_a\upsilon^a$ the projection of \eqref{bfgeo:metric-variation} onto a cut then yields
\begin{equation}
\bm\theta\big|_{\mathscr H}
=\frac1{8\pi G_N}\delta(\kappa\bm\eta)+\dd\bm b+\bm{\mathcal E},
\label{bfgeo:complete-null-potential}
\end{equation}
with
\begin{equation}
16\pi\GN\,\bm{\mathcal E}
=\bm\eta\left[
\sigma^{ab}h_{ab}
-\left(\kappa+\frac12\Theta_{(\ell)}\right)q^{ab}h_{ab}
-2(\varpi_a+\Theta_{(\ell)} n_a)\upsilon^a+2\Theta_{(\ell)} \delta\psi\right],
\label{bfgeo:null-potential-flux}
\end{equation}
where $16\pi\GN\,\bm b=(h-2\delta\psi)\bm\mu+i_\upsilon\bm\eta$.
For $\left.h_{ab}\ell^a\ell^b\right|_{\mathscr H}=0$,
\begin{equation}
 \delta\kappa-\kappa\,\delta\psi-\ell^a\nabla_a(\delta\psi)=\upsilon^a\nabla_a\psi+
\frac{1}{2}n^a\nabla_a(h_{bc}\ell^b\ell^c).\label{bfgeo:variation-of-inaffinity}
\end{equation}
The boundary action thus cancels the exact variation of $\kappa\bm\eta$ with the sign appropriate to the orientation of the null boundary, yielding the symplectic current $\bm\omega=\delta\bm{\mathcal E}+\dd\delta\bm b$ \cite{CF}.

\subsection{Corners and edge modes}

We saw in Section~\ref{bfgeo:null-initial-data} that the corner charge $\mathscr K(y)$ is conjugate to a relative boost.\footnote{In first order gravity one can also derive the brackets of fields at a corner from the algebra of constraints on the boundary~\cite{Canepa:2022uii,Cattaneo:2026jvj}. For four dimensional BF theory the resulting corner algebra has been quantized~\cite{Canepa:2026jfi}.} To compute the brackets of these variables with the radiative modes in~\eqref{bfgeo:null-potential-flux}, we first need to know what freedom remains in identifying the null generators on the two null boundaries once their induced metrics agree at the corner. For $\mathscr H_L$ and $\mathscr H_R$ intersecting at $S$, we write the induced metric on $\mathscr H_j$ as
\begin{equation}
q_{AB}=\varrho_S(y)^2 v_j^2\bar q_{AB},
\label{bfapp:cut-metric-on-null-boundaries}
\end{equation}
where $j=L,R$, $v_j|_S=1$ and $\varrho_S(y)=\varrho|_S$. In these coordinates the induced metrics agree when $\bar q_L|_S=\bar q_R|_S$, while the relative normalization of the two null generators can still vary through $\lambda=-\log|\partial_{v_L}\cdot\partial_{v_R}|$.

Varying the angular identification also displaces the null generators on one null boundary relative to those on the other. The momenta for these displacements depend on the normal connections $\varpi_L,\varpi_R$. The angular coordinates $s_j^i(y)$ specify where the null generator labeled by $y$ meets the cut, and their gradients form the Jacobian of $s_j$. Choosing the left sided null generators as coordinates sets $s_L=\mathrm{id}$, so the relative angular identification $Y=s_R$ has momentum $p_i$,
\begin{subequations}\label{bfgeo:twist-densities-at-the-cuts}
\begin{align}
&t_L=\varrho_S(y)^2(\dd\lambda+\varpi_L-\varpi_R),\label{bfgeo:twist-densities-at-the-cuts:a}\\
&t_R=\varrho_S(y)^2(\dd\lambda-\varpi_L+\varpi_R),\label{bfgeo:twist-densities-at-the-cuts:b}\\
&\widetilde t_{ji}\mathring{\nabla}_a s_j^i=t_{ja},\label{bfgeo:twist-densities-at-the-cuts:c}\\
&p_i=\frac{\widetilde t_{Ri}}{16\pi G_N}.\label{bfgeo:twist-densities-at-the-cuts:d}
\end{align}
\end{subequations}
 Together with the pair $(\mathscr K(y),\lambda)$, the two angular pairs $(Y^i,p_i)$ account for the three corner degrees of freedom and their momenta. Their coupling to the conformal metric on each null boundary follows from the symplectic form in these coordinates.

On each null boundary we use the coordinates $z_j$ of~\eqref{bfq:canonical-conformal-metric:a} and define $\xi_j=\delta s_j\circ s_j^{-1}$, so
\begin{equation}
 \delta^yz_j=\delta z_j-\Lie_{\xi_j}z_j,\ \delta^oz_j=\delta^yz_j-\frac{v_j}{2}\delta^y\log(\varrho_S(y)^2)\,\partial_{v_j}z_j.\label{bfgeo:variations-at-fixed-labels}
\end{equation}
Here $\delta^yz_j$ removes the variation induced by changing $s_j$, while $\delta^oz_j$ compares variations at the same area. The Lie derivatives act on the conformal metric, then on its coordinates $z_j$. Substitution in $\Omega$ yields \cite{Reisenberger07122541,Reisenberger12113880}
\begin{equation}
\begin{aligned}
&\Omega=\frac1{16\pi G_N}\sum_{j=L,R}\int_S\bm\epsilon_0\bigg[
\delta\lambda\wedge\delta(\varrho_S(y)^2)+
\delta\widetilde t_{ji}\wedge\delta s_j^i
-\frac12\delta^y(\varrho_S(y)^2)\wedge \mathsf g_{\alpha\beta}\,\partial_{v_j}z_j^\alpha\,\delta^y z_j^\beta\\
&+\varrho_S(y)^2\int_1^{b_j}v_j^2
\mathsf g_{\alpha\beta}\,\nabla_{v_j}(\delta^o z_j)^\alpha\wedge(\delta^o z_j)^\beta\dd v_j
\bigg].
\end{aligned}
\label{bfgeo:coupled-form}
\end{equation}
Along a curve $z^\alpha(v)$ the covariant derivative of a variation $X^\alpha$ is
\begin{equation}
\nabla_v X^\alpha=\partial_v X^\alpha+\Gamma^\alpha{}_{\beta\gamma}(\partial_v z^\beta)X^\gamma,
\label{bfapp:covariant-derivative-along-curve}
\end{equation}
with $\Gamma^\alpha{}_{\beta\gamma}$ determined by $\mathsf g_{\alpha\beta}$.
For the three corner pairs we obtain~\cite{CarlipTeitelboim1995}
\begin{equation}
\Omega\supset\int_S\bm\epsilon_0\bigl[\delta\lambda\wedge\delta\mathscr K(y)+\delta p_i\wedge\delta Y^i\bigr],
\label{bfapp:corner-symplectic-form}
\end{equation}
while the mixed terms in~\eqref{bfgeo:coupled-form} couple $\varrho_S(y)^2$ and $s_j$ to $z_j$.

At an outer cut $C_j=\{v_j=b_j\}$ the variation depends on how we specify its location. Fixing $\varrho_{C_j}^2$ determines $b_j$. Holding the affine coordinate $u(b,z)$ fixed instead determines the displacement of the cut and hence its area variation,
\begin{subequations}\label{bfgeo:variations-of-the-inner-area}
\begin{align}
&b_j^2=\frac{\varrho_{C_j}(s_j)^2\lvert\det(\mathring{\nabla}_a s_j^i)\rvert}{\varrho_S(y)^2},\label{bfgeo:variations-of-the-inner-area:a}\\
&\delta b=\left.(\partial_u v)\right|_{v=b}[\delta[u(b,z)]-(\delta_z u)(b)],\label{bfgeo:variations-of-the-inner-area:b}\\
&\delta\bm\mu_C=[b^2\delta(\varrho_S(y)^2)+2\varrho_S(y)^2b\,\delta b]\bm\epsilon_0.\label{bfgeo:variations-of-the-inner-area:c}
\end{align}
\end{subequations}
 The symplectic potential on the spacelike cap bounded by $C$ contains
\begin{equation}
-\frac{\epsilon_C}{8\pi G_N}\int_C\log(-n_C\cdot\ell)\,\delta\bm\mu_C,
\label{bfapp:outer-cut-potential}
\end{equation}
where $n_C$ is the future unit normal of the cap and $\epsilon_C$ is its oriented sign.

A cut at fixed affine parameter can move in the area coordinate as we vary the fields. Its displacement $Y_u$ enters the variation $X^\alpha$ of the conformal metric,
\begin{equation}
Y_u=\delta^yu-\frac{v\,\delta^y\log(\varrho_S(y)^2)}{2\partial_u v},\
X^\alpha=\delta^y_u z^\alpha+Y_u\partial_u z^\alpha.
\label{bfgeo:affine-mixed-form}
\end{equation}
For a second variation $\delta'$ we write $X'{}^\alpha$ and $Y'_u$ for the same quantities. The contribution of the null boundaries to the symplectic form is
\begin{equation}
\Omega_{\mathscr H}(\delta,\delta')
=\frac1{16\pi G_N}\sum_j\int_S\bm\epsilon_0\int \varrho_S(y)^2v^2
\left[X'_{\alpha}\nabla_u X^\alpha-X_{\alpha}\nabla_u X'{}^\alpha\right]\dd u.
\label{bfgeo:affine-mixed-form:c}
\end{equation}
In addition to terms linear in the displacements $Y_u$ and $Y'_u$, the integrand in~\eqref{bfgeo:affine-mixed-form:c} contains the mixed term
\begin{equation}
\mathsf g_{\alpha\beta}Y'_u(\partial_u z^\beta)\nabla_u(Y_u\partial_u z^\alpha)-(\delta\leftrightarrow\delta').
\label{bfapp:affine-displacement-term}
\end{equation}
At the corner, varying~\eqref{bfgeo:capped-clock} yields the displacement entering~\eqref{bfgeo:affine-mixed-form},
\begin{equation}
\delta u(1,y)=\delta u(\epsilon,y)+\delta\left[\frac{u_*(y)}{p_{\epsilon,y}}\right].
\label{bfapp:affine-corner-displacement}
\end{equation}

To find how the affine parameter of a cut at fixed $v$ depends on the conformal metric, we vary the null constraint $\mathfrak C_\ell=0$. Writing $u=u(1,y)+\exp(\Gamma) s$, we separate off changes in affine origin and normalization by introducing $f$. Holding the area parameter fixed under variations, we obtain
\begin{subequations}\label{bfgeo:variations-along-the-null-generator}
\begin{align}
&f=(\partial_u v)[\delta u-\delta u(1,y)-(u-u(1,y))\delta\Gamma],\label{bfgeo:variations-along-the-null-generator:a}\\
&\delta_vz=\delta_u^y z+f\partial_v z,\label{bfgeo:variations-along-the-null-generator:b}\\
&\delta\log(\partial_u v)=-\delta\Gamma-\partial_v f-\kappa_vf,\label{bfgeo:variations-along-the-null-generator:c}\\
&\kappa_v=-\partial_v\log(\partial_u v)=\frac{v}{4}\mathsf g_{\alpha\beta}(\partial_v z^\alpha)(\partial_v z^\beta).\label{bfgeo:variations-along-the-null-generator:d}
\end{align}
\end{subequations}
Linearizing $\mathfrak C_\ell=0$ yields
\begin{equation}
\left(\partial_v^2-\kappa_v\partial_v+\frac{\kappa_v}{v}\right)f
=\frac v2\mathsf g_{\alpha\beta}(\partial_v z^\alpha)\nabla_v(\delta_u^y z)^\beta.
\label{bfgeo:clock-linearization}
\end{equation}
The corner data $f(1)$ and $(\partial_v f)(1)$ fix the coefficients of the two homogeneous solutions
\begin{equation}
v,\ v\int_1^v\frac{\dd w}{w^2(\partial_u v)(w)}.
\label{bfapp:clock-homogeneous-solutions}
\end{equation}
Including variations of radiative modes, we obtain
\begin{equation}
f(v)=vf(1)+G(v,1)[(\partial_v f)(1)-f(1)]
+\int_1^vG(v,w)\frac{w}{2}\mathsf g_{\alpha\beta}(\partial_w z^\alpha)\nabla_w(\delta_u^y z)^\beta\,\dd w,
\label{bfgeo:clock-green}
\end{equation}
where
\begin{equation}
G(v,w)=vw(\partial_u v)(w)\int_w^v\frac{\dd s}{s^2(\partial_u v)(s)}.
\label{bfgeo:clock-green:b}
\end{equation}
On a smooth null cone, $(\partial_v\bar q)^{AB}\delta\bar q_{AB}=2\mathsf g_{\alpha\beta}(\partial_v z^\alpha)\delta z^\beta$ expresses the shear terms of~\eqref{bfgeo:cone-form-main} in these coordinates. Varying the angular identification also yields the terms involving $s_j$ in~\eqref{bfgeo:coupled-form}. 

For the null normals $\ell_p,\ell_q$ on the intersecting null cones, we write $k_q=\exp(\Gamma_q)\ell_q$. Then $\Gamma_q=\log(-\ell_p\cdot k_q)-\eta$, where $\eta=\log(-\ell_p\cdot\ell_q)$. At fixed $\ell_p,\ell_q$, $k_q\mapsto\exp(s)k_q$ implies $\Gamma_q\mapsto\Gamma_q+s$. The two corner terms therefore combine into one canonical pair, $(\mathscr K(y),\lambda-\gamma)$,
\begin{equation}
 -\mathscr K(y)\delta\lambda+\mathscr K(y)\delta\gamma=-\mathscr K(y)\delta(\lambda-\gamma).\label{bfgeo:matched-area-boost-variation}
\end{equation}
 As for the angular coordinates, $Y$ identifies the copies of $S$ on $\mathscr H_p$ and $\mathscr H_q$, so we match their area two forms after pullback by $Y$. The clock on each null cone fixes its own $\varrho_{S,p}$ or $\varrho_{S,q}$ through~\eqref{bfgeo:integrated-focusing-main:d}. With inward null generators, the area elements and normal connections obey
\begin{equation}
 Y^*(\varrho_{S,q}^{\,2}\bm\epsilon_{0,q})=\varrho_{S,p}^{\,2}\bm\epsilon_{0,p},\ \varpi_p+Y^*\varpi_q=\dd\eta.\label{bfgeo:matching-of-area-and-twist}
\end{equation}

\subsection{The null constraints}

At a given affine parameter the area element depends on the shear integrated from the vertex. In using this relation to define a clock it matters whether we vary the null data or merely reparametrize the same null generator. For a choice of parameter $v$ the constraint $\mathfrak C_v=0$ generates the latter transformation, while the momentum conjugate to $\varrho^2$ in the symplectic potential is proportional to $\mu=\kappa+(\partial_v\varrho^2)/(2\varrho^2)$. Including the terms at the cuts in $\Theta_\partial$, we have~\cite{HopfmuellerFreidel2018}
\begin{subequations}
\label{bfgeo:area-constraint}
\begin{align}
&\Theta=\frac1{8\pi G_N}\int\left[-\mu\,\delta(\varrho^2)+
\frac{\varrho^2}{2}\mathsf g_{\alpha\beta}\,\partial_v z^\alpha\,\delta z^\beta\right]\dd v+\Theta_\partial,\\
&\mathfrak C_v=\partial_v^2\varrho^2-\mu(\partial_v\varrho^2)+\frac{\varrho^2}{2}\mathsf g_{\alpha\beta}(\partial_v z^\alpha)(\partial_v z^\beta)=0 .
\end{align}
\end{subequations}
For affine $u$, $\mathfrak C_u=\varrho^2\mathfrak C_\ell$. A reparametrization generated by $(8\pi G_N)^{-1}\int f\mathfrak C_v\,\dd v$ has charge $[\varrho^2\partial_vf-f\partial_v\varrho^2]/(8\pi G_N)$ at a cut~\cite{HopfmuellerFreidel2018}.

We choose a parameter $V$ satisfying $\mu=\partial_v\log(\partial_v V)$~\cite{CiambelliFreidelLeigh2024DressingTime,CiambelliFreidelLeigh2024QuantumNull}, and write
$\varrho^2=\varrho(V)^2$ and $z=\zeta(V)$. With $\dot\zeta^\alpha=\partial_V\zeta^\alpha$ and $\dot\zeta_\alpha=\mathsf g_{\alpha\beta}\dot\zeta^\beta$, the constraint involves a second order differential operator,
\begin{equation}
\mathcal D_{\varrho^2}=\partial_V^2+\frac{1}{2}\dot\zeta^\alpha\dot\zeta_\alpha,\ \mathfrak C_v=(\partial_v V)^2\mathcal D_{\varrho^2}(\varrho^2).
\label{bfgeo:clock-reduction}
\end{equation}
In the coordinate $V$ the symplectic potential is
\begin{equation}
\begin{aligned}
&\Theta=
\int\frac{\mathfrak C_v}{8\pi G_N(\partial_v V)}\,\delta V\,\dd v
+\frac1{16\pi G_N}\int \varrho^2\,\mathsf g_{\alpha\beta}\,\partial_V\zeta^\alpha\,\delta\zeta^\beta\dd V
+\Theta_\partial\\
&{}+\frac1{8\pi G_N}\left[
\varrho^2\,\delta\log(\partial_v V)-\frac{\partial_v\varrho^2}{\partial_v V}\delta V
\right]_{\partial}.
\end{aligned}
\label{bfgeo:clock-reduction:c}
\end{equation}
Thus $\mathfrak C_v=0$ sets the momentum conjugate to $V$ to zero in the interior, while the terms at the cuts still enter the symplectic potential. To determine the variation of the area element we hold $\varrho(V_0)^2$ and $(\partial_V \varrho^2)(V_0)$ fixed. The linearized constraint then determines the area variation,
\begin{subequations}\label{bfgeo:area-green}
\begin{align}
&\mathcal D_{\varrho^2}\,\delta(\varrho^2)=-\frac{1}{2}\varrho^2\delta(\dot\zeta^\alpha\dot\zeta_\alpha),\label{bfgeo:area-green:a}\\
&\delta(\varrho^2)(V)=-\frac12\int_{V_0}^VG_{\varrho^2}(V,s)\varrho(s)^2\delta(\dot\zeta^\alpha\dot\zeta_\alpha)(s)\dd s,\label{bfgeo:area-green:b}\\
&G_{\varrho^2}(V,s)=\varrho(V)^2\varrho(s)^2\int_s^V\frac{\dd t}{\varrho(t)^4}.\label{bfgeo:area-green:c}
\end{align}
\end{subequations}

The affine parameter $u$ satisfies $\dd V/\dd u=c\varrho$ for a positive normalization constant $c$, and hence
$\Delta u=c^{-1}\int\dd V/\varrho$.

At fixed area element, the momenta $\pi_\alpha$ conjugate to $z^\alpha$ obey second class constraints whose Poisson brackets act as a differential operator on $\delta z^\alpha$~\cite{Reisenberger07122541,AlexandrovSpeziale2015},
\begin{equation}
\pi_\alpha-\frac{\varrho^2}{16\pi G_N}\,\mathsf g_{\alpha\beta}\partial_v z^\beta=0,\ (\Delta_{\mathfrak C}\delta z)_\alpha=-\frac{\varrho}{8\pi G_N}\,\mathsf g_{\alpha\beta}\nabla_v(\varrho\,\delta z^\beta).
\label{bfgeo:dirac-inverse}
\end{equation}
The inverse along a null generator involves parallel transport $\mathsf P(v,w)$ from $z(w)$ to $z(v)$. Integration by parts yields the boundary term
\begin{equation}
-\frac{1}{16\pi G_N}\bigl[\varrho^2\mathsf g_{\alpha\beta}\,\delta z^\beta\,\delta' z^\alpha\bigr]_{\partial},
\label{bfapp:dirac-boundary-term}
\end{equation}
which enters the boundary conditions together with $\delta\Theta_\partial$ from the cuts. With these contributions included the kernel is
\begin{equation}
(\Delta_{\mathfrak C}^{-1})^{\alpha\beta}(v,w)
=-\frac{8\pi G_N\operatorname{sgn}(v-w)}
{2\varrho(v)\varrho(w)}\,\mathsf P^\alpha{}_\gamma(v,w)\mathsf g^{\gamma\beta}(z(w))
+\mathsf k_\partial^{\alpha\beta}(v,w).
\label{bfgeo:dirac-inverse:c}
\end{equation}
The conditions at the cuts and corners in \eqref{bfgeo:coupled-form} fix $\mathsf k_\partial$. On a null cone, $\varrho^2=\varrho_S(y)^2v^2$, so the coefficient of parallel transport is $-\operatorname{sgn}(v-w)/(2\mathscr K(y)vw)$.
Configurations of the conformal metric at separated points on one null generator thus have a nonzero reduced bracket \cite{Reisenberger07122541}. To allow the area element to vary as well, we must include $\delta(\varrho^2)$ from~\eqref{bfgeo:area-green:b} in the symplectic form before taking its inverse.

For a variation $U$ of the bulk degrees of freedom between cuts $V_-$ and $V_+$, we solve for the area variation with homogeneous data at $V_-$ and for $b$ with homogeneous data at $V_+$,
\begin{subequations}\label{bfgeo:coupled-area-operator}
\begin{align}
&\delta_U(\varrho^2)=-\mathcal D_{\varrho^2,-}^{-1}[\varrho^2\,\dot{\zeta}_\alpha\nabla_V U^\alpha],\label{bfgeo:coupled-area-operator:a}\\
&\delta_U(\varrho^2)(V_-)=(\partial_V\delta_U(\varrho^2))(V_-)=0,\label{bfgeo:coupled-area-operator:initial-data}\\
&b=\mathcal D_{\varrho^2,+}^{-1}[\dot{\zeta}_\alpha U^\alpha],\label{bfgeo:coupled-area-operator:b}\\
&b(V_+)=(\partial_Vb)(V_+)=0.\label{bfgeo:coupled-area-operator:final-data}
\end{align}
\end{subequations}
In terms of these solutions, the operator acting on $U$ is
\begin{equation}
(\mathcal D_z U)^\alpha=(2\varrho^2\nabla_V+\partial_V \varrho^2)U^\alpha+\dot{\zeta}^\alpha\delta_U(\varrho^2)-\nabla_V(\varrho^2\dot{\zeta}^\alpha b).
\label{bfgeo:coupled-area-operator:c}
\end{equation}
Pairing $\mathcal D_zU$ with a second variation $\eta$ yields the bulk integral in the symplectic form,
\begin{equation}
\Omega(U,\eta)=\frac1{16\pi G_N}\int_{V_-}^{V_+}\eta_\alpha(\mathcal D_z U)^\alpha\dd V
+\Omega_\partial(U,\eta).
\label{bfgeo:coupled-area-operator:d}
\end{equation}
The contribution $\Omega_\partial$ includes all terms at $V_-,V_+$ from $\delta\Theta$ in~\eqref{bfgeo:clock-reduction:c}. In terms of
\begin{equation}
w=U-\frac{b}{2}\,\dot{\zeta},\ 
y=\delta_U(\varrho^2)-\frac{b}{2}\,\partial_V(\varrho^2),
\label{bfgeo:coupled-area-variables}
\end{equation}
the equation $\mathcal D_z U=f$ then reduces to
\begin{equation}
\partial_V^2y=-\frac12\dot{\zeta}_\alpha f^\alpha,\ (2\varrho^2\nabla_V+\partial_V\varrho^2)w^\alpha=f^\alpha-\dot{\zeta}^\alpha y,\ \partial_V^2b=\dot{\zeta}_\alpha w^\alpha,
\label{bfgeo:coupled-area-inverse}
\end{equation}
with $U=w+(b/2)\dot{\zeta}$. The initial conditions in~\eqref{bfgeo:coupled-area-operator:initial-data} then imply
\begin{equation}
y(V_-)=-\frac{1}{2}(\partial_V\varrho^2)(V_-)b(V_-),\ (\partial_Vy)(V_-)=-\frac{1}{2}\bigl[(\partial_V^2\varrho^2)(V_-)b(V_-)+(\partial_V\varrho^2)(V_-)(\partial_Vb)(V_-)\bigr].
\label{bfgeo:initial-conditions-for-area-variation}
\end{equation}
 Together with the boundary contribution $\Omega_\partial$ these conditions fix $y(V_-)$, $(\partial_Vy)(V_-)$ and $w(V_-)$. For nonzero $\delta(\varrho^2)(V_-)$ or $(\partial_V\delta(\varrho^2))(V_-)$ the homogeneous solution enters the boundary symplectic form $\Omega_\partial$ at $V_\pm$. This form also depends on variations of $\mathfrak e|_{V_-}$. It is thus the full symplectic form in \eqref{bfgeo:coupled-form} that determines the brackets of bulk degrees of freedom and edge modes for arbitrary angular dependence, and hence the normalization of the canonical variables used in Section~\ref{sec:bf-quantization}.

\subsection{Canonical pairs on an expanding null cone}
\label{bfapp:incoming-canonical-pairs}

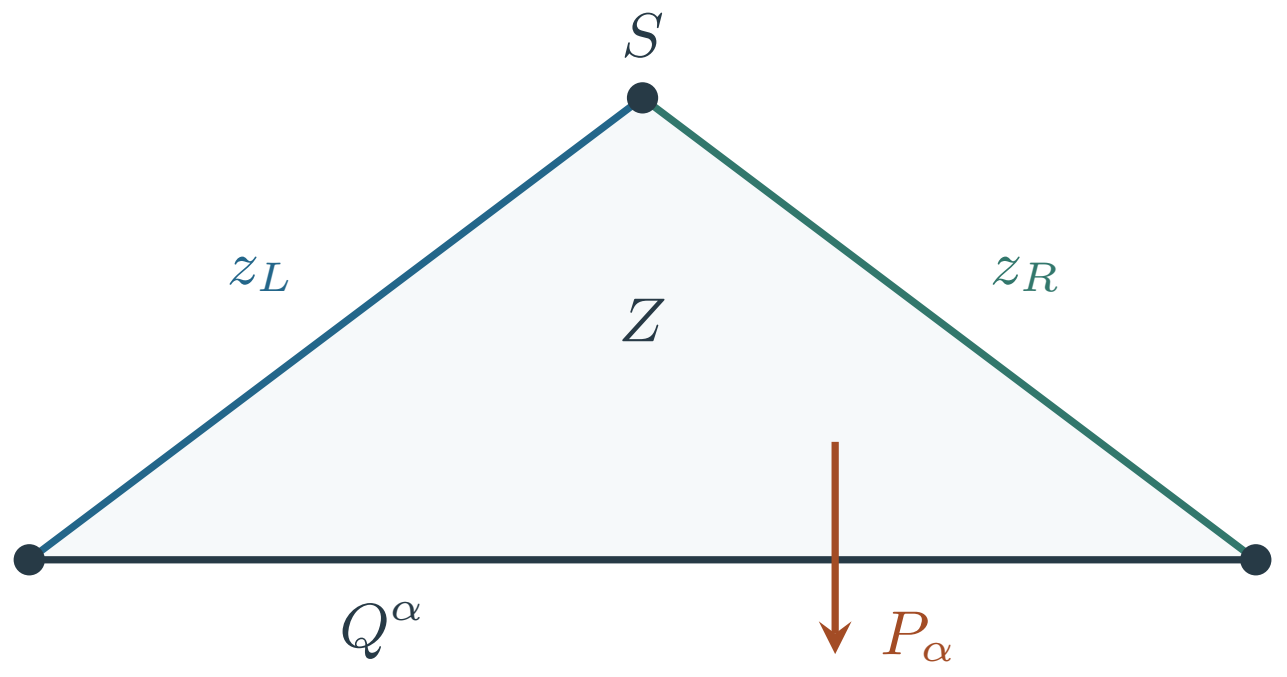
\begin{figure}[!htb]
\centering
\resizebox{.55\linewidth}{!}{%
\begingroup
\definecolor{charLeft}{HTML}{23668A}
\definecolor{charRight}{HTML}{32776C}
\definecolor{charInk}{HTML}{273A46}
\definecolor{charNormal}{HTML}{A34C25}
\begin{tikzpicture}[x=1cm,y=1cm,text=charInk,font=\small,>=stealth,
    profile/.style={line width=1.05pt},cut/.style={circle,fill=charInk,inner sep=1.6pt}]
  \coordinate (S) at (0,2.35);
  \coordinate (L) at (-3.12,0);
  \coordinate (R) at (3.12,0);
  \fill[charLeft!4] (S) -- (L) -- (R) -- cycle;
  \draw[profile,charLeft] (S)--(L);
  \draw[profile,charRight] (S)--(R);
  \draw[profile,charInk] (L)--(R);
  \node[cut,label=above:{$S$}] at (S) {};
  \node[anchor=south east,text=charLeft,inner sep=4pt] at (-1.62,1.22) {$z_L$};
  \node[anchor=south west,text=charRight,inner sep=4pt] at (1.62,1.22) {$z_R$};
  \node at (0,1.22) {$Z$};
  \node[cut] at (L) {};
  \node[cut] at (R) {};
  \node[anchor=north,inner sep=6pt] at (-1.32,0) {$Q^\alpha$};
  \draw[->,charNormal,line width=1.05pt] (.98,.60)--(.98,-.48);
  \node[text=charNormal,anchor=west,inner sep=4pt] at (1.06,-.39) {$P_\alpha$};
\end{tikzpicture}
\endgroup%
}
\caption[Canonical data for two null cone profiles]{At fixed $y$ we prescribe $z_L(1-r,y)$ on the blue side and $z_R(1-l,y)$ on the green side, with their common value at $S$. On the side $r+l=a$ joining the outer cuts, $Z^\alpha$ specifies $Q^\alpha$ and its derivative along the orange arrow, $\partial_r+\partial_l$, determines $P_\alpha$ by~\eqref{bfapp:characteristic-pairs}.}
\label{fig:characteristic-data}
\end{figure}

A change of variables motivated by the integrability of cylindrically symmetric gravity simplifies the Poisson brackets of free null initial data~\cite{FuchsReisenberger170406992}. For the Gaussian wavefunction in Section~\ref{bfq:algebra-states} we seek independent canonical pairs at fixed area. The inverse in~\eqref{bfgeo:coupled-area-inverse} determines variations along each null generator from the data at its cuts. To construct the pairs while preserving the matching at $S$, we introduce an auxiliary field $Z$ on the triangle in Figure~\ref{fig:characteristic-data} and prescribe one conformal metric profile on each of two sides. At the vertex where these sides meet, $Z$ takes their common value at $S$. On the third side, the field value serves as a coordinate and the boundary variation determines its conjugate momentum from the derivative of $Z$ across that side. Solving the field equation in either direction relates this canonical pair to the two original profiles.

We first fix the area density and angular identification $x=(\mathscr K(y),Y^i(y))$, so $\delta^o z_j=\delta z_j$ in~\eqref{bfgeo:coupled-form}. For outer cuts at $v_j=\epsilon$, $0<\epsilon<1$, the triangle has $r,l\ge0$ and $r+l\le a=1-\epsilon$. At each angular label $y$ we prescribe the left sided profile $Z(r,0,y)=z_L(1-r,y)$ and the right sided profile $Z(0,l,y)=z_R(1-l,y)$. To reproduce the weight $v_j^2$ in the radiative symplectic form on both sides, we use the action
\begin{equation}
 \int_S\frac{\varrho_S^2\bm\epsilon_0}{16\pi G_N}
 \int_{r,l\ge0,\ r+l\le a}(1-r-l)^2\mathsf g_{\alpha\beta}(Z)
 \partial_rZ^\alpha\,\partial_lZ^\beta\,\dd r\,\dd l .
 \label{bfapp:characteristic-action}
\end{equation}
With $\nabla_r$ the covariant derivative of $\mathsf g$ along $Z$, the interior variation yields
\begin{equation}
 \nabla_r\partial_lZ^\alpha
 =\frac{\partial_r Z^\alpha+\partial_l Z^\alpha}{1-r-l}.
 \label{bfapp:characteristic-evolution}
\end{equation}
On $l=a-r$ we read off $Q^\alpha$ and its conjugate momentum $P_\alpha$ from the boundary variation of~\eqref{bfapp:characteristic-action},
\begin{subequations}\label{bfapp:characteristic-boundary}
\begin{align}
 &\int_S\frac{\epsilon^2\varrho_S^2\bm\epsilon_0}{16\pi G_N}
 \int_0^a\left[\mathsf g_{\alpha\beta}(Z)\,(\partial_r+\partial_l)Z^\beta\,\delta Z^\alpha\right]_{l=a-r}\dd r
 =\int_S\bm\epsilon_0\int_0^a P_\alpha\,\delta Q^\alpha\,\dd r ,
 \label{bfapp:characteristic-side-variation}\\
 &Q^\alpha(r,y)=Z^\alpha(r,a-r,y),\ P_\alpha(r,y)=\left.\frac{\epsilon^2\varrho_S^2}{16\pi G_N}
 \mathsf g_{\alpha\beta}(Q)(\partial_r+\partial_l)Z^\beta\right|_{l=a-r}.
 \label{bfapp:characteristic-pairs}
\end{align}
The two null cone profiles lie at the lower limits $l=0$ and $r=0$ in~\eqref{bfapp:characteristic-action}. Taking $v_j$ from $1$ at $S$ to $\epsilon$ at each outer cut, we obtain their contribution to the boundary variation,
\begin{equation}
 -\int_S\frac{\varrho_S^2\bm\epsilon_0}{16\pi G_N}\sum_{j=L,R}\int_1^\epsilon
 v_j^2\mathsf g_{\alpha\beta}(z_j)\,\partial_{v_j}z_j^\alpha\,\delta z_j^\beta\,\dd v_j .
 \label{bfapp:characteristic-axis-variation}
\end{equation}
\end{subequations}
The sum of~\eqref{bfapp:characteristic-side-variation} and~\eqref{bfapp:characteristic-axis-variation} is the variation of the on shell action, so its exterior derivative vanishes. At fixed $x$ we therefore recover the radiative part of~\eqref{bfgeo:coupled-form} as the canonical symplectic form for $(Q,P)$.

To recover the original profiles from $(Q,P)$ we first determine $\partial_r Z$ and $\partial_l Z$ on $l=a-r$. Differentiating $Q(r,y)=Z(r,a-r,y)$ with respect to $r$ fixes their difference. Equation~\eqref{bfapp:characteristic-pairs} fixes their sum from $P$ since $\epsilon\varrho_S>0$ and $\mathsf g$ is positive definite. To bound both derivatives as we integrate back to the axes, we write $w=1-r-l$ in~\eqref{bfapp:characteristic-evolution}. Since $\nabla_l\partial_r Z=\nabla_r\partial_l Z$, we obtain
\begin{equation}
 \nabla_l(w\partial_r Z)=\partial_l Z,\ \nabla_r(w\partial_l Z)=\partial_r Z .
 \label{bfapp:characteristic-transport}
\end{equation}
Let $M_a$ be the largest norm of either derivative on $r+l=a$, measured with $\mathsf g$, and let $M$ be their largest norm on a closed strip extending back from this side. To reach $(r,l)$ we integrate the first equation in~\eqref{bfapp:characteristic-transport} from $l=a-r$ at fixed $r$ and the second from $r=a-l$ at fixed $l$. Each integral has length $a-r-l=w-\epsilon$. Since parallel transport preserves the norm, at a point where either derivative attains $M$ we obtain
\begin{equation}
 wM\le\epsilon M_a+(w-\epsilon)M,\ M\le M_a .
 \label{bfapp:characteristic-inverse-estimate}
\end{equation}
Conversely, let $M_0$ bound $|\partial_r Z|$ on $l=0$ and $|\partial_l Z|$ on $r=0$. For $r+l\le s$ let $m(s)$ be the largest of $w|\partial_r Z|$ and $w|\partial_l Z|$. Integrating~\eqref{bfapp:characteristic-transport} from the axes yields $m(s)\le M_0+\int_0^s m(s')\,\dd s'/(1-s')$, hence $m(s)\le M_0/(1-s)$. Both derivatives are therefore bounded by $M_0/\epsilon^2$ on the triangle. The metric~\eqref{bfq:canonical-conformal-metric:c} is the complete metric on the hyperbolic plane, so the derivative bounds keep $Z$ in a compact metric ball. The local solution of~\eqref{bfapp:characteristic-evolution} thus extends across the triangle in either direction. Differentiating~\eqref{bfapp:characteristic-transport} yields smooth dependence on the data and on $y$.

For smooth $(Q,P)$ the two recovered profiles agree at $S$, where $z_L(1,y)=z_R(1,y)=Z(0,0,y)$. At their outer cuts $v_j=\epsilon$ we have $z_R(\epsilon,y)=Q(0,y)$ and $z_L(\epsilon,y)=Q(a,y)$. The recovered profiles have finite $\Xi_y(v)$, so the null generators remain expanding by~\eqref{bfgeo:integrated-focusing-main:b} for $p_{\epsilon,y}>0$. For a single incoming null cone we prescribe its profile on both axes. Uniqueness makes $Q$ and $P$ symmetric under $r\mapsto a-r$, so integrating over half the side yields the symplectic form of one copy.

To allow $x$ to fluctuate in~\eqref{bfq:fixed-area-gaussian}, we need corner coordinates that can range over the real line. At fixed $\mathscr A$ every linear mode of the area density obeys $|\int_S g\,\dd A|\le\mathscr A\|g\|_\infty$. We therefore use the modes of its logarithm $h$ with the constant part removed. For the angular identification we choose a reference $Y_0$ and smooth vector harmonics $e_k$ along $Y_0$ for a fixed reference metric on $S$. Using the exponential map of this metric, we write
\begin{subequations}\label{bfapp:gaussian-corner-data}
\begin{align}
 &\left.\dd A\right|_S=\frac{\mathscr A\exp(h)\bm\epsilon_0}{\int_S\exp(h)\bm\epsilon_0},\ \int_S h\bm\epsilon_0=0,\label{bfapp:gaussian-corner-data:area}\\
 &Y(y)=\exp_{Y_0(y)}\!\left(\sum_k\eta_k e_k(y)\right),\ \eta_k=\frac{\xi_k}{[1+(\xi_k/b_k)^2]^{1/2}}.\label{bfapp:gaussian-corner-data:angular}
\end{align}
\end{subequations}
We choose the harmonics orthogonal in $H^3(S)$ and the $b_k>0$ to decrease polynomially in their angular eigenvalues, with $\sum_k b_k^2\|e_k\|_{H^3(S)}^2<R^2$. Since the $H^3(S)$ norm controls the $C^1$ norm on the two dimensional cut, we can take $R$ small enough that displacements of norm at most $2R$ have invertible differential after applying the exponential map. The resulting maps have the same degree as $Y_0$ and hence are diffeomorphisms. Since $|\eta_k|<b_k$, every real choice of the $\xi_k$ specifies a diffeomorphism in this chart. Likewise, arbitrary finite variations of the modes of $h$ preserve both the positivity of the area density and its integral.

When $x$ varies, the radiation also contributes to the corner momenta $\pi$, with the relative boost as in~\eqref{bfgeo:matched-area-boost-variation}. Subtracting $\int_S\bm\epsilon_0\,\delta\pi\wedge\delta x$ and the radiative canonical form from~\eqref{bfgeo:coupled-form} yields a closed two form in $(x,z)$ that vanishes on pairs of variations at fixed $x$. At each point we join a smooth shear free reference profile $z_*(y)$ to $z$ along the unique geodesic of~\eqref{bfq:canonical-conformal-metric:c}. Contracting the two form with the tangent to this path and integrating its pullback along the path yields a one form linear in $\delta x$. The part at $z_*$ has a primitive in $x$ alone on the contractible corner chart. Angular integration by parts on the closed surface $S$ thus puts their sum in the form $\int_S\bm\epsilon_0\,A(x,z)\,\delta x$. Shifting $\pi$ by $A$ absorbs the mixed terms, and subtracting $A(x,z)$ after recovering the profiles inverts the shift.

In the coordinates~\eqref{bfapp:gaussian-corner-data} we transform the shifted momenta by preserving the one form $\int_S\bm\epsilon_0(\pi+A)\,\delta x$. The momentum for $\xi_k$ is $(\dd\eta_k/\dd\xi_k)$ times the momentum for $\eta_k$. This derivative is positive for every finite $\xi_k$, so the conjugate momentum ranges over the full real line as well. Inverting the momentum shift preserves the common conformal metric at $S$ and keeps both outer cuts at $v_j=\epsilon$. Thus $\mathscr K(y)$ Poisson commutes with both radiative canonical fields, and at fixed $\mathscr A$ the uniform relative boost acts by the phase in Section~\ref{bfq:algebra-states}.

\section{Boundary terms in the Hamiltonian}
\label{app:bf-hamiltonian}

Although we evaluate the Hamiltonian $H_\Sigma$ at the cuts in Section~\ref{sec:bf-action}, the spin connection there depends on the radiative modes between them. To use the Poisson brackets of Appendix~\ref{app:bf-geometry} to generate time evolution, we make this dependence explicit by solving the null constraints for the connection in terms of the conformal metric and corner data. Translating a cut varies its embedding and hence its normal frame in the boundary action. At an interior cut the two contributions cancel when the fields and normal frames match, since the two sides induce opposite orientations.

\subsection{The Hamiltonian}

On a null cone, $q_{AB}(v,y)$ fixes the shear, while the charge~\eqref{bfgeo:canonical-hamiltonian} for a displacement in $t$ also depends on derivatives along the other null normal, which we determine from corner data using the constraints. With the tetrad of Appendix~\ref{app:bf-geometry} in coordinates $(t,v,y^A)$ adapted to the intersecting null cones, we write the spacetime metric near $S$ as
\begin{equation}
\dd s^2=-F\,\dd t^2-2N\,\dd t\,\dd v
+q_{AB}(\dd y^A+V^A\dd t)(\dd y^B+V^B\dd t).
\label{bfgeo:full-metric}
\end{equation}
Where the null boundary is expanding, we use area per unit solid angle to define a radial coordinate,
$\varrho^2=\varrho_S(t,y)^2v^2$, with $\varrho_S(t,y)=\varrho(t,1,y)$. Since $\sqrt{\det q_{AB}}\,\dd^2y=\varrho^2\bm\epsilon_0$, we can choose local angular coordinates for which $\bm\epsilon_0=\dd^2y$. In these coordinates it remains to determine $N,V^A,F$ from the null constraints.

With $\mathring{\nabla}_A$ the derivative operator compatible with $q_{AB}$ and $f=-\log N$, we write $K^{(n)}_{AB}$ for the extrinsic curvature along the other null normal $n$ and obtain
\begin{subequations}
\label{bfgeo:transverse-data}
\begin{align}
&K_{AB}=\frac12\partial_v q_{AB}=\frac12\Theta_{(\ell)} q_{AB}+\sigma_{AB},\displaybreak[3]\label{bfgeo:transverse-data:a}\\
&NK^{(n)}_{AB}=\frac12(\partial_t-\Lie_V)q_{AB}-\frac{F}{2N}K_{AB},\displaybreak[3]\label{bfgeo:transverse-data:b}\\
&\eta_{AB}=NK^{(n)}_{AB}-\frac12\vartheta q_{AB},\label{bfgeo:transverse-data:c}\\
&w_A=-\frac1{2N}q_{AB}\partial_v V^B,\label{bfgeo:transverse-data:d}\\
&\mu_c=(\partial_t-V^A\mathring{\nabla}_A)f+\frac{\partial_v F}{2N},\label{bfgeo:transverse-data:e}\\
&n=\frac{1}{N}\left[\partial_t-V^A\partial_A-\frac{F}{2N}\partial_v\right],\label{bfgeo:transverse-data:f}\\
&\vartheta=Nq^{AB}K^{(n)}_{AB},\label{bfgeo:transverse-data:g}\\
&\varpi_A=w_A-\frac{1}{2}\mathring{\nabla}_Af,\label{bfgeo:transverse-data:h}
\end{align}
\end{subequations}
where $\Theta_{(\ell)}=2/v$.
The tensors $\sigma_{AB},\eta_{AB}$ are trace free. For $\ell=\partial_v$ the inaffinity is $(\partial_vN)/N$, and we normalize the transverse null normal by $\ell\cdot n=-1$. Here $\mu_c$ denotes a transverse component of the connection, whereas $\mu=\kappa+(\partial_v\varrho^2)/(2\varrho^2)$ in~\eqref{bfgeo:area-constraint} involves derivatives along the null generators.

The $\ell\ell$ projection of $F^{\mu\nu}=\Phi^{\mu\nu}{}_{\rho\sigma}B^{\rho\sigma}$ is $R_{vv}=0$. Since $\Theta_{(\ell)}=2/v$ implies $\partial_v\Theta_{(\ell)}+\tfrac12\Theta_{(\ell)}^2=0$, the Raychaudhuri equation reduces to
\begin{equation}
\partial_v f=-\frac{\sigma_{AB}\sigma^{AB}}{\Theta_{(\ell)}}.
\label{bfgeo:lapse-constraint}
\end{equation}
The integration constant fixes the normalization of the affine parameter at the corner. Once $f$ is known, the constraint $\mathfrak C_{\ell a}=0$ in Section~\ref{sec:metric-bulk-observables} determines the momentum $\Pi_A=-2\varrho^2 w_A$ and angular shift $V^A$ along each null generator,
\begin{subequations}\label{bfgeo:shift-constraint}
\begin{align}
&\partial_v\Pi_A=\varrho^2\bigl(2\mathring{\nabla}_B\sigma^B{}_A\mathbin{+}\mathring{\nabla}_A\partial_v f\mathbin{-}\mathring{\nabla}_A\Theta_{(\ell)}\mathbin{-}\Theta_{(\ell)}\mathring{\nabla}_Af\bigr),\label{bfgeo:shift-constraint:a}\\
&V^A(v)=V^A(v_0)+\int_{v_0}^v \frac{N(s)}{\varrho(s)^2}q^{AB}(s)\Pi_B(s)\dd s.\label{bfgeo:shift-constraint:b}
\end{align}
\end{subequations}
Here $\Pi_A(v_0)$ and $V^A(v_0)$ specify the momentum and angular frame at the corner, respectively, with angular derivatives taken with respect to $q_{AB}$.

The other normal and transverse projections of $F^{\mu\nu}=\Phi^{\mu\nu}{}_{\rho\sigma}B^{\rho\sigma}$ propagate the connection components $\eta_{AB},\vartheta,\mu_c$~\cite{Hayward1994Dynamics},
\begin{subequations}
\label{bfgeo:complementary-constraints}
\begin{align}
&\partial_v(\varrho^2\vartheta)=-N\varrho^2
\left[\frac{1}{2}R[q]-\mathring{\nabla}_A\varpi^A-\varpi_A\varpi^A-\Lambda\right],\\*
&(\partial_v\eta_{AB})^{\rm TF}
=\frac{1}{2}\Theta_{(\ell)}\eta_{AB}-\frac{1}{2}\vartheta\sigma_{AB}
+N(\mathring{\nabla}_{(A}\varpi_{B)}+\varpi_A\varpi_B)^{\rm TF},\\*
&\partial_v\mu_c=\sigma^{AB}\eta_{AB}-\frac{1}{2}\Theta_{(\ell)}\vartheta
+N\left[-\frac{1}{2}R[q]-\frac{1}{4}\mathring{\nabla}_Af\mathring{\nabla}^Af
+3w_Aw^A+w^A\mathring{\nabla}_Af\right].
\end{align}
\end{subequations}
Here the intersecting null boundary and its normal supply the initial values. Since $R_{ab}\ell^an^b=-\Lambda$, $\Lambda$ enters the equation for $\varrho^2\vartheta$ directly and the one for $\mu_c$ through $\vartheta$.

We can now use \eqref{bfgeo:transverse-data} to read off the coefficient $F$ together with the time derivatives of $q_{AB}$ and $f$,
\begin{subequations}
\label{bfgeo:temporal-reconstruction}
\begin{align}
&F=Nv\,[\partial_t\log(\varrho_S(y)^2)-\mathring{\nabla}_AV^A-\vartheta],\\
&\partial_tq_{AB}
=\Lie_Vq_{AB}+2\eta_{AB}+\vartheta q_{AB}
+\frac FNK_{AB},\\
&\partial_tf=\mu_c-\frac{\partial_v F}{2N}+V^A\mathring{\nabla}_Af .
\end{align}
\end{subequations}
Once $q_{AB},N,V^A$ are known, equations~\eqref{bfgeo:complementary-constraints} for $\varrho^2\vartheta,\eta_{AB},\mu_c$ are linear along each null generator. Their values on one cut thus suffice to determine them along the adjacent null boundary.

On a solution of the constraints the terms proportional to $N\varrho^2$ in~\eqref{bfgeo:canonical-hamiltonian} form a divergence. With $\varpi_A=w_A-\mathring{\nabla}_Af/2$, the first equation in~\eqref{bfgeo:complementary-constraints} reads
\begin{equation}
N\varrho^2\left[-R[q]-\frac12\mathring{\nabla}_Af\mathring{\nabla}^Af+2w_Aw^A+2\Lambda\right]
=2\partial_v(\varrho^2\vartheta)-2\partial_A(N\varrho^2\varpi^A).
\label{bfgeo:hamiltonian-divergence}
\end{equation}
For a closed cut $C$, the angular divergence integrates to zero, so only the terms at $v_\pm$ survive. Including $\dd\bm\alpha$ from~\eqref{bfgeo:potential-improvement}, we thus recover the surface charge~\eqref{bfgeo:eh-hamiltonian}. Translating a cut also varies the boundary action $I_{\partial \mathscr M}$ through its boundary frame $\mathfrak e$.

\subsection{Translating the boundary}

We now evaluate the charge on the displaced null boundary. In the coordinates of~\eqref{bfgeo:full-metric} its new location $v=b(t,y)$ fixes the normal $s_a$, which must satisfy $s_as^a=0$ there. The corresponding null generator $m^a$ has $m^t=1$ and $m^b\nabla_bm^a=\kappa_m m^a$. For $n^a=d\,m^a$ to be affinely parametrized the normalization must compensate the inaffinity. In terms of the metric on the displaced boundary, these conditions read
\begin{subequations}
\label{bfgeo:area-side}
\begin{align}
&s_a=N\nabla_a(v-b),\label{bfgeo:area-side:a}\\
&m^a=-g^{ab}s_b,\label{bfgeo:area-side:b}\\
&m^a\nabla_a\log d=-\kappa_m,\label{bfgeo:area-side:c}\\
&s_as^a=F+2N(\partial_tb-V^A\mathring{\nabla}_Ab)+N^2q^{AB}\mathring{\nabla}_A b\mathring{\nabla}_B b,\displaybreak[0]\label{bfgeo:area-side:d}\\
&m=\partial_t+(N\mathring{\nabla}^Ab-V^A)\partial_A
+\frac{N^2q^{AB}\mathring{\nabla}_A b\mathring{\nabla}_B b-F}{2N}\partial_v,\label{bfgeo:area-side:e}\\
&\kappa_m=m^a\nabla_a\log N-\frac{\partial_v(s_as^a)}{2N},\label{bfgeo:area-side:f}\\
&m^a\nabla_a\log(dN)=\frac{\partial_v(s_as^a)}{2N}.\label{bfgeo:side-normalization}
\end{align}
\end{subequations}
Here $\partial_v(s_as^a)$ is evaluated at fixed $b$, $\partial_t b$ and $\partial_A b$. Specifying $d$ on one cut fixes the normalization of $n^a$ along each null generator of the displaced boundary. If $c_L$ fixes the scale of the other null normal, the boost angle is $\log(c_L dN/2)$. For the affine coordinate $r$ used below, the metric has the form~\eqref{bfgeo:full-metric} with $r$ in place of $v$ and $N=1$.

Translating a cut also changes the limits of integration in the action. To see which surface terms this entails, we evaluate the bulk action in the null coordinates of~\eqref{bfgeo:full-metric}, keeping the total derivatives that arise upon integration by parts. For the expansions and shears along the two coordinate directions we set
\begin{subequations}\label{bfgeo:strip-expansions-and-shears}
\begin{align}
&\Theta_{(t)}=\frac{(\partial_t-\Lie_V)\varrho^2}{\varrho^2},\label{bfgeo:strip-expansions-and-shears:a}\\
&\Theta_{(r)}=\frac{\partial_r\varrho^2}{\varrho^2},\label{bfgeo:strip-expansions-and-shears:b}\\
&\sigma_{t\,AB}=\frac{1}{2}[(\partial_t-\Lie_V)q_{AB}-\Theta_{(t)}q_{AB}],\label{bfgeo:strip-expansions-and-shears:c}\\
&\sigma_{r\,AB}=\frac{1}{2}(\partial_rq_{AB}-\Theta_{(r)}q_{AB}).\label{bfgeo:strip-expansions-and-shears:d}
\end{align}
\end{subequations}
Substituting the spin connection determined by $\nabla_{[a}e^\mu{}_{b]}=0$ into the bulk action and writing the total derivatives as $\partial_a L^a$, we obtain
\begin{equation}
B_{\mu\nu}\curlywedge F^{\mu\nu}
-\frac{1}{2}B^{\mu\nu}\curlywedge\Phi_{\mu\nu\rho\sigma}B^{\rho\sigma}
=(L_0+\Delta L_0+\partial_a L^a)\,
\dd t\curlywedge\dd r\curlywedge\dd^2y,
\label{bfgeo:bf-density-evaluation}
\end{equation}
where
\begin{subequations}
\label{bfgeo:first-derivative-action}
\begin{align}
&L_0=\varrho^2\left[
2\sigma_t^{AB}\sigma_{r\,AB}-\Theta_{(t)}\Theta_{(r)}
+R[q]+2w_Aw^A-2\Lambda\right],\\
&\Delta L_0=(\partial_r\varrho^2)(\partial_r F)+
F\left[\frac{(\partial_r\varrho^2)^2}{2\varrho^2}-\varrho^2\,\sigma_r^{AB}\sigma_{r\,AB}\right],\\
&L^t=2(\partial_r\varrho^2),\\
&L^r=2(\partial_t\varrho^2)-2\partial_A(\varrho^2 V^A)-\varrho^2 (\partial_r F)-2F(\partial_r\varrho^2),\\
&L^A=-2(\partial_r\varrho^2)V^A.
\end{align}
\end{subequations}
Here $w_A=-q_{AB}(\partial_rV^B)/2$. On $r=b(t,y)$ we use the null condition $s_as^a=0$ to evaluate the boundary term from the divergence,
\begin{equation}
s_a L^a=2(\partial_t\varrho^2)-2\partial_A(\varrho^2 V^A)-\varrho^2(\partial_r F)-F(\partial_r\varrho^2)+(\partial_r\varrho^2)q^{AB}\mathring{\nabla}_A b\mathring{\nabla}_B b.
\label{bfgeo:side-divergence}
\end{equation}
Since $g^{tt}=0$ in~\eqref{bfgeo:full-metric}, we describe each spacelike cap $\Sigma_i$ locally by $t=\tau_i(r,y)$. We choose $w_i=t-\tau_i$ to increase toward the future. With its future unit normal fixed by this choice, pulling the divergence in~\eqref{bfgeo:bf-density-evaluation} back to this cap brings in all three components,
\begin{equation}
 W_i=(-\nabla_aw_i\nabla^aw_i)^{1/2}>0,\ n_{i\,a}=-\frac{\partial_aw_i}{W_i},\ L^a\partial_aw_i=L^t-L^r\partial_r\tau_i-L^A\partial_A\tau_i.\label{bfgeo:cap-unit-normal}
\end{equation}
Writing $\bm\lambda=L^a i_{\partial_a}(\dd t\curlywedge\dd r\curlywedge\dd^2y)$ for the corresponding three form, we thus obtain
\begin{subequations}
\label{bfgeo:strip-action}
\begin{align}
&I=\frac1{16\pi G_N}\bigg[\int_{\mathscr M}(L_0+\Delta L_0)\dd t\,\dd r\,\dd^2y
+\int_{\mathscr H}\bm\lambda
+\left[\int_{\Sigma_i}L^a\partial_aw_i\,\dd r\,\dd^2y\right]_{i=0}^{1}\bigg]+I_{\partial\mathscr M},\label{bfgeo:strip-action-cuts}\\
&I_{\mathscr H}=\frac1{8\pi\GN}
\int\dd t\sum_s\epsilon_s\int_{C_s}
\varrho^2[\kappa_m+m^a\nabla_a\log d]\dd^2y.
\end{align}
\end{subequations}
We orient $\mathscr H$ outward in the first line while in the second we use coordinates $r=b(t,y)$ on each null boundary and integrate over its cuts $C_s$ at fixed $t$. In the boundary action~\eqref{bfgeo:boundary-action:a}, the cap volume form and extrinsic curvature are
\begin{equation}
 \bm\epsilon_\Sigma=\varrho^2 W_i\,\dd r\,\dd^2y,\ K_i=\frac{\partial_a(\varrho^2n_i^a)}{\varrho^2}.\label{bfgeo:cap-volume-and-extrinsic-curvature}
\end{equation}
The two cap integrals therefore combine with the divergence to yield the density $L^a\partial_aw_i-2\varrho^2W_iK_i$ in the boundary terms. At $\partial\Sigma_0$ and $\partial\Sigma_1$ we then evaluate $-n_i\cdot\ell_s=\ell_s^a\partial_aw_i/W_i$ for $s=p,q$, respectively, where the joint at $S$ preserves the relative normalization of the two null normals in~\eqref{bfgeo:boundary-action:a}.

When $\Sigma_i$ deforms, we vary the pullback of each surface or joint integrand $\bm a$ by its embedding $X$, so
\begin{equation}
\delta(X^*\bm a)=X^*(\delta\bm a+\Lie_\chi\bm a),
\label{bfapp:surface-integrand-variation}
\end{equation}
with $\chi=\delta X\circ X^{-1}$. This includes the displacement of $\partial\Sigma_i$. We vary $I_{\mathscr H}$ before imposing \eqref{bfgeo:side-normalization}. If two subregions share this null boundary, matching identifies the integrands and their contributions then cancel by opposite orientations.

For $\chi=\delta X\circ X^{-1}$ with $\chi^t=0$ near the cut $C$ at $r=b(t,y)$, the torsion free Noether charge of Section~\ref{bfgeo:reduced-hamiltonian-main} yields
\begin{equation}
\bm Q_\chi\big|_C
=\frac1{16\pi G_N}\varrho^2\left[
\partial_r\chi^r-q_{AB}(\partial_rV^A)\chi^B
-q^{AC}\mathring{\nabla}_Cb\,\partial_r(q_{AB}\chi^B)
\right]\dd^2y .
\label{bfgeo:reference-charge}
\end{equation}
The term $\partial_r\chi^r$ generates a boost, while $\chi^A$ is the angular displacement of $X$ along $C$. When we vary $X$ at fixed $t$, the symplectic potential contains the surface charge
\begin{equation}
\int_C\bm Q_{\delta X\circ X^{-1}}.
\label{bfapp:embedding-surface-charge}
\end{equation}
For a translation of the spacelike cap $X_T(\Sigma)$ at fixed canonical data $\phi$, the charge enters the Hamiltonian with the opposite sign,
\begin{equation}
-\int_C\bm Q_{\partial_T X_T\circ X_T^{-1}}.
\label{bfapp:cap-translation-charge}
\end{equation}
For this deformation we use the covariant charge~\eqref{bfgeo:torsion-free-noether-charge}, since $\partial_T X_T$ can have a component along $\partial_t$.

Equivalently, the pulled back first order fields vary as
\begin{equation}
\delta(X^*B^{\mu\nu})=X^*\bigl(\delta B^{\mu\nu}+\Lie_\chi B^{\mu\nu}\bigr),\ 
\delta(X^*\omega^{\mu\nu})=X^*\bigl(\delta\omega^{\mu\nu}+\Lie_\chi\omega^{\mu\nu}\bigr).
\label{bfapp:pulled-back-field-variations}
\end{equation}
The extended potential is
\begin{equation}
\Theta_\Sigma^{(X)}
=\int_\Sigma(\bm\theta+i_\chi\bm L)
+\int_{\partial\Sigma}\bm Q_\chi,
\label{bfgeo:extended-potential}
\end{equation}
Here $\bm\theta$ is the bulk symplectic potential with the improvement~\eqref{bfgeo:potential-improvement}, and $\bm L$ is the first order Lagrangian. In writing the potential this way we include $\delta I_{\partial \mathscr M}$ only once \cite{Speranza}. Since $\delta^2=0$ on the pulled back fields, $\delta\chi+(1/2)[\chi,\chi]=0$, and the last integral contains the charge $\bm Q_\chi$, which reduces to~\eqref{bfgeo:reference-charge} for the displacement at fixed $t$.

After gluing, we integrate the divergence in~\eqref{bfgeo:bf-density-evaluation} in the original coordinate $u_s$ of each null boundary segment $s$. Its contribution to the symplectic potential is
\begin{subequations}
\label{bfgeo:inherited-intervals}
\begin{equation}
\frac{1}{16\pi G_N}\,\delta\!\left[
\sum_s\epsilon_s\int_{u_{s,-}}^{u_{s,+}}\!
\left(2\partial_{u_s}\varrho_s^2-\varrho_s^2\partial_{u_s}f_s\right)\dd u_s\,\dd^2y\right].
\end{equation}
When the cuts $C_s$ move, their velocities $\dot u_s$ in the coordinates $u_s$ enter the Hamiltonian through
\begin{equation}
-\frac{1}{16\pi G_N}\sum_s\epsilon_s\int_{C_s}
\dot u_s\left(2\partial_{u_s}\varrho_s^2-\varrho_s^2\partial_{u_s}f_s\right)\dd^2y .
\end{equation}
\end{subequations}
Equation~\eqref{bfgeo:inherited-intervals} accounts for the null segments, while the terms on the spacelike caps and their joints enter through~\eqref{bfgeo:strip-action-cuts}. The terms involving $\delta\mathfrak e$ on adjacent null boundaries remain when $\partial_{u_s}f_s=0$.

\subsection{The reduced Hamiltonian}

The term $-\Theta_T(\partial_T\iota_T)$ in~\eqref{bfq:hcan:b} accounts for the change in the embedding $\iota_T$ of the reduced data as we deform $\Sigma_T=X_T(\Sigma)$ at fixed $\phi$. On this slice the spin connection enters~\eqref{bfgeo:canonical-hamiltonian} through
\begin{equation}
\omega_T=i_{\partial_T X_T\circ X_T^{-1}}\omega.
\label{bfapp:translated-spin-connection}
\end{equation}
We include the cap and joint terms in~\eqref{bfgeo:strip-action-cuts} and vary the embedding as in~\eqref{bfgeo:extended-potential} before setting $\mathfrak C_A=0$ or $\kappa_s=0$. At fixed $\phi$, the boundary terms $[B_S]_{T_i}^{T_f}$ in the action shift $\Theta_T$ by $\delta B_S$ and $H$ by $-\partial_T B_S$ in~\eqref{bfq:hcan:b}. For two null boundaries joined at $S$, the shift $\delta B_S$ appears explicitly in the symplectic potential~\eqref{bfgeo:glued-potential-main}.

After subtracting $[B_S]_{T_i}^{T_f}$ from the action, we have $\Upsilon=\tfrac12\delta\log|c_L/d|$ in~\eqref{bfgeo:cone-form-main:a}, where the normalization factors $c_L$ and $d$ of the null normals depend on the metric and on the location $r=b(t,y)$ of the boundary.

When we translate an intermediate cut, the displacements on the two sides must also preserve the matching there. Thus the map $\iota_t$ in~\eqref{bfq:hcan:b} is constrained even after we identify configurations related by simultaneous gauge transformations. For an angular displacement, the counterpart of the boost charge in Section~\ref{sec:bf-action} for the canonical pairs $(q^i,\Pi_i)$ and $(\mathscr K(y),\Pi_{\mathscr K})$ is
\begin{equation}
 \int_S\bm\epsilon_0\left[\sum_i\Pi_i\Lie_\xi q^i+\Pi_{\mathscr K}\partial_A(\mathscr K(y)\xi^A)\right].\label{bfgeo:angular-relabeling-generator}
\end{equation}

\section{Representations of the algebra of observables}
\label{app:bf-quantization}

The relations of Section~\ref{sec:bf-quantization} determine products of $S_I(F)$ before we specify any state. Choosing a positive normalized state $\omega$ supplies the inner product $\langle a|b\rangle=\omega(a^\dagger b)$ with which we represent $S_I(F)$ on physical states and sum over intermediate states when we glue. For a causal diamond we can then identify which states enter the equilibrium trace once we impose the boundary condition on the clock.

\subsection{Physical inner product}
\label{app:physical-inner-product}

A change of boundary variables acts on a preparation by a unitary that includes the phase from the boundary action. To follow this phase through two successive shifts at fixed cosmological constant, we restrict the variation of the Lagrange multiplier by $\epsilon^{\mu\nu\rho\sigma}\varphi_{\mu\nu\rho\sigma}=0$ and include $I_{\partial\mathscr M}$ in the action whenever a shift reaches a cut. For states on the cap $\Sigma_t$, the unitary $V_{ji}(t)$ implements the change from boundary variables $i$ to $j$. Since $I+F^s+\Delta_s I=(I+F)^s$, changing variables on $\Sigma_{t_a}$ and $\Sigma_{t_b}$ yields
\begin{equation}
U_{I+F^s+\Delta_s I}(t_b,t_a)
=V_{ji}(t_b)\,U_{I+F}(t_b,t_a)\,V_{ji}(t_a)^{-1}.
\label{bfq:ward-central}
\end{equation}
In particular, \eqref{bfq:relations:d} sets $S_I(\Delta_sI)=\widehat{1}$ for a shift supported in the interior, whereas a shift at an outer boundary acts on the edge modes there and thus enters the identification of boundary states. For successive shifts $s$ from $i$ to $j$ and $s'$ from $j$ to $k$,
\begin{equation}
\Delta_{s+s'}I=(\Delta_s I)^{s'}+\Delta_{s'}I.
\label{bfq:successive-shifts}
\end{equation}
Applying~\eqref{bfq:ward-central} twice yields
\begin{equation}
U_{I+F^{s+s'}+\Delta_{s+s'}I}(t_b,t_a)
=V_{kj}(t_b)V_{ji}(t_b)\,U_{I+F}(t_b,t_a)\,V_{ji}(t_a)^{-1}V_{kj}(t_a)^{-1}.
\label{bfq:ward-successive}
\end{equation}

For a change of reference action by $V$, the admissible deformations of $I+V$ are precisely those $F$ for which $V+F$ is admissible for $I$. The relative $S$ matrix $R_V$ of Section~\ref{sec:bf-quantization} identifies the two algebras~\cite{DutschFredenhagen2001,BuchholzFredenhagen190206062},
\begin{equation}
\beta_{I,V}(S_{I+V}(F))=R_V(F).
\label{bfq:reference-action-maps}
\end{equation}
For $\phi\mapsto\phi+s$ with $s$ vanishing near the boundary, applying~\eqref{bfq:relations:d} to $V+F$ yields
\begin{equation}
 \Delta_s(I+V)=\Delta_sI+V^s-V,\ R_V(F^s+\Delta_s(I+V))=R_V(F).\label{bfq:reference-action-causal-relations}
\end{equation}
 When $s$ reaches an outer boundary we include its action on the boundary state, along with its phase, in comparing amplitudes via~\eqref{bfq:reference-action-maps}. Thus a state $\omega_I$ for the original action corresponds to the pullback state $\omega_{I+V}$ for the deformed action.

Consequently, for a normalized state $\omega$, the change of reference action in~\eqref{bfq:reference-action-maps} preserves the physical inner product~\eqref{bfq:pairing} and hence its null states~\cite{Haag1996},
\begin{equation}
 \omega_{I+V}=\omega_I\circ\beta_{I,V},\ 
 \langle a|b\rangle_{\omega_{I+V}}
=\langle\beta_{I,V}(a)|
\beta_{I,V}(b)\rangle_{\omega_I}.\label{bfq:reference-null-transport}
\end{equation}
Since $\beta_{I,V}$ preserves products and is onto, $|a\rangle\mapsto|\beta_{I,V}(a)\rangle$ extends to a unitary between the two Hilbert spaces, with $a$ and $\beta_{I,V}(a)$ acting identically on corresponding states. Time evolution likewise preserves null states by~\eqref{bfq:cap-evolution-norm}. In a causal diamond, $\rho_{\mathscr D}$ is strictly positive on $\mathcal H_{\mathscr D}$ by~\eqref{bfent:density-main}, so $\operatorname{Tr}(\rho_{\mathscr D}Q^\dagger Q)=0$ precisely when $Q$ annihilates every state in $\mathcal H_{\mathscr D}$. For a product of insertions $Q$ this criterion applies to the product itself, with its intermediate states summed as in~\eqref{bfq:compressed-source-products}.

Under a change of boundary variables, a state on $\Sigma_t$ acquires the phase $\exp(i\chi_t/\hbar)$ from~\eqref{bfq:hcan:a}. At an intermediate cap the adjacent factors in~\eqref{bfq:ward-central} cancel as $V_{ji}(t)^{-1}V_{ji}(t)=\widehat{1}$ and only the identifications at the outer boundaries survive gluing.

\subsection{Weyl ordering and contact terms}
\label{bfq:canonical-representation}

To obtain the contact terms in~\eqref{bfq:operator-insertion-second-derivative} we vary $U_j$ with respect to $j$, holding the canonical coordinates $\phi$ fixed. The boundary data $\mathfrak h_{j}=\iota_{j,t}(\phi)$ still varies on the constraint surface since $\iota_{j,t}$ depends on $j$. We therefore evaluate the tetrad and metric on the same $\mathfrak h_j$ when differentiating $H_c[j]$ in~\eqref{bfq:hcan:b}.\footnote{For two tetrad variations the contact term in~\eqref{bfq:operator-insertion-second-derivative} is $-\mathcal Q(\partial_{j_r}\partial_{j_s}H_c)|_0$ in the interaction picture~\cite{DutschFredenhagen2001,HollandsWald2005}.}

For a deformation $F=\int f_t\,\dd t$ of $I_0$, suppose $\mathcal Q(f_t)$ is bounded, self adjoint and strongly continuous with $\int\|\mathcal Q(f_t)\|\,\dd t<\infty$. The time evolution operator on $\mathcal K_{\mathscr A}$ is
\begin{equation}
U(t_b,t_a)=\mathrm T\exp\!\left[\frac{i}{\hbar}\int_{t_a}^{t_b}\mathcal Q(f_t)\,\dd t\right].
\label{bfapp:time-ordered-source}
\end{equation}
 Differentiating $U_j$ at $j=0$ thus inserts the operator
\begin{equation}
\widehat{\mathcal O}_1(t)=-\left.\partial_j\widehat H[j](t)\right|_{j=0}
\label{bfapp:first-source-insertion}
\end{equation}
between $U_0(t_f,t)$ and $U_0(t,t_i)$. A second source derivative can act at two different times or twice at the same time. With $\widehat{\mathcal O}_2(t)=-\partial_j^2\widehat H[j](t)|_{j=0}$ and the subscript $\mathrm I$ denoting the interaction picture of $\widehat H=\mathcal Q(H_c[0])$, we obtain
\begin{equation}
\left.\partial_j^2(U_0^\dagger U_j)\right|_{j=0}
=\frac{i}{\hbar}\int\dd t\,\widehat{\mathcal{O}}_{2,\mathrm I}(t)
-\frac{2}{\hbar^2}\int_{t>s}\dd t\,\dd s\,
\widehat{\mathcal{O}}_{1,\mathrm I}(t)\widehat{\mathcal{O}}_{1,\mathrm I}(s).
\label{bfq:operator-insertion-second-derivative}
\end{equation}
The first term is the contact term~\cite{HollandsWald2001,HollandsWald2005}. For polynomial interactions we evaluate the derivatives order by order with the same smearings in products and contractions. In perturbative quantum gravity one can define renormalized time ordered products using the Batalin Vilkovisky formalism~\cite{Fredenhagen:2011mq,BrunettiFredenhagenRejzner2016}. For functions $f,g$ of smeared canonical variables, Weyl ordering expresses the product of their operators as~\cite{Moyal1949},
\begin{equation}
\mathcal Q(f)\mathcal Q(g)=\mathcal Q(f\star g),\
f\star g=f\exp\!\left[\frac{i\hbar}{2}\overleftarrow{\delta}_I J^{IJ}\overrightarrow{\delta}_J\right]g.
\label{bfq:canonical-products}
\end{equation}
To evaluate $e^{\widehat\Xi_y(v)}$, let $x$ denote the phase space function obtained by multiplying the factors in~\eqref{bfq:canonical-clock-integral} with the indicated ordering using~\eqref{bfq:canonical-products}, so that $\widehat\Xi_y(v)=\mathcal Q(x)$. Order by order,
\begin{equation}
 e^{\widehat\Xi_y(v)}=\mathcal Q(1+x+\frac{1}{2} x\star x+\frac{1}{3!}x\star x\star x+\cdots).\label{bfq:weyl-exponential-of-focusing}
\end{equation}

To now justify the norm limit in~\eqref{bfq:clock-filter-definition}, let $Q_{\Xi_*}$ denote the quadratic form~\eqref{bfent:clock-form} with $\widehat\Xi_y(v)$ replaced by $\min(\widehat\Xi_y(v),\Xi_*)$. At positive corner area, $\widehat{\mathsf a}$ has trivial kernel. Indeed, if $\widehat{\mathsf a}\psi=0$, positivity of $\widehat\varrho_S(y)$ implies $\widehat\varrho_S(y)\psi=0$ for almost every $y$. The area normalization in~\eqref{bfq:canonical-area-density:e} then yields $\mathscr A\|\psi\|^2=0$, hence $\psi=0$. It follows that $\widehat{\mathsf a}^{-1/2}$ has dense domain, on which the same normalization yields
\begin{equation}
(1-\epsilon)\mathscr A\|\widehat{\mathsf a}^{-1/2}\psi\|^2
\le Q_{\Xi_*}[\psi]\le(1-\epsilon)\mathscr A e^{\Xi_*}\|\widehat{\mathsf a}^{-1/2}\psi\|^2 .
\label{bfq:capped-clock-form}
\end{equation}
Thus $Q_{\Xi_*}$ is closed and defines the positive self adjoint operator $\widehat{\bar u}_*(\Xi_*)$. As we raise $\Xi_*$, the quadratic forms $Q_{\Xi_*}$ increase to the integral $Q$ in~\eqref{bfent:clock-form}.

For any normalized incoming state $|\psi\rangle$, write $\psi(u)$ and $\Psi(u)$ for the wavefunctions in the spectral representation of $\widehat{\bar u}_*(\Xi_*)$. Completing the square in the integrand of~\eqref{bfq:incoming-clock-functional} yields
\begin{equation}
 \|\Psi(u)-\psi(u)\|^2+\frac{u}{L}\|\Psi(u)\|^2
 =\left(1+\frac{u}{L}\right)\left\|\Psi(u)-\frac{L}{L+u}\psi(u)\right\|^2
 +\frac{u}{L+u}\|\psi(u)\|^2 .\label{bfq:clock-variational-equation}
\end{equation}
 At fixed $L$, monotone convergence for closed quadratic forms determines the norm limit as $\Xi_*\to\infty$. If $M$ is the closure of the domain of $Q$, with $P_M$ the projection onto $M$ and the clock acting on $M$, the resulting wavefunction is
\begin{equation}
 L(L+\widehat{\bar u}_*(\Xi_*))^{-1}\psi \longrightarrow \Psi_L=L(L+\widehat{\bar u}_*)^{-1}P_M\psi.\label{bfq:clock-resolvent-limit}
\end{equation}
 Its projections onto finite readings extend by zero on $M^\perp$~\cite[Theorem~4.2]{BehrndtHassiDeSnooWietsma2010}.

Integrating $uL^2/(L+u)^2\le L/4$ over the clock readings in $P_M\psi$ bounds $Q[\Psi_L]$. Taking the inner product of $(\widehat 1+\widehat{\bar u}_*/L)\Psi_L=P_M\psi$ with $\Psi_L$ then relates the integral to the overlap with $\psi$,
\begin{equation}
Q[\Psi_L]\le\frac L4,\ \|\psi-\Psi_L\|^2+\frac{Q[\Psi_L]}{L}=1-\langle\psi|\Psi_L\rangle .
\label{bfq:clock-form-approximation}
\end{equation}
As $L\to\infty$, $|\Psi_L\rangle\to P_M|\psi\rangle$, so $1-\langle\psi|\Psi_L\rangle$ tends to $\|(\widehat 1-P_M)|\psi\rangle\|^2$.

For the finite set $\mathcal L$ of reduced canonical pairs, Fourier transformation of the density matrix in~\eqref{bfq:fixed-area-gaussian} with respect to the difference of its configuration space arguments at fixed midpoint $q$ yields
\begin{subequations}\label{bfq:clock-semiclassical-estimates}
\begingroup
\begin{equation}
 W_{\Omega_{\mathscr A}}(q,p)
 =\prod_{j\in\mathcal L}\frac{1}{\pi\hbar}
 \exp\!\left[-\frac{(q_j-q_j^0)^2}{2(\Delta q_j)^2}
             -\frac{2(\Delta q_j)^2(p_j-p_j^0)^2}{\hbar^2}\right].
 \label{bfq:incoming-wigner-density}
\end{equation}
For almost every $(q,p)$ with respect to~\eqref{bfq:incoming-wigner-density}, the reduced fields $\iota(q,p)$ in~\eqref{bfgeo:constraint-surface-pullback} have a conformal metric smooth in $v$ up to $S$. Thus $\bar u_*(\Xi_*)$ increases to the finite reading~\eqref{bfq:finite-classical-clock}. Monotone convergence in $\Xi_*$ followed by dominated convergence in $L$, with majorant $W_{\Omega_{\mathscr A}}$, yields
\endgroup
\begin{equation}
\sup_{\Xi_*>0}\int\dd q\,\dd p\,W_{\Omega_{\mathscr A}}(q,p)\,
\frac{\bar u_*(\Xi_*)}{L+\bar u_*(\Xi_*)}
=\int\dd q\,\dd p\,W_{\Omega_{\mathscr A}}(q,p)\,
\frac{\bar u_*}{L+\bar u_*}\to0 .
\label{bfq:classical-clock-expectation-limit}
\end{equation}
For higher orders at fixed $G_N$ we choose a closed ball about the spherical data $\phi_0=(q^0,p^0)$ on which the incoming null cones remain smooth with $\partial_u\varrho>0$ up to $S$. At fixed resolution $\Xi_y(v)$ is bounded on this ball, so once $\Xi_*$ exceeds its maximum the focusing limit drops out of the local expansion. At fixed $L$ let $u_n$ and $f_n$ denote the coefficients of $\hbar^n$ in the Weyl expansions of the clock and of $f$ in~\eqref{bfq:overlap-weyl-identity} on this ball. With $u_0=\bar u_*$, the Weyl product~\eqref{bfq:canonical-products} in that identity yields
\begingroup
\begin{align}
&f_1=\frac{L\,u_1}{(L+\bar u_*)^2},\label{bfq:clock-mode-derivatives}\\
&f_n=\frac{1}{L+\bar u_*}\left(u_n-
 \left[\left(\sum_{m=0}^{n}\hbar^m u_m\right)\star\sum_{j=0}^{n-1}\hbar^j f_j\right]_{\hbar^n}\right),\
 n\ge1 .\label{bfq:clock-local-weyl-limit}
\end{align}
Here $[\cdot]_{\hbar^n}$ denotes the coefficient of $\hbar^n$. Every derivative of $(L+\bar u_*)^{-1}$ contains an inverse power of $L+\bar u_*$, and every fixed derivative of $u_n$ is bounded on the ball. Induction in~\eqref{bfq:clock-local-weyl-limit} therefore bounds $f_n$ and each of its fixed derivatives there by a constant times $1/L$.

The Taylor coefficient with $2m$ fluctuations in $f_{n-m}$ carries $1/(2m)!$. After contraction with its symmetric derivatives, the $(2m)!/(2^m m!)$ Wick contractions have equal weight. With $\partial_I=\partial/\partial\phi^I$ and the propagator $\Sigma^{IJ}$ in~\eqref{bfq:clock-canonical-propagator}, this yields
\begin{equation}
 \big[1-\langle\Omega_{\mathscr A}|\Psi_L\rangle\big]_{\hbar^n}
 =\sum_{m=0}^n\frac{1}{2^m m!}
 \left[\left(\frac{\Sigma^{IJ}}{\hbar}\partial_I\partial_J\right)^m
 f_{n-m}\right]_{\phi=\phi_0}
 =O(L^{-1}).
 \label{bfq:clock-gaussian-coefficients}
\end{equation}
\endgroup
\end{subequations}
\begingroup
For the first term in~\eqref{bfq:clock-finite-hbar-estimate}, we use operator monotonicity of $u/(L+u)=1-L/(L+u)$ for $u\ge0$ in~\eqref{bfq:exact-corner-clock}. This yields
\begin{subequations}\label{bfq:clock-focusing-estimates}
\begin{equation}
 \left\langle\frac{\widehat{\bar u}_*(\Xi_*)}{L+\widehat{\bar u}_*(\Xi_*)}
 \right\rangle_{\Omega_{\mathscr A}}
 \le
 \left\langle\frac{(1-\epsilon)\mathscr A}
 {(1-\epsilon)\mathscr A+Le^{-\Xi_*}\widehat{\mathsf a}}
 \right\rangle_{\Omega_{\mathscr A}} .
 \label{bfq:clock-corner-resolvent}
\end{equation}
At $\Xi_*=\tfrac12\log(L/b)$ the integrand against the spectral measure of $\widehat{\mathsf a}$ is bounded by one and tends to zero for every $a>0$. Since $\widehat{\mathsf a}$ has trivial kernel at positive $\mathscr A$ by the argument for~\eqref{bfq:capped-clock-form}, dominated convergence applies even when the corner area density is correlated with radiative modes.

Substituting the conformal metric~\eqref{bfq:canonical-conformal-metric} on the reduced fields $\mathfrak h=\iota(\phi)$ into~\eqref{bfgeo:integrated-focusing-main:c}, then using~\eqref{bfgeo:integrated-focusing-main:b}, we obtain
\begin{equation}
 \Xi_y(1;\phi)
 =\frac14\int_\epsilon^1 v\left[(\partial_v z^1)^2
       +e^{2z^1}(\partial_v z^2)^2\right]\dd v
 =\log\!\left(\frac{p_{\epsilon,y}}{\left.\partial_u\varrho\right|_{v=1}}\right).
 \label{bfq:clock-full-focusing}
\end{equation}
As we increase $\Xi_*$ above $\tfrac12\log(L/b)$ at fixed $L$, the classical clock~\eqref{bfgeo:capped-clock},~\eqref{bfgeo:area-and-clock-average} is unchanged unless the focusing on some null generator exceeds $\tfrac12\log(L/b)$. Monotonicity of the reading in $\Xi_*$ then implies
\begin{equation}
 0\le \frac{L}{L+\bar u_*(\tfrac12\log(L/b))}-\frac{L}{L+\bar u_*(\Xi_*)}
 \le \mathbf 1_{\{\max_{y\in S}\Xi_y(1;\phi)>\tfrac12\log(L/b)\}} .
 \label{bfq:clock-classical-focusing-difference}
\end{equation}
For any $\delta>0$ we choose a compact set $K$ in the regular domain of $\iota(q,p)$ carrying at least $1-\delta$ of the measure~\eqref{bfq:incoming-wigner-density}. At fixed resolution $\Xi_y(1;\phi)$ in~\eqref{bfq:clock-full-focusing} attains a finite maximum on $K\times S$ by continuity. Once $\tfrac12\log(L/b)$ reaches this maximum, the integral over the subset in~\eqref{bfq:clock-classical-focusing-difference} receives contributions only from outside $K$, so
\begin{equation}
 L\ge b\exp\!\left[2\max_{\phi\in K,\,y\in S}\Xi_y(1;\phi)\right]
 \ \Longrightarrow\
 \int_{\max_{y\in S}\Xi_y(1)>\tfrac12\log(L/b)}\dd q\,\dd p\,W_{\Omega_{\mathscr A}}(q,p)\le\delta .
 \label{bfq:clock-radiation-probability}
\end{equation}
\end{subequations}
Since $K$ is independent of $\Xi_*$ and $\delta$ is arbitrary, the classical integral of the difference in~\eqref{bfq:clock-classical-focusing-difference} tends to zero uniformly for $\Xi_*\ge\tfrac12\log(L/b)$.
\endgroup

For~\eqref{bfq:state-clock-weight} we need this estimate only for $\Omega_{\mathscr A}$. To see how it extends to occupation number states, we displace the center of mode $j$ by a real valued coherent amplitude $\alpha$. Choosing the phase so that $\langle\Omega_{\mathscr A}|\Omega_\alpha\rangle>0$, we have
\begin{equation}
e^{\alpha^2/2}|\Omega_\alpha\rangle=e^{\alpha\mathfrak a_j^\dagger}|\Omega_{\mathscr A}\rangle.
\label{bfapp:displaced-occupation-state}
\end{equation}
 The finite differences defining the $n$th derivative at $\alpha=0$ converge in norm to $(\mathfrak a_j^\dagger)^n|\Omega_{\mathscr A}\rangle$. If $1-\langle\psi|\Psi_L\rangle$ tends to zero as $L$ increases for each nearby displaced wavefunction $\psi$,~\eqref{bfq:clock-form-approximation} implies that each belongs to $M$. Since $M$ is closed, it also contains their derivatives. Mixed derivatives in finitely many modes then yield every occupation number state in~\eqref{bfq:fixed-area-fock:a}, so under the same estimate the domain of~\eqref{bfent:clock-form} is dense in $\mathcal K_{\mathscr A}$.

\subsection{Boundary operators}
\label{bfq:fixed-area-source-algebra}

To construct the rank one operators of Section~\ref{bfq:algebra-states}, we work at fixed corner area on the Hilbert space $\mathcal K_{\mathscr A}$ before imposing $P_{\mathscr D}$. It suffices to realize a rank one operator on $n$ canonical pairs tensored with the identity on the remaining modes, since the projections in~\eqref{bfq:source-vacuum-projection} converge on every state in $\mathcal K_{\mathscr A}$ as $n\to\infty$. In the canonical coordinates $\phi^I$ of Section~\ref{sec:bf-quantization}, with $\mathfrak C_A\circ\iota_t=0$, we therefore deform the reduced action by $F_{\lambda,f}$ for real valued $f\in\mathcal S(\mathbb R^{2n})$ and a smooth real valued $j$ of compact support. Appendix~\ref{bfapp:incoming-canonical-pairs} reconstructs smooth expanding null data for every value of these $n$ canonical pairs. Since the first derivatives of $f$ are bounded, its Hamiltonian flow exists for every finite source strength and leaves the other modes fixed. For any angular mode entering $f$, the derivatives of $\xi_k$ with respect to $\eta_k$ in~\eqref{bfapp:gaussian-corner-data:angular} grow only as powers of $|\xi_k|$ as $|\eta_k|$ approaches $b_k$. Schwartz decay therefore makes the source and its derivatives vanish at these limits. To define the deformation on the same reduced phase space as $I$, we extend the source by zero outside the angular chart about $Y_0$ with a smooth cutoff equal to one for every value of the Gaussian coordinates.

 To isolate the operator due to $f$ when $I=I_0+V$, we form $Y_f(\lambda)$ from the two relative $S$ matrices. The factors of $S_{I_0}(V)^{-1}$ in~\eqref{bfq:reference-action-maps} cancel in this product. On the incoming Fock space $\mathcal K_{\mathscr A}$, the time ordered exponential of Appendix~\ref{bfq:canonical-representation} represents $S_{I_0}(F_{\lambda,f})$. We denote this operator by $\widehat Y_f(\lambda)$. Since $j(t)$ multiplies the same bounded operator $\mathcal Q(f)$ at every $t$, time ordering is immaterial and $\int j(t)\,\dd t=1$ yields
\begin{subequations}\label{bfq:bounded-source-word}
\begin{align}
&F_{\lambda,f}=\lambda\int j(t)f(q(t),p(t))\,\dd t,\label{bfq:bounded-source-word:a}\\
&Y_f(\lambda)=S_I(-V)^{-1}S_I(F_{\lambda,f}-V),\label{bfq:bounded-source-word:b}\\
&\beta_{I_0,V}(Y_f(\lambda))=S_{I_0}(F_{\lambda,f}),\label{bfq:bounded-source-word:c}\\
&\widehat Y_f(\lambda)
=\exp\!\left[\frac{i\lambda}{\hbar}\mathcal Q(f)\right].\label{bfq:bounded-source-word:d}
\end{align}
\end{subequations}
To realize $|u\rangle\langle v|$ we take $u,v$ to be Schwartz functions of the first $n$ configuration variables. Weyl quantization then yields
\begin{subequations}
\label{bfq:source-matrix-symbol}
\begin{align}
&f_{uv}(q,p)=\int\mathrm d^ny\,e^{-ip\cdot y/\hbar}
u(q+y/2)v^*(q-y/2),\\
&\mathcal Q(f_{uv})=|u\rangle\langle v|\otimes\widehat{1}_{>n},\\
&|u\rangle\langle u|\otimes\widehat{1}_{>n}
=\frac{\widehat{1}-\widehat Y_{f_{uu}}(\pi\hbar)}{2},\; \|u\|=1.
\end{align}
\end{subequations}
To realize the generally complex $f_{uv}$ using real valued deformations, we couple $\operatorname{Re}f_{uv}$ and $\operatorname{Im}f_{uv}$ separately. For the Gaussian state $|\Omega_{\mathscr A}\rangle$ of Section~\ref{bfq:algebra-states} and operators $\widehat a,\widehat b$ in the algebra generated by these deformations,
\begin{equation}
E_N=\bigotimes_{j\le N}|\Omega_j\rangle\langle\Omega_j|\otimes\widehat{1}_{>N}\longrightarrow|\Omega_{\mathscr A}\rangle\langle\Omega_{\mathscr A}|,\ \overline{\operatorname{span}\{\widehat a|\Omega_{\mathscr A}\rangle\}}=\mathcal K_{\mathscr A}.
\label{bfq:source-vacuum-projection}
\end{equation}
On a product vector with only finitely many factors different from $|\Omega_j\rangle$, $E_N$ acts as $|\Omega_{\mathscr A}\rangle\langle\Omega_{\mathscr A}|$ for all sufficiently large $N$. Such products span a dense subspace of $\mathcal K_{\mathscr A}$, and $\|E_N\|=1$, so the convergence holds in norm on every state. The products $\widehat a|\Omega_{\mathscr A}\rangle\langle\Omega_{\mathscr A}|\widehat b^\dagger$ thus generate the rank one operators used in~\eqref{bfq:fixed-area-type-i-main}.

On $\mathcal K_{\mathscr A}$, the reflected amplitude in~\eqref{bfent:full-gluing-main} assigns zero norm precisely to $\ker P_{\mathscr D}(\mathscr A)$, since the weight $\exp[-\mathscr A/(4G_N\hbar)]$ is strictly positive. Each class $[\xi]\in\mathcal K_{\mathscr A}/\ker P_{\mathscr D}(\mathscr A)$ is thus represented by $P_{\mathscr D}(\mathscr A)\xi$. For the pure state on $\mathcal A$ subsequently chosen in Section~\ref{bfq:state-extension}, the inner product~\eqref{bfq:pairing} embeds $\mathcal K_{\mathscr A}$ isometrically in $\mathcal H_\omega$. For boundary operators $X,Y\in\pi_\omega(\mathcal A)$, we restrict clock readings only on external states when evaluating a matrix element of their product $XY$. Between $Y$ and $X$ we therefore insert a resolution of identity on $\mathcal K_{\mathscr A}$ for insertions that preserve area and on $\mathcal H_\omega$ in general, so that
\begin{equation}
\begin{aligned}
&P_{\mathscr D}XY P_{\mathscr D}=(P_{\mathscr D}XP_{\mathscr D})(P_{\mathscr D}YP_{\mathscr D})
+P_{\mathscr D}X(\widehat{1}-P_{\mathscr D})YP_{\mathscr D}.
\end{aligned}
\label{bfq:compressed-source-products}
\end{equation}
Since $E_N$ converges to $|\Omega_{\mathscr A}\rangle\langle\Omega_{\mathscr A}|$ on every state, multiplication by the bounded operators $P_{\mathscr D}\widehat a$ and $\widehat b^\dagger P_{\mathscr D}$ preserves this limit, so
\begin{equation}
\lim_{N\to\infty}P_{\mathscr D}\widehat a E_N\widehat b^\dagger P_{\mathscr D}=P_{\mathscr D}\widehat a|\Omega_{\mathscr A}\rangle\langle\Omega_{\mathscr A}|\widehat b^\dagger P_{\mathscr D}.
\label{bfq:subregion-type-i}
\end{equation}
These rank one operators generate all bounded operators on $P_{\mathscr D}(\mathscr A)\mathcal K_{\mathscr A}$.

\subsection{Focusing bound and two loop divergence for radiative overlaps at fixed area}
\label{bfq:fixed-area-radiative-estimate}

To express $\Pi_b$ as an integral of the shear, we determine the smearing on the null cone from the wave packet $b$ on the canonical line of Appendix~\ref{bfapp:incoming-canonical-pairs}. In that construction the two axes $l=0$ and $r=0$ carry copies of the same conformal metric $z(v,y)$ at the same $v$ and $y$. We specify $b$ on $0\le r\le(1-\epsilon)/2$ and extend it to the other half of the line by $b(1-\epsilon-r,y)=b(r,y)$. With $w=1-r-l$ and the connection of $\mathsf g_{\alpha\beta}$ in~\eqref{bfapp:characteristic-action}, we transport the polarization to the axes by solving
\begin{subequations}\label{bfq:radiative-polarization-transport}
\begin{equation}
 \nabla_l(wH_r)=-H_l,\ \nabla_r(wH_l)=-H_r,\ H_r|_{w=\epsilon}=H_l|_{w=\epsilon}=b .
 \label{bfq:radiative-polarization-transport:equations}
\end{equation}
Pairing these equations with the field equation for $Z$ in~\eqref{bfapp:characteristic-transport} yields
\begin{equation}
 \partial_l\!\left[w^2\langle\partial_r Z,H_r\rangle_{\mathsf g}\right]
 =w\!\left(\langle\partial_l Z,H_r\rangle_{\mathsf g}
           -\langle\partial_r Z,H_l\rangle_{\mathsf g}\right)
 =-\partial_r\!\left[w^2\langle\partial_l Z,H_l\rangle_{\mathsf g}\right].
 \label{bfq:radiative-polarization-transport:flux}
\end{equation}
Integrating over the triangle expresses the integral on $w=\epsilon$ as the sum of the two axis integrals. The symmetry under $r\leftrightarrow l$ makes these contributions equal, so either axis yields the integral on half the canonical line. Setting $H_b(v,y)=H_r(1-v,0,y)$ and substituting $\partial_r Z=-\partial_v z$ on $l=0$ and $P_\alpha$ from~\eqref{bfapp:characteristic-pairs} on $w=\epsilon$, we obtain
\begin{equation}
 -\frac{1}{16\pi G_N}\int_S\dd A_S\int_\epsilon^1 v^2\langle\partial_v z,H_b\rangle_{\mathsf g}\,\dd v
 =\int_S\bm\epsilon_0\int_0^{(1-\epsilon)/2} b^\alpha P_\alpha\,\dd r=\Pi_b ,
 \label{bfq:radiative-polarization-transport:momentum}
\end{equation}
\end{subequations}
which proves~\eqref{bfq:affine-radiative-preparation:momentum}. Integrating the first equation in~\eqref{bfq:radiative-polarization-transport:equations} at fixed $r$ and the second at fixed $l$, as in~\eqref{bfapp:characteristic-inverse-estimate}, also bounds $|H_b|_{\mathsf g}$ by the largest $|b|_{\mathsf g(Q)}$ on the line.

Let $D=(1-\epsilon)/2$, $c_y=16\pi G_N/(\epsilon^2\varrho_S^2)$ and $\Xi=\Xi_y(1)$. The difference between the two preparations depends on the range of $v$ where $\tau(d_y(v))-1$ is nonzero. To bound this range, we first use~\eqref{bfapp:characteristic-evolution} to integrate the divergence of $w^4|\partial_r Z|_{\mathsf g}^2$ and $w^4|\partial_l Z|_{\mathsf g}^2$ on the triangle. The divergence is $-2w^3|\partial_rZ-\partial_lZ|_{\mathsf g}^2$, so the integral on the axes bounds that on the line. Substituting the canonical data~\eqref{bfapp:characteristic-pairs} yields
\begin{equation}
 \Xi\ \geq\ \frac{\epsilon^4}{8}\int_0^D
       \left(|\partial_r Q|_{\mathsf g(Q)}^2+c_y^2|P|_{\mathsf g(Q)^{-1}}^2\right)\dd r .
 \label{bfq:radiative-focusing-energy}
\end{equation}
Writing $\bar Q^1=D^{-1}\int_0^D Q^1\dd r$, this inequality bounds $|Q^1-\bar Q^1|$ by $C_\epsilon\sqrt\Xi$. The largest value of a sum of $N+1$ normalized harmonics is at most $C_\epsilon\sqrt{N+1}$ times its $L^2$ norm. Applying~\eqref{bfapp:characteristic-inverse-estimate} to $\partial_r Z$ and $\partial_l Z$ on the axes yields $\sup_v|\partial_v z|_{\mathsf g}^2\le C_\epsilon(N+1)\Xi\,e^{C_\epsilon\sqrt\Xi}$. Along a null generator $|\partial_{d_y}e^{\Xi_y(v)}|=\partial_v\Xi_y(v)$. On the fixed affine support of $1-\tau$ this bounds $|\partial v/\partial d_y|$ by $e^{-\Xi}(1+d_y\sup_v\partial_v\Xi_y)$. Cauchy Schwarz in the integral over $v$ in~\eqref{bfq:affine-radiative-preparation} then bounds the square of its change at each angle by $C(N+1)(1+e^{2\bar Q^1})e^{-c\Xi}$, with $c>0$.

We Weyl order the entire source, including its affine support, before evaluating the norm on the Gaussian. At a fixed angle each Fourier mode of either polarization selects one linear combination of angular harmonics for its position and momentum. Their commutator is $i\hbar B_L^2$ with $B_L^2=(L+1)^2/(4\pi)$. Applying Plancherel's theorem to the integral in~\eqref{bfq:source-matrix-symbol}, we bound the squared norm by integrating over $(q,p)$ the squared modulus of the contribution at this angle to $\Pi_{b,\tau}-\Pi_b$. For the second polarization, keep the two lowest nonconstant positions $u=(q_{2,1},q_{2,2})$ outside this integral first. Bounding their incoming wavefunction by its supremum and applying weighted Cauchy Schwarz costs $\int_{\mathbb R^2}(1+e^{2\bar Q^1}|u|^2)^{-2}\dd^2u=\pi e^{-2\bar Q^1}$. This cancels the growth of $|H_b|_{\mathsf g}^2$ for that polarization. Equation~\eqref{bfq:radiative-focusing-energy} absorbs the inverse weight into a fraction of $e^{-c\Xi}$. For the first polarization $|b|_{\mathsf g(Q)}$ is already bounded independently of $\bar Q^1$.

Each nonconstant position in~\eqref{bfq:radiative-focusing-energy} contributes the inverse Fourier frequency to the phase space integral and each conjugate momentum an inverse $c_y$. For the second polarization the two Gram determinants have weights $e^{2Q^1}$ on the sines and $e^{-2Q^1}$ on the cosines. Jensen's inequality on their normalized eigenvectors bounds the logarithm of their product from below by $2\int Q^1\sum_{n=1}^N(\psi_n^2-\varphi_n^2)\dd r$, where $\psi_n,\varphi_n$ are the normalized sines and cosines. Its absolute value is bounded by $C_\epsilon\|\partial_rQ^1\|_2$ since only the cosine coefficients of $Q^1$ at $2n$ enter. The first polarization therefore absorbs this factor. The product of inverse frequencies in both polarizations supplies $(N!)^{-2}$. The constant position in $z^2$ acts by a target isometry. For the remaining constant pair one uses the same Weyl kernel with the Gaussian position density and the Gaussian bound on its momentum from~\eqref{bfq:radiative-focusing-energy}.

We choose the widths of the angular density modes so that $\langle\|h\|_{H^4(S)}^2\rangle<\infty$ in~\eqref{bfapp:gaussian-corner-data:area}. Integrating $e^h$ about a maximum on $S$ bounds $\sup\varrho_S^2$ by $C\mathscr A(1+\|h\|_{H^4})$. The $m$th root of the $m$th moment of this Gaussian norm grows at most as $C\sqrt m$. After the area integration, the resulting bound on the same incoming Gaussian wavefunction $\chi$, or on any fixed finite translation of it, is
\begin{equation}
 \big\|[\mathcal Q(\Pi_{b,\tau,N,L})-\widehat{\Pi}_{b,N,L}]\chi\big\|^2
 \le C_0(C_1/N)^N .
 \label{bfq:radiative-gaussian-tail}
\end{equation}
The constants are uniform in $N,L$ for fixed source, state, $\epsilon$, $G_N$ and $\hbar$. For a smooth wave packet with $\sum_j |b_j|^2(\Delta p_j)^2<\infty$, the tail of $\widehat{\Pi}_{b,N,L}\chi$ tends to zero separately. Multiplying the limiting preparations by finite phases in the independent corner coordinates of~\eqref{bfapp:gaussian-corner-data} extends their dense span to $\mathcal K_{\mathscr A}$.

For the expansion in~\eqref{bfq:fixed-area-uv-expansion:series}, take the source constant in angle and choose $b=\partial_r(\partial_r^2+\omega_1^2)\zeta$ in the second polarization, with $\zeta$ smooth and supported in $(D/2,3D/4)$ and $\omega_n=\pi n/D$. Then $b_0=b_1=0$. On the shear free solution the transported polarization vanishes near both ends of the physical null cone. We turn the source on near $S$ with $\tau_S(d)=(1-d)^2 u'(d)$ for smooth $u$ supported in $(\delta/4,\delta/2)$, $0<\delta<D/4$. A second smooth term rises from zero to one for $|d-(1-\epsilon)|<\delta/2$ and equals one thereafter. These profiles are held fixed throughout both limits.

On the shear free solution $d=1-v=r$. Transporting the first variation of the prescribed polarization in~\eqref{bfq:radiative-polarization-transport} yields $2(-M^2,M^1)/v^2$ near $S$, with $M^\alpha=\int_0^D bQ^\alpha\dd r$, and zero near the other end. Thus the quadratic source is $(8\pi G_N)^{-1}\int_S(M^1 V^2-M^2 V^1)\dd A_S$, where $V^\alpha=\int\tau_S(r)\partial_r z^{\alpha(1)}(1-r)\dd r$. Its coefficient of $Q^\alpha(0)$ is proportional to $\int\tau_S(r)\dd r/(1-r)^2=\int u'=0$. All other radial coefficients decrease rapidly. The momentum coefficient of mode one is proportional to $\omega_1\int u(r)(1-r)\sin(\omega_1r)\dd r>0$ for $u\ge0$, so the one loop pair sum is finite and nonzero at fixed $L$.

The first change of the affine reading is $d_y^{(2)}(r)=\frac14\int_0^{1-\epsilon}\min(r,x)(1-x)\sum_\alpha|\partial_xz^{\alpha(1)}(1-x)|^2\dd x$. Contracting its two fields on the Gaussian yields
\begin{equation}
 \langle d_y^{(2)}(r)\rangle
 =\frac{k_y}{s}F_\epsilon(r)\sum_{n=1}^N n+O_L(N/s),
 \ F_\epsilon(r)=\int_0^{1-\epsilon}\frac{\min(r,x)}{1-x}\,\dd x,
 \ k_y=\frac{\epsilon^2B_L^2}{16D}\left[\omega_1^2+(c_y\hbar s)^2\right].
 \label{bfq:radiative-affine-contraction}
\end{equation}
Indeed the leading derivative is $\epsilon[Q'+c_yP]/[2(1-x)]$ by~\eqref{bfapp:characteristic-pairs}, and the two polarizations contribute equally. In the quartic source this contraction multiplies $\tau'_S$ inside $V^\alpha$. Pairing with the quadratic source therefore yields the variation of the one loop sum under $\tau_S\mapsto\tau_S+k_yF_\epsilon\tau'_S\,\sum_{n=1}^N n/s$.

The other quartic contractions can pair the derivative at $x$ with the derivative at $r$. Integration against $\min(r,x)$ then gains one inverse frequency. In~\eqref{bfapp:characteristic-evolution} the two differentiated fields carry phases in $r$ and $l$, respectively. Integrating at fixed $r$ or fixed $l$ likewise gains one inverse frequency. These contractions cost at most $O_L(N)$ after summing. In the cubic source the two derivatives in the product of the linear shear and the quadratic transported polarization also occur in distinct integration variables. An integration by parts leaves at most one differentiated mode. For two fixed undifferentiated indices $m,k$, Bessel's inequality in that variable bounds the squared coefficient summed over the differentiated modes by $C N/(mk)$. Internally contracting two undifferentiated fields costs at most $C(1+\log N)$. A derivative contracted with an undifferentiated field sums sines of $\omega_n(r\pm l)$, with $L^1$ norm at most $C(1+\log N)$. The squared cubic contribution therefore grows at most as $O_L(N(1+\log N)^2+(1+\log N)^3)$. Only~\eqref{bfq:radiative-affine-contraction} contributes to the quadratic power of the cutoff.

To fix the sign, take $u=\psi(1+\cos\lambda r)/\lambda$ with $\psi\ge0$ smooth of the stated support and $\lambda=J\pi/D$ large but fixed. For $h_S=\tau_S/v$, the leading position and momentum contributions to the one loop sum are positive multiples of $\langle h_S,|\partial_r|h_S\rangle$. Their variations replace the second $h_S$ by $F_\epsilon h_S'-F_\epsilon h_S/v$. The symmetric Fourier kernel of the derivative term contains $|k|k'-k|k'|$, which vanishes for frequencies of the same sign. For opposite signs the smooth coefficient decreases rapidly in $|k-k'|$. This term and the lower order contributions are bounded as $J$ increases. The remaining term equals $-(\lambda/2)\int(F_\epsilon/v)v^2\psi^2\dd r+O(1)$, which is strictly negative for sufficiently large fixed $J$. The interference thus has the sign in~\eqref{bfq:fixed-area-uv-expansion:two-loop}.

\section{Fluctuations about the classical solution}
\label{app:bf-classical}

The backreaction of gravitons in Section~\ref{bfq:graviton-backreaction} depends on the state on $\Sigma_0$ through the boundary conditions for the graviton two point function. We obtain this propagator here from the first order action, including the boundary term for that state. With a radiative insertion on each side of $S$, the same calculation lets us check the boost dependence of $\mathcal G_\sigma$ after gluing in Section~\ref{bfent:subregion-boost-main}. At one loop we also bound the number of graviton states allowed by focusing. In Section~\ref{app:replica-area} we derive the logarithmic correction to the entropy of a Schwarzschild black hole at fixed horizon area and at fixed mass.

\subsection{One loop determinant}
\label{app:semiclassical-determinant}

Evaluating~\eqref{bfq:hcan:a} at nonzero $j$, we write the action in canonical coordinates on the Schwinger Keldysh contour $\mathcal C$. Its value at the stationary point $\phi_j$ determines the phase of the semiclassical amplitude in terms of the boundary data, as in the Hamilton Jacobi formulation of constrained systems~\cite{Cattaneo:2020cwj}. With $\Omega_0=\delta\theta_0$, the action and its Hessian at $\phi_j$ take the form
\begin{subequations}\label{semi:canonical-hessian}
\begin{align}
&I_j[\phi]=\int_{\mathcal C}\dd t\,[p_k\dot q^k-H_c[j](\phi,t)]+I_{\partial,j},\label{semi:canonical-hessian:a}\\
&(K_j)_{IJ}(t,t')
=\left[(\Omega_0)_{IJ}\partial_t
-\delta_I\delta_J H_c[j]\right]
\delta_{\mathcal C}(t-t')+(K_{\partial,j})_{IJ}(t,t'),\label{semi:canonical-hessian:b}
\end{align}
\end{subequations}
Symmetrizing $p_k\dot q^k$ on each branch, we include $\tfrac12[p_kq^k]_{t_a}^{t_b}$ in the boundary action. The boundary action also includes the initial density operator and the inner product on $\Sigma_1$, where we join the two branches. Its second variation $K_{\partial,j}$, together with the surface terms from integration by parts, thus fixes the boundary conditions for $K_j^{-1}$. The contour and the state fix the phase of the determinant, so normalizing at $j=0$ yields
\begin{equation}
\mathcal W[j_+,j_-]
=\Delta_j\left(I_j[\phi_j]
+\frac{i\hbar}{2}\operatorname{Tr}_{\mathcal C}\log K_j
-i\hbar\log\mu_{0,j}\right)+O(\hbar^2).
\label{semi:one-loop}
\end{equation}
Here $\Delta_j f=f[j]-f[0]$ and $\mu_{0,j}=\mu_j[\phi_j]$. For independent sources on the two branches,
\begin{equation}
\mathcal W[j_+,j_-]=-i\hbar\log Z[j_+,j_-].
\label{semi:connected-generating-functional}
\end{equation}
Unitarity implies $Z[j,j]=1$. Weyl ordering introduces no correction linear in $\hbar$ to the classical function inserted in the integral.

To evaluate the one loop determinant in first order variables, we vary the independent fields about a classical solution according to $(B,\omega,\Phi)\mapsto(B,\omega,\Phi)+\varepsilon(b,\delta\omega,\varphi)$ and keep the coefficient of $\varepsilon^2$,
\begin{equation}
\begin{aligned}
&I_{\mathscr M}^{(2)}=\frac{1}{64\pi G_N}\int_{\mathscr M}
\bigl[2b_{\mu\nu ab}\nabla_{[c}\delta\omega_{d]}^{\mu\nu}
+2B_{\mu\nu ab}\delta\omega_{[c}{}^\mu{}_{\rho}\delta\omega_{d]}{}^{\rho\nu}\\
&-\frac{1}{2}b_{\mu\nu ab}\Phi^{\mu\nu}{}_{\rho\sigma}b^{\rho\sigma}_{cd}
-b_{\mu\nu ab}\varphi^{\mu\nu}{}_{\rho\sigma}B^{\rho\sigma}_{cd}\bigr]
\dd x^a\curlywedge\dd x^b\curlywedge\dd x^c\curlywedge\dd x^d
+I_{\partial\mathscr M}^{(2)}.
\end{aligned}
\label{semiapp:bf-quadratic-action}
\end{equation}
With $\epsilon^{\mu\nu\rho\sigma}\varphi_{\mu\nu\rho\sigma}=0$, $\delta_\varphi I_{\mathscr M}^{(2)}=0$ imposes the linearization of~\eqref{bfgeo:simplicity}. We can therefore eliminate $\delta\omega$ using the linearized torsion equation of Section~\ref{bfq:graviton-backreaction} and write the result as $I_{\mathscr M}^{(2)}[h,\delta\mathfrak e]$ with $h_{ab}=2e_{\mu(a}\delta e^\mu{}_{b)}$. The boundary frame $\mathfrak e$ must vary along with the bulk fields.

 For two independent variations $h_{ab}=\delta_h g_{ab}$ and $k_{ab}=\delta_k g_{ab}$, with $k=g^{ab}k_{ab}$ and $E_{ab}=G_{ab}+\Lambda g_{ab}$, we have
\begin{equation}
\delta_k\delta_hI
 =-\frac1{16\pi G_N}\int_{\mathscr M}\bm\epsilon\,h_{ab}
       \left(g^{ac}g^{bd} E^{(1)}_{cd}[k]
           +\frac{1}{2}k E^{ab}
           -2k^{c(a} E^{b)}{}_c\right)
+\delta_k\!\left[\int_{\partial\mathscr M}\bm\theta_g(h)+\delta_h I_{\partial\mathscr M}\right].
\label{semiapp:metric-hessian}
\end{equation}
Thus the boundary conditions for the inverse of the bulk operator in~\eqref{semiapp:metric-hessian} follow from the surface integral together with the second variation of $I_{\partial\mathscr M}$ in~\eqref{bfgeo:boundary-action:a}. This includes the displacement of each boundary as in~\eqref{semiapp:pullback-second}. Then, on a vacuum solution $R_{ab}=\Lambda g_{ab}$, only the term with $E^{(1)}_{ab}[k]$ contributes to the bulk Hessian in~\eqref{semiapp:metric-hessian}~\cite{DeWitt1967II}. 
On a classical solution with $j\ne0$, the terms proportional to $E_{ab}$
in \eqref{semiapp:metric-hessian}, variations of $F[j]$ and the
constraint terms also contribute. Furthermore, at fixed $j$, the fields $\mathfrak h=\iota_{j,t}(\phi)$ satisfy $\mathfrak C_A\circ\iota_j=0$, so their variations are tangent to the constraint surface. At a stationary point of $I\circ\iota_j$ we have $\delta I=-\lambda^A\delta\mathfrak C_A$ so for fixed canonical variations $\delta,\delta'$, differentiating $\mathfrak C_A\circ\iota_j=0$ twice then yields
\begin{equation}
\delta'\delta(I\circ\iota_j)
=\left(\frac{\delta^2 I}{\delta \mathfrak h^2}+\lambda^A\frac{\delta^2\mathfrak C_A}{\delta \mathfrak h^2}\right)(\delta\mathfrak h,\delta'\mathfrak h).
\label{semiapp:hessian-pullback}
\end{equation}
Here $I$ includes $F[j]$ and the boundary state. Evaluating the quadratic form in parentheses on $\delta_I\mathfrak h$ and $\delta_J\mathfrak h$ yields the components $(K_j)_{IJ}$ of the canonical Hessian.

It remains to include the measure in these coordinates. On a smooth nondegenerate region satisfying~\eqref{bfgeo:simplicity} it takes the form
\begin{equation}
\dd\mu_{\phi}=\mathcal D \mathfrak h\,
\delta(\mathfrak C^{(2)})\sqrt{|\det\{\mathfrak C^{(2)},\mathfrak C^{(2)}\}|}\,
\delta(\mathfrak C^{(1)})\delta(\mathfrak G)|\det\{\mathfrak G,\mathfrak C^{(1)}\}_D|.
\label{semiapp:bf-reduced-measure}
\end{equation}
Here $\mathfrak C^{(1)}$ and $\mathfrak C^{(2)}$ denote first class constraints and second class constraints, $\mathfrak G=0$ is the gauge condition, and $\{\ ,\ \}_D$ is the Dirac bracket after imposing $\mathfrak C^{(2)}=0$~\cite{Senjanovic1976}. In the canonical coordinates $\phi$ this reduces to the Liouville measure, including the pairs at the corners. What the change of variables to $\phi$ preserves is the combination $\mu_{0,j}(\det_{\mathcal C}K_j)^{-1/2}$, since the Jacobians in the measure and determinant cancel. In particular, integrating out $\eta$ with invertible Hessian $K_{\eta\eta}$ yields the Hessian for the remaining fluctuations and a factor in $\mu_{0,j}$, respectively,
\begin{equation}
 K_j=K_{\phi\phi}-K_{\phi\eta}K_{\eta\eta}^{-1}K_{\eta\phi},\ (\det K_{\eta\eta})^{-1/2}.\label{semiapp:reduced-hessian}
\end{equation}
 For the spin connection this is precisely the step that produced $I_{\mathscr M}^{(2)}[h,\delta\mathfrak e]$, including the variation of the boundary frame.

To insert the metric at $p$ we quantize $g_{ab}(p)$, including its dependence on the boundary frame $\mathfrak e(\phi)$, and use~\eqref{bfq:canonical-products} for products. The $j$ dependence of $\iota_{j,t}$ and $\chi_{j,t}$ at fixed $\phi$ enters the contact term $\widehat{\mathcal O}_2$ in~\eqref{bfq:operator-insertion-second-derivative}. The one loop contribution to $\delta'\delta\mathcal W$ in~\eqref{semi:one-loop} is
\begin{equation}
\frac{i\hbar}{2} \operatorname{Tr}_{\mathcal C}\left( K_j^{-1}\delta'\delta K_j -K_j^{-1}\delta'K_j K_j^{-1}\delta K_j\right) -i\hbar\delta'\delta\log\mu_{0,j}.
\label{semiapp:source-contact-terms}
\end{equation}
Here the derivatives with respect to $j$ also act on $\phi_j$ and $I_{\partial,j}$. In particular $(i\hbar/2)\operatorname{Tr}_{\mathcal C}(K_j^{-1}\delta'\delta K_j)$ is the contact term required by~\eqref{bfq:operator-insertion-second-derivative}.

To obtain $\langle\Delta\phi^I\rangle$ to first order in $\hbar$, we need $I_j$ through cubic order and $\mu_j[\phi]$ through first order about $\phi_j$. With integrals over $\mathcal C$ understood,
\begin{equation}
\langle\Delta\phi^I\rangle
=i\hbar(K_j^{-1})^{IJ}
\left[\frac{\delta\log\mu_j[\phi]}{\delta\phi^J}
-\frac12(K_j^{-1})^{KL}\frac{\delta^3 I_j}{\delta\phi^J\,\delta\phi^K\,\delta\phi^L}\right]_{\phi_j}
+O(\hbar^2).
\label{semiapp:one-loop-shift}
\end{equation}
Consequently the expectation value of a Weyl ordered observable is
\begin{equation}
 \langle\mathcal Q(\mathcal{O})\rangle=\mathcal{O}(\phi_j)+(\delta_I \mathcal{O})(\phi_j)\langle\Delta\phi^I\rangle+\frac{i\hbar}{2}(\delta_I\delta_J \mathcal{O})(\phi_j)(K_j^{-1})^{IJ}+O(\hbar^2).\label{semiapp:second-order-observable-expectation}
\end{equation}
The derivative of $\log\mu_j$ in~\eqref{semiapp:one-loop-shift} accounts for the variation of the reduced measure on the contour.

The Schwinger Keldysh effective action~\cite{CalzettaHu1987}, obtained by
the Legendre transform in sources for the canonical variables, has the expansion
\begin{equation}
\Gamma_{\mathcal C}[\bar\phi]=\left.I_j[\bar\phi]\right|_{j=0}+\hbar\Gamma_{\mathcal C}^{(1)}+O(\hbar^2),\ \delta\Gamma_{\mathcal C}^{(1)}=\frac i2\operatorname{Tr}_{\mathcal C}(K^{-1}\delta K)-i\delta\log\mu_0.
\label{semiapp:effective-action}
\end{equation}
With~\eqref{semi:equation-variations}, stationarity of the effective action, $\delta\Gamma_{\mathcal C}=0$, thus recovers the semiclassical Einstein equations~\eqref{semi:semiclassical-einstein}.

\subsection{Corner boosts and two sided graviton correlators}
\label{app:collar-return}

Section~\ref{bfent:subregion-boost-main} uses the fact that boosting the normal frame on just one side of $S$ changes a graviton two point function across the cut. To determine this change, we evaluate the boost generated by $\widehat{\mathscr A}/(8\pi G_N)$ first on $\sigma_{AB}$ in a freely falling local Lorentz frame and then on the intermediate states over which we sum when we close the cut.

To glue the normal frames $(\ell_R^a,n_R^a)$ and $(\ell_L^a,n_L^a)$ at $S$, we integrate over the rapidity $\alpha$ of the intermediate normal frame. On a state $|s\rangle$ with $\widehat{\mathscr A}|s\rangle=4G_N\hbar s|s\rangle$, a boost of rapidity $\Gamma$ acts by $\exp[i\Gamma s/(2\pi)]$. Composing the boosts from $\Gamma_L$ to $\alpha$ and from $\alpha$ to $\Gamma_R$ yields
\begin{equation}
 \int_{\mathbb R}\frac{\dd\alpha}{4\pi^2}
 e^{\,i[s_R(\Gamma_R-\alpha)+s_L(\alpha-\Gamma_L)]/(2\pi)}
 =\delta(s_R-s_L)\,e^{\,is_R(\Gamma_R-\Gamma_L)/(2\pi)} .
 \label{eq:collar-boost-match}
\end{equation}
Thus the two areas must agree, $s_R=s_L$, but the amplitude still depends on the relative boost $\Gamma_R-\Gamma_L$. To determine how this enters $\mathcal G_\sigma$, we take a trace free radiative perturbation $h_{AB}$ of the shear free congruence in Section~\ref{bfent:subregion-boost-main}. In the observer's affine parameter $u$ we have $\sigma_{AB}^{(1)}=\partial_u h_{AB}/2$ to linear order, so to leading order in $d/b$ (see~\eqref{bfent:shear-preparation} and~\eqref{bfent:graviton-correlator}),
\begin{equation}
 \mathcal G_\sigma(u,u')=\frac1{8}\partial_u\partial_{u'}\,
 \langle\{h_{AB}(u),h^{AB}(u')\}\rangle .
 \label{bfapp:shear-contraction}
\end{equation}
In a transverse basis parallel transported along each null generator, the smearing functions of Section~\ref{bfent:subregion-boost-main} are $f_{L,R}^{AB}(u,y)=f^{AB}(u\pm d,y)$, with $\supp f$ in $|u|<\Delta u$. To compute the two point function of $\widehat\sigma_{f_L}$ and $\widehat\sigma_{f_R}$ at leading order, we rescale $\Delta\phi_a\mapsto\lambda\Delta\phi_a$ on each half $a=L,R$ and keep the coefficient $I_a^{(2)}$ of $\lambda^2$ in the action and the coefficient of $\lambda$ in each insertion $\sigma_{f_a}$. By~\eqref{bfq:diamond-cyclic-packet}, the states $P_{\mathscr D}\pi_\omega(X)|\psi\rangle$ span a dense subspace of $\mathcal H_{\mathscr D}$. Using an orthonormal basis obtained from these states, we insert a resolution of identity at the cut. The external preparation $C_b$ fixes the weight at each area, including the factor $f_b(\hat s)^{1/2}$. In the unnormalized amplitude at quadratic order we integrate once over the data $\Delta\mathfrak h$ on $\Sigma_1$ and $S$ in~\eqref{bfent:full-gluing-main}. For the ordering $\exp(ij_L\widehat\sigma_{f_L})\exp(ij_R\widehat\sigma_{f_R})$ this yields
\begin{equation}
 Z(j_L,j_R)=\int\mathcal D\Delta\mathfrak h\,\prod_{a=L,R}\big[\int_{\Delta\mathfrak h}\mathcal D\Delta\phi_a\,e^{\,iI_a^{(2)}/\hbar+ij_a(\delta_I \sigma_{f_a})\Delta\phi_a^I}\big].\label{bfapp:conditional-shear-source-integral}
\end{equation}
After matching at $S$, the bulk and boundary terms on the two oriented halves add to a single quadratic action,
\begin{equation}
I_L^{(2)}+I_R^{(2)}=\frac{1}{2}\Delta\phi^I(K_0)_{IJ}\Delta\phi^J.
\label{bfapp:shear-source-integral:a}
\end{equation}
Completing the square with the Hessian $K_0$ in~\eqref{semi:canonical-hessian:b} at $j_L=j_R=0$ yields the same factor $(\det_{\mathcal C}K_0)^{-1/2}$ in $Z(j_L,j_R)$ and $Z(0,0)$. Since shear vanishes on the background, the logarithm of the normalized amplitude is quadratic in $j_L,j_R$,
\begin{equation}
\log\frac{Z(j_L,j_R)}{Z(0,0)}
 =\frac12\sum_{a,c=L,R}j_a j_c
  \left.\frac{\partial^2\log Z}{\partial j_a\,\partial j_c}\right|_{0}
 =-\frac{i\hbar}{2}\sum_{a,c}j_a j_c(\delta_I \sigma_{f_a})(K_0^{-1})^{IJ}(\delta_J \sigma_{f_c}).
\label{bfapp:shear-source-integral:b}
\end{equation}
Here we evaluate the derivatives at $j_L=j_R=0$. For real valued smearing functions the two orderings are complex conjugates, so taking the real part yields $\mathcal G_\sigma$,
\begin{equation}
 \frac{1}{2}\langle\{\widehat\sigma_{f_L},\widehat\sigma_{f_R}\}\rangle
 =-\left.\operatorname{Re}\frac{\partial^2\log Z}{\partial j_L\,\partial j_R}\right|_{0}
 =\operatorname{Re}\!\left[i\hbar(\delta_I \sigma_{f_L})(K_0^{-1})^{IJ}(\delta_J \sigma_{f_R})\right].
 \label{bfapp:shear-source-derivative}
\end{equation}
We implement $\ell_R^a\mapsto e^\eta\ell_R^a$ in~\eqref{bfapp:shear-source-integral:b} by transforming $\sigma_{f_R}$ while keeping $\sigma_{f_L}$ fixed.
On a flat null plane the radiative symplectic form fixes the positive frequency measure of $\sigma_{AB}$ to be proportional to $\omega\,\dd\omega$~\cite{Reisenberger12113880,CiambelliFreidelLeigh2024QuantumNull}. With the transverse smearing of Section~\ref{bfent:subregion-boost-main}, the freely falling vacuum at $u\ne u'$ therefore has
\begin{equation}
 \mathcal G_{\sigma,0}(u,u')=c_\perp\lim_{\tau\downarrow0}\int_0^\infty
 \omega e^{-\tau\omega}\cos[\omega(u-u')]\,\dd\omega
 =-\frac{c_\perp}{(u-u')^2},
 \label{bfapp:shear-singularity}
\end{equation}
where $c_\perp>0$ includes the transverse smearing and both polarizations. We choose a polarization for which the correlator in~\eqref{bfent:shear-boost-ratio} changes under the boost. For this choice, the two radiative insertions commute since their supports are disjoint and $[\widehat\sigma^{(1)}_{AB}(u),\widehat\sigma^{(1)}_{CD}(u')]$ is proportional to $\partial_u\delta(u-u')$ at this order. We may therefore use~\eqref{bfapp:shear-source-integral:b} at fixed $j_L,j_R$ to evaluate~\eqref{bfent:bounded-graviton-observable} in the freely falling vacuum. Writing $G_{ac}=\tfrac12\langle0|\{\widehat\sigma_{f_a},\widehat\sigma_{f_c}\}|0\rangle$ for $a,c\in\{L,R\}$, we obtain
\begin{equation}
 \langle0|\widehat{\mathcal O}|0\rangle
 =\exp[-(j_L^2G_{LL}+j_R^2G_{RR})/2]\sinh(j_Lj_RG_{LR}) .
 \label{bfapp:shear-bounded-functions}
\end{equation}
A relative boost preserves $G_{LL}$ and $G_{RR}$ but deforms $G_{LR}$ by~\eqref{bfent:shear-boost-ratio}, so $\delta_\eta\ne0$ for suitable $j_L,j_R$.

Since the background is shear free, this same contraction determines $\langle\sigma_{AB}\sigma^{AB}\rangle$ in the Raychaudhuri equation~\eqref{bfgeo:focusing:e} through one loop. In particular the derivatives in~\eqref{bfapp:shear-contraction} act at separated points as in~\eqref{semiapp:propagator-derivatives}. We subtract the local vacuum singularity before taking coincidence in both the affine and transverse coordinates, then integrate against the transverse smearing function~\cite{Wald1994QFT,HollandsWald2001}. The local curvature terms agree under both approaches to $S$. If boosting $\ell_R^a$ at fixed $\ell_L^a$ preserved $\mathcal G_\sigma$, subtracting~\eqref{bfapp:shear-singularity} would require
\begin{equation}
 d^2\big[(\mathcal G_\sigma-\mathcal G_{\sigma,0})(-d,d)-e^\eta(\mathcal G_\sigma-\mathcal G_{\sigma,0})(-d,e^\eta d)\big]=\frac{c_\perp}{4}\tanh^2(\eta/2).
 \label{bfapp:shear-focusing}
\end{equation}
Both subtracted two point functions have a finite coincidence limit at $S$, so the left hand side vanishes as $d\to0$ for fixed $\eta\ne0$. This is incompatible with the positive right hand side. At finite separation, the ratio in~\eqref{bfent:shear-boost-ratio} measures the same change in $\mathcal G_\sigma$.

To evaluate the same boost on intermediate states in~\eqref{bfent:shear-finite-observable}, we pair $|i\rangle$ with its conjugate $|i^*\rangle$ on the oppositely oriented copy of $S$, so the sum computes an expectation value in
\begin{equation}
 |\Psi_P\rangle=\frac{\sum_i C_b|i\rangle\otimes|i^*\rangle}{\sqrt{\Tr(PC_b^\dagger C_b)}} .
 \label{bfapp:shear-open-sum}
\end{equation}
Since an overall phase drops out of $\langle\widehat{\mathcal O}\rangle$, we are free to multiply the boosted state by $e^{-i\eta s_0/(2\pi)}$ for any $s_0$. Writing $\hat s^*$ for the area in units of $4G_N\hbar$ on the conjugate copy of $S$, we can therefore use $|e^{ix}-1|\le|x|$ to bound the distance between the two states,
\begin{equation}
 \left\|\bigl(\widehat{1}\otimes e^{i\eta(\hat s^*-s_0)/(2\pi)}-\widehat{1}\bigr)|\Psi_P\rangle\right\|^2
 \le\frac{\eta^2}{4\pi^2}\Tr[\rho_P(\hat s-s_0)^2].
 \label{bfapp:boosted-state-distance}
\end{equation}
The corresponding change in $\langle\widehat{\mathcal O}\rangle$ is at most $2\|\widehat{\mathcal O}\|$ times this distance in norm, which proves~\eqref{bfent:boost-spread}.

Under the hypothesis $\dim\mathcal H_{\mathscr D}=\infty$, infinitely many orthogonal states have support in the interval $I$ of width at most $\Delta s$ chosen in Section~\ref{bfent:subregion-boost-main}, whether the spectrum in $I$ is discrete or continuous. Given any $k$ orthonormal intermediate states we can therefore add $N$ more from $I$, orthogonal to the first $k$. For $s_0\in I$, $\Tr[P(\widehat{1}-\mathbf1_I(\hat s))]\le k$, and~\eqref{bfent:collar-comparison} implies $PC_b^\dagger C_bP\le P$ and $\Tr(PC_b^\dagger C_b)\ge(k+N)e^{-s_*}m_{b,\mathscr D}$. It follows that
\begin{equation}
 \Tr[\rho_P(\hat s-s_0)^2]\le (\Delta s)^2+
 \frac{s_*^2 k}{(k+N)e^{-s_*}m_{b,\mathscr D}} .
 \label{bfapp:shear-band}
\end{equation}

To keep the fraction of states outside $I$ in~\eqref{bfapp:shear-band} small for every finite union of complete cycles, it suffices to include at most one such state in each cycle of length $r$. We can arrange this for an infinite basis of any cardinality by partitioning $[0,s_*]$ into finitely many intervals of width at most $\Delta s$. Since their spectral subspaces span $\mathcal H_{\mathscr D}$, at least one interval $I$ satisfies $\dim(\mathbf1_I(\hat s)\mathcal H_{\mathscr D})=\dim\mathcal H_{\mathscr D}$. A basis adapted to $\mathbf1_I(\hat s)$ thus has enough vectors in $I$ to group $r-1$ of them with each vector outside $I$ and to form any further cycles entirely within $I$. For the resulting permutation $V$, every projection $P$ onto a finite union of complete cycles then satisfies $VPV^\dagger=P$. By~\eqref{bfapp:shear-band} the fraction of states outside the interval bounds the spread in area,
\begin{equation}
 \Tr[P(\widehat{1}-\mathbf1_I(\hat s))]\leq\frac{\rank(P)}{r},\ \Tr[\rho_P(\hat s-s_0)^2]\leq(\Delta s)^2+\frac{s_*^2}{r e^{-s_*}m_{b,\mathscr D}}.\label{bfapp:area-interval-concentration}
\end{equation}
 For a fixed boost with $\delta_\eta\ne0$ in~\eqref{bfapp:shear-bounded-functions}, we may thus choose $\Delta s$ and $r$ so that the upper bound in~\eqref{bfent:boost-spread} is less than $|\delta_\eta|/2$ for every finite union of cycles.

The sum over cycles can thus suppress $\langle\Delta_\eta\widehat{\mathcal O}\rangle_P$ even though $\delta_\eta\ne0$. To determine what this would cost in focusing along the null generators through $S$, we use the transverse smearing $f_\perp^{AB}$ of~\eqref{bfent:transversely-smeared-graviton}. Writing $f_L^{AB}=p_L(u)f_\perp^{AB}(y)$ and $f_R^{AB}=p_R(u)f_\perp^{AB}(y)$, we have $p_{R,\alpha}(u)=p_R(e^{-\alpha}u)$ for a relative boost of rapidity $\alpha$. For smearing in the affine parameter we write
\begin{equation}
\widehat\sigma_\perp[h]=\int h(u)\widehat\sigma_\perp(u)\,\dd u.
\label{bfapp:affinely-smeared-graviton}
\end{equation}
 To include both insertions before and after the boost, we choose a real valued function $g(u)=g_0(u/B)$ with
\begin{subequations}\label{bfapp:shear-interval-window}
\begin{equation}
g_0\in C_c^\infty((-1,1)),\ g_0(v)=1\ (|v|\le\frac{1}{2}),\label{bfapp:shear-interval-window:a}
\end{equation}
\begin{equation}
2\max(1,e^\eta)(d+\Delta u)<B\ll b .\label{bfapp:shear-interval-window:b}
\end{equation}
\end{subequations}
To the order used in~\eqref{bfapp:shear-singularity}, $\dd A=b^2\dd\Omega$ at $S$. With $c_f$ the coefficient of that two point function for $f_\perp^{AB}$, subtracting the freely falling vacuum before taking $u'\to u$ yields
\begin{equation}
 E_g=\int g^2\langle\normalorder{\widehat\sigma_\perp^2}\rangle\,\dd u
       +\frac{c_f}{2}\|\partial_u g\|_2^2
     =\frac1\pi\int_0^\infty
       \langle\widehat\sigma_\perp[g e^{iku}]^\dagger
              \widehat\sigma_\perp[g e^{iku}]\rangle\,\dd k .
 \label{bfapp:shear-interval-square}
\end{equation}
Indeed, the commutator cancels out upon subtraction, so $\langle\normalorder{\widehat\sigma_\perp(u)\widehat\sigma_\perp(u')}\rangle=\langle\normalorder{\widehat\sigma_\perp(u')\widehat\sigma_\perp(u)}\rangle$ and integrating over $k>0$ takes $u'$ to $u$. For $\widetilde g(k)=\int e^{-iku}g(u)\,\dd u$, the integral on the right in $|0\rangle$ is
\begin{equation}
 \frac{c_f}{\pi}\int_0^\infty\dd k\int_0^\infty\dd\xi\,\xi|\widetilde g(k+\xi)|^2=\frac{c_f}{2}\|\partial_u g\|_2^2.\label{bfapp:shear-smearing-energy}
\end{equation}

We can use the positive integral in~\eqref{bfapp:shear-interval-square} to bound the change in $\langle\widehat{\mathcal O}\rangle$ because the two radiative insertions commute. Before and after the relative boost,
\begin{equation}
 2\widehat{\mathcal O}_\alpha
 =\cos\widehat\sigma_\perp[h_{\alpha,-}]-\cos\widehat\sigma_\perp[h_{\alpha,+}],\label{bfapp:shear-interval-cosines}
\end{equation}
where
\begin{equation}
 h_{\alpha,\pm}=j_L p_L\pm j_Rp_{R,\alpha},\ \alpha\in\{0,\eta\}.\label{bfapp:shear-interval-profiles}
\end{equation}
For each such $h$ we use its positive frequency part to define the mode operator. Since $g h=h$, the operator and its adjoint sum to the smeared radiative operator,
\begin{equation}
 h_+(u)=\frac{1}{2\pi}\int_0^\infty\widetilde h(k)e^{iku}\,\dd k,\ \mathfrak{a}_h=\widehat\sigma_\perp[g h_+],\ \mathfrak{a}_h+\mathfrak{a}_h^\dagger=\widehat\sigma_\perp[h].\label{bfapp:positive-frequency-shear}
\end{equation}
Where $g$ differs from the identity, $h_+$ is purely imaginary, so $\operatorname{Im}(\overline{h_+}\partial_u h_+)=0$ there and
\begin{equation}
 [\mathfrak{a}_h,\mathfrak{a}_h^\dagger]=v_h\widehat{1},\ v_h=c_f\int_0^\infty k|\widetilde h(k)|^2\,\dd k.\label{bfapp:shear-commutator-normalization}
\end{equation}

The Fock space vacuum $\mathfrak{a}_h|0_h\rangle=0$ reproduces $\langle0|\cos\widehat\sigma_\perp[h]|0\rangle=e^{-v_h/2}$ since $v_h=\langle0|\widehat\sigma_\perp[h]^2|0\rangle$. For a state $\rho_h$ of this mode, the expectation value of the occupation number bounds the probability of a nonzero occupation and hence the trace distance to the vacuum. Using $\|\cos\widehat\sigma_\perp[h]\|\le1$ in the trace distance bound and applying Cauchy Schwarz to the integral for $\mathfrak{a}_h$, we obtain
\begin{subequations}\label{bfapp:shear-interval-cosine}
\begin{align}
&1-\langle0_h|\rho_h|0_h\rangle\le\frac{\langle\mathfrak{a}_h^\dagger\mathfrak{a}_h\rangle}{v_h},\label{bfapp:shear-interval-cosine:a}\\
&\frac{1}{2}\|\rho_h-|0_h\rangle\langle0_h|\|_1\le(1-\langle0_h|\rho_h|0_h\rangle)^{1/2},\label{bfapp:shear-interval-cosine:b}\\
&\left|\langle\cos\widehat\sigma_\perp[h]\rangle-e^{-v_h/2}\right|^2
 \le\frac{4}{v_h}\langle\mathfrak{a}_h^\dagger\mathfrak{a}_h\rangle
 \le\frac{\|h\|_2^2}{v_h}E_g .\label{bfapp:shear-interval-cosine:c}
\end{align}
\end{subequations}

For any state on $(-B,B)$, we can now sum~\eqref{bfapp:shear-interval-cosine:c} over the four functions $h$ in~\eqref{bfapp:shear-interval-profiles}. Using $E_g$ from~\eqref{bfapp:shear-interval-square}, we find
\begin{equation}
 \int g^2\langle\normalorder{\widehat\sigma_\perp^2}\rangle\,\dd u
 \ge \frac{4\big|\langle\Delta_\eta\widehat{\mathcal O}\rangle-\delta_\eta\big|^2}
 {\left(\sum_h\frac{\|h\|_2}{\sqrt{v_h}}\right)^2}
 -\frac{c_f}{2}\|\partial_u g\|_2^2 .
 \label{bfapp:shear-interval-reading}
\end{equation}
At fixed $\sqrt{c_f}\,j_L p_L$ and $\sqrt{c_f}\,j_Rp_R$ as functions of $u/d$, $\delta_\eta$ is independent of $d$ and $(\sum_h\|h\|_2/\sqrt{v_h})^2\propto d/c_f$. Since $\|\partial_u g\|_2^2=B^{-1}\|\partial_v g_0\|_2^2$, this proves the lower bound in~\eqref{bfent:shear-interval-cost:a} for $d\ll B$. For the uncut preparation $C_b$ we have $\langle\normalorder{\widehat\sigma_\perp^2}\rangle_{C_b}=O(c_f/b^2)$ on the smooth background as in Section~\ref{bfent:subregion-boost-main}. Applying~\eqref{bfapp:shear-interval-reading} to this state therefore yields
\begin{equation}
 \big|\langle\Delta_\eta\widehat{\mathcal O}\rangle_{C_b}-\delta_\eta\big|^2
 \lesssim \frac dB+\frac{dB}{b^2} .
 \label{bfapp:shear-interval-contrast}
\end{equation}
At fixed $\ell_P\ll\Delta u\ll d$, we can thus choose $B$ and $b$ so that $|\langle\Delta_\eta\widehat{\mathcal O}\rangle_{C_b}-\delta_\eta|<|\delta_\eta|/2$, as required in~\eqref{bfent:boost-deformation-bounds}.

\subsection{Focusing and the number of graviton states}
\label{app:focusing-graviton-states}

We can also check the finiteness result of Section~\ref{bfent:subregion-boost-main} at one loop by bounding the number of graviton states allowed by focusing at fixed corner area. To that end we expand about a causal diamond with a spherical corner, for which $4\pi\varrho_S^2=\mathscr A$ and corrections to the flat spacetime calculation are of order $\varrho_S^2/R_c^2$. The dependence on $\Lambda$ is especially simple. For a spherical cut $C_u$ at affine parameter $u$ we have
\begin{equation}
\int_{C_u} R[q]\,\dd A=8\pi,
\label{eq:collar-cut-curvature}
\end{equation}
so integrating~\eqref{bfgeo:transverse-constraints-main:d} over $C_u$ yields
\begin{equation}
 \partial_u\int_{C_u}\Theta_{(n)}\,\dd A
 =-4\pi+\int_{C_u}|\varpi|^2\,\dd A+\Lambda\,\mathscr A(C_u).
 \label{eq:collar-cross-focusing}
\end{equation}
With the affine parameter fixed at $\partial\Sigma_i$ as in~\eqref{bfgeo:capped-clock}, the $\Lambda$ term is suppressed by $O(|\Lambda|\varrho_S^2)$ relative to the spherical term, while $R_{ab}\ell^a\ell^b=0$ on the original null boundary.

Where $\partial_u\varrho>0$, we write a regular perturbation of the conformal metric in tensor harmonics normalized on the unit sphere, with coefficients $h_I(v)$. Here $I=(j,m,p)$, with $p$ distinguishing the two polarizations. With $t_j^2=(j-1)(j+2)/2$ the coefficient $\varpi_I$ of the induced normal connection obeys the linearized angular constraint
\begin{equation}
 \partial_v(v^2\varpi_I)=-\frac{t_j}{2}v^2\partial_v h_I .
 \label{eq:collar-angular}
\end{equation}
The regular data at $\partial\Sigma_i$ fixes the integration constant for the normal connection. In terms of $v$, the change in $\Theta_{(n)}$ in~\eqref{eq:collar-cross-focusing} contains the following positive contribution at second order in $h_I$,
\begin{equation}
 Q[h]=\sum_I\int_\epsilon^1\dd v\,
 \left(|\varpi_I|^2+\frac{v^2}{8}|\partial_v h_I|^2\right).
 \label{eq:collar-radiative-cost}
\end{equation}
The $|\varpi_I|^2$ term comes from transverse focusing~\cite{Hayward1994Dynamics}, while $v^2|\partial_vh_I|^2/8$ accounts for the change in affine slope on the original null boundary. With the corner data fixed, specifying $\int_{C_u}\Theta_{(n)}\,\dd A$ at $C_u=\partial\Sigma_i$ in~\eqref{eq:collar-cross-focusing} bounds $Q[h]$ by $Q_*$. 

The short wavelength behavior of $Q[h]$ already indicates how focusing limits the number of states. For one graviton of radial frequency $|k|$, the symplectic form~\eqref{bfgeo:cone-form-main:b} fixes $s|k|\,|h_I|^2$ at a given radial profile. By~\eqref{eq:collar-angular} the shear and normal connection terms in~\eqref{eq:collar-radiative-cost} then scale at large $j$ as
\begin{equation}
k^2|h_I|^2\propto\frac{|k|}{s},\ j(j+1)|h_I|^2\propto\frac{j(j+1)}{s|k|}.
\label{bfapp:radial-angular-focusing-scales}
\end{equation}
 Increasing the radial frequency thus suppresses the second contribution at the expense of the first.

To bound the number of radiative modes, we hold the data on $\Sigma_i$ and at $S$ fixed in the spherically symmetric causal diamond. The symplectic pairing of two perturbations $h,g$ in one harmonic depends on their first derivatives along a null generator. Since $g(\epsilon)=g(1)=0$, integration by parts in~\eqref{eq:collar-angular} expresses this pairing in terms of the normal connection $\varpi_I[h]$ induced by $h$. Cauchy Schwarz in~\eqref{eq:collar-radiative-cost} therefore bounds the pairing,
\begin{subequations}\label{eq:collar-symplectic-overlap}
\begin{align}
&\Omega(h,g)\propto \frac{\varrho_S^2}{G_N}\int_\epsilon^1 v^2[(\partial_vh)g-(\partial_vg)h]\,\dd v,\label{eq:collar-symplectic-overlap:a}\\
&\int_\epsilon^1v^2(\partial_v h)g\,\dd v=\frac{2}{t_j}\int_\epsilon^1 v^2\varpi_I[h]\partial_v g\,\dd v,\label{eq:collar-symplectic-overlap:b}\\
&|\Omega(h,g)|\le C_\epsilon\frac{\varrho_S^2}{G_N t_j}\sqrt{Q[h]Q[g]}.\label{eq:collar-symplectic-overlap:c}
\end{align}
\end{subequations}

The radial dependence constrains the same pairing. With $x=vh$ and $y=vg$ the integral in~\eqref{eq:collar-symplectic-overlap:a} becomes
\begin{equation}
\int_\epsilon^1\bigl[(\partial_v x)y-(\partial_v y)x\bigr]\,\dd v,
\label{eq:radial-symplectic-integral}
\end{equation}
where $\|\partial_v x\|_2^2\le C_\epsilon Q[h]$ by~\eqref{eq:collar-radiative-cost}. On the complement of the first $N-1$ radial sine modes we also have
\begin{equation}
 \|x\|_2\le \frac{(1-\epsilon)\|\partial_vx\|_2}{\pi N},\label{eq:radial-poincare-bound}
\end{equation}
and similarly for $y$, so the pairing there is suppressed by $1/N$.

Combining the angular and radial bounds, the two polarizations with $j\le J$ in the first $J$ radial modes account for $2J\sum_{j=2}^J(2j+1)=O(J^3)$ directions. On their complement the pairing is suppressed by order $1/J$ in the $Q$ norm, whereas quantization requires a symplectic area of order $\hbar$ per excitation. If a graviton in mode $i$ increases $Q[h]$ by $\nu_i$, these estimates bound the number of modes with $\nu_i\le\nu$. The radiative harmonics start at $j=2$, and a radially constant perturbation that vanishes at $\partial\Sigma_i$ must vanish everywhere. At fixed $s$, $d(s)$ includes only states whose graviton occupation numbers $n_i$ satisfy $\sum_i n_i\nu_i\le Q_*$. For every $\beta>0$ we therefore have
\begin{subequations}\label{eq:collar-occupation}
\begin{align}
&\sum_{i:\,\nu_i\leq\nu}1\leq C_\epsilon(s\nu)^3,\label{eq:collar-occupation:a}\\
&\log d(s)
 \le \beta Q_*-\sum_i\log\!\left(1-e^{-\beta\nu_i}\right)
 \ \lesssim_\epsilon\ \beta Q_*+ (s/\beta)^3+1 .\label{eq:collar-occupation:b}
\end{align}
\end{subequations}
For the logarithm in~\eqref{eq:collar-occupation:b}, we bound the sum using the number of modes, then choose $\beta$ to minimize the bound for $Q_*>0$,
\begin{subequations}\label{eq:collar-occupation-bound}
\begin{align}
&\sum_{m\ge1}m^{-1}\sum_i e^{-m\beta\nu_i}\lesssim_\epsilon(s/\beta)^3\sum_{m\ge1}m^{-4},\label{eq:collar-occupation-bound:a}\\
&\beta=(s^3/Q_*)^{1/4},\label{eq:collar-occupation-bound:b}\\
&\log d(s)\lesssim_\epsilon 1+(sQ_*)^{3/4}.\label{eq:collar-occupation-bound:c}
\end{align}
\end{subequations}
 On the expanding null cone in flat spacetime, $Q_*\lesssim_\epsilon 1+[\ell(\mathscr D)/\varrho_S-(1-\epsilon)]^{1/3}$, so $sQ_*$ stays bounded even as $\varrho_S$ tends to zero. On a weakly curved background the correction from $\Lambda$ is controlled by~\eqref{eq:collar-cross-focusing}.

We can now use the bound on $d(s)$ to compare the two ways of closing the thermal trace. We first sum over a finite dimensional subspace $\mathcal V$ of intermediate states, with $\widetilde\rho_{\mathcal V}=P_{\mathcal V}C^\dagger C P_{\mathcal V}$ and $Z_{\mathcal V}=\Tr_{\mathcal V}\widetilde\rho_{\mathcal V}$. Closing $n$ copies separately then yields $Z_{\mathcal V}^n$, whereas cyclic gluing carries a single state index through all $n$ copies. For $0\le C^\dagger C\le\widehat{1}$ and $n>1$, this bounds their ratio on $\mathcal V$. If the ratio has a positive limit as we include all intermediate states, then
\begin{equation}
 R_{\mathcal V}(n)=\frac{\Tr_{\mathcal V}(\widetilde\rho_{\mathcal V}^{\,n})}{Z_{\mathcal V}^n}\le Z_{\mathcal V}^{1-n},\ R_{\mathcal V}(n)\to R(n)>0,\ 
 \Tr(C^\dagger C)\le R(n)^{-1/(n-1)} .\label{bfent:cyclic-trace-bound}
\end{equation}
For $C=C_{\mathscr D}$, we have $R(n)=\Tr(\rho_{\mathscr D}^{\,n})$, so $\dim\mathcal H_{\mathscr D}<\infty$ (Section~\ref{bfent:subregion-boost-main}) implies $R(2)>0$ and $\lim_{n\downarrow1}R(n)=1$.

To evaluate both choices of gluing for $C_b$ at one loop, we sum over the $d(s)$ states allowed by focusing at each $s$. Their weights for one copy are the eigenvalues $k_a(s)\ge0$ of the reflected amplitude obtained from the Hessian~\eqref{semiapp:hessian-pullback}. In addition to $f_b$, each $k_a(s)$ includes a factor $e^s$ from a disk normal to $S$. Indeed, by the Gauss Bonnet theorem the Euclidean action of this disk tends to $-\hbar s$ as it shrinks, once its boundary term is included~\cite{CarlipTeitelboim1995}. With two copies glued cyclically there is only one such disk, so $k_a(s)^2$ enters with a factor $e^s/e^{2s}=e^{-s}$,
\begin{equation}
Z_1^{\rm reg}=\int\dd s\,\sum_{a=1}^{d(s)}k_a(s),\ Z_{2,\rm cyc}^{\rm reg}=\int\dd s\,e^{-s}\sum_{a=1}^{d(s)}k_a(s)^2.
\label{eq:collar-regular-integrals}
\end{equation}
The area integral runs over the regular family of saddles within $0\le s\le s_*$. Applying Cauchy Schwarz to $\exp(s/2)$ and $\exp(-s/2)k_a(s)$ under the integral and the sum over $a$ yields
\begin{equation}
 R_2^{\rm reg}=\frac{Z_{2,\rm cyc}^{\rm reg}}{(Z_1^{\rm reg})^2}
 \ \ge\ \frac{1}{\int\dd s\,e^s d(s)}>0 .
 \label{eq:collar-regular-return}
\end{equation}
Since~\eqref{eq:collar-occupation:b} bounds the number of graviton states throughout the allowed range of $s$, this lower bound stays positive when we include them all.

Even if shear is concentrated on a region $P$ of area $A_P$, the clock and the radiative symplectic form weight it by the same area element $\dd A=\varrho_S^2\bm\epsilon_0$. Positivity in~\eqref{bfgeo:capped-clock} and the bound on area in~\eqref{bfent:clock-area-main:d} imply
\begin{equation}
\int_P\dd A\int_\epsilon^1e^{\Xi_y}\,\dd v\le(1-\epsilon)\mathscr A_*(\mathscr D),\ \mathscr A_*(\mathscr D)=4G_N\hbar s_*.
\label{bfapp:focusing-on-a-patch}
\end{equation}
Along null generator $y$ we write the total variation of the conformal metric as
\begin{equation}
D(y)=\int_\epsilon^1\bigl[(\partial_v\bar q)^{AB}\partial_v\bar q_{AB}\bigr]^{1/2}\,\dd v.
\label{bfapp:conformal-variation-on-generator}
\end{equation}
 With $t=[1+\log(\mathscr A_*(\mathscr D)/A_P)]^{-1}$, Cauchy Schwarz along each null generator and H\"older's inequality on $P\times[\epsilon,1]$ yield
\begin{subequations}\label{eq:collar-patch-clock}
\begin{align}
&D(y)^2=8\left(\int_\epsilon^1(\partial_v\Xi_y/v)^{1/2}\dd v\right)^2
\le\frac{8[1-e^{-t\Xi_y(1)}]}{\epsilon t}\int_\epsilon^1e^{t\Xi_y}\dd v,\label{eq:collar-patch-clock:a}\\
&\int_P D(y)^2\dd A\le\frac{8(1-\epsilon)A_P}{\epsilon t}
\left(\frac{\mathscr A_*(\mathscr D)}{A_P}\right)^t
\le\frac{8e(1-\epsilon)A_P}{\epsilon}\left[1+\log\!\left(\frac{\mathscr A_*(\mathscr D)}{A_P}\right)\right].\label{eq:collar-patch-clock:b}
\end{align}
\end{subequations}

To see whether deformations confined to $P$ can enclose a symplectic area of order $\hbar$, we compare four radiative profiles that agree outside $P$ and at $\partial\Sigma_i$ and $S$. We vary only the conformal metric, keeping $\varrho_S(y)$, the null normals $(\ell^a,n^a)$ at $\partial\Sigma_i$ and at $S$, and the angular coordinates $y^A$ fixed. At fixed $(v,y)$, $z_t$ lies a fraction $t$ of the way along the geodesic between successive profiles $z_0$ and $z_1$, with distance measured by $\mathsf g_{\alpha\beta}$ in~\eqref{bfq:canonical-conformal-metric:b}. Since the profiles agree at $\partial\Sigma_i$ and distances between geodesics are convex in $t$,
\begin{equation}
|\partial_t z_t|\le \frac{D_0+D_1}{\sqrt{2}},\ |\partial_v z_t|\le(1-t)|\partial_v z_0|+t|\partial_v z_1|,
\end{equation}
where $D_i=D(y)$ on profile $i$. By~\eqref{bfgeo:capped-clock} the entire loop lies within the allowed range of the gravitational clock,
\begin{equation}
\bar u_*[z_t]\le(1-t)\bar u_*[z_0]+t\bar u_*[z_1]\le\ell(\mathscr D).
\label{bfapp:clock-along-radiative-loop}
\end{equation}
 Since the corner data is fixed, only the radiative term of~\eqref{bfgeo:cone-form-main:b} contributes to its symplectic area. The exact variations integrate to zero around the loop, and~\eqref{eq:collar-patch-clock:b} along each edge $(i,j)$ implies
\begin{equation}
 \left|\oint\Theta_{\mathrm{rad}}\right|{}\le\frac{1}{64\pi G_N}\sum_{(i,j)}\int_P(D_i+D_j)^2\,\dd A\lesssim_\epsilon
 \frac{A_P}{G_N}\left[1+\log\!\left(\frac{\mathscr A_*(\mathscr D)}{A_P}\right)\right].
 \label{eq:collar-patch-cell}
\end{equation}
At fixed $\ell(\mathscr D)$ the symplectic area enclosed by these four deformations falls below $\hbar$ for sufficiently small $A_P$, even if the polarization changes along each null generator.

If only $\mathscr A$ is fixed, the classical clock attains the lower bound in~\eqref{bfent:clock-area-main:c} for vanishing shear and $\varrho_S(y)=\sqrt{\mathscr A/(4\pi)}$.

\subsection{Fluctuations of the horizon area and angular momentum}
\label{app:replica-area}

A displacement of $S$ and a perturbation of the metric can compensate one another in the variation of the horizon area. In computing the density $W_\ell(s)$ in~\eqref{bfent:clock-smoothed-density} we integrate over both, fixing only the total area. Since the area generates boosts at $S$, we can impose this constraint in the path integral~\eqref{bfq:reduced-path-integral:d} by integrating over the boost~\eqref{eq:collar-boost-match}. For $n$ interiors glued cyclically at $S$ we have
\begin{equation}
 \int_{\mathbb R}\frac{\dd\beta}{2\pi}\,
 \Tr\!\left[P_\ell\exp\!\left(-n\hat s-\nu(\hat s-s)\right)\right]
 =\sum_{\bar u_{*,i}\le\ell}\exp(-ns_i)\,\delta(s-s_i),\ 
 \nu=1-n+i\beta .
 \label{app:replica-area-multiplier}
\end{equation}

To expand this integral about Schwarzschild, consider how the solution varies with the horizon radius $b$. At fixed dimensionless coordinates the metric and the classical clock reading in the area band~\eqref{bfent:one-loop-band-probability} satisfy
\begin{equation}
 g_{ab}=b^2\bar g_{ab},\ 
 s=\frac{\pi b^2}{\ell_P^2},\ 
 \bar u_*=(1-\epsilon)b=\ell\sqrt{\frac{s}{s_*}}\le\ell\sqrt{a_2}<\ell .
 \label{app:replica-radius-family}
\end{equation}
Under a rescaling of $b$, the area elements at $\partial\Sigma_1$ and $S$ scale in the same proportion~\eqref{bfgeo:diamond-boundary-conditions}, as do $\varrho$ and the affine parameter $u$~\eqref{bfgeo:capped-clock}. We can thus vary $b$ while satisfying the boundary conditions on the null cone and the bound on the clock. Matching the induced metrics at $S$ requires the same $\delta b$ on every copy.

Denote this family of solutions, including the embeddings of the boundaries, by $\phi_{n,s}$. The surface terms at $\Sigma_1$ in the variation of the action cancel between the two branches. At $S$ the excess angle multiplies the charge of a unit boost $\xi$~\eqref{bfgeo:charge-main},~\eqref{bfgeo:unit-corner-boost-charge}. Including the multiplier in~\eqref{app:replica-area-multiplier}, we obtain
\begin{equation}
 \left.\delta_\phi\big[I_n+\nu s[\phi]\big]\right|_{\phi_{n,s}}
 =\frac{2\pi(n-1)}{\hbar}\int_S
   \left(\delta\bm Q_\xi-\bm Q_{\delta\xi}-i_\xi\bm\theta\right)
   +\nu\,\delta s[\phi]=(n-1+\nu)\,\delta s[\phi].
 \label{app:replica-boundary-variation}
\end{equation}
The stationary value $\nu=1-n$ thus cancels the term proportional to the second variation of the area in~\eqref{bfent:replica-corner-hessian}, including the variation of the embedding in $\delta'\delta q_{AB}$~\eqref{semiapp:pullback-second},
\begin{equation}
 4G_N\hbar\,\delta'\delta s[\phi]=
 \int_S\!\left[\frac12 q^{AB}\delta'\delta q_{AB}
 +\frac14(q^{AB}\delta q_{AB})(q^{CD}\delta' q_{CD})
 -\frac12 q^{AC}q^{BD}\delta q_{AB}\delta' q_{CD}\right]\dd A .
 \label{app:replica-area-second-variation}
\end{equation}
Since $\nu$ is independent of $s$, all the solutions $\phi_{n,s}$ extremize the same action. Differentiating~\eqref{app:replica-boundary-variation} along this family, we find for the Hessian $H_n$ in~\eqref{semiapp:hessian-pullback} and any variation $h$,
\begin{equation}
 H_n(v_n,h)=\partial_s\!\left(\left.\delta_h\big[I_n+(1-n)s[\phi]\big]\right|_{\phi_{n,s}}\right)=0,\ 
 v_n=\partial_s\phi_{n,s},\ \delta_{v_n}s[\phi]=1 .
 \label{app:replica-area-tangent}
\end{equation}

To express the measure for this zero mode in terms of $\dd s$ we use the same inner product as in the graviton determinant~\cite{BastianelliBonezziMelis2022}. The metric perturbation along $v_n$ obeys de Donder gauge by~\eqref{app:replica-radius-family},
\begin{equation}
 h_{ab}=\partial_s g_{ab}=\frac{g_{ab}}{s},\ 
 \nabla^a h_{ab}-\frac12\nabla_b h^a{}_a=0 .
 \label{app:replica-radius-gauge}
\end{equation}
With $\dd V$ the positive volume element on each branch, the squared norm in Planck units is
\begin{equation}
 \mathcal J_n^2=\frac{1}{4\ell_P^4}\left|\int_{\mathscr D_n}
 \left[h_{ab}h^{ab}-\frac12(h^a{}_a)^2\right]\dd V\right|
 =\frac{\operatorname{Vol}(\mathscr D_n)}{\ell_P^4s^2}
 =\frac{1}{\pi^2}\int_{\mathscr D_n}\dd V_{\bar g}.
 \label{app:replica-area-norm}
\end{equation}
Thus the measure $\mathcal J_n\,\dd s$ in~\eqref{bfent:replica-radius-measure} is uniform in $s$ at fixed $\epsilon$. As in~\eqref{semiapp:bf-reduced-measure}, the product of the measure and the determinant factor is invariant under the change from reduced to metric variables.

A fluctuation orthogonal to $v_n$ can itself change the area of $S$. To include this change in the integral over $\beta$, we write
\begin{equation}
 \delta\phi=\frac{z}{\mathcal J_n}v_n+h_\perp,\ 
 \delta s[\phi]=\frac{z}{\mathcal J_n}+Lh_\perp,\ 
 Lh_\perp=\delta_{h_\perp}s[\phi].
 \label{app:replica-area-coordinates}
\end{equation}
By~\eqref{app:replica-area-tangent}, $z$ couples only to $\beta$ in the quadratic action. In the coordinates $(z,h_\perp,\beta)$ the Hessian of the constrained action and its determinant are
\begin{equation}
 \mathbb H_n=
 \begin{pmatrix}
  0 & 0 & i/\mathcal J_n\\
  0 & H_{n,\perp} & iL^{\mathsf T}\\
  i/\mathcal J_n & iL & 0
 \end{pmatrix},\ 
 \det\mathbb H_n=\mathcal J_n^{-2}\det H_{n,\perp}.
 \label{app:replica-area-determinant}
\end{equation}
We can see why the result is independent of $L$ by performing two of the integrals explicitly,
\begin{equation}
 \int_{\mathbb R}\dd z\int_{\mathbb R}\frac{\dd\beta}{2\pi}
 \exp\!\left[-i\beta\!\left(\frac{z}{\mathcal J_n}+Lh_\perp\right)\right]
 =\int_{\mathbb R}\dd z\,\delta\!\left(\frac{z}{\mathcal J_n}+Lh_\perp\right)=\mathcal J_n .
 \label{app:replica-area-integral}
\end{equation}

In the ratio~\eqref{bfent:replica-area-density} we perform this integral once in the numerator and $n$ times in the denominator. The same counting enters the determinants through the heat kernel of $H_n$,
\begin{equation}
 \exp(-\tau H_n)v_n=v_n\ \implies\
 k_{n,\perp}(\tau)-n k_{1,\perp}(\tau)=k_n(\tau)-n k_1(\tau)+n-1 .
 \label{app:replica-area-heat-trace}
\end{equation}
Lengths in~\eqref{app:replica-radius-family} scale with $b$, so in the proper time integral of Eq.~(17) of~\cite{BastianelliBonezziMelis2022} we set $t=\tau/b^2$. For the term $n-1$ in~\eqref{app:replica-area-heat-trace}, the part of the integral that depends on $b$ is
\begin{equation}
 \frac{n-1}{2}\int_{\ell_P^2/b^2}^{1}\frac{\dd t}{t}
 =(n-1)\log\!\left(\frac{b}{\ell_P}\right).
 \label{app:replica-area-proper-time}
\end{equation}
Together with~\eqref{app:replica-area-norm}, this yields the following term in $\log W_\ell(s)$,
\begin{equation}
 -\left.\partial_n\!\left[(n-1)\log\!\left(\frac{b}{\ell_P}\right)
   +\log\!\left(\frac{\mathcal J_n}{\mathcal J_1^n}\right)\right]\right|_{n=1}
 =-\frac12\log s+O(1).
 \label{app:replica-density-shift}
\end{equation}

We can also obtain the curvature term~\eqref{bfent:schwarzschild-trace-one-loop} from the Euler characteristic, since on the Schwarzschild background the squared Riemann curvature equals the Euler density. We continue the metric perturbations and ghosts smoothly across the disk normal to $S$. Outside a tubular neighborhood of $S$ both gluings have $n$ copies of the interior. The Gauss Bonnet boundary terms on the two sides of the tube cancel, and each tube retracts to $S$. We have one such tube in the cyclic trace and $n$ in the product of traces (Figure~\ref{fig:summary-replica}). Thus~\cite{FursaevPatrushevSolodukhin2013}
\begin{equation}
 32\pi^2\big[\chi(\mathscr D_n)-n\chi(\mathscr D_1)\big]
 =32\pi^2(1-n)\chi(S)=8\pi(1-n)\int_S R[q]\,\dd A .
 \label{bfent:schwarzschild-trace-one-loop:a}
\end{equation}

Both terms in~\eqref{bfent:clock-smoothed-density} thus follow at fixed $s$. To see how the width $\sigma_s$ of a preparation affects $Z_{b,1}$, we set $s=s_b+\sigma_s t$ in~\eqref{bfent:preparation-log-coefficient}. At one loop,
\begin{equation}
 \frac{Z_{b,1}}{\sqrt{2\pi}\,\sigma_s}=
 \int_{-s_b/\sigma_s}^{(s_*-s_b)/\sigma_s}
 \frac{\dd t}{\sqrt{2\pi}}\,\exp(-t^2/2)\, W_\ell(s_b+\sigma_s t).
 \label{app:replica-width-integral}
\end{equation}
We hold $s_b/s_*$ fixed between $a_1$ and $a_2$ and take $\sigma_s\propto s_b^\alpha$ for $0<\alpha<1$. Within the band, $s/s_b$ stays bounded above and away from zero. The Gaussian suppresses contributions from outside the band exponentially in $(s_b/\sigma_s)^2$. Thus by~\eqref{bfent:clock-smoothed-density}
\begin{equation}
 \log\!\left(\frac{Z_{b,1}}{\sqrt{2\pi}\,\sigma_s}\right)
 =\frac{167}{90}\log s_b+O(1).
 \label{app:replica-width-average}
\end{equation}

At fixed mass $M$ the area of the horizon decreases with $J^2$~\eqref{bfent:clock-kerr-reweight}. Consequently, on factoring $\exp(s_M)$ out of the density of states per unit $s_M$, the weight of a state relative to its Boltzmann weight in $Z(\ell)$ is
\begin{equation}
 \frac{\exp(-s_M)}{\exp(-s)}
 =\exp\!\left[-\frac{4\pi^2 J^2}{\hbar^2s}\right].
 \label{app:replica-spin-ratio}
\end{equation}
To average~\eqref{app:replica-spin-ratio} we need the fluctuations of the normal connection $\varpi_A$, which we held fixed in the integral over angular frames~\eqref{bfent:replica-rotation-integral}. Indeed, let $X^A$ generate a rotation of a spherically symmetric cut $C_v$. The angular momentum $J_X$ about this axis varies as~\eqref{bfgeo:eh-hamiltonian},~\eqref{bfgeo:transverse-data:h}
\begin{equation}
 \delta J_X(C_v)=-\frac{1}{16\pi G_N}\int_{C_v} X^A\delta\Pi_A\,\bm\epsilon_0
 =\frac{1}{8\pi G_N}\int_{C_v} X^A\delta\varpi_A\,\dd A .
 \label{app:replica-spin-charge}
\end{equation}
Since $X^A$ is divergence free, the angular gradients in the linearized null constraint~\eqref{bfgeo:shift-constraint:a} integrate to zero. Integrating the divergence of the shear by parts and using the Killing equation, we obtain
\begin{equation}
 \partial_v\delta J_X(C_v)
 =\frac{1}{8\pi G_N}\int_{C_v}\delta\sigma^{AB}\mathring\nabla_{(A}X_{B)}\,\dd A=0
 \implies \delta J_X(S)=\delta J_X(\partial\Sigma_1).
 \label{bfent:clock-spin-conservation}
\end{equation}

Thus only the three $j=1$ harmonics of $\varpi_A$ along the rotational Killing fields on $\partial\Sigma_1$ enter the variance of $J$ at this order. To determine its dependence on $s$, let $X_i$ generate rotations about three orthogonal axes and write $J_i=J_{X_i}$. In terms of the perturbation $\bar h_{ab}$ of $\bar g_{ab}$ in~\eqref{app:replica-radius-family}, the quadratic action~\eqref{semiapp:metric-hessian} and the charges~\eqref{app:replica-spin-charge} take the form
\begin{equation}
 I_1^{(2)}=\frac{s}{2}\langle\bar h,\bar H_\perp\bar h\rangle,\ 
 \frac{\delta J_i}{\hbar}=\frac{s}{8\pi^2}\int_S X_i^A\delta\varpi_A\,\dd\Omega=s\,B_i\bar h .
 \label{app:replica-spin-scaling}
\end{equation}
Here $\bar H_\perp$ includes the boundary terms on the cap and $B_i$ acts linearly on $\bar h$. Both are independent of $b$ for the family~\eqref{app:replica-radius-family}. Contracting the two variations of the charge with the graviton propagator at fixed $s$, we find
\begin{equation}
 \frac{\langle\delta J_i\,\delta J_j\rangle_{\ell,s}}{\hbar^2 s}
 =B_i\bar H_\perp^{-1}B_j^{\mathsf T}\equiv \mathsf C_{ij}.
 \label{app:replica-spin-covariance}
\end{equation}
Equation~\eqref{bfent:clock-spin-scale} thus follows provided the quadratic action for fluctuations of these three components of the angular momentum on the cap is nondegenerate.

Fixing the mass also changes the argument of the delta function in~\eqref{app:replica-area-multiplier}. Since the angular momentum vanishes along the entire Schwarzschild family,~\eqref{bfent:clock-kerr-reweight} implies
\begin{equation}
 \delta_{v_1}J_i=0,\ 
 \delta s_M=\delta s,\ 
 \delta^2(s_M-s)=\frac{8\pi^2}{\hbar^2s}\sum_{i=1}^3(\delta J_i)^2 .
 \label{app:replica-spin-hessian}
\end{equation}
At $n=1$ the multiplier vanishes at the saddle, so only the first variation of the constraint enters~\eqref{app:replica-area-determinant}. The integral over $z$ and $\beta$ is therefore the same at fixed $s_M$ and at fixed $s$. In the remaining integral we include the factor~\eqref{app:replica-spin-ratio}. With $B$ the matrix whose rows are $B_i$, the ratio of the two densities at one loop is
\begin{equation}
 \frac{\exp(-s_M)}{W_\ell(s_M)}\frac{\dd N_\ell}{\dd s_M}
 =\left[\frac{\det\bar H_\perp}{\det(\bar H_\perp+8\pi^2 B^{\mathsf T}B)}\right]^{1/2}
 =\frac{1}{\sqrt{\det_{\mathbb R^3}(\mathbf 1+8\pi^2\mathsf C)}} .
 \label{app:replica-spin-determinant}
\end{equation}
Since $\mathsf C$ is independent of $b$, the change in the logarithmic coefficient in~\eqref{bfent:band-density-per-mass} comes entirely from the Jacobian $\dd s_M/\dd M$.

\section{Axiomatic formulation of Lorentzian quantum gravity}
\label{sec:lorentzian-algebras}
\label{app:lorentzian-algebras}

The overlap of a connected preparation with a disconnected one was determined entirely by closing their boundaries in Section~\ref{topology:relation}. More generally, amplitudes on closed spacetimes can tell us which states propagate across a cut and with what multiplicities they enter a trace. We would like to recover that Hilbert space and its operator algebra from the amplitudes themselves.

In Euclidean gravity these properties follow from axioms for amplitudes on closed manifolds~\cite{ColafranceschiDongMarolfWang2024}. Reflection positivity ensures a positive inner product after quotienting by null states, while the full set of axioms determines type~I algebras of boundary observables with positive integer weights in their traces. These traces identify entropy computed by the replica trick with von Neumann entropy at finite coupling, without assuming a dual field theory.

\subsection{Conditions on amplitudes}
\label{sec:lorentzian-assumptions}

\noindent Gluing a Lorentzian preparation to its time reflection pairs the time evolution operator with its adjoint. We denote a cut with compact spatial sections and specified boundary sources $j$ by $\Sigma$. Reversing the orientation of a cobordism $\mathscr M$ and conjugating $j$ defines $\mathscr M^\dagger$. Once every boundary has been closed, $\zeta_L(\mathscr M)$ is its unnormalized amplitude.

\begin{enumerate}
\item \emph{Diffeomorphism invariance.} We require invariance under diffeomorphisms that preserve the boundary conditions and under changes of intermediate cuts. For three adjacent subregions we may glue either neighboring pair first, and in a closed configuration we may start gluing at any cut. We allow adjoints, disjoint unions and every pairing of labelled boundaries with matching data, including pairings between different components.

\item \emph{Reality and positivity.} Let $|\mathscr M_i\rangle_\Sigma$ denote the states prepared by a finite collection of cobordisms with matching boundaries in the domain of condition~\ref{ax:lorentzian-finiteness}. For every allowed pairing of any number of boundary components and all coefficients $c_i$ we require
\begin{subequations}
\label{eq:lorentzian-positivity}
\begin{align}
&\zeta_L(\mathscr M^\dagger)=\zeta_L(\mathscr M)^*,\\
&\langle \mathscr M_1|\mathscr M_2\rangle_{\Sigma}=\zeta_L(\mathscr M_1^\dagger\cup_{\Sigma} \mathscr M_2),\\
&\sum_{i,j}c_i^*c_j \langle \mathscr M_i|\mathscr M_j\rangle_{\Sigma}\geq0.
\end{align}
\end{subequations}

\item \emph{Factorization.} For disjoint closed configurations, we require
\begin{equation}
\zeta_L(\mathscr M\sqcup \mathscr N)=\zeta_L(\mathscr M)\zeta_L(\mathscr N),\;
\zeta_L(\varnothing)=1.
\label{eq:lorentzian-factorization}
\end{equation}

\item \emph{Continuous unitary time evolution.} Real time evolution $U$ from $\Sigma$ to $\Sigma'$ has inverse $U^\dagger$ and preserves the inner product.\footnote{In algebraic QFT one can likewise describe time evolution in terms of Lorentzian cobordisms between Cauchy surfaces~\cite{Bunk:2023ley}. For an additive theory satisfying the time slice axiom one can also encode locality by allowing several causally disjoint incoming regions~\cite{Bunk:2025siq} or by using time ordered products in a factorization algebra~\cite{Benini:2019ujs}.} At fixed $\Sigma$ the translations form a one parameter unitary group with continuous matrix elements between preparations,
\begin{equation}
T_{\Sigma}(t+s)=T_{\Sigma}(t)T_{\Sigma}(s),\;
T_{\Sigma}(t)^\dagger=T_{\Sigma}(-t),\;
T_{\Sigma}(0)=\widehat{1}.
\label{eq:lorentzian-translations}
\end{equation}

\item \emph{Spectrum and finiteness.}\label{ax:lorentzian-finiteness} To restrict boundary energies to bounded intervals we superpose time translations using the Fourier transforms of smooth functions $g(E)$ of compact support,
\begin{subequations}
\label{eq:lorentzian-smearing}
\begin{align}
&F_{\Sigma,g}=\int_{\mathbb R}\dd t\,\check g(t)T_{\Sigma}(t),\\
&\check g(t)=\frac{1}{2\pi\hbar}\int_{\mathbb R}\dd E\,g(E)e^{iEt/\hbar}.
\end{align}
\end{subequations}
where $t$ and $E$ denote time and energy for time translations, or rapidity and boost charge at a corner. We evaluate these integrals in matrix elements between preparations of finite norm. Gluing and conjugation respect superposition.

For a cobordism $\mathscr M$ we allow $F_{\Sigma,g}\mathscr M F_{\Sigma,h}$, their finite linear combinations, products and adjoints. Every compatible pairing that closes their boundaries must have finite amplitude. Denoting the identification of the last two matching boundaries by $\operatorname{cl}_\Sigma$, we require
\begin{equation}
\zeta_L\!\left(\operatorname{cl}_{\Sigma}(F_{\Sigma,g}^\dagger F_{\Sigma,h})\right)
=\int g^*(E)h(E)\,\dd\nu_{\Sigma}(E).
\label{eq:lorentzian-spectral-measure}
\end{equation}
where $\nu_\Sigma$ is a positive measure supported in $[E_{0,\Sigma},\infty)$.
\end{enumerate}

\begingroup

\subsection{Equivalence with Euclidean formulation}
\label{app:lorentzian-euclidean-equivalence}

To continue a boundary time translation to a Euclidean cylinder, we first construct the Hilbert space on which it acts. For two matching boundaries $\Sigma$, let $\mathcal A_0$ be the span of the preparations $F_{\Sigma,g}\mathscr M F_{\Sigma,h}$. Closing the boundaries of $a\in\mathcal A_0$ fixes its trace, and gluing $b$ to the time reflection of $a$ yields their inner product,
\begin{equation}
\tau_0(a)=\zeta_L(\operatorname{cl}_{\Sigma}a),\ \langle a|b\rangle_{\mathrm L}=\tau_0(a^\dagger b).
\label{eq:lorentzian-source-pairing}
\end{equation}
We quotient by null states and complete to obtain $\mathcal K_\Sigma$. Gluing on the two open boundaries acts by $L(a)|b\rangle=|ab\rangle$ and $R(a)|b\rangle=|ba\rangle$. Moving the cut yields $\tau_0(ab)=\tau_0(ba)$, so reversal is isometric. Exchanging the left boundaries of $a\sqcup b$ and applying Cauchy Schwarz yields $\|L(a)\|\le\|a\|_2$~\cite{ColafranceschiDongMarolfWang2024}. For $e_m=F_{\Sigma,\chi_m}$ choose $0\le\chi_m\le1$ smooth of compact support, $\chi_{m+1}=1$ on $\supp\chi_m$, and $\chi_m(E)\to1$. Composing the smearings in~\eqref{eq:lorentzian-smearing} yields $e_m=e_mT_\Sigma(0)e_{m+1}\in\mathcal A_0$ and $e_ma=ae_m=a$ for all sufficiently large $m$. The preparations thus form a Hilbert algebra, whose trace extends to a faithful normal semifinite trace $\tau_{\mathrm L}$ on the left algebra~\cite{Takesaki1970}.

The boundary translations are strongly continuous by unitarity and continuity of their matrix elements. Smearing their self adjoint generators yields $L(F_{\Sigma,g})=g(\widehat H_L)$ and $R(F_{\Sigma,g})=g(\widehat H_R)$. A smearing supported below $E_{0,\Sigma}$ has zero norm by~\eqref{eq:lorentzian-spectral-measure}, so both generators are bounded below. With $Q_m^{L,R}=\chi_m(\widehat H_{L,R})\uparrow\widehat 1$, normality and~\eqref{eq:lorentzian-spectral-measure} yield, for $A\ge0$ and any closed bounded interval $I$,
\begin{subequations}
\begin{align}
&\tau_{\mathrm L}(A)=\sup_m\langle e_m|A|e_m\rangle_{\mathrm L},\label{eq:lorentzian-trace-definition}\\
&\tau_{\mathrm L}(\mathbf1_I(\widehat H_L))=\nu_\Sigma(I)<\infty.\label{eq:lorentzian-thermal-trace}
\end{align}
\end{subequations}

Continuing the left boundary translation to $t=-is$ inserts a Euclidean cylinder between the two preparations. Cutting it in half yields reflection positivity,
\begin{equation}
C(s)=e^{-s\widehat H_L/\hbar},\ \langle\psi|C(s)|\psi\rangle_{\mathrm L}=\|C(s/2)|\psi\rangle\|^2.
\label{eq:equivalence-filtered-cylinder}
\end{equation}
The lower bound $\widehat H_L\ge E_{0,\Sigma}$ ensures norm continuity for $s>0$, with $\|C(s)\|\le e^{-sE_{0,\Sigma}/\hbar}$. As $s\downarrow0$, strong convergence to the identity recovers $\langle a|b\rangle_{\mathrm L}$.

For a bounded insertion $A$ and smooth $g$ of compact support, the spectral bound~\eqref{eq:lorentzian-thermal-trace} lets us close the cylinder,
\begin{equation}
\big|\tau_{\mathrm L}(A g(\widehat H_L)C(s))\big|
 \le \|A\|\,\|g\|_\infty e^{-s E_{0,\Sigma}/\hbar}
       \tau_{\mathrm L}(\mathbf1_{\supp g}(\widehat H_L))<\infty.
\label{eq:equivalence-filtered-cylinder:c}
\end{equation}
Without an energy restriction, $|\tau_{\mathrm L}(AC(s))|\le\|A\|\tau_{\mathrm L}(C(s))$ suffices whenever the cylinder has finite trace, as at a reflecting AdS boundary in~\eqref{bfent:ads-joint-bound}. Reversal conjugates the amplitude and disjoint closures multiply. At fixed insertions we require all decompositions of a given closed configuration to be related by the diffeomorphisms and changes of cut of axiom~1.

Conversely, for cylinder amplitudes satisfying the Euclidean axioms~\cite{ColafranceschiDongMarolfWang2024}, gluing and reflection yield
\begin{equation}
C(s+t)=C(s)C(t),\ C(s)=C(s/2)^\dagger C(s/2)\ge0.
\label{bfapp:euclidean-cylinder-composition}
\end{equation}
If $C(s_0)\psi=0$, composition gives $C(s)^m\psi=0$ for $ms>s_0$, and positivity implies $C(s)\psi=0$ for all $s>0$. The ranges of short cylinders span densely because every preparation admits one at its boundary, so $\psi=0$. Subdivision and continuity then determine
\begin{equation}
 C(s)=C(s_0)^{s/s_0},\ \widehat H=-\frac{\hbar}{s_0}\log C(s_0).\label{eq:euclidean-cylinder-generator}
\end{equation}
The Hamiltonian is self adjoint, bounded below and independent of $s_0$.

By~\cite{ColafranceschiDongMarolfWang2024} the Euclidean boundary algebra is a sum of type~I factors $\mathcal B(\mathcal H_\mu)$ with positive integer weights $m(\mu)$ in the trace. In this trace each energy $E\le R$ carries weight at least $e^{-s_0R/\hbar}$. Closing the cylinder thus bounds the spectral weight in condition~\ref{ax:lorentzian-finiteness},
\begin{equation}
 \tau_{\mathrm E}(\mathbf1_{(-\infty,R]}(\widehat H))
 \le e^{s_0R/\hbar}\tau_{\mathrm E}(C(s_0))<\infty ,\ 
 \nu_\Sigma=\sum_{\mu,j}m(\mu)\delta_{E_{\mu j}}.\label{eq:equivalence-energy-measure}
\end{equation}
Since $m(\mu)\ge1$, the spectral projection below $R$ has finite rank across all sectors. As $R$ increases these subspaces span the $\mathcal H_\mu$. Smearing multiplies each eigenvector by $g(E_{\mu j})$, with weight $m(\mu)$ in~\eqref{eq:lorentzian-spectral-measure}. With compact energy support each gluing is a finite contraction of the indices fixed by the pairing. Euclidean reflection ensures positivity and conjugation, and disjoint components factorize.

For $a,b\in\mathcal A_0$, let $\iota(a),\iota(b)$ denote their Euclidean preparations. Choose smooth $g$ of compact support equal to one on their left boundary energy intervals. Smearing with $g_s(E)=g(E)e^{-sE/\hbar}$ inserts the cylinder,
\begin{equation}
 L(F_{g_s})=g(\widehat H_L)C_{\mathrm L}(s),\ 
 \langle\iota(a)|C_{\mathrm E}(s)|\iota(b)\rangle_{\mathrm E}=\langle a|L(F_{g_s})|b\rangle_{\mathrm L}.\label{eq:equivalence-cylinder-insertion}
\end{equation}
Taking $s\downarrow0$ equates the two inner products, since $C_{\mathrm E}(s)\to\widehat{1}$ strongly and $g_s\to g$ uniformly. Thus $V|a\rangle_{\mathrm L}=|\iota(a)\rangle_{\mathrm E}$ extends by completion to an isometry.

For any finite norm Euclidean preparation built from these insertions and cylinders of positive length, $Q_m^LQ_m^R|\mathscr M\rangle_{\mathrm E}\to|\mathscr M\rangle_{\mathrm E}$ in norm. At fixed $m$, replace each internal cylinder in turn by the smearing with $\chi_n(E)e^{-sE/\hbar}$. Uniform convergence for $E\ge E_{0,\Sigma}$ approximates the restricted preparation in norm by $\iota(a)$ with $a\in\mathcal A_0$. Thus $V$ maps onto this Euclidean sector. Gluing identifies $\iota(ab)=\iota(a)\iota(b)$, so $V$ intertwines both boundary algebras and their cylinders. Smearing also identifies $V|e_m^{\mathrm L}\rangle=|e_m^{\mathrm E}\rangle$, hence for $A_{\mathrm E}=VAV^\dagger\ge0$ we have
\begin{subequations}\label{eq:equivalence-preserved-trace}
\begin{align}
&V|a\rangle_{\mathrm L}=|\iota(a)\rangle_{\mathrm E},\label{eq:equivalence-preserved-trace:a}\\
&VC_{\mathrm L}(s)V^\dagger=C_{\mathrm E}(s),\label{eq:equivalence-preserved-trace:b}\\
&\tau_{\mathrm L}(A)=\sup_m\langle e_m^{\mathrm E}|A_{\mathrm E}|e_m^{\mathrm E}\rangle
=\sup_m\tau_{\mathrm E}(A_{\mathrm E}^{1/2}\chi_m(\widehat H)^2 A_{\mathrm E}^{1/2})
=\tau_{\mathrm E}(A_{\mathrm E})\label{eq:equivalence-preserved-trace:c}
\end{align}
\end{subequations}
by normality, including for infinite trace. The same trace therefore determines the positive integer sector weights and identifies the normalized replica moments with those of an ordinary density operator, as in the Euclidean theory~\cite{ColafranceschiDongMarolfWang2024}.

\endgroup

\subsection{Realization in first order gravity}
\label{sec:lorentzian-construction}

In first order gravity we can evaluate the same amplitude by closing a cobordism in the path integral or by summing over the states already constructed on the cut. For either $\mathcal H_\Sigma=\mathcal H_{\mathscr D}$ or $\mathcal H_\Sigma=\mathcal H_\partial$ of Section~\ref{bfent:subregion-boost-main}, this comparison lets us verify the axioms directly. If $\widehat a$ represents a linear combination $a$ of cobordisms, closing $a$ sums its diagonal matrix elements. Gluing $b$ to the reflection of $a$ then computes their overlap,
\begin{subequations}
\label{eq:lorentzian-physical-closing}
\begin{align}
&\zeta_L(\operatorname{cl}_{\Sigma} a)=\operatorname{Tr}_{\mathcal H_{\Sigma}}\widehat a,\\
&\langle a|b\rangle_{\Sigma}=\operatorname{Tr}_{\mathcal H_{\Sigma}}(\widehat a^\dagger\widehat b).
\end{align}
\end{subequations}
The boundary identifications in~\eqref{bfq:ward-central} act by unitaries $V$ whose factors $V^{-1}V$ cancel at a cut, including the phases from the boundary action, and for a causal diamond, the map $Y$ in~\eqref{bfgeo:geometric-matching} identifies the points on the two copies of $S$ that label the null generators. Changing variables by $Y$ in~\eqref{bfent:clock-form} and conjugating each operator by $V$, we thus obtain
\begin{equation}
 \widehat{\bar u}_*'=V\widehat{\bar u}_*V^\dagger,\ P_{\mathscr D}'=VP_{\mathscr D}V^\dagger.\label{eq:lorentzian-clock-conjugation}
\end{equation}
Therefore, $V$ identifies the allowed subspaces and maps null states to null states. Relabeling several boundaries permutes their matching conditions and clock projectors together, preserving the pairing in~\eqref{eq:lorentzian-physical-closing}. 

The trace in~\eqref{eq:lorentzian-physical-closing} is positive on $\widehat a^\dagger\widehat a$, respects conjugation and factorizes on tensor products. Unitary time evolution preserves this inner product, and self adjoint generators yield strongly continuous unitary groups.

Finally, in~\eqref{bfent:trace-bound-main}, we proved $d_{\mathscr D}=\dim\mathcal H_{\mathscr D}<\infty$ for a causal diamond with a fixed range of the clock. We can therefore choose a smooth function $g$ of compact support equal to unity on the bounded spectrum of the boost generator:
\begin{equation}
F_{\Sigma,g}=g(\widehat{\mathscr K})=\widehat{1},\ \left|\operatorname{Tr}_{\mathcal H_{\mathscr D}}\widehat a\right|\le d_{\mathscr D}\|\widehat a\|.
\label{eq:lorentzian-diamond-finiteness}
\end{equation}
Every closed contraction of finitely many allowed operators is consequently finite, as required by condition~\ref{ax:lorentzian-finiteness} on the full physical Hilbert space.

On the AdS boundary sector, the thermal trace bounds the number of states in each bounded energy interval by~\eqref{bfent:ads-energy-count}. It follows that $g(\widehat H)$ has finite rank for every smooth $g$ of compact support, and all compatible contractions of $g(\widehat H)\widehat a\,h(\widehat H)$ are finite. This lets us approximate any state of finite norm in~\eqref{eq:lorentzian-physical-closing} by states prepared with boundary sources. Indeed, $\|\widehat a\|_2^2=\operatorname{Tr}(\widehat a^\dagger\widehat a)$ is a sum of squared matrix elements. When it converges, $\|P_\Lambda\widehat aP_{\Lambda'}-\widehat a\|_2\to0$ as $\Lambda,\Lambda'\to\infty$, where $P_\Lambda=\mathbf1_{[E_0,E_0+\Lambda]}(\widehat H)$. Since the energy levels are isolated, we can choose a smooth $g$ of compact support with $g(\widehat H)=P_\Lambda$. The time smearings in~\eqref{eq:lorentzian-smearing} thus restrict the approximation to finitely many matrix elements in these two intervals.

In Section~\ref{bfent:subregion-boost-main} we also showed that boundary sources generate $\mathcal B(\mathcal H_\partial)$, so we can choose uniformly bounded $\widehat a_i$ in the algebra of boundary insertions with $\widehat a_i|\psi\rangle\to\widehat a|\psi\rangle$ for every $|\psi\rangle$. Between the selected intervals this convergence controls the error in the physical inner product,
\begin{equation}
 \|P_\Lambda(\widehat a_i-\widehat a)P_{\Lambda'}\|_2^2
 =\sum_{\substack{E_m\le E_0+\Lambda\\ E_n\le E_0+\Lambda'}}
 |\langle m|(\widehat a_i-\widehat a)|n\rangle|^2
 \xrightarrow{i\to\infty}0.
 \label{eq:lorentzian-construction-source-density}
\end{equation}
Gluing a bounded insertion $\widehat b$ on either side preserves finite norm even when $\widehat b$ changes area, since $\max\{\|\widehat b\widehat a\|_2,\|\widehat a\widehat b\|_2\}\le\|\widehat b\|\,\|\widehat a\|_2$. This verifies condition~\ref{ax:lorentzian-finiteness} even when $\mathcal H_\partial$ is infinite dimensional.

\section{Gravitational amplitudes and extensions of the fixed area state}
\label{app:dressed-graviton-area}

The amplitude for emission and reabsorption of a radiative excitation across \(S\) depends on the extension \(\omega\) chosen in Section~\ref{bfq:state-extension}, since the corner area can change between the two insertions. On the uncut preparation \(C_b\) of~\eqref{bfent:shear-preparation} we compare the sum over intermediate areas with the two point function obtained by continuing a freely falling graviton mode through \(S\) to the reflected source. We then use the null constraints to estimate how short wavelength virtual gravitons contribute to this sum. Since the finiteness argument in Section~\ref{bfent:subregion-boost-main} uses \(C_b\) after projection onto the allowed clock readings, we also bound the change in the overlap of the one graviton wave packet with its boosted copy by the probability of an excluded reading.

\subsection{A radiative insertion on the incoming state}
\label{app:dressed-graviton-open}

In Section~\ref{bfq:state-extension} we chose an incoming wavefunction that depends on both the radiation and the relative boost angle at \(S\). We first determine how a source specified in the right sided null frame acts on its radiative modes. On a shear free background of area \(\mathscr A=4\pi b^2\) the source couples to a graviton through \(\sigma^{(1)}_{AB}=\partial_u h_{AB}/2\). We measure the affine parameter \(u\) from \(S\) and center the real valued smearing \(p_R\) at \(u=d>0\) with width \(\ell_P\ll\Delta u\ll d\ll b\), as in~\eqref{bfent:shear-scale-hierarchy}. Holding \(f_\perp^{AB}\) fixed in the local Lorentz frame parallel transported along each null generator, we vary the reduced canonical data \(\phi^I=(q_j,p_j)\) to obtain
\begin{equation}
 \lambda_I[p_R]=\delta_I\sigma_\perp[p_R]
 =\frac12\int_0^\infty \dd u\,p_R(u)
     \int_S \dd A\, f_\perp^{AB}\,\partial_u(\delta_I h_{AB})(u,y).
 \label{app:gravarea-source}
\end{equation}
Here \(\delta_I h_{AB}\) is the linearized solution for a unit variation of \(\phi^I\). Boosting the right sided local Lorentz frame through rapidity \(\gamma\) while holding the left one fixed replaces \(p_R(u)\) by \(p_{R,\gamma}(u)=p_R(e^{-\gamma}u)\). Before we glue the two sides we therefore keep the uniform relative boost angle conjugate to \(\widehat{\mathscr A}/(8\pi G_N)\) in~\eqref{bfgeo:two-sided-boost-potential} as an argument of the incoming wavefunction. For changes in dimensionless area small compared with \(s_b=\mathscr A/(4G_N\hbar)\), we evaluate the normalization of its canonical modes at \(s_b\). Centering the radiative modes on the shear free solution, we represent a source of strength \(j\) by the Weyl displacement~\eqref{bfads:rep-weyl}. With the corner phase~\eqref{eq:collar-boost-match} factored out, the source and boost act on the remaining wavefunction \(\chi(\gamma)\) by
\begin{subequations}\label{app:gravarea-operators}
\begin{align}
 &\big(e^{ij\widehat\sigma_\perp[p_R]}\chi\big)(\gamma)
     =D\big(j\lambda[p_{R,\gamma}]\big)\chi(\gamma),\label{app:gravarea-source-action}\\
 &\big(B_\eta\chi\big)(\gamma)
     =e^{i\eta s_b/(2\pi)}\chi(\gamma+\eta),\
       B_\eta=e^{i\eta\hat s/(2\pi)}.\label{app:gravarea-boost-action}
\end{align}
\end{subequations}
Conjugation by \(B_\eta\) thus boosts the smearing of the source. Taking Fock space inner products at each \(\gamma\) and integrating over \(\dd\gamma\) defines a positive state on products of these unitaries. To check its restriction to the incoming observables, we take \(\chi(\gamma)=g(\gamma)|\Omega_{\mathscr A}\rangle\) with \(g(\gamma)\propto\exp[-\gamma^2/(4(\Delta\gamma)^2)]\) normalized in \(\dd\gamma\). The area preserving displacement \(D(\lambda)\) acts on \(|\Omega_{\mathscr A}\rangle\) independently of \(\gamma\), so the boost angle and the radiative modes contribute separately,
\begin{equation}
 \langle\chi|D(\lambda)B_\eta|\chi\rangle
 =e^{i\eta s_b/(2\pi)-\eta^2/[8(\Delta\gamma)^2]}
     \langle\Omega_{\mathscr A}|D(\lambda)|\Omega_{\mathscr A}\rangle .
 \label{app:gravarea-restriction}
\end{equation}
The incoming fluctuations are therefore those of~\eqref{bfq:fixed-area-gaussian}, and the boost phase approaches that at fixed area for \(|\eta|\ll\Delta\gamma\). On a finite null cone we choose \(\Delta\gamma\) small compared with both \(\log(b/d)\) and \(\log(\Delta u/\ell_P)\), so that the part of the wavefunction of appreciable weight keeps the insertion near \(S\) and its width large compared with \(\ell_P\). The resulting spread \(\Delta s=\pi/\Delta\gamma\) sets the area resolution of this approximation.

\subsection{Reflection across the corner}
\label{app:dressed-graviton-reflection}

The width \(\Delta\gamma\) fixes how finely the incoming state resolves area. To determine the weights at intermediate areas, we now pair the source with its reflection and vary their relative boost. We compute the unnormalized two point amplitude on the uncut \(C_b\) of~\eqref{bfent:shear-preparation}, centering its area weight on the smooth solution of Appendix~\ref{app:collar-return}. A freely falling observer uses the same affine parameter on a null generator through \(S\) to define positive frequency for both sources. At affine frequency \(k\omega_1\) the metric variance in the incoming Gaussian wavefunction used in~\eqref{bfq:translated-pair-state} is \(1/(2s_b k)\). Since \(\sigma^{(1)}_{AB}=\partial_u h_{AB}/2\), we differentiate the metric two point function once at each insertion along the null generator. The two insertions multiply the metric variance by \((k\omega_1)^2/4\), yielding a spectral weight proportional to \(k\) as in~\eqref{bfapp:shear-singularity} at short wavelengths. With \(c_\perp\) fixed by the transverse smearing, the graviton created on side \(a=L,R\) by \(p_a\) has wavefunction \(\psi_a(\omega)\). A boost on the right rescales its affine frequency,
\begin{equation}
 \psi_a(\omega)=\sqrt{c_\perp\omega}\int_{\mathbb R}e^{i\omega u}p_a(u)\,\dd u,\
 \psi_{R,\eta}(\omega)=e^{\eta/2}\psi_R(e^\eta\omega),\ \omega>0.
 \label{app:gravarea-affine}
\end{equation}
In \(\log(\omega/\omega_1)\) this boost is a translation. Fourier transforming in that variable resolves the one graviton wave packet by boost frequency \(\nu\), which fixes the change in area by~\eqref{bfq:source-boost-phase:a},
\begin{equation}
 m_a(\nu)=\frac{1}{\sqrt{2\pi}}\int_0^\infty\frac{\dd\omega}{\sqrt{\omega}}
      \left(\frac{\omega}{\omega_1}\right)^{-i\nu}\psi_a(\omega),\
 \nu=\frac{s'-s_b}{2\pi}.
 \label{app:gravarea-mellin}
\end{equation}
Indeed \(m_{R,\eta}(\nu)=e^{i\nu\eta}m_R(\nu)\), and the transform preserves the norm. More explicitly, the intermediate area in a pair of weak right sided insertions has characteristic function
\begin{equation}
 \frac{\langle\chi|\widehat\sigma_\perp[p_R] B_\eta
                    \widehat\sigma_\perp[p_R]|\chi\rangle}
      {\langle\chi|\widehat\sigma_\perp[p_R]^2|\chi\rangle}
 \simeq
 e^{i\eta s_b/(2\pi)-\eta^2/[8(\Delta\gamma)^2]}
 \frac{\int_{\mathbb R}e^{i\nu\eta}|m_R(\nu)|^2\,\dd\nu}
      {\int_{\mathbb R}|m_R(\nu)|^2\,\dd\nu}.
 \label{app:gravarea-intermediate}
\end{equation}
In deriving~\eqref{app:gravarea-intermediate}, the two smearings in~\eqref{app:gravarea-source-action} have rapidities \(\gamma\) and \(\gamma+\eta\). Their common dilation drops out of the scalar product in~\eqref{app:gravarea-affine}. Thus \(|m_R(\nu)|^2\,\dd\nu\) determines how much of the one graviton norm the source assigns to each change in area, to the resolution of~\eqref{app:gravarea-restriction}.

To compare this distribution with the uncut two point function, we reflect the smearing about \(S\), so \(p_L(-u)=p_R(u)\) for \(u>0\). The Fourier integrals in~\eqref{app:gravarea-mellin} then have opposite phases. Inserting \(\exp(-\tau\omega)\) with \(\tau>0\) yields a factor \(\Gamma(1-i\nu)(\tau\mp iu)^{-1+i\nu}\) on the right and left, respectively. Taking \(\tau\) to zero, we obtain
\begin{subequations}\label{app:gravarea-reflection}
\begin{align}
 &m_R(\nu)=i\sqrt{\frac{c_\perp}{2\pi}}\,
   e^{\pi\nu/2}\Gamma(1-i\nu)
      \int_0^\infty\frac{\dd u}{u}\,(\omega_1 u)^{i\nu}p_R(u),\label{app:gravarea-right}\\
 &m_L(\nu)=-e^{-\pi\nu}m_R(\nu),\
      e^{-\pi\nu}=\frac{\exp(-s'/2)}{\exp(-s_b/2)}.\label{app:gravarea-left}
\end{align}
\end{subequations}
The factor \(e^{-\pi\nu}\) from continuing the mode across \(u=0\) thus agrees with the ratio of the weights at \(s'\) and \(s_b\) obtained by analytically continuing the corner boost through \(i\pi\) in~\eqref{bfent:full-gluing-main}. The additional factor from \(f_b^{1/2}\) in~\eqref{bfent:shear-preparation} is \(\exp[-\pi^2\nu^2/\sigma_s^2]\). We choose the width of the area band so that \(|s'-s_b|=2\pi|\nu|\ll\sigma_s\ll s_b\) for the transitions that contribute appreciably to the two point function.

We hold the left sided insertion fixed and boost the right sided insertion to obtain the symmetrized two point function \(G_{LR}(\eta)\). For a smearing with \(\int p_R(u)\,\dd u=1\) and \(\Delta u\ll d\), the integral over \(u\) in~\eqref{app:gravarea-right} approaches \((\omega_1 d)^{i\nu}/d\). Using \(|\Gamma(1-i\nu)|^2=\pi\nu/\sinh(\pi\nu)\), we find
\begin{subequations}\label{app:gravarea-correlator}
\begin{align}
 &G_{LR}(\eta)=-\operatorname{Re}\int_{\mathbb R}\dd\nu\,
          e^{i\nu\eta} e^{-\pi\nu}|m_R(\nu)|^2,\label{app:gravarea-correlator-weight}\\
 &G_{LR}(\eta)\simeq-\frac{c_\perp}{2d^2}\int_{\mathbb R}
       \frac{\nu\cos(\nu\eta)}{\sinh(\pi\nu)}\,\dd\nu
       =-\frac{c_\perp}{4d^2\cosh^2(\eta/2)}.\label{app:gravarea-correlator-local}
\end{align}
For example, at \(\eta=\log 3\) the magnitude of the cross contraction falls by a quarter, as in~\eqref{bfent:shear-boost-ratio}. Since the boost preserves both variances, it increases the expectation value of the bounded observable \(\widehat{\mathcal O}\) in~\eqref{bfapp:shear-bounded-functions} for \(j_Lj_R>0\). The same integral as in~\eqref{app:gravarea-correlator-local} appears in~\eqref{topology:horizon-correlator} when gravitons on the two horizons of the eternal black hole are paired by time reflection. Here its frequency resolves the change in corner area produced by a dressed radiative insertion.

In the interacting theory a boost frequency \(\nu\) still fixes the change in the eigenvalue of \(\widehat{\mathscr A}\) to \(8\pi G_N\hbar\nu\)~\eqref{bfq:source-boost-phase:a}. Interactions correct the weights of these transitions and allow the dressed source to create several radiative excitations. To include them in the same preparation, take the quadratic action \(I^{(2)}\) as reference, with the boundary phases included in the incoming and outgoing states. Let \(V\) contain the cubic and higher terms of the reduced action in the slab and \(F\) the full dressed source. The change of reference action~\eqref{bfq:reference-action-maps} then assigns to \(F\) the relative \(S\) matrix
\begin{equation}
 R_V(F)=S_{I^{(2)}}(V)^\dagger S_{I^{(2)}}(V+F),\ \
 R_V(F_{\rm in})=S_{I^{(2)}}(F_{\rm in}) .
 \label{app:gravarea-interacting-source}
\end{equation}
Here \(F_{\rm in}\) inserts an area preserving observable before the interaction in the slab. Causal factorization~\eqref{bfq:relations:e} yields the second equality, so the matrix elements on the incoming Gaussian wavefunction are independent of \(V\) to the area resolution of~\eqref{app:gravarea-restriction}. For the amplitude~\eqref{bfq:pairing} with a dressed source on each side of \(S\), we take
\begin{equation}
 \omega\!\left(S_{I^{(2)}+V}(F_L)^\dagger S_{I^{(2)}+V}(F_R)\right)
   =\langle\chi|R_V(F_L)^\dagger R_V(F_R)|\chi\rangle ,
 \label{app:gravarea-interacting-overlap}
\end{equation}
\end{subequations}
where \(\chi(\gamma)=g(\gamma)|\Omega_{\mathscr A}\rangle\) as in~\eqref{app:gravarea-restriction}. The same prescription applies to ordered products of dressed sources. For the later insertions we expand the relative \(S\) matrices about the uncut saddle using the vertices in both \(V\) and \(F\). At each fixed order, an additional graviton loop carries a factor \(G_N\hbar/(\Delta u)^2\) up to logarithms after renormalization~\eqref{bfent:shear-fluctuation-estimate}.

At finite mode resolution the probability to measure a dimensionless area in a band \(J\) after applying the source is \(\|\mathbf1_J(\hat s)R_V(F)|\chi\rangle\|^2\). These probabilities sum to one for a partition of the spectrum, also for sources that prepare several radiative excitations. Equation~\eqref{app:gravarea-intermediate} computes the leading distribution for the one graviton component.

To determine whether the reflected overlap approaches a definite value as we include shorter wavelengths, we hold \(\chi\), both sources and their strengths, the relative boost, \(G_N\) and \(\hbar\) fixed. At each resolution \(N\) both preparations evolve from the same incoming state up to the first insertion in~\eqref{app:gravarea-interacting-overlap}.

To see what must converge, we express the reflected overlap as a sum over area transitions on the cut of \(C_b\). Each state \(|i\rangle\) of dimensionless area \(s_i\) enters with weight \(w_i=e^{-s_i/2}\sqrt{f_b(s_i)}\) by~\eqref{bfent:full-gluing-main},~\eqref{bfent:shear-preparation}. For the transition \(i\to j\) write \(\sigma_{ji}=\langle j|\widehat{\sigma}_{f_R}|i\rangle\). The right sided insertion acts on \(|i\rangle\) with amplitude \(m_{R,ji}=w_i\sigma_{ji}\). Reflection transposes the insertion and reverses the null tangent, so \(m_{L,ji}=-w_j\sigma_{ji}\). Boosting the right sided insertion therefore yields
\begin{subequations}\label{app:gravarea-transition-overlap}
\begin{align}
 &m_{L,ji}=-e^{-\pi\nu_{ji}}\sqrt{\frac{f_b(s_j)}{f_b(s_i)}}\,m_{R,ji},\
 \nu_{ji}=\frac{s_j-s_i}{2\pi},\label{app:gravarea-reflection-interacting}\\
 &\operatorname{Re}\sum_{i,j}\overline{m_{L,ji}}\,e^{i\eta\nu_{ji}}m_{R,ji}
 =-\sum_{i,j}w_iw_j|\sigma_{ji}|^2\cos(\eta\nu_{ji}).\label{app:gravarea-reflected-sum}
\end{align}
The factor \(e^{-\pi\nu_{ji}}\) in~\eqref{app:gravarea-reflection-interacting} comes from analytically continuing the corner boost through \(i\pi\). Transitions at \(\nu=0\) cancel in \(G_{LR}(\eta)-G_{LR}(0)\). With the boost weight alone, interchanging \(i\) and \(j\) multiplies \(|m_{R,ji}|^2\) by \(e^{-2\pi\nu_{ji}}\). The commutator therefore weights each transition by \(1-e^{-2\pi\nu_{ji}}\) times \(|m_{R,ji}|^2\), so for \(\nu\ne0\) we obtain
\begin{equation}
 \int_{\mathbb R}e^{-i\nu\eta}G_{LR}(\eta)\,\dd\eta
 =-\frac{1}{2\sinh(\pi\nu)}\int_{\mathbb R}e^{-i\nu\eta}
 \big\langle[\widehat\sigma_{f_R},B_\eta\widehat\sigma_{f_R}B_\eta^\dagger]\big\rangle\,\dd\eta .
 \label{app:gravarea-reflected-commutator}
\end{equation}
\end{subequations}
The reduced null symplectic form fixes this commutator to leading order at short wavelengths~\eqref{bfgeo:cone-form-main:b},~\eqref{bfapp:shear-singularity}. For \(C_b\) we include the area profile \(f_b\) in~\eqref{app:gravarea-reflected-sum}.

As we include shorter wavelengths, new intermediate states enter~\eqref{app:gravarea-reflected-sum} and virtual gravitons can change \(\sigma_{ji}\) even for fixed \(i,j\). A smooth source of width \(\Delta u\) suppresses direct creation of two short wavelength gravitons because its Fourier transform enters at the sum of their positive frequencies. A virtual pair can instead enter through a frequency difference smaller than \(1/\Delta u\). To see how such a pair enters the null constraints, consider its contribution to \(\sigma^{AB}\eta_{AB}\) in~\eqref{bfgeo:complementary-constraints}. Over one short wavelength the transverse expansion \(\vartheta\) is slowly varying. In a parallel transported transverse basis the term \(-\vartheta\sigma_{AB}/2\) in the constraint for \(\eta_{AB}\) then yields
\begin{equation}
 \delta\sigma_{AB}\simeq\tfrac12\partial_v h_{AB},\ 
 \delta\eta_{AB}\simeq-\tfrac14\vartheta h_{AB},\ 
 \delta\sigma^{AB}\delta\eta_{AB}\simeq-\tfrac1{16}\vartheta\,\partial_v(h^{AB}h_{AB}).
 \label{app:gravarea-virtual-transverse}
\end{equation}
Integrating along the null generator from $\partial\Sigma_0$ to $S$ yields
\begin{equation}
 \frac1{16}\int_\epsilon^1\!\vartheta\,\partial_v(h^{AB}h_{AB})\,\dd v
 =\frac1{16}\left.(\vartheta h^{AB}h_{AB})\right|_{S}
 -\frac1{16}\left.(\vartheta h^{AB}h_{AB})\right|_{\partial\Sigma_0}
 -\frac1{16}\int_\epsilon^1\!(\partial_v\vartheta)h^{AB}h_{AB}\,\dd v .
 \label{app:gravarea-virtual-transverse-cuts}
\end{equation}
Each term remains finite as the difference of the two frequencies tends to zero.

For the angular term \((\varpi_A\varpi_B)^{\rm TF}\) the single Gaussian contraction is pure trace on the spherical preparation after summing over angular orientations. At fixed transverse momentum~\eqref{bfgeo:shift-constraint:a} makes the linearized \(\varpi_A\) proportional to an angular derivative of \(h_{AB}\) at short wavelengths. Its variance thus has the same \(1/k\) dependence as the incoming metric variance \(1/(2s_b k)\). For the connected contraction with fixed affine frequency difference \(r\omega_1\), the modes added on doubling \(N\) contribute, apart from fixed angular factors,
\begin{equation}
 \frac{1}{s_b^2}\sum_{\substack{N<k\le2N\\N<k+r\le2N}}
 \frac{1}{k(k+r)}
 =\frac{1}{2s_b^2N}+O\!\left(\frac{1+|r|}{s_b^2N^2}\right).
 \label{app:gravarea-virtual-angular}
\end{equation}
There are at most \(N\) terms, each bounded by \(N^{-2}\), for any \(r\). We can therefore sum against the Fourier coefficients of the fixed smooth smearings. Since the leading coefficient is independent of the frequency difference, it acts locally along the null generator and its contribution is summable over successive doublings of \(N\), including at \(r=0\). The change in area under a null translation is likewise suppressed by focusing where the shear is large~\eqref{bfq:displaced-corner-overlap}. Once this coefficient is fixed, a term that resolves the variation of a smooth source over a wavelength \(\lambda\) carries at least one power of \(\lambda/\Delta u\) in a local expansion. Applying this expansion to the remaining vertices leads us to expect convergence of the weighted overlap for the two fixed sources at finite \(G_N\) and \(\hbar\), even as interactions redistribute weight among intermediate states. We next estimate the effect of the clock projection on the uncut preparation.

\subsection{Clock readings in a one graviton wave packet}
\label{app:dressed-graviton-clock}

The reflected two point function in~\eqref{app:gravarea-correlator-local} receives most of its weight from \(|\nu|\) of order one, so the change in a macroscopic \(s_b\) is small. Focusing by this graviton can increase the reading of the gravitational clock \(\widehat{\bar u}_*\) in~\eqref{bfq:canonical-clock}, so projecting onto readings below \(\ell(\mathscr D)\) can change the overlap of the wave packet with its boosted copy. Projecting before the unitary source acts costs at most the norm in~\eqref{bfq:clock-projection-estimate}. To estimate the effect of projecting afterwards we keep the term linear in each source in~\eqref{bfent:graviton-mixed-source}. At fixed mode resolution the Fourier transform in \(\gamma\) then selects a one graviton wave packet in the nearby area sector \(\mathscr A'=4G_N\hbar s'\), which after normalization takes the form
\begin{equation}
 |1\rangle=\sum_j v_j\mathfrak a_j^\dagger|\Omega_{\mathscr A'}\rangle,\
 \sum_j|v_j|^2=1.
 \label{app:gravarea-one}
\end{equation}
To apply the estimate of Section~\ref{bfq:diamond-state-existence} to \(|1\rangle\), we express its Wigner function in terms of that of the incoming Gaussian state. The complex oscillator coordinate \(\alpha\) of the excited mode follows from the annihilation operators in~\eqref{bfq:gaussian-ladder}. Taking the Wigner transform of the one quantum oscillator wavefunction along this mode yields
\begin{subequations}\label{app:gravarea-wigner}
\begin{align}
 &\alpha=\sum_j\overline v_j\left(\frac{q_j-q_j^0}{2\Delta q_j}
        +\frac{i\Delta q_j}{\hbar}(p_j-p_j^0)\right),\label{app:gravarea-oscillator}\\
 &W_1=(4|\alpha|^2-1)W_{\Omega_{\mathscr A'}},\
 \int\dd q\,\dd p\,W_{\Omega_{\mathscr A'}}(4|\alpha|^2-1)^2=5.\label{app:gravarea-wigner-square}
\end{align}
\end{subequations}
The numerical factor is independent of the coefficients \(v_j\) because both quadratures of \(\alpha\) have variance \(1/4\) in~\eqref{bfq:incoming-wigner-density}. At leading order in \(\hbar\) and fixed \(G_N\), the loss of overlap in~\eqref{bfq:clock-overlap-limit} is determined by the full nonlinear classical clock. Its contribution for \(|1\rangle\) obeys, by Cauchy Schwarz,
\begin{equation}
 \left|\int\dd q\,\dd p\,W_1 f_0\right|
 \le\left(5\int\dd q\,\dd p\,W_{\Omega_{\mathscr A'}}f_0\right)^{1/2},\
 f_0=\frac{\bar u_*}{L+\bar u_*}.
 \label{app:gravarea-clock-classical}
\end{equation}
The upper bound tends to zero with increasing \(L\) by~\eqref{bfq:classical-clock-expectation-limit}. Since the only property of \(f_0\) used in~\eqref{app:gravarea-clock-classical} was \(0\le f_0\le1\), the same estimate applies to the indicator of strong focusing in~\eqref{bfq:clock-finite-hbar-estimate}. The corner contribution there tends to zero even at finite \(\hbar\), since \(\widehat{\mathsf a}\) has trivial kernel at positive area.

At finite \(\hbar\) we still need the overlap with the wavefunction \(|\Psi_L\rangle\) of~\eqref{bfq:clock-filter-definition}, now with \(|1\rangle\) as its input. By~\eqref{bfq:clock-resolvent-limit} this wavefunction approaches the projection of \(|1\rangle\) onto the closure of the domain of the clock form as \(L\) grows. If the finite \(\hbar\) estimate of Section~\ref{bfq:diamond-state-existence} holds for Gaussian states displaced slightly along the excited mode, they belong to this closed subspace. Their derivative in norm at zero displacement is \(|1\rangle\) by~\eqref{bfapp:displaced-occupation-state}, so the same conclusion holds for the graviton. For a finite maximum reading,~\eqref{bfq:clock-form-approximation} then implies
\begin{equation}
 \big\|(\widehat 1-P_{\mathscr D})|1\rangle\big\|
 \le \big(1-\langle1|\Psi_L\rangle\big)^{1/2}
       +\frac12\left(\frac{L}{\ell(\mathscr D)}\right)^{1/2}.
 \label{app:gravarea-clock-quantum}
\end{equation}
For the area measurement in~\eqref{app:gravarea-intermediate}, let \(|1_R\rangle\) be the normalized excitation restricted to the band of nearby positive areas. Since the clock preserves area, its excluded probability is the average of the excluded probabilities in this band. Under the above hypothesis each tends to zero as we raise \(\ell(\mathscr D)\). A boost preserves both the allowed and excluded subspaces, so for \(|1_R\rangle\)
\begin{subequations}
\begin{equation}
 \big|\langle1_R|B_\eta|1_R\rangle
      -\langle1_R|P_{\mathscr D}B_\eta P_{\mathscr D}|1_R\rangle\big|
 \le \big\|(\widehat 1-P_{\mathscr D})|1_R\rangle\big\|^2 .
 \label{app:gravarea-projected-boost}
\end{equation}
For the two point function we compare the overlap with the left sided graviton state before and after the boost. We center \(f_b\) on the corner area of a smooth shear free solution \(g_0\) and choose spherically symmetric cap data so that the expectation value of a single trace free radiative insertion vanishes. Let \(Z_\eta(j_L,j_R)\) denote the unnormalized amplitude of this uncut \(C_b\) with the right sided insertion boosted through \(\eta\). At \(\eta=\log 3\) the insertion moves from \(d\) to \(3d\), so~\eqref{bfapp:shear-source-derivative} and~\eqref{app:gravarea-correlator-local} yield the change in the two point function and its value at one loop,
\begin{align}
 &\Delta_\eta G_{LR}^{C_b}
  =-\left.\operatorname{Re}\frac{\partial^2}{\partial j_L\,\partial j_R}
       \log\frac{Z_\eta(j_L,j_R)}{Z_0(j_L,j_R)}\right|_{j_L=j_R=0},
       \label{app:gravarea-mixed-derivative}\\
 &\big(\Delta_\eta G_{LR}^{C_b}\big)^{(1)}
  =\frac{c_\perp}{16d^2}
       \left[1+O\!\left(\frac{(\Delta u)^2}{d^2}+\frac{d^2}{b^2}\right)\right].
       \label{app:gravarea-saddle-contrast}
\end{align}
With the cap data and transverse smearing fixed in units of \(b\), the two corrections in~\eqref{app:gravarea-saddle-contrast} come from the width of the insertions and the regular part of the correlator on \(g_0\)~\eqref{bfapp:shear-focusing}. To estimate the effect of the clock projection, let \(p\) be the largest probability of a reading above \(\ell(\mathscr D)\) in the normalized one graviton states created by \(p_L,p_R,p_{R,\eta}\). For the wavefunctions in~\eqref{app:gravarea-affine}, the part of each overlap omitted by \(P_{\mathscr D}\) is the inner product of its two excluded components. Cauchy Schwarz therefore yields
\begin{equation}
 \left|\langle\psi_L|(\widehat 1-P_{\mathscr D})
        \big(|\psi_{R,\eta}\rangle-|\psi_R\rangle\big)\right|
 \le 2p\sqrt{G_{LL}G_{RR}}
 =O\!\left(\frac{p\,c_\perp}{(\Delta u)^2}\right).
 \label{app:gravarea-projected-overlaps}
\end{equation}
We can thus choose \(L\) for all three states in~\eqref{app:gravarea-clock-quantum} and then raise \(\ell(\mathscr D)\) at that \(L\) until \(p\ll(\Delta u/d)^2\). The clock projection then changes their contraction about \(g_0\) by much less than \(c_\perp/(16d^2)\).

Closing \(C_b\) sums over every state admitted by the clock, with the same weight \(e^{-s}f_b(s)\) at a given area~\eqref{bfent:full-gluing-main},~\eqref{bfent:shear-preparation}. The focusing bound~\eqref{bfapp:shear-interval-reading} applies also to a mixed state of the polarization common to both insertions, so halving the increase under a boost would require focusing of order \(c_f/d\) in this sum. We can estimate that focusing with both sources switched off. Since the shear vanishes on \(g_0\), at one loop we contract the two fields in \(\widehat\sigma_\perp^2\) with \(K_0^{-1}\), integrating all quadratic fluctuations of the uncut \(C_b\)~\eqref{bfapp:conditional-shear-source-integral}. With \(b\) the scale set by the cap data, the vacuum subtracted contraction is \(O(c_f/b^2)\) near \(S\). Its integral over a width \(B\) containing both insertions is therefore \(O(c_f B/b^2)\)~\eqref{bfent:shear-interval-cost:b}.

Sustained shear delays the gravitational clock~\eqref{bfgeo:capped-clock}, and \(P_{\mathscr D}\) excludes readings above \(\ell(\mathscr D)\). For the focusing near \(S\) at finite \(G_N\) we use the saddle approximation on the uncut \(C_b\). We choose the source separation \(d\) independently of this preparation, so for \(d\ll B\ll b\) the integral could reach \(c_f/d\) only if it exceeded its one loop scale by a factor of order \(b^2/(Bd)\).\footnote{For the modes \(p_L\pm p_{R,\alpha}\),~\eqref{bfapp:shear-interval-cosine:c} bounds the occupation number expectation value \(n\) by \(O(d/B+dB/b^2)\) at fixed radial profiles in units of \(d\), including the vacuum term in~\eqref{bfapp:shear-interval-square}. Cauchy Schwarz then bounds the fractional change of each second moment from its vacuum value by \(2n+2\sqrt{n(n+1)}\). Subtracting the second moments for the two signs yields \(4G_{LR}^{C_b}(\alpha)\).}

To include the nonlinear terms in the mixed derivative we use the size \(h_{AB}=O(\ell_P/\Delta u)\) of the fluctuations about \(g_0\)~\eqref{bfent:shear-fluctuation-estimate}. A cubic term in \(V\) multiplying the two linear insertions is odd in \(h\), as is a quadratic term in either source multiplying the other linear insertion. Their expectation values vanish on the centered Gaussian saddle. The first nonzero correction thus carries two powers of \(\ell_P/\Delta u\). Estimating the error of the expansion by this order in units of \(G_{aa}\sim c_\perp/(\Delta u)^2\) yields, up to logarithms after renormalization,
\begin{equation}
 \frac{\left|\Delta_\eta G_{LR}^{C_b}-\big(\Delta_\eta G_{LR}^{C_b}\big)^{(1)}\right|}
      {\big(\Delta_\eta G_{LR}^{C_b}\big)^{(1)}}
 \ \lesssim\ \frac{G_N\hbar\,d^2}{(\Delta u)^4}.
 \label{app:gravarea-relative-correction}
\end{equation}
For \(\sqrt{\ell_P d}\ll\Delta u\ll d\ll B\ll b\), we therefore expect the mixed derivative on this clock projected \(C_b\) to reproduce more than half the increase in the freely falling vacuum at finite \(G_N\). Taking \(j_L=j_R=j\) in~\eqref{bfent:bounded-graviton-observable}, finite second moments of the smeared radiative operators imply
\begin{equation}
 \lim_{j\to0}\frac{\operatorname{Re}\langle\Delta_\eta\widehat{\mathcal O}\rangle_{C_b}}{j^2}
   =\Delta_\eta G_{LR}^{C_b}.
 \label{app:gravarea-weak-sources}
\end{equation}
\end{subequations}
Hence sufficiently weak, fixed sources on this one uncut preparation suffice for the right hand inequality in~\eqref{bfent:boost-deformation-bounds}.\footnote{For \(j_a^2G_{aa}=1\),~\eqref{bfapp:shear-bounded-functions} yields \(\delta_\eta\sim(\Delta u/d)^2\), with the one loop error on \(C_b\) bounded by~\eqref{bfapp:shear-interval-contrast}. With \(V[\lambda h]/(\hbar\lambda^2)\) and \(F_a[\lambda h]/(\hbar\lambda)\) in the exponent of~\eqref{app:gravarea-interacting-source}, the first variation of \(\langle\Delta_\eta\widehat{\mathcal O}\rangle\) at \(\lambda=0\) vanishes by the same parity. If the renormalized connected moments of the smeared vertices in variables of unit Gaussian variance stay of order one for \(0\le\lambda\le1\), integrating the second variation with weight \(1-\lambda\) yields a correction of order \(G_N\hbar/(\Delta u)^2\) up to logarithms. Dividing by \(\delta_\eta\) then recovers~\eqref{app:gravarea-relative-correction} for the bounded insertions.}

\subsection{Comparison with matter on a fixed background}
\label{app:dressed-graviton-scalar}

In~\eqref{app:gravarea-correlator-weight} we obtained the radiative two point function $G_{LR}(\eta)$ by boosting the right sided insertion while holding $S$ and the left sided insertion fixed. To compare the change in area fluctuations under a null translation~\eqref{bfq:null-translation-area-variance} with the corresponding change in a nonrenormalizable field theory and in QED, we translate a spherical cut on a fixed Minkowski background. The boost generator for the ball of radius $b$ is
\begin{equation}
 K_b(t)=\int_{r<b}w_b(r)\,T_{00}(t,\mathbf x)\,\dd^3x,\ 
 w_b(r)=\frac{b^2-r^2}{2b}.
 \label{app:matter-ball-boost}
\end{equation}
The weight vanishes at the cut and has unit inward derivative there. An inward null translation takes $(t,b)=(0,b)$ to $(d,b-d)$. Since the incoming free vacuum $\chi$ is invariant under time reversal, the time derivative at fixed $b$ vanishes and
\begin{equation}
 \left.\partial_d\operatorname{Var}_\chi K_{b-d}(d)\right|_{d=0}
 =\left.(\partial_t-\partial_b)\operatorname{Var}_\chi K_b(t)\right|_{t=0}
 =-\partial_b\operatorname{Var}_\chi K_b(0).
 \label{app:matter-null-translation}
\end{equation}

We use the Weyl ordered Hamiltonian in each theory and hold $\chi$, $\hbar$ and the bare couplings fixed as we raise the spatial momentum cutoff $\Lambda_{\rm UV}$. Inserting the occupation number basis at $t=0$ yields
\begin{equation}
 \operatorname{Var}_\chi K_b=\sum_{\boldsymbol n\ne\boldsymbol 0}
       \big|\langle\boldsymbol n|K_b|\chi\rangle\big|^2 .
 \label{app:matter-boost-states}
\end{equation}
The interaction $g\phi^6$ creates six scalar quanta, while the transverse coupling $e A_i\bar\psi\gamma^i\psi$ creates a photon and an electron positron pair. Let $\widetilde w_b$ be the Fourier transform of $w_b$ extended by zero outside the ball. Up to constants independent of $b$ and the cutoff, the contributions of these states are
\begin{subequations}\label{app:matter-boost-channels}
\begin{align}
 &\left.\operatorname{Var}_\chi(K_b)\right|_{6\phi}
 \propto g^2\int^{\Lambda_{\rm UV}}\!
      \left|\widetilde w_b\!\left(\sum_{i=1}^6\mathbf p_i\right)\right|^2
      \prod_{i=1}^6\frac{\dd^3p_i}{2E_i},\label{app:matter-boost-six}\\
 &\left.\operatorname{Var}_\chi(K_b)\right|_{\gamma e^-e^+}
 \propto e^2\sum_{\lambda,\sigma,\sigma'}\int^{\Lambda_{\rm UV}}\!
     \frac{\dd^3 k}{2|\mathbf k|}\,
     \frac{\dd^3p\,\dd^3p'}{4E_{\mathbf p}E_{\mathbf p'}}
     \big|\widetilde w_b(\mathbf k+\mathbf p+\mathbf p')\big|^2
     \big|\varepsilon^\lambda_i(\mathbf k)\,
          \bar u_\sigma(\mathbf p)\gamma^i v_{\sigma'}(\mathbf p')\big|^2 .
 \label{app:matter-boost-photon}
\end{align}
\end{subequations}
Here the cutoff bounds each spatial momentum, $E$ denotes its on shell energy, and $\lambda,\sigma,\sigma'$ label the transverse photon polarization and the two fermion spins.

At fixed total momentum \(\mathbf P\), we set \(\mathbf p_6=\mathbf P-\sum_{i=1}^5\mathbf p_i\) in~\eqref{app:matter-boost-six} and integrate over \(\mathbf p_i/\Lambda_{\rm UV}\) for \(i=1,\ldots,5\). The five momentum measures and the six inverse energies yield \(\Lambda_{\rm UV}^{15-6}\). In~\eqref{app:matter-boost-photon} we set \(\mathbf p'=\mathbf P-\mathbf k-\mathbf p\) and integrate over \(\mathbf k/\Lambda_{\rm UV}\) and \(\mathbf p/\Lambda_{\rm UV}\). For noncollinear momenta the sum over the photon polarizations and the fermion spins yields
\begin{equation}
 0<\lim_{\Lambda_{\rm UV}\to\infty}
 \sum_{\lambda,\sigma,\sigma'}\frac{
 \big|\varepsilon_i^\lambda(\mathbf k)\,\bar u_\sigma(\mathbf p)\gamma^i v_{\sigma'}(\mathbf p')\big|^2}
 {4E_{\mathbf p}E_{\mathbf p'}}
 =2\big[1-(\hat{\mathbf k}\cdot\hat{\mathbf p})(\hat{\mathbf k}\cdot\hat{\mathbf p}')\big]\le4 .
 \label{app:matter-boost-spin-sum}
\end{equation}
Here we hold the rescaled integration variables fixed, and the hats denote unit vectors. The two momentum measures thus yield \(\Lambda_{\rm UV}^{6}\) and the factor \(1/(2|\mathbf k|)\) in the photon measure yields \(\Lambda_{\rm UV}^{-1}\). In both integrals \(\mathbf P\) now enters through \(\mathbf P/\Lambda_{\rm UV}\), so their leading terms at fixed \(\mathbf P\) are independent of the total momentum. The remaining integral over $\mathbf P$ yields
\begin{equation}
 \int\frac{\dd^3P}{(2\pi\hbar)^3}\,|\widetilde w_b(\mathbf P)|^2
 =4\pi\int_0^b r^2\left(\frac{b^2-r^2}{2b}\right)^2\dd r
 =\frac{8\pi b^5}{105}.
 \label{app:matter-boost-weight}
\end{equation}
Differentiating with respect to $b$, we obtain positive constants $C_\phi,C_\gamma$ independent of $b$ and $\Lambda_{\rm UV}$ such that
\begin{subequations}\label{app:matter-boost-powers}
\begin{align}
 &\partial_b\left.\operatorname{Var}_\chi(K_b)\right|_{6\phi}
   \sim C_\phi\,g^2 b^4\,\Lambda_{\rm UV}^{9},\label{app:matter-boost-scalar-power}\\
 &\partial_b\left.\operatorname{Var}_\chi(K_b)\right|_{\gamma e^-e^+}
   \sim C_\gamma\,e^2 b^4\,\Lambda_{\rm UV}^{5}.\label{app:matter-boost-photon-power}
\end{align}
\end{subequations}
In gravity the factor $e^{-\Xi_y(1)}$ in~\eqref{bfq:null-translation-area} reduces the area lost along a null generator with strong focusing. On the fixed background the cut advances independently of the fluctuating fields.

\bibliographystyle{JHEP}
\begingroup
\let\TopologyOriginalBibliography\thebibliography
\renewcommand{\thebibliography}[1]{\TopologyOriginalBibliography{#1}\setlength{\itemsep}{0pt}\addtolength{\parskip}{-2.85pt}\interlinepenalty=10000 }
\bibliography{references,cobordism_extra,christodoulou-1970}
\endgroup
\end{document}